\documentclass[reqno,12pt]{article}
\usepackage{amsfonts,amssymb,
amsmath,amscd, bm, mathrsfs, 
mathrsfs,delarray,subfigure}
\usepackage{hyperref,cite}
\usepackage{xypic}
\usepackage[dvips]{color}
\usepackage[dvips]{graphicx}

\DeclareMathOperator{\gr}{gr}
\DeclareMathOperator{\ad}{ad}
\DeclareMathOperator{\Ad}{Ad}

\DeclareMathOperator{\im}{im}
 
\DeclareMathOperator{\id}{id}

\DeclareMathOperator{\Hom}{Hom}
 
\DeclareMathOperator{\Fun}{Fun}
\DeclareMathOperator{\Map}{Map}

\DeclareMathOperator{\Vect}{Vect}
\DeclareMathOperator{\Gau}{Gau}
\DeclareMathOperator{\UGau}{UGau}

\DeclareMathOperator{\Aut}{Aut}
\DeclareMathOperator{\Inn}{Inn}
\DeclareMathOperator{\UAut}{UAut}

\DeclareMathOperator{\Lie}{Lie}

\DeclareMathOperator{\Ima}{Im}
\DeclareMathOperator{\Rea}{Re}

\numberwithin{equation}{subsection} 
\numberwithin{subsection}{section} 

\font\sansserif=cmss12
\font\scriptsansserif=cmss12 at 7 truept
\font\scriptscriptsansserif=cmss10 at 5 truept
\font\euler=eusm10 at 12.8 truept
\font\scripteuler=eusm7
\font\scriptscripteuler=eusm5 
\def\eul{\fam=12}
\newcommand{\matheul}[1]{{{\eul #1}}}

\newcommand{\mathbfs}[1]{{\boldsymbol{#1}}}
\newcommand{\bfdot}{{\boldsymbol{\,\cdot\,}}}

\newcommand{\ul}[1]{{\underline{#1}}{}}

\begin{document}

\hrule\vskip.5cm
\hbox to 14.5 truecm{November 2012 \hfil DFUB 12}
\vskip.5cm\hrule
\vskip.7cm
\centerline{\textcolor{blue}{\bf AKSZ MODELS OF SEMISTRICT HIGHER GAUGE THEORY}}   
\vskip.2cm
\centerline{by}
\vskip.2cm
\centerline{\bf Roberto Zucchini}
\centerline{\it Dipartimento di Fisica, Universit\`a degli Studi di Bologna}
\centerline{\it V. Irnerio 46, I-40126 Bologna, Italy}
\centerline{\it I.N.F.N., sezione di Bologna, Italy}
\centerline{\it E--mail: zucchinir@bo.infn.it}
\vskip.7cm
\hrule
\vskip.6cm
\centerline{\bf Abstract} 
\par\noindent
In the first part of this paper, we work out a perturbative Lagrangian formulation of semistrict higher gauge theory, that
avoids the subtleties of the relationship between Lie $2$--groups and 
algebras by relying exclusively on the structure semistrict Lie $2$--algebra
$\mathfrak{v}$ and its automorphism $2$--group $\Aut(\mathfrak{v})$. 
Gauge transformations are defined and shown to form a strict
$2$--group depending on $\mathfrak{v}$. Fields are $\mathfrak{v}$--valued and their global
behaviour is controlled by appropriate gauge transformation gluing
data organized as a strict $2$--groupoid. In the second part,   
using the BV quantization method in the AKSZ geometrical version,
we write down a $3$--dimensional semistrict higher BF gauge theory generalizing ordinary BF theory, 
carry out its gauge fixing and obtain as end result a semistrict higher topological gauge 
field theory of the Witten type. 
We also introduce a related $4$--dimensional semistrict higher Chern--Simons 
gauge theory. 

\vfill\eject

\tableofcontents

\vfill\eject



\section{\normalsize \textcolor{blue}{Introduction}}\label{sec:intro}


\hspace{.5cm} 
{\it Higher gauge theory} is a generalization of ordinary gauge theory in which 
the gauge potentials have form degree $p\geq 1$ and, correspondingly, 
their gauge curvatures have degree $p+1\geq 2$.  
Parallel transport, as a consequence, is defined along $p$--dimensional 
submanifolds.
Higher gauge theory can be both Abelian and non Abelian 
and may require several gauge potentials of different form degree 
for its consistency. 

The origin of Abelian higher gauge theory can be traced back to the early days 
of supergravity. Expectedly, Abelian higher gauge theory enters also in string 
and $M$--theory, which have supergravity as their low energy energy limit 
\cite{Polchinski:1998rr,Becker:2007zj}. 
The Kalb--Ramond $B$--field of the $10$--dimensional type II 
theories is a $2$--form field governed by a higher form of electrodynamics, 
$2$--form electrodynamics, in which fundamental strings act as sources. 
Likewise, the Ramond–-Ramond fields are $p$--form fields 
described by $p$--form electrodynamics whose sources are $D$--branes.
Analogously, the $11$--dimensional theory $C$--field is a $3$--form field
described by $3$--form electrodynamics and sourced by $M2$ branes. 

The physics of stacks of coinciding branes is encoded in non Abelian higher 
gauge theory \cite{Johnson:2003gi}. While stacks of $D$--branes are governed by 
ordinary Yang--Mills theory, those of $M5$--branes are believed to be described by a higher 
non Abelian gauge theory whose details are still not completely understood. 

Non Abelian higher gauge theory, especially in its integral version,
has also found application in the theories of quantum gravity 
alternative to string theory such as loop quantum gravity and, in particular, 
spin foam models \cite{Baez:1999sr,Rovelli:2004tv}. 

In general, higher gauge theory is expected to be play a basic role whenever 
charged higher--dimensional extended objects are involved. For this reason, 
higher gauge theory has been the object of independent analysis since quite early 
\cite{Savit:1977fw,Freedman:1980us,Nepomechie:1982rb,
Teitelboim:1985ya,Henneaux:1986ht,Blau:1989bq, Solodukhin:1992jg}.
However, its intensive study began only in the last decade or so
\cite{Lahiri:2001di,Lahiri:2001ks,Chepelev:2001mg,Baez:2002jn,
Hofman:2002ey,Pfeiffer:2003je}.
See ref. \cite{Baez:2010ya} for an up--to--date 
review of this subject and extensive referencing. 

From a mathematical point of view, higher gauge theory is intimately related to higher
algebraic structures, such as 2--categories, 2--groups \cite{Baez5,Baez:2003fs} 
and strong homotopy Lie or $L_\infty$ algebras \cite{Lada:1992wc,Lada:1994mn} 
and higher geometrical structures such as gerbes \cite{Brylinski:1993ab,Breen:2001ie}.

Since its inception, higher gauge theory has always had a topological flavour.
In fact, higher gauge theory has intersected topological field theory
at relevant points. In particular, the so called 
BF theory \cite{Schwarz:1978,Horowitz:1989ng} and its variants
have played an important part \cite{Guo:2002yc,Freidel:2005ak,
Girelli:2003ev,Girelli:2007tt}.

Higher gauge theory has been developed by promoting 
the basic topological and geometrical structures of ordinary 
gauge theory to a higher level in category theory by a procedure 
called {\it internalization} \cite{Baez:2004in,Baez:2005qu}.
In the same way as ordinary gauge theory relies on Lie groups, Lie algebras
and principal bundles over manifolds with connections, 
higher gauge theory does on their next level categorified 
counterparts, viz Lie $2$--groups, Lie $2$--algebras and principal $2$--bundles
on $2$--manifolds with $2$--connections.  
This outline, nicely intuitive as it is, hides the complexity of the matter however.
Lie $2$--groups and $2$--algebras come in two broad varieties. 
They are either {\it strict},
when the basic Lie group and algebra theoretic relations hold strictly 
as identities, or {\it non strict}, when those relations are allowed to hold 
only up to isomorphism. In the non strict case, the so--called
{\it weak} or {\it coherent $2$---groups} and {\it semistrict Lie $2$--algebras} %
have been studied, e. g. in \cite{Baez5,Baez:2003fs}. 
We thus distinguish strict Lie $2$--group based higher gauge theory, 
henceforth referred to as {\it strict higher gauge theory}, 
from coherent Lie $2$--group based higher gauge theory,
which we christen {\it semistrict higher gauge theory}.
While there exists a large body of literature about the strict theory,
a comparatively smaller amount of work has been devoted to the study and 
exemplification of the semistrict theory.
The present paper is a further modest contribution to the latter. 
We are going to propose a partially original formulation of semistrict higher gauge 
theory, candidly assessing its merits and shortcomings and illustrating it 
with a number of sample calculations. A summary description of it is given next. 
 
Simply speaking, a {\it semistrict Lie $2$--algebra}
is a $2$--vector space $\mathfrak{v}$ equipped with a bilinear functor
$[\cdot, \cdot]:\mathfrak{v}\times\mathfrak{v}\mapsto\mathfrak{v}$, the Lie bracket,
that is antisymmetric and satisfies the Jacobi identity up to a possibly trivial completely antisymmetric
trilinear natural isomorphism, the {\it Jacobiator}, which in turn is required to satisfy 
a certain coherence relation, the {\it Jacobiator identity}. 

The finite counterpart of a Lie $2$--algebra is a $2$--group. 
A {\it $2$--group} is a category equipped with a multiplication, 
a unit and an inversion functor analogous to group operations but
satisfying the associativity, unit and inverse law only up to possibly trivial natural 
isomorphisms satisfying coherence relations. 
A {\it Lie $2$--group} is a $2$ group, whose underlying category is a smooth one.

An {\it $L_\infty$ algebra} is a chain complex $\mathfrak{v}$ of vector spaces equipped
with a bilinear antisymmetric operation
$[\cdot, \cdot]:\mathfrak{v}\times\mathfrak{v}\mapsto\mathfrak{v}$,
which satisfies the Jacobi identity up to an infinite tower of chain homotopies.
When the complex is non trivial only in degree $0,\dots,n-1$, we have an {\it $n$--term $L_\infty$ 
algebra}. $2$--term $L_\infty$ algebras and semistrict Lie $2$--algebras constitute $2$--categories
which can be shown to be equivalent in the appropriate sense. 
Therefore, we can formulate the theory of semistrict Lie $2$--algebras
in the language of that of $2$--term $L_\infty$ algebras. This is indeed
the way we shall proceed and, for this reason, in the following we shall often refer to semistrict 
higher gauge theory as {\it $2$--term $L_\infty$ algebra gauge theory}. 
The basic notions and properties of $2$--term $L_\infty$ algebras and their manifold relations
to Lie $2$--groups are reviewed in sect. \ref{sec:gtwogr}.

A conventional formulation of semistrict gauge theory modelled on that of ordinary gauge 
theory along the lines of refs. \cite{Baez:2004in,Baez:2005qu} (see also ref. \cite{Aschieri:2003mw}) 
would presumably require a principal 
$2$--bundle $P(M,V)$ on a $2$--manifold $M$ with a structure coherent $2$--group $V$. The structure 
Lie $2$--algebra $\mathfrak{v}$ would be only a derived secondary object. 
(For the sake of simplicity, but oversimplifying a bit, 
we leave aside the subtle issues involved in the relationship between $V$ 
and $\mathfrak{v}$ in the non strict case.)
This type of approach is the most powerful in theory, but its concrete implementation
appears to be presently beyond our reach in practice and, for this reason, we follow another route.

Consider an ordinary gauge theory with structure 
group $G$. The topological background of the theory is then a principal $G$--bundle $P$ 
represented by an equivalence class of $G$--valued $1$--cocycles $\gamma=\{\gamma_{ij}\}$
with respect to an open covering $U=\{U_i\}$ of the base manifold $M$. 
Since in gauge theory all fields are in the adjoint of $G$, 
the effective structure group is the adjoint group $\Ad G=G/Z(G)$ rather than $G$.  
Can we, then, replace $G$ by $\Ad G$ in our gauge theory? 
The answer to this question is positive if, from the knowledge of the data $g_{ij}=\Ad\gamma_{ij}$,
it is possible to reconstruct the data $\sigma_{ij}=\gamma_{ij}{}^{-1}d\gamma_{ij}$ 
which control the global definition of the gauge fields. This can be done only 
$G$ is semisimple, e. g. $G=\mathrm{SU}(n)$.
However, we can still work with $\Ad G$ rather than $G$, if we give up the condition 
that the $\sigma_{ij}$ be determined by the $g_{ij}$ and regard the $g_{ij}$ and $\sigma_{ij}$ 
as a whole set of data satisfying the relations
\begin{subequations}
\label{intro1}
\begin{align}
&d\sigma_{ij}+\frac{1}{2}[\sigma_{ij},\sigma_{ij}]=0,
\vphantom{\Big]}
\label{intro1a}
\\
&g_{ij}{}^{-1}dg_{ij}(x)-[\sigma_{ij},x]=0, \qquad x\in\mathfrak{g},
\vphantom{\Big]}
\label{intro1b}
\end{align}
\end{subequations}
together the cocycle conditions 
\begin{subequations}
\label{intro2}
\begin{align}
&g_{ij}g_{jk}=g_{ik},
\vphantom{\Big]}
\label{intro2a}
\\
&\sigma_{ik}-\sigma_{jk}-g_{jk}{}^{-1}(\sigma_{ij})=0.
\vphantom{\Big]}
\label{intro2b}
\end{align}
\end{subequations}
As $\Ad G \simeq\Inn(\mathfrak{g})\subset \Aut(\mathfrak{g})$, however, 
proceeding in this way we are generalizing gauge theory, since now 
$g_{ij}$ is allowed to take values in the full automorphism group $\Aut(\mathfrak{g})$
rather than the inner one $\Inn(\mathfrak{g})$. 

This leads to a formulation of gauge \pagebreak theory that can be summarized as follows. 
The basic datum is a finite dimensional structure Lie algebra $\mathfrak{g}$. At finite level, instead 
of a Lie group $G$ integrating $\mathfrak{g}$, we rely on the automorphism group
$\Aut(\mathfrak{g})$ of $\mathfrak{g}$. A gauge transformation on a neighborhood $O$ consists of a map
$g\in\Map(O,\Aut(\mathfrak{g}))$ together with a flat connection $\sigma$ on $O$
satisfying certain relations. 
Gauge transformations form an infinite dimensional group $\Gau(O,\mathfrak{g})$. 
A left action of $\Gau(O,\mathfrak{g})$ on fields on $O$ is defined. 
Given an open covering $U=\{U_i\}$,  
the global definedness of the fields is controlled by a $\Gau(\cdot,\mathfrak{g})$--valued
cocycle, which comprises an $\Aut(\mathfrak{g})$--valued cocycle $\{g_{ij}\}$
and a set of flat connection data $\{\sigma_{ij}\}$ satisfying \eqref{intro1}, \eqref{intro2}. 
These latter constitute the $0$--cells of a groupoid $\check{\mathcal{P}}(U,\mathfrak{g})$ 
which describes the underlying topology.
 
The theory, so, can be formulated to a large extent relying on the Lie algebra $\mathfrak{g}$ only.
It is clear that the gauge theoretic framework outlined above can only work in 
{\it perturbative Lagrangian field theory}. 
As it is, it is unsuitable for the analysis 
of parallel transport, a central issue in gauge theory. Further, as it is well--known, 
important non perturbative effects are attached to the center $Z(G)$ of $G$, 
information about which is lost. The reason why the approach is nevertheless useful 
is that it can be directly generalized to semistrict higher gauge theory.

Our formulation of semistrict higher gauge theory follows basically the same lines. 
The basic datum is a finite dimensional structure Lie $2$--algebra $\mathfrak{v}$,  
conveniently viewed as a $2$--term $L_\infty$ algebra. 
At finite level, instead of a Lie $2$--group $V$ integrating $\mathfrak{v}$, which
may be infinite dimensional or may be something more general than a mere coherent
$2$--group, we rely on the automorphism $2$--group $\Aut(\mathfrak{v})$ of $\mathfrak{v}$,
which is always finite dimensional and strict. A gauge transformation on a neighborhood $O$ consists of a map
$g\in\Map(O,\Aut(\mathfrak{v}))$ together with a flat connection doublet $(\sigma,\varSigma)$ on $O$
and other form data satisfying a set of relations. Gauge transformations form an infinite dimensional group 
$\Gau_1(O,\mathfrak{v})$, which is the $1$--cell group of a strict $2$--group $\Gau(O,\mathfrak{v})$. 
A left action of $\Gau_1(O,\mathfrak{v})$ on fields on $O$ is defined. 
Given an open covering $U=\{U_i\}$,  
the global definedness of the fields is controlled by a $\Gau(\cdot,\mathfrak{g})$--valued
cocycle, which comprises an $\Aut(\mathfrak{v})$--valued cocycle $\{g_{ij},W_{ijk}\}$,
a set of flat connection doublet data $\{\sigma_{ij},\varSigma_{ij}\}$ and other form data 
satisfying relations generalizing \eqref{intro1}, \eqref{intro2}. 
These latter constitute the $0$--cells of a strict $2$--groupoid $\check{\mathcal{P}}_2(U,\mathfrak{v})$ 
which describes the underlying topology. Given its novelty and its relevance in the subsequent constructions, 
this matter is expounded in great detail in sect. \ref{sec:linfgauge}.

The strict case where $\mathfrak{v}$ is the differential Lie crossed module $(\mathfrak{g},\mathfrak{h})$ 
associated with a Lie crossed module $(G,H)$, which is widely discussed in the literature 
and very well understood, can be described in our framework. Indeed, one can show that $i)$ gauge transformations
on $O$, as customarily defined in this case (see e. g. \cite{Baez:2004in,Baez:2005qu}), can be organized 
in an infinite dimensional strict $2$--group $\Gau(O,G,H)$ and $ii)$ that there is a natural
$2$--group morphism to $\Gau(O,G,H)\rightarrow\Gau(O,\mathfrak{v})$, which translates the familiar strict 
higher gauge theoretic framework into ours. A recent very general formulation of higher gauge theory, 
proposed in refs. \cite{Fiorenza2011,Schreiber2011}, is also   
related to ours, though non manifestly so. See again sect. \ref{sec:linfgauge}.
 
Our approach has its advantages and disadvantages. On the differential 
side, it is very efficient and provides a powerful algorithm for the construction of local 
semistrict higher gauge models in perturbative Lagrangian field theory. On the integral side, 
as its counterpart of ordinary gauge 
theory, it is apparently not suitable for the study and efficient computation of higher parallel transport, 
even in the strict theory. With its admitted limitations, this is anyway the line of thought we follow. 

With the suitable differential geometric tools available to us, the construction 
of semistrict higher gauge field theories becomes possible. Indeed, there is an elegant
methodology for working out consistent quantum field theories relying only on a given set of 
differential geometric data based on the Batalin--Vilkovisky (BV) quantization algorithm 
\cite{BV1,BV2} and known as Alexandrov--Kontsevich--Schwartz--Zaboronsky (AKSZ) construction
\cite{AKSZ,Ikeda:2012pv}. Following this path and borrowing ideas from previous work 
\cite{Zucchini6,Zucchini7}, we are able to write down a $3$--dimensional 
semistrict higher BF gauge theory, carry out its gauge fixing and obtain eventually 
a semistrict higher topological gauge theory of the Witten type
\cite{Witten:1988ze}. The BF gauge field theory we get 
differs at significant points from the ones which have appeared in the literature
mentioned earlier and is interesting on its own. We also outline briefly 
a related $4$--dimensional semistrict higher Chern--Simons gauge theory \cite{Witten:1988hf}, here 
understood as a gauge theory whose field equations are flatness conditions. 
These models are illustrated in sect. \ref{sec:bvlinf}. We have found that they belong 
to the class of models covered by the general analysis of though not explicitly studied in 
refs. \cite{Fiorenza:2011jr,Kotov:2007nr}.

Many problems remain open and many issues require further investigation. 
They are discussed briefly in sect. \ref{sec:concl}.

Finally, in the appendices, we provide explicit formulae 
for the action and BV symmetry variations in components for the BV field theories
studied in the main body of the paper.

\vfil\eject

\section{\normalsize \textcolor{blue}{Lie $2$--algebras and $2$--groups}}\label{sec:gtwogr}


The symmetries of higher gauge theory are believed to be encoded 
in Lie $2$--algebras and $2$--groups \cite{Baez:2010ya}. 
So, we may begin our discussion from these.
Below, we review the basic notions of the theory of Lie $2$--algebras  
and $2$--groups, a subject still largely unknown among non experts. 
Our presentation is admittedly incomplete, leaving as it does the 
important categorical aspects in the background, and occasionally
lacking in mathematical rigour. An exhaustive treatment would go beyond the scope 
of the present paper. The one given below furnishes the reader 
with all the basic definitions and results required for the understanding of the second 
half of the paper. It also sets our notation and terminology once and for all.
Though with  limitations, it is tailor made for our purposes.

\subsection{\normalsize \textcolor{blue}{Lie $2$--algebras}}\label{sec:twolie}

\hspace{.5cm} 
Lie $2$--algebras are the next higher analog of ordinary Lie algebras.
Let us recall briefly the definition of this notion.  
A {\it $2$--vector space} is a category internal to the category $\Vect$ of
vector spaces, that is a category whose objects and morphisms are
vector spaces and whose source, target, identity and composition maps
are all linear \cite{Baez:2003fs}.   
A {\it semistrict Lie $2$--algebra}, or more concisely a Lie $2$--algebra,  
is a $2$--vector space equipped with a bilinear and
antisymmetric bracket functor, which satisfies the Jacobi identity up
to a natural isomorphism, called the Jacobiator. This latter in turn
satisfies a coherence law, the Jacobiator identity. 

In \cite{Baez:2003fs}, it is shown that there is a one--to--one correspondence between 
equivalence classes of Lie $2$--algebras and isomorphism classes of 
the following data:
\begin{enumerate}

\item a Lie algebra $\mathfrak{g}$;

\item an Abelian Lie algebra $\mathfrak{h}$;

\item a homomorphism $\rho:\mathfrak{g}\rightarrow\mathfrak{der(h)}$;

\item an element $[j]\in H^3(\mathfrak{g},\mathfrak{h})$ of the Lie algebra cohomology of $\mathfrak{g}$
with values in $\mathfrak{h}$.

\end{enumerate}
The correspondence hinges on the close relationship between Lie $2$--algebras and 
$2$--term $L_\infty$ algebras, as we explain next.

$L_\infty$ algebras were originally introduced by Schlessinger and Stasheff
in \cite{Stasheff1}. Since then, they have found several applications 
in field and string theory. (See \cite{Lada:1992wc} for a readable 
self contained account.) 
A $L_\infty$ algebra is a higher generalization of a Lie algebra, in which 
the Jacobi identity holds only up to higher coherent homotopy.
A {\it semistrict $2$--term $L_\infty$ algebra}, or more briefly a $2$--term $L_\infty$ algebra, 
is a special, particularly simple kind of $L_\infty$ algebra. It generalizes 
a Lie algebra and a differential Lie crossed module by allowing the Lie bracket to have
a non trivial Jacobiator. 
In \cite{Baez:2003fs}, it is proven that Lie $2$--algebras form a $2$--category 
which is $2$--equivalent to the $2$--category of $2$--term $L_\infty$ algebras. 
In practice this means that we can think of every Lie $2$--algebra equivalently as a
$2$--term $L_\infty$ algebra. 

The proof of the classification theorem quoted above in outline goes as follows. 
First, one proves that a given Lie $2$--algebra $\mathfrak{v}$ is equivalent to 
a {\it skeletal} Lie $2$--algebra $\mathfrak{v}_s$, that is one in which all isomorphic objects are equal.
Next, one demonstrates that under the equivalence of the categories of Lie $2$--algebras 
and $2$--term $L_\infty$ algebras, the isomorphism classes of skeletal Lie $2$--algebras are in one--to--one 
correspondence with those of $2$--term $L_\infty$ algebras with vanishing differential. Finally, one shows that 
these latter classes are in one--to--one correspondence with the isomorphism classes of the above data. 

Though from a categorical point of view is more natural
to work with  Lie $2$--algebras, in field theoretic
applications it is definitely more convenient to deal with 
$2$--term $L_\infty$ algebras, because these lend themselves 
to rather explicit calculations. We thus turn to them.

\vfill\eject

\subsection{\normalsize \textcolor{blue}{$2$--term $L_\infty$ algebras}}\label{sec:linfty}

\hspace{.5cm} A {\it $2$--term $L_\infty$ algebra} 
consists of the following set of data: 
\begin{enumerate}

\item a pair of vector spaces on the same field
$\mathfrak{v}_0,\mathfrak{v}_1$;

\item a linear map $\partial:\mathfrak{v}_1\rightarrow\mathfrak{v}_0$;

\item a linear map $[\cdot,\cdot]:\mathfrak{v}_0\wedge \mathfrak{v}_0\rightarrow \mathfrak{v}_0$;

\item a linear map $[\cdot,\cdot]:\mathfrak{v}_0\otimes \mathfrak{v}_1\rightarrow \mathfrak{v}_1$;

\item a linear map 
$[\cdot,\cdot,\cdot]:\mathfrak{v}_0\wedge \mathfrak{v}_0\wedge \mathfrak{v}_0\rightarrow \mathfrak{v}_1$
\footnote{$\vphantom{\bigg[}$ We denote by $[\cdot,\cdot]$ both 
$2$--argument brackets. It will be clear from context which is which.}.

\end{enumerate}
These are required to satisfy the following axioms:
\begin{subequations}
\label{2tlinalg}
\begin{align}
&[x,\partial X]-\partial[x,X]=0,\hspace{5.1cm}
\vphantom{\Big]}
\label{2tlinalga}
\\
&[\partial X,Y]+[\partial Y,X]=0,
\vphantom{\Big]}
\label{2tlinalgb}
\\
&[x,[y,z]]+[y,[z,x]]+[z,[x,y]]-\partial[x,y,z]=0,
\vphantom{\Big]}
\label{2tlinalgc}
\\
&[x,[y,X]]-[y,[x,X]]-[[x,y],X]-[x,y,\partial X]=0,
\vphantom{\Big]}
\label{2tlinalgd}
\\
&[x,[y,z,w]]-[y,[z,w,x]]+[z,[w,x,y]]-[w,[x,y,z]]
\vphantom{\Big]}
\label{2tlinalge}
\\
&-[x,y,[z,w]]-[x,z,[w,y]]-[x,w,[y,z]]
\vphantom{\Big]}
\nonumber
\\
&+[y,z,[w,x]]+[w,y,[z,x]]+[z,w,[y,x]]=0,
\vphantom{\Big]}
\nonumber
\end{align}
\end{subequations}
where $x,y,z,\ldots\in \mathfrak{v}_0$, $X,Y,Z,\ldots\in \mathfrak{v}_1$ 
here and below. 
In the following, we shall denote a $2$--term $L_\infty$ algebra such as the above 
by $\mathfrak{v}$ or, more explicitly, by 
$(\mathfrak{v}_0,\mathfrak{v}_1,\partial,[\cdot,\cdot],[\cdot,\cdot,\cdot])$
to emphasize its underlying structure. 
When dealing with several such algebras, we shall use apexes to distinguish them.

A $2$--term $L_\infty$ algebra $\mathfrak{v}$ is frequently thought of as a 2--term 
chain complex $\xymatrix{0\ar[r]&\mathfrak{v}_1\ar[r]^{\partial}&\mathfrak{v}_0\ar[r]&0}$
(whence its name) equipped with a degree $0$ graded antisymmetric bilinear bracket
$[\cdot,\cdot]$ and a degree $1$ graded antisymmetric trilinear bracket $[\cdot,\cdot,\cdot]$
enjoying the following properties. First, the boundary $\partial$ satisfies the graded Leibniz
identity with respect to $[\cdot,\cdot]$ (cf. eqs. \eqref{2tlinalga}, \eqref{2tlinalgb}). 
Second, $[\cdot,\cdot]$ does not satisfy the graded Jacobi identity and, so,  
is not a Lie bracket in general. 
The Jacobiator of $[\cdot,\cdot]$ equals the Leibnizator of $\partial$ with respect to 
$[\cdot,\cdot,\cdot]$ (cf. eqs. \eqref{2tlinalgc}, \eqref{2tlinalgd}). 
Third, the brackets $[\cdot,\cdot]$, $[\cdot,\cdot,\cdot]$ must satisfy a
certain consistency condition (cf. eq. \eqref{2tlinalge}).
Thus, $\mathfrak{v}$ is a graded differential Lie algebra up to coherent homotopy
(whence the name $2$--term strong homotopy Lie algebra frequently used.)


\subsection{\normalsize \textcolor{blue}{$2$--term $L_\infty$ 
algebra morphisms}}\label{sec:linftymorph}

\hspace{.5cm} The notion of $2$--term $L_\infty$ algebra morphism is expected to play 
an important role in the theory of $2$--term $L_\infty$ algebras on general grounds. 
It also underlies the basic fact that $2$--term $L_\infty$ algebras form a $2$--category. 
Our treatment follows closely that given in \cite{Baez:2003fs}.

Let $\mathfrak{v}$, $\mathfrak{v}'$ be $2$--term $L_\infty$ algebras. A {\it $2$--term $L_\infty$ algebra
$1$--morphism} from $\mathfrak{v}$ to $\mathfrak{v}'$ consists of the following data:
\begin{enumerate}

\item a vector space morphism  $\phi_0:\mathfrak{v}_0\rightarrow\mathfrak{v}'{}_0$;

\item a vector space morphism $\phi_1:\mathfrak{v}_1\rightarrow\mathfrak{v}'{}_1$; 

\item a vector space morphism $\phi_2:\mathfrak{v}_0\wedge\mathfrak{v}_0\rightarrow\mathfrak{v}'{}_1$. 

\end{enumerate}
These are required to satisfy the following relations:
\begin{subequations}
\label{mor2tlinalg}
\begin{align}
&\phi_0(\partial X)-\partial'\phi_1(X)=0,
\vphantom{\Big]}
\label{mor2tlinalga}
\\
&\phi_0([x,y])-[\phi_0(x),\phi_0(y)]'-\partial'\phi_2(x,y)=0,
\vphantom{\Big]}
\label{mor2tlinalgb}
\\
&\phi_1([x,X])-[\phi_0(x),\phi_1(X)]'-\phi_2(x,\partial X)=0,
\vphantom{\Big]}
\label{mor2tlinalgc}
\end{align}
\begin{align}
&[\phi_0(x),\phi_2(y,z)]'+[\phi_0(y),\phi_2(z,x)]'+[\phi_0(z),\phi_2(x,y)]'
+\phi_2(x,[y,z])
\vphantom{\Big]}
\label{mor2tlinalgd}
\\
&+\phi_2(y,[z,x])+\phi_2(z,[x,y])-\phi_1([x,y,z])+
[\phi_0(x),\phi_0(y),\phi_0(z)]'=0.
\vphantom{\Big]}
\nonumber
\end{align}
\end{subequations}
In the following, we shall denote a $1$--morphism 
such as the above one by $\phi$ or, more explicitly, by 
$(\phi_0,\phi_1,\phi_2)$ to emphasize its constituent components. 
We shall also write $\phi:\mathfrak{v}\rightarrow\mathfrak{v}'$ to indicate 
the source and target algebras of $\phi$.

To make the notion of $1$--morphism more transparent, it is necessary to 
think of $\mathfrak{v}$ as a 2--term chain complex 
$\xymatrix{0\ar[r]&\mathfrak{v}_1\ar[r]^{\partial}&\mathfrak{v}_0\ar[r]&0}$
equipped with a degree $0$ graded antisymmetric bilinear bracket
$[\cdot,\cdot]$ and a degree $1$ graded antisymmetric trilinear bracket $[\cdot,\cdot,\cdot]$
with certain properties and similarly for $\mathfrak{v}'$ (cf. sect. \ref{sec:linfty}). 
If $\phi:\mathfrak{v}\rightarrow\mathfrak{v}'$ is a $1$--morphism, then the maps $\phi_0,~\phi_1$ 
are the components of 
a chain map of $\tilde \phi:\mathfrak{v}\rightarrow \mathfrak{v}'$ (cf. eq. \eqref{mor2tlinalga}). Further, 
$[\tilde\phi(\cdot), \tilde\phi(\cdot)], \tilde\phi([\cdot, \cdot]):\mathfrak{v}\otimes \mathfrak{v}\rightarrow
\mathfrak{v}'$ are chain maps, $\mathfrak{v}\otimes \mathfrak{v}$ being the tensor square of the 
chain complex $\mathfrak{v}$, and $\phi_2$ is a chain homotopy of such chain maps
(cf. eqs. \eqref{mor2tlinalgb}, \eqref{mor2tlinalgc}). Finally, $\phi_0,~\phi_1,~\phi_2$ 
must satisfy a coherence relation ensuring their compatibility with the basic relations
\eqref{2tlinalg} satisfied by the brackets of $\mathfrak{v}$ (cf. eq. \eqref{mor2tlinalgd}).

For any two $2$--term $L_\infty$ algebra 
$1$--morphisms $\phi,\psi:\mathfrak{v}\rightarrow\mathfrak{v}'$, 
a {\it $2$--term $L_\infty$ algebra $2$--morphism} from $\phi$ to $\psi$ consists of a single datum:
\begin{enumerate}

\item a linear map $\varPhi:\mathfrak{v}_0\rightarrow\mathfrak{v}'_1$. 

\end{enumerate}
This must satisfy the following relations
\begin{subequations}
\label{mor0tlinalg}
\begin{align}
&\phi_0(x)-\psi_0(x)=\partial'\varPhi(x),
\vphantom{\Big]}
\label{mor0tlinalga}
\\
&\phi_1(X)-\psi_1(X)=\varPhi(\partial X),
\vphantom{\Big]}
\label{mor0tlinalgb}
\\
&\phi_2(x,y)-\psi_2(x,y)+[\phi_0(x),\varPhi(y)]'-[\psi_0(y),\varPhi(x)]'-\varPhi([x,y])=0.
\vphantom{\Big]}
\label{mor0tlinalgc}
\end{align}
\end{subequations}
We shall write a $2$--morphism such as this as $\varPhi$ or as $\varPhi:\phi\Rightarrow\psi$
to emphasize its source and target.  

To clarify the notion of $2$--morphism, one must  
regard again $\mathfrak{v}$ as a 2--term chain complex 
$\xymatrix{0\ar[r]&\mathfrak{v}_1\ar[r]^{\partial}&\mathfrak{v}_0\ar[r]&0}$
equipped with a graded antisymmetric multilinear brackets
$[\cdot,\cdot]$, $[\cdot,\cdot,\cdot]$ and similarly $\mathfrak{v}'$ 
and view $\phi,\psi:\mathfrak{v}\rightarrow \mathfrak{v}'$ as chain maps
$\tilde\phi,\tilde\psi:\mathfrak{v}\rightarrow \mathfrak{v}'$
as explained earlier. Then, a $2$--morphim $\varPhi:\phi\Rightarrow \psi$
is a chain homotopy of the chain maps $\tilde\phi,\tilde\psi$.
(cf. eqs. \eqref{mor0tlinalga}, \eqref{mor0tlinalgb}). 
$\varPhi$ must satisfy further a coherence relation ensuring its compatibility with the 
 basic relations \eqref{mor2tlinalg} satisfied by $\phi,\psi$ (cf. eq. \eqref{mor0tlinalgc}).

Next, we shall define a composition law 
and a unit for $1$--morphisms and horizontal and vertical composition laws 
and units for $2$--morphisms.  

The composition of two $2$--term $L_\infty$ algebra $1$--morphisms
$\phi:\mathfrak{v}\rightarrow\mathfrak{v}'$, $\psi:\mathfrak{v}'\rightarrow\mathfrak{v}''$
is the 
$1$--morphism $\psi\circ\phi:\mathfrak{v}\rightarrow\mathfrak{v}''$
defined componentwise by 
\begin{subequations}
\label{mor3tlinalg}
\begin{align}
&(\psi\circ \phi)_0(x)=\psi_0\phi_0(x),
\vphantom{\Big]}
\label{mor3tlinalga}
\\
&(\psi\circ \phi)_1(X)=\psi_1\phi_1(X),
\vphantom{\Big]}
\label{mor3tlinalgb}
\\
&(\psi\circ \phi)_2(x,y)=\psi_1\phi_2(x,y)+\psi_2(\phi_0(x),\phi_0(y)).
\vphantom{\Big]}
\label{mor3tlinalgc}
\end{align}
\end{subequations}
For any $2$--term $L_\infty$ algebra $\mathfrak{v}$, the identity of $\mathfrak{v}$
is the $2$--term $L_\infty$ algebra $1$--morphism $\id_{\mathfrak{v}}:\mathfrak{v}\rightarrow\mathfrak{v}$
defined componentwise by 
\begin{subequations}
\label{mor3/1tlinalg}
\begin{align}
&\id_{\mathfrak{v}0}(x)=x,
\vphantom{\Big]}
\label{mor3tlinalgg}
\\
&\id_{\mathfrak{v}1}(X)=X,
\vphantom{\Big]}
\label{mor3tlinalgh}
\\
&\id_{\mathfrak{v}2}(x,y)=0.
\vphantom{\Big]}
\label{mor3tlinalgi}
\end{align}
\end{subequations}

The horizontal composition of two $2$--term $L_\infty$ algebra $2$--morphisms 
$\varPhi:\lambda\Rightarrow\mu$, $\varPsi:\phi\Rightarrow\psi$, with 
$\lambda,\mu:\mathfrak{v}\rightarrow\mathfrak{v}'$, $\phi,\psi:\mathfrak{v}'\rightarrow\mathfrak{v}''$ 
$2$--term $L_\infty$ algebra $1$--morphisms, is the $2$--morphism 
$\varPsi\circ \varPhi:\phi\circ\lambda\Rightarrow\psi\circ\mu$ defined by
\begin{equation}
\varPsi\circ \varPhi(x)=\varPsi\lambda_0(x)+\psi_1\varPhi(x)=\varPsi\mu_0(x)+\phi_1\varPhi(x).
\label{mor4tlinalga}
\end{equation}
The vertical composition of two $2$--term $L_\infty$ algebra $2$--morphisms 
$\varPi:\lambda\Rightarrow\mu$, $\varLambda:\mu\Rightarrow\nu$, with 
$\lambda,\mu,\nu:\mathfrak{v}\rightarrow\mathfrak{v}'$ $2$--term $L_\infty$ algebra $1$--morphisms, 
is the $2$--morphism $\varLambda\bfdot\varPi:\lambda\Rightarrow\nu$ defined by
\begin{equation}
\varLambda\bfdot \varPi(x)=\varPi(x)+\varLambda(x). 
\label{mor4tlinalgb}
\end{equation}
Finally, for any $2$--term $L_\infty$ algebra $1$--morphism
$\phi:\mathfrak{v}\rightarrow\mathfrak{v}'$, the identity of $\phi$  
is the  $2$--term $L_\infty$ algebra $2$--morphism $\mathrm{Id}_\phi:\phi\Rightarrow\phi$ given by 
\begin{equation}
\mathrm{Id}_\phi(x)=0.
\label{mor4tlinalgc}
\end{equation}

The composition of $1$--morphisms, the unit of an $L_\infty$ algebra, the horizontal and vertical composition
of $2$--morphisms and the unit of a $1$--morphism satisfy the following basic relations
\begin{subequations}
\label{twocat1}
\begin{align} 
&(\nu\circ \mu)\circ \lambda=\nu\circ(\mu\circ \lambda),
\vphantom{\Big]}
\label{twocat1a}
\\
&\lambda\circ \id_{\mathfrak{v}}=\id_{\mathfrak{v}'}\circ \lambda=\lambda,
\vphantom{\Big]}
\label{twocat1c}
\\
&(\varLambda\circ \varPsi)\circ \varPhi=\varLambda\circ(\varPsi\circ \varPhi),
\vphantom{\Big]}
\label{twocat1d}
\\
&\varPhi\circ \mathrm{Id}_{\id_{\mathfrak{v}}}=\mathrm{Id}_{\id_{\mathfrak{v}'}}\circ \varPhi=\varPhi,
\vphantom{\Big]}
\label{twocat1f}
\\
&(\varLambda\bfdot \varPsi)\bfdot \varPhi=\varLambda\bfdot(\varPsi\bfdot \varPhi),
\vphantom{\Big]}
\label{twocat1g}
\\
&\varPhi\bfdot \mathrm{Id}_\lambda=\mathrm{Id}_\mu\bfdot \varPhi=\varPhi,
\vphantom{\Big]}
\label{twocat1i}
\\
&(\varTheta\bfdot \varLambda)\circ(\varPsi\bfdot \varPhi)=(\varTheta\circ \varPsi)\bfdot(\varLambda\circ \varPhi),
\vphantom{\Big]}
\label{twocat1j}
\end{align}
\end{subequations}%
holding whenever the various instances of morphism composition are defined. 
\eqref{twocat1} are precisely the relations which render the class of $2$--term $L_\infty$
algebras a (strict) $2$--category. This fact has a great mathematical salience,
though it will matter only marginally in the field theoretic applications
treated later. 

With the appropriate $2$--term $L_\infty$-algebra morphism structure at one's disposal, 
it is possible to define the notion of equivalence of $2$--term $L_\infty$-algebras.
Two such algebras $\mathfrak{v}$, $\mathfrak{v}'$ are said {\it equivalent} if there are $1$--morphisms
$\phi:\mathfrak{v}\rightarrow\mathfrak{v}'$, $\psi:\mathfrak{v}'\rightarrow\mathfrak{v}$
and vertically invertible \pagebreak $2$--morphisms $\varPhi:\psi\circ\phi\Rightarrow \id_{\mathfrak{v}}$
and $\varPsi:\phi\circ\psi\Rightarrow \id_{\mathfrak{v}'}$ \footnote{ $\vphantom{\dot{\dot{\dot{\dot{x}}}}}$
A $2$--morphism 
$\varLambda:\lambda\Rightarrow \mu$ is said vertically invertible if there is a  $2$--morphism 
$\varPi:\mu\Rightarrow \lambda$ such that $\varPi\bfdot\varLambda=\mathrm{Id}_\lambda$,
$\varLambda\bfdot\varPi=\mathrm{Id}_\mu$. $\varPhi,\varPsi$ here 
can be shown to be automatically vertically invertible.}. Isomorphism implies equivalence but not viceversa. 


\subsection{\normalsize \textcolor{blue}{Strict Lie $2$--algebras and differential Lie crossed modules}}
\label{sec:difliecr}

\hspace{.5cm} 
Strict Lie $2$--algebras form a distinguished subclass of the class of Lie
$2$--algebras, which is well understood and appears in many important applications. 
Further, they are intimately related to differential Lie crossed modules. 

A $2$--term $L_\infty$ algebra $(\mathfrak{v}_0,\mathfrak{v}_1,\partial,[\cdot,\cdot],[\cdot,\cdot,\cdot])$
is {\it strict} if $[\cdot,\cdot,\cdot]=0$ identically. 

Inspecting \eqref{2tlinalg}, we realize that then $\mathfrak{v}_0$ is an ordinary Lie algebra,
$\mathfrak{v}_1$ is a $\mathfrak{v}_0$ Lie module and $\partial$ is a Casimir for the latter. 


A {\it differential Lie crossed module} \cite{Gerstenhaber:1964}
consists in the following elements.
\begin{enumerate}

\item A pair of Lie algebras $\mathfrak{g}$, $\mathfrak{h}$.

\item A Lie algebra morphism $\tau:\mathfrak{h}\rightarrow\mathfrak{g}$.

\item A Lie algebra morphism $\mu:\mathfrak{g}\rightarrow\mathfrak{der}(\mathfrak{h})$, where
$\mathfrak{der}(\mathfrak{h})$ is the Lie algebra of 
derivations of $\mathfrak{h}$.

\end{enumerate}
Further, the following conditions are verified,
\begin{subequations}
\label{liecross}
\begin{align}
&\tau(\mu(x)(X))=[x,\tau(X)]_{\mathfrak{g}},
\vphantom{\Big]}
\label{liecrossa}
\\
&\mu(\tau(X))(Y)=[X,Y]_{\mathfrak{h}},
\vphantom{\Big]}
\label{liecrossb}
\end{align}
\end{subequations}
where $x,y,\ldots\in\mathfrak{g}$, $X,Y,\dots\in\mathfrak{h}$.
We shall denote a differential Lie crossed module such as this
by $(\mathfrak{g},\mathfrak{h})$ or $(\mathfrak{g},\mathfrak{h},\tau,\mu)$ to explicitly indicate 
its underlying structure. 

There exists a one--to--one
correspondence between strict $2$--term $L_\infty$ algebras and differential Lie crossed modules.
With any differential Lie crossed module $(\mathfrak{g},\mathfrak{h})$, 
there is associated a strict $2$--term $L_\infty$ algebra $\mathfrak{v}$ as follows.
$\vphantom{\ul{\ul{\ul{x}}}}$
\begin{enumerate}

\item $\mathfrak{v}_0=\mathfrak{g}$;

\item $\mathfrak{v}_1=\mathfrak{h}$;

\item $\partial X=\tau(X)$;

\item $[x,y]=[x,y]_{\mathfrak{g}}$; 

\item $[x,X]=\mu(x)(X)$;

\item $[x,y,z]=0$.

\end{enumerate}
Conversely, with any strict $2$--term $L_\infty$ algebra $\mathfrak{v}$, 
there is associated a differential Lie crossed module $(\mathfrak{g},\mathfrak{h})$ 
as follows.  
\begin{enumerate}

\item $\mathfrak{g}=\mathfrak{v}_0$;

\item $\mathfrak{h}=\mathfrak{v}_1$;

\item $[x,y]_{\mathfrak{g}}=[x,y]$;

\item $[X,Y]_{\mathfrak{h}}=[\partial X,Y]$;

\item $\tau(X)=\partial X$;

\item $\mu(x)(X)=[x,X]$.

\end{enumerate}

Let $\mathfrak{v}$, $\mathfrak{v}'$ be strict $2$--term $L_\infty$ algebras. 
A {\it strict $2$--term $L_\infty$ algebra $1$--morphism} from $\mathfrak{v}$ to 
$\mathfrak{v}'$ is a $2$--term $L_\infty$ algebra $1$--morphism
$\phi:\mathfrak{v}\rightarrow\mathfrak{v}'$ such that $\phi_2=0$.
For two strict $2$--term $L_\infty$ algebra $1$--morphisms
$\phi,\psi:\mathfrak{v}\rightarrow\mathfrak{v}'$, a strict $2$--term $L_\infty$ algebra $2$--morphism
from $\phi$ to $\psi$ is an ordinary $2$--term $L_\infty$ algebra $2$--morphism 
$\varPhi:\phi\Rightarrow \psi$.
With the strict morphism structure defined above, 
strict $2$--term $L_\infty$ algebras form a sub--$2$--category of the $2$--category 
of $2$--term $L_\infty$ algebras.




Let $(\mathfrak{g},\mathfrak{h})$, $(\mathfrak{g}',\mathfrak{h}')$ be 
differential Lie crossed modules. 
A {\it differential Lie crossed module morphism} from $(\mathfrak{g},\mathfrak{h})$ to 
$(\mathfrak{g}',\mathfrak{h}')$ is a pair of 
\begin{enumerate}

\item a Lie algebra morphism $\beta:\mathfrak{g}\rightarrow\mathfrak{g}'$, 

\item a Lie algebra  morphism $\gamma:\mathfrak{h}\rightarrow\mathfrak{h}'$ 

\end{enumerate}
preserving the crossed module relations,
\begin{subequations}
\label{difliecr3}
\begin{align}
&\beta(\tau(X))=\tau'(\gamma(X)),
\vphantom{\Big]}
\label{difliecr3a}
\\
&\gamma(\mu(x)(X))=\mu'(\beta(x))(\gamma(X)).
\vphantom{\Big]}
\label{difliecr3b}
\end{align}
\end{subequations}
We shall denote a crossed module morphism like the above 
as $(\beta,\gamma)$ or $(\beta,\gamma):(\mathfrak{g},\mathfrak{h})\rightarrow (\mathfrak{g}', \mathfrak{h}')$.

With the morphism structure just defined, differential Lie crossed modules form a category. 

There is a obvious one--to--one correspondence between strict $2$--term $L_\infty$ algebra $1$--morphisms 
$\phi:\mathfrak{v}\rightarrow \mathfrak{v}'$  and crossed module morphism
$(\beta,\gamma):(\mathfrak{g},\mathfrak{h})\rightarrow (\mathfrak{g}',\mathfrak{h}')$, 
obtained by viewing the strict $2$--term $L_\infty$ algebras $\mathfrak{v}$, $\mathfrak{v}'$ 
as the differential Lie crossed modules $(\mathfrak{g},\mathfrak{h})$, $(\mathfrak{g}',\mathfrak{h}')$, 
as described above. Explicitly, 
\begin{enumerate}

\item $\phi_0(x)=\beta(x)$;

\item $\phi_1(X)=\gamma(X)$. 

\end{enumerate}
In this way, the category of differential Lie crossed modules can be extended to a $2$--category
which is identified with the $2$--category of strict $2$--term $L_\infty$ Lie algebras. 


\subsection{\normalsize \textcolor{blue}{Examples of  Lie $2$--algebras}}\label{sec:linftyex}

\hspace{.5cm} Below, we shall illustrate some simple but important examples of 
Lie $2$--algebras.


{\it 1. Lie algebras}

Every Lie algebra $\mathfrak{l}$ can be regarded as a strict $2$--term $L_\infty$ algebra, 
denoted by the same symbol. As a differential crossed module, \pagebreak $\mathfrak{l}$ is 
defined by the data $(\mathfrak{g},\mathfrak{h},\tau,\mu)$, where 
$\mathfrak{g}=\mathfrak{l}$, $\mathfrak{h}=0$,   
$\tau:\mathfrak{h}\rightarrow\mathfrak{g}$ vanishes and
$\mu:\mathfrak{g}\rightarrow\mathfrak{der(h)}$ is trivial. 

{\it 2. Inner derivation Lie $2$--algebras} 

With any Lie algebra $\mathfrak{l}$, there is associated canonically 
a strict $2$--term $L_\infty$ algebra $\mathfrak{inn(l)}$ defined as follows.
As a differential crossed module, $\mathfrak{inn(l)}$ is the quadruple of data 
$(\mathfrak{g},\mathfrak{h},\tau,\mu)$, where $\mathfrak{g}=\mathfrak{l}$, $\mathfrak{h}=\mathfrak{l}$, 
$\tau:\mathfrak{h}\rightarrow\mathfrak{g}$ is the identity $\id_{\mathfrak{l}}$
and $\mu:\mathfrak{g}\rightarrow\mathfrak{der(h)}$ is the adjoint action $\ad_{\mathfrak{l}}$
of $\mathfrak{l}$ on itself. $\mathfrak{inn(l)}$ is 
called the {\it inner derivation Lie $2$--algebra of $\mathfrak{l}$}.

{\it 3. Derivation Lie $2$--algebras} 

The derivations of a Lie  algebra $\mathfrak{l}$, $\mathfrak{der(l)}$ 
form a Lie algebra and thus also a strict $2$--term $L_\infty$ algebra, by example 1. 
However, $\mathfrak{der(l)}$ has a second strict $2$--term $L_\infty$ 
algebra structure defined as follows. Viewed again as a differential crossed module,  
$\mathfrak{der(l)}$ is specified by the data $(\mathfrak{g},\mathfrak{h}, \tau,\mu)$, 
where $\mathfrak{g}=\mathfrak{der(l)}$, $\mathfrak{h}=\mathfrak{l}$, 
$\tau:\mathfrak{h}\rightarrow\mathfrak{g}$ is the adjoint Lie algebra morphism $\ad_{\mathfrak{l}}$
and $\mu:\mathfrak{g}\rightarrow\mathfrak{der(h)}$ is the identity $\id_{\mathfrak{der(l)}}$. 
$\mathfrak{aut(l)}$ is called the {\it derivation Lie $2$--algebra of $\mathfrak{l}$}.

{\it 4. Central extension Lie $2$--algebras}

Consider a central extension of a Lie algebra $\mathfrak{l}$ by an Abelian Lie
algebra $\mathfrak{a}$, that is a third Lie algebra $\mathfrak{e}$
fitting in  a short exact sequence of Lie algebras 
\begin{equation}
\xymatrix{0\ar[r]&\mathfrak{a}\ar[r]&\mathfrak{e}\ar[r]&\mathfrak{l}\ar[r]&0,}
\label{2tliex1}
\end{equation}
with the image of $\mathfrak{a}$ contained in the center of $\mathfrak{e}$.
With the extension, there is associated a canonical differential crossed module
$(\mathfrak{g},\mathfrak{h},\tau,\mu)$, hence 
a strict $2$--term $L_\infty$ algebra, as follows. $\mathfrak{g}=\mathfrak{l}$, $\mathfrak{h}=\mathfrak{e}$.
$\tau:\mathfrak{h}\rightarrow\mathfrak{g}$ is the third morphism in the sequence 
\eqref{2tliex1}. $\mu:\mathfrak{g}\rightarrow\mathfrak{der(h)}$ is defined by choosing a linear mapping
$\sigma:\mathfrak{l}\rightarrow\mathfrak{e}$ such that $\tau\circ\sigma=\id_{\mathfrak{l}}$
and setting $\mu(x)(X)=[\sigma(x),X]_{\mathfrak{e}}$. As $\sigma$ is defined mod $\ker\tau$ which is 
contained in the center of $\mathfrak{e}$, $\mu$ is well--defined. The resulting Lie $2$--algebra
$\mathfrak{c}_{\mathfrak{e}}$ is called the {\it central extension Lie $2$--algebra of $\mathfrak{e}$}. 

Next we consider a few examples of non strict Lie $2$--algebras. 

{\it 5. Jacobiator Lie 2--algebras}

A {\it pre--Lie algebra} is a vector space $\mathfrak{l}$ equipped with a linear map 
$[\cdot,\cdot]_{\mathfrak{l}}:\mathfrak{l}\wedge \mathfrak{l}\rightarrow \mathfrak{l}$.
It is {\it not} assumed that $[\cdot,\cdot]_{\mathfrak{l}}$ satisfies the Jacobi identity. 

Let $\mathfrak{l}$ be a pre--Lie algebra.
Let $\mathfrak{v}_0=\mathfrak{l}$, $\mathfrak{v}_1=\mathfrak{l}$  and let
$\partial:\mathfrak{v}_1\rightarrow \mathfrak{v}_0$ be the identity map $\id_{\mathfrak{l}}$.
Further, let $[\cdot,\cdot]:\mathfrak{v}_0\wedge \mathfrak{v}_0\rightarrow \mathfrak{v}_0$
and $[\cdot,\cdot]:\mathfrak{v}_0\otimes \mathfrak{v}_1\rightarrow \mathfrak{v}_1$
be the bracket $[\cdot,\cdot]_{\mathfrak{l}}$ of $\mathfrak{l}$ and 
$[\cdot,\cdot,\cdot]:\mathfrak{v}_0\wedge \mathfrak{v}_0\wedge \mathfrak{v}_0\rightarrow \mathfrak{v}_1$
be defined by
\begin{equation}
[x,y,z]=[x,[y,z]_{\mathfrak{l}}]_{\mathfrak{l}}+[y,[z,x]_{\mathfrak{l}}]_{\mathfrak{l}}+[z,[x,y]_{\mathfrak{l}}]_{\mathfrak{l}},
\label{jacobor}
\end{equation}
that is the Jacobiator of $[\cdot,\cdot]_{\mathfrak{l}}$. 
Then, $(\mathfrak{v}_0,\mathfrak{v}_1,\partial,[\cdot,\cdot],[\cdot,\cdot,\cdot])$ is a $2$--term $L_\infty$ algebra
canonically associated with the pre--Lie algebra $\mathfrak{l}$, which we shall denote by
$\mathfrak{j}_{\mathfrak{l}}$ and call the {\it Jacobiator Lie $2$--algebra
of $\mathfrak{l}$}. When $\mathfrak{l}$ is a Lie algebra, $\mathfrak{j}_{\mathfrak{l}}$ reduces to $\mathfrak{inn(l)}$. 

An important illustration of this is furnished by the imaginary octonions $\Ima\mathbb{O}\simeq\mathbb{R}^7$.
In this case, $[\cdot,\cdot]_{\Ima\mathbb{O}}$ is the customary octonionic commutator 
\begin{equation}
[x,y]_{\Ima\mathbb{O}}=xy-yx,
\label{octcom}
\end{equation}
$x,y\in\Ima\mathbb{O}$. The Jacobiator algebra $\mathfrak{j}_{\Ima\mathbb{O}}$ is therefore defined. 
Since octonionic multiplication is not associative, the associated $3$ argument bracket is non trivial. 
Remarkably, the $2$ argument and $3$ argument brackets of $\mathfrak{j}_{\Ima\mathbb{O}}$
are related in simple manner to 
the associative $3$--form $\phi$ and coassociative $4$--form $\psi$ of $\Ima\mathbb{O}$:
\begin{align}
&\phi(x,y,z)=-\frac{1}{2}\Rea(x[y,z]),	
\vphantom{\Big]}
\label{assform}
\\
&\psi(x,y,z,w)=-\frac{1}{12}\Rea(x[y,z,w]).	
\vphantom{\Big]}
\label{coassform}
\end{align}

{\it 6. The string Lie $2$--algebra}

The {\it string Lie $2$--algebra} $\mathfrak{string}_k(\mathfrak{g})$, where $\mathfrak{g}$
is a simple Lie algebra of com\-pact type and $k\in\mathbb{R}$,  is an important example of non strict
Lie $2$--algebra. For $\mathfrak{g}=\mathfrak{so}(n)$, it is relevant in string theory. 
As a $2$--term $L_\infty$ algebra, it can be presented as the set of data
$(\mathfrak{v}_0,\mathfrak{v}_1,\partial,[\cdot,\cdot],[\cdot,\cdot,\cdot])$ defined as follows. 
$\mathfrak{v}_0=\mathfrak{g}$, $\mathfrak{v}_1=\mathbb{R}$.  
$\partial:\mathfrak{v}_1\rightarrow\mathfrak{v}_0$ vanishes. 
$[\cdot,\cdot]:\mathfrak{v}_0\wedge \mathfrak{v}_0\rightarrow \mathfrak{v}_0$
is the Lie bracket $[\cdot,\cdot]_{\mathfrak{g}}$ of $\mathfrak{g}$,
$[\cdot,\cdot]:\mathfrak{v}_0\otimes \mathfrak{v}_1\rightarrow \mathfrak{v}_1$ vanishes and
$[\cdot,\cdot,\cdot]:\mathfrak{v}_0\wedge \mathfrak{v}_0\wedge \mathfrak{v}_0\rightarrow \mathfrak{v}_1$
is defined by $[x,y,z]=k\langle x,[y,z]_{\mathfrak{g}}\rangle$, where $\langle\cdot,\cdot\rangle$
is a suitably normalized invariant symmetric non singular bilinear form on $\mathfrak{g}$.

Another non strict example is provided by weak Courant--Dorfman algebras \cite{Dorfman:1987,Courant:1990}, 
in particular by Courant algebroids \cite{Roytenberg:1998vn}. 

\subsection{\normalsize \textcolor{blue}{$2$--groups}}\label{sec:weaktwogr} 

\hspace{.5cm} 
Algebraically, the finite counterpart of a Lie $2$--algebra should be 
a $2$--group. {\it Weak} or {\it coherent $2$--groups}, 
or $2$--groups for short, have been studied in depth in \cite{Baez5}, which addresses various 
notions of $2$--groups appeared in the literature giving a synthesis. 
A coherent $2$--group is a category equipped with a multiplication, a unit and an inversion 
functor analogous to group operations but
satisfying the associativity, unit and inverse law only up to coherent natural isomorphisms. 
As this definition already suggests, there are remarkable structural similarities between the theory of $2$--groups and 
that of Lie $2$--algebras. In particular, the classification theorem of Lie $2$--algebras
stated in subsect. \ref{sec:twolie} has a close $2$--group analog.
In \cite{Baez5}, it is shown that there is a one--to--one correspondence between 
equivalence classes of $2$--groups and isomorphism classes of 
the following data:
\begin{enumerate}

\item a group $G$,

\item an Abelian group $H$, 

\item a homomorphism $\alpha:G\rightarrow \Aut(H)$, 

\item an element $[a]\in H^3(G,H)$ of group cohomology of $G$ with values in $H$.

\end{enumerate}
The proof of the theorem also follows a similar course.
The $2$--group counterpart of a $2$--term $L_\infty$ algebra is a 
{\it special}  $2$--group, a $2$--group which is skeletal, that is  
all isomorphic objects are equal, and such that the unit and inverse laws hold strictly.
Every coherent $2$--group $V$ is equivalent to a special $2$--group $V_s$, 
Isomorphism classes of these latter can then be shown to be in 
one--to--one correspondence with the isomorphism classes of the above data. 

A {\it Lie $2$--group} is a $2$--group, in which objects and morphisms are smooth manifolds
and the multiplication, unit and inversion functors are smooth.
In spite of the close formal similarities 
noticed above, a relationship between Lie $2$--algebras 
and Lie $2$--groups analogous to that existing between ordinary Lie algebras 
and Lie groups does not appear to exist. 
In fact, unlike what happens for groups, 
in general Lie $2$--algebras do not straightforwardly integrate to Lie $2$--groups. 
We illustrate this situation with the following classical example taken from ref. \cite{BCSS}.

Suppose that $\mathfrak{g}$ is a simple Lie algebra of compact type. 
Let us look for a coherent Lie $2$--group integrating the string Lie $2$--algebra
$\mathfrak{string}_k(\mathfrak{g})$  introduced at the end of subsect. \ref{sec:linftyex}. 
According to the Lie $2$--algebras classification theorem of subsect. \ref{sec:twolie}, 
$\mathfrak{string}_k(\mathfrak{g})$ corresponds to the Lie algebra $\mathfrak{g}$,
the Abelian Lie algebra $\mathfrak{u}(1)$, the trivial homomorphism $\mathfrak{g}
\rightarrow\mathfrak{der}(\mathfrak{u}(1))$ and the suitably normalized canonical 
$\mathfrak{u}(1)$--valued $\mathfrak{g}$--$3$--cocycle $j=\langle\cdot,[\cdot,\cdot]_{\mathfrak{g}}\rangle$. 
To build a $2$--group $G_k$ integrating 
$\mathfrak{string(g)}_k$, we need somehow to map $H^3(\mathfrak{g},\mathfrak{u}(1))$ into 
$H^3(G,\mathrm{U}(1))$. $H^3 (\mathfrak{g},\mathfrak{u}(1))$ contains a lattice $\Lambda$	
consisting of the integer multiples of $[j]$. Chern--Simons \cite{CS}
and Cheeger--Simons \cite{CheegSim} construct an inclusion $\iota:\Lambda\rightarrow H^3(G,\mathrm{U}(1))$.
Thus, by the $2$--group classification theorem recalled above, 
when $k\in\mathbb{Z}$, we can build a special $2$--group $G_k$ corresponding to 
the group $G$, the Abelian group $\mathrm{U}(1)$, the trivial homomorphism $G\rightarrow \Aut(\mathrm{U}(1))$ 
and the cohomology class $k\iota[j]\in H^3(G,\mathrm{U}(1))$.
Unfortunately, for $k\not=0$, $G_k$ is not and cannot be a Lie $2$--group, as there is no continuous representative 
of the cohomology class $k\iota[j]$ if $G$ and $\mathrm{U}(1)$ are given the usual topology, 
except for the trivial case  $k=0$. More on this in subsect. \ref{sec:2grex}.

The abstract categorical setting, in which $2$--groups are defined, albeit 
very elegant and powerful from a mathematical perspective,  
makes it difficult to manipulate them in detailed field theoretic applications.
For this and other reasons, in this paper, we shall base 
our formulation of semistrict higher gauge theory not directly on $2$--group theory,
but on Lie $2$--algebra. However, when dealing with global 
issues in higher gauge theory, it is not possible to restrict oneself to the infinitesimal 
Lie $2$--algebraic level. A $2$--group structure is bound to emerge in a way or another.
Our proposal is to make reference to the automorphism $2$--group of the
underlying Lie $2$--algebra viewed as a $2$--term $L_\infty$ 
algebra. This is a $2$--group of a special sort, called strict.

{\it Strict $2$--groups} form a distinguished subclass of the class of $2$--groups
for which matters are much simpler. 
Strict Lie $2$--groups integrate strict Lie $2$--algebra 
algebras much as ordinary Lie groups integrate ordinary Lie algebras.
Hence, they are of a special interest.
There are other reasons why they are relevant for us.
 As already recalled, the automorphisms of a general $2$--term $L_\infty$ 
algebra form a strict $2$--group. Further, 
the gauge transformation group of our version of higher gauge theory
is an infinite dimensional strict $2$--group. 
For these reasons,  we shall concentrate exclusively on strict $2$--groups 
in the next few sections.


\subsection{\normalsize \textcolor{blue}{Strict $2$--groups}}\label{sec:twogr}

\hspace{.5cm} 
The theory of strict $2$--groups is phrased most efficiently in the 
language of higher category theory. We shall restrict ourselves 
to providing only the basic definitions and properties. 
See ref. \cite{Baez5} for a comprehensive categorical treatment. 
Strict $2$--groups are also intimately related to crossed modules 
and are so amenable to a more conventional Lie algebraic 
treatment.

\hspace{.5cm} A {\it strict $2$--group} (in delooped form)
consists of the following set of data: 
\begin{enumerate}

\item a set of $1$-cells $V_1$;

\item a composition law of $1$--cells $\circ: V_1\times V_1\rightarrow V_1$;

\item a inversion law of $1$--cells ${}^{-1_\circ}: V_1\rightarrow V_1$;

\item a distinguished unit $1$--cell $1\in V_1$;

\item for each pair of $1$--cells $a,b\in V_1$, a set of $2$--cells $V_2(a,b)$;

\item for each quadruple of $1$--cells $a,b,c,d\in V_1$, a horizontal composition law of $2$--cells
$\circ:V_2(a,c)\times V_2(b,d)\rightarrow V_2(b\circ a,d\circ c)$;

\item for each pair of $1$--cells $a,b\in V_1$, 
a horizontal inversion law of $2$--cells ${}^{-1_\circ}: V_2(a,b)\rightarrow V_2(a^{-1_\circ},b^{-1_\circ})$;

\item for each triple of $1$--cells $a,b,c\in V_1$, a vertical composition law of $2$--cells
$\bfdot:V_2(a,b)\times V_2(b,c)\rightarrow V_2(a,c)$;

\item for each pair of $1$--cells $a,b\in V_1$, 
a vertical inversion law of $2$--cells ${}^{-1_\bfdot}\!\!: V_2(a,b)\rightarrow V_2(b,a)$;

\item for each $1$--cell $a$, a distinguished 
unit $2$--cell $1_a\in V_2(a,a)$.
\end{enumerate}
 These are required to satisfy the following axioms. 
\begin{subequations}
\label{twogr1}
\begin{align}
&(c\circ b)\circ a=c\circ(b\circ a),
\vphantom{\Big]}
\label{twogr1a}
\\
&a^{-1_\circ}\circ a=a\circ a^{-1_\circ}=1,
\vphantom{\Big]}
\label{twogr1b}
\\
&a\circ 1=1\circ a=a,
\vphantom{\Big]}
\label{twogr1c}
\\
&(C\circ B)\circ A=C\circ(B\circ A),
\vphantom{\Big]}
\label{twogr1d}
\\
&A^{-1_\circ}\circ A=A\circ A^{-1_\circ}=1_1,
\vphantom{\Big]}
\label{twogr1e}
\\
&A\circ 1_1=1_1\circ A=A,
\vphantom{\Big]}
\label{twogr1f}
\\
&(C\bfdot B)\bfdot A=C\bfdot(B\bfdot A),
\vphantom{\Big]}
\label{twogr1g}
\\
&A^{-1_\bfdot}\!\bfdot A=1_a,\qquad A\bfdot A^{-1_\bfdot}=1_b,
\vphantom{\Big]}
\label{twogr1h}
\\
&A\bfdot 1_a=1_b\bfdot A=A,
\vphantom{\Big]}
\label{twogr1i}
\\
&(D\bfdot C)\circ(B\bfdot A)=(D\circ B)\bfdot(C\circ A).
\vphantom{\Big]}
\label{twogr1j}
\end{align}
\end{subequations}%
Here and in the following, $a,b,c,\dots\in V_1$, $A,B,C,\dots\in V_2$, where 
$V_2$ denotes the set of all $2$-cells. For clarity, we often denote $A\in V_2(a,b)$
as $A:a\Rightarrow b$. 
All identities involving the vertical composition and inversion hold whenever defined.  
Relation \eqref{twogr1j} is called interchange law. 
In the following, we shall denote a $2$--group such as the above as $V$ or $(V_1,V_2)$
or $(V_1,V_2,\circ,{}^{-1_\circ},\bfdot,{}^{-1_\bfdot},1_-)$ to emphasize the underlying structure.

If $(V_1,V_2,\circ,{}^{-1_\circ},\bfdot,{}^{-1_\bfdot},1_-)$ is a strict $2$--group, then 
$(V_1,\circ,{}^{-1_\circ},1)$ is an ordinary group and $(V_1,V_2,\bfdot,{}^{-1_\bfdot},1_-)$ is a groupoid. 
Viewing this as a category $V$, $\circ:V\times V\rightarrow V$ 
and ${}^{-1_\circ}:V\rightarrow V$ are both functors. Indeed, 
$V$ is a strict monoidal category in which every morphism is invertible 
and every object has a strict inverse. $V$ can also be viewed as a one--object
strict $2$--category in which all $1$--morphisms are invertible  and all $2$--morphisms
are both horizontal and vertical invertible, that is a one--object strict $2$--groupoid. 

A {\it crossed module} \cite{Whitehead:1946} 
consists in the following elements.
\begin{enumerate}

\item a pair of groups $G$, $H$;

\item a group morphism $t:H\rightarrow G$;

\item a group morphism $m:G\rightarrow\Aut(H)$, where
$\Aut(H)$ is the group of automorphisms of $H$.

\end{enumerate}
Further, the following conditions are met.
\begin{subequations}
\label{twogr3}
\begin{align}
&t(m(a)(A))=a t(A)a^{-1},
\vphantom{\Big]}
\label{twogr3a}
\\
&m(t(A))(B)=ABA^{-1},
\vphantom{\Big]}
\label{twogr3b}
\end{align}
\end{subequations}
where here and in the following $a,b,c,\dots\in G$, $A,B,C,\dots\in H$.
We shall denote a crossed module such as this
by $(G,H)$ or $(G,H,t,m)$ to explicitly indicate its underlying structure. 

There exists a one--to--one \pagebreak 
correspondence between strict $2$--groups and crossed modules \cite{Brown:1976}.
With any crossed module $(G,H)$, there is associated a strict $2$--group $V$
as follows. 
\begin{enumerate}

\item  $V_1=G$;

\item  
$b\circ a=ba$;

\item  
$a^{-1_\circ}=a^{-1}$;

\item $1=1_G$;

\item  
$V_2(a,b)$ is the set of pairs $(a,A) \in G\times H$ 
such that $b = t(A)a$;

\item 
$(b,B)\circ(a,A)=(ba, Bm(b)(A))$;

\item  
$(a,A)^{-1_\circ}=(a^{-1},m(a^{-1})(A^{-1}))$;

\item  for composable 
$(a,A)$, $(b,B)$, $(b,B)\bfdot(a,A)=(a, BA)$;

\item  
$(a,A)^{-1_\bfdot}=(t(A)a,A^{-1})$;

\item 
$1_a=(a,1_H)$.

\end{enumerate}
Conversely, with any strict $2$--group $V$ there is associated a crossed module $(G,H)$, 
as follows. 
\begin{enumerate}

\item $G=V_1$;

\item 
$ba=b\circ a$;

\item 
$a^{-1}=a^{-1_\circ}$;

\item $1_G=1$;

\item $H$ is the set of all $2$--cells of the form $A:1\Rightarrow a$
for some $a$; 

\item 
$BA=B\circ A$;

\item 
$A^{-1}=A^{-1_\circ}$;

\item $1_H=1_1$;

\item 
$t(A)=a$ if $A:1\Rightarrow a$. 

\item 
$m(a)(A) = 1_a \circ A\circ 1_{a^{-1_\circ}}$.
\end{enumerate}

A strict Lie $2$--group is a strict $2$--group $(V_1,V_2,\circ,{}^{-1_\circ},\bfdot,{}^{-1_\bfdot},1_-)$ such that 
$V_1$, $V_2$ are smooth manifolds and $\circ,{}^{-1_\circ},\bfdot,{}^{-1_\bfdot},1_-$ are smooth mappings.
Similarly, a Lie crossed module is a crossed module $(G,H,t,m)$ such that 
$G$, $H$ are Lie groups and $t$, $m$ are smooth mappings.



Let $V$, $V'$ be strict $2$--groups.
A {\it strict $2$--group $1$--morphism} from $V$ to $V'$ is a pair of 
\begin{enumerate}

\item a mapping $\theta_1:V_1\rightarrow V'{}_1$,

\item for any two $1$--cells $a,b\in V_1$, a mapping $\theta_2:V_2(a,b)\rightarrow V'{}_2(\theta_1(a),\theta_1(b))$ 

\end{enumerate}
preserving the $2$--group structure:
\begin{subequations}
\label{twogr2}
 \begin{align}
&\theta_1(b\circ a)=\theta_1(b)\circ\theta_1(a),
\vphantom{\Big]}
\label{twogr2a}
\\
&\theta_1(a^{-1_\circ})=\theta_1(a)^{-1_\circ},
\vphantom{\Big]}
\label{twogr2b}
\\
&\theta_1(1)=1',
\vphantom{\Big]}
\label{twogr2c}
\\
&\theta_2(B\circ A)=\theta_2(B)\circ\theta_2(A),
\vphantom{\Big]}
\label{twogr2d}
\\
&\theta_2(A^{-1_\circ})=\theta_2(A)^{-1_\circ},
\vphantom{\Big]}
\label{twogr2e}
\\
&\theta_2(B\bfdot A)=\theta_2(B)\bfdot\theta_2(A),
\vphantom{\Big]}
\label{twogr2f}
\\
&\theta_2(A^{-1_\bfdot})=\theta_2(A)^{-1_\bfdot},
\vphantom{\Big]}
\label{twogr2g}
\\
&\theta_2(1_a)=1'{}_{\theta_1(a)}.
\vphantom{\Big]}
\label{twogr2h}
\end{align}
\end{subequations} 
We shall denote such a $1$--morphism as $\theta$ or, more explicitly, as $(\theta_1,\theta_2)$. 
We shall also write $\theta:V\rightarrow V'$ to emphasize the source and target $2$--groups. 

If $\theta:V\rightarrow V'$ is a strict $2$--group $1$--morphism, 
then $\theta_1:V_1\rightarrow V'{}_1$ is a group morphism and 
$\theta:(V_1,V_2)\rightarrow (V'{}_1,V'{}_2)$ is a groupoid morphism. 
$\theta:V\rightarrow V'$ can also be viewed as a $2$--functor 
of the $2$--categories $V$, $V'$.


It is also possible to introduce the notion of $2$--morphism
from a $1$--morphism to another. For any two  strict $2$--group 
$1$--morphisms $\theta,\upsilon:V\rightarrow V'$,
a strict $2$--group $2$--morphism from $\theta$ to $\upsilon$ consists of a full set of data 
of the form 
\begin{enumerate}

\item for any two $a\in V_1$, an element $\varTheta(a) \in V'{}_2(\theta_1(a),\upsilon_1(a))$ 

\end{enumerate}
such that the following relations are satisfied 
\begin{subequations}
\label{twogr5}
 \begin{align}
&\varTheta(b\circ a)=\varTheta(b)\circ \varTheta(a),
\vphantom{\Big]}
\label{twogr5a}
\\
&\varTheta(a^{-1_\circ})=\varTheta(a)^{-1_\circ},
\vphantom{\Big]}
\label{twogr5b}
\\
&\varTheta(1)=1'{}_1,
\vphantom{\Big]}
\label{twogr5c}
\\
&\varTheta(b)\bfdot \theta_2(A)=\upsilon_2(A)\bfdot\varTheta(a),
\vphantom{\Big]}
\label{twogr5d}
\end{align}
\end{subequations} 
where $A:a\Rightarrow b$. We shall denote a morphism such as this 
as $\varTheta$ or more explicitly as $\varTheta:\theta \Rightarrow \upsilon$.

If $\theta,\upsilon:V\rightarrow V'$ are $1$--morphisms and 
$\varTheta:\theta\Rightarrow\upsilon$ is a $2$--morphism, 
then $\varTheta$ is a pseudonatural 
transformation of the $2$--functors
$\theta,\upsilon$.

Next, having in mind a $2$--categorical structure, 
we shall define a composition law and a unit for $1$--morphisms and horizontal 
and vertical composition laws and units for $2$--morphisms.

The composition of two strict $2$--group $1$--morphisms 
$\theta:V\rightarrow V'$, $\upsilon:V'\rightarrow V''$
is the 
$1$--morphism $\upsilon\circ\theta:V\rightarrow V''$, 
defined componentwise by  
\begin{subequations}
\label{twogr6}
\begin{align}
&(\upsilon\circ \theta)_1(a)=\upsilon_1(\theta_1(a)),
\vphantom{\Big]}
\label{twogr6a}
\\
&(\upsilon\circ \theta)_2(A)=\upsilon_2(\theta_2(A)).
\vphantom{\Big]}
\label{twogr6b}
\end{align}
\end{subequations}
For any strict $2$--group $V$, the identity of $V$ \pagebreak 
is the strict $2$--group $1$--morphism $\id_{V}:V\rightarrow V$
defined componentwise by 
\begin{subequations}
\label{twogr7}
\begin{align}
&\id_{V1}(a)=a,
\vphantom{\Big]}
\label{twogr7a}
\\
&\id_{V2}(A)=A.
\vphantom{\Big]}
\label{twogr7b}
\end{align}
\end{subequations}

The horizontal composition of two strict $2$--group $2$--morphisms 
$\varLambda:\phi\Rightarrow\psi$, $\varTheta:\lambda\Rightarrow\mu$, with 
$\phi,\psi:V\rightarrow V'$, $\lambda,\mu:V'\rightarrow V''$
strict $2$--group $1$--morphisms, is the $2$--morphism 
$\varTheta\circ\varLambda:\lambda\circ\phi\Rightarrow\mu\circ\psi$ defined by
\begin{equation}
\varTheta\circ\varLambda(a)
=\varTheta(\psi_1(a))\bfdot \lambda_2(\varLambda(a))
=\mu_2(\varLambda(a))\bfdot\varTheta(\phi_1(a)).
\label{twogr8}
\end{equation}
The vertical composition of two strict $2$--group $2$--morphisms 
$\varPi:\lambda\Rightarrow\mu$, $\varLambda:\mu\Rightarrow\nu$, with 
$\lambda,\mu,\nu:V\rightarrow V'$ strict $2$--group $1$--morphisms, 
is the $2$--morphism $\varLambda\bfdot\varPi:\lambda\Rightarrow\nu$ defined by
\begin{equation}
\varLambda\bfdot\varPi(a)=\varLambda(a)\bfdot\varPi(a). 
\label{twogr9}
\end{equation}
Finally, for any strict $2$--group $1$--morphism
$\theta:V\rightarrow V'$, the identity of $\theta$  
is the  $2$--group $2$--morphism $\mathrm{Id}_\phi:\theta\Rightarrow\theta$ given by 
\begin{equation}
\mathrm{Id}_\theta(a)=1'{}_{\theta_1(a)}.
\label{twogr10}
\end{equation}

The composition of $1$--morphisms, the unit of an $L_\infty$ algebra, the horizontal and vertical composition
of $2$--morphisms and the unit of a $1$--morphism satisfy the following basic relations
\begin{subequations}
\label{twogr11}
\begin{align} 
&(\nu\circ \mu)\circ \lambda=\nu\circ(\mu\circ \lambda),
\vphantom{\Big]}
\label{twogr11a}
\\
&\lambda\circ \id_{V}=\id_{V'}\circ \lambda=\lambda,
\vphantom{\Big]}
\label{twogr11c}
\\
&(\varPi\circ \varLambda)\circ \varTheta=\varPi\circ(\varLambda\circ \varTheta),
\vphantom{\Big]}
\label{twogr11d}
\\
&\varTheta\circ \mathrm{Id}_{\id_{V}}=\mathrm{Id}_{\id_{V'}}\circ \varTheta=\varTheta,
\vphantom{\Big]}
\label{twogr11f}
\\
&(\varPi\bfdot \varLambda)\bfdot \varTheta=\varPi\bfdot(\varLambda\bfdot \varTheta),
\vphantom{\Big]}
\label{twogr11g}
\\
&\varTheta\bfdot \mathrm{Id}_\lambda=\mathrm{Id}_\mu\bfdot \varTheta=\varTheta,
\vphantom{\Big]}
\label{twogr11i}
\end{align}
\begin{align} 
&(\varXi\bfdot \varPi)\circ(\varLambda\bfdot \varTheta)=(\varXi\circ \varLambda)\bfdot(\varPi\circ \varTheta),
\vphantom{\Big]}
\vphantom{\dot{\dot{\dot{\dot{\dot{\dot{x}}}}}}}
\label{twogr11j}
\end{align}
\end{subequations}%
holding whenever the various instances of morphism composition are defined. 
\eqref{twogr11} 
are precisely the relations which render the class of strict $2$--groups
a (strict) $2$--category. 

Let  $(G,H)$, $(G',H')$ be crossed modules. 
A {\it crossed module morphism} from $(G,H)$ to $(G',H')$ is a pair of 
\begin{enumerate}

\item a group morphism $\rho:G\rightarrow G'$, 

\item a group morphism $\sigma:H\rightarrow H'$ 

\end{enumerate}
preserving the crossed module relations, 
\begin{subequations}
\label{twogr4}
\begin{align}
&\rho(t(A))=t'(\sigma(A)),
\vphantom{\Big]}
\label{twogr4a}
\\
&\sigma(m(a)(A))=m'(\rho(a))(\sigma(A)).
\vphantom{\Big]}
\label{twogr4b}
\end{align}
\end{subequations}
We shall denote a crossed module morphism like the above 
as $(\rho,\sigma)$ or $(\rho,\sigma):(G,H)\rightarrow (G',H')$.


With the morphism structure just defined, strict $2$--groups form a category. 

There is a obvious one--to--one correspondence between $2$--group $1$--morphism 
$\theta:V\rightarrow V'$  and crossed module morphism
$(\rho,\sigma):(G,H)\rightarrow (G',H')$, obtained by viewing 
the strict $2$--groups $V$, $V'$ as the 
crossed modules $(G,H)$, $(G',H')$, as indicated above. 
\begin{enumerate}

\item $\theta_1(a)=\rho(a)$;

\item $\theta_2(a,A)=(\rho(a),\sigma(A))$. 

\end{enumerate}
In this way, the category of crossed modules can be extended to a $2$--category
which is identified with the $2$--category of strict $2$--groups in the way explained in detail above. 

If $V$, $V'$ are  strict Lie $2$--groups,
a strict Lie $2$--group $1$--morphism $\theta:V\rightarrow V'$ is a strict $2$--group 
$1$ morphism such that $\theta_1$, $\theta_2$ are both smooth.
One can similarly define strict Lie $2$--group $1$--morphism $\varTheta:\lambda\Rightarrow\mu$.
Similarly, if $(G,H)$, $(G',H')$ are Lie crossed modules, a Lie crossed module morphism 
$(\rho,\sigma):(G,H)\rightarrow (G',H')$ is a crossed module morphism such that
$\rho$, $\sigma$ are both smooth.


\subsection{\normalsize \textcolor{blue}{Strict Lie $2$--groups and strict Lie $2$--algebras}}\label{sec:st2grst2lie}

\hspace{.5cm} 
Much as with any Lie group $G$, there is associated a Lie algebra $\mathfrak{g}$, 
with any strict Lie $2$--group $V$, there is associated a strict Lie $2$--algebra $\mathfrak{v}$. 
Showing this is straightforward, if one sees the former as 
a Lie crossed module $(G,H)$ and the latter as a differential Lie crossed module 
$(\mathfrak{g},\mathfrak{h})$.

Let $(G,H)$ be a Lie crossed module. With the group morphism $t:H\rightarrow G$, there
is associated the Lie algebra morphism $\dot t:\mathfrak{h}\rightarrow \mathfrak{g}$ defined by 
\begin{equation}
\dot t(X)=\frac{d t(C(s))}{ds}\Big|_{s=0},
\label{st2grst2lie1}
\end{equation} 
with $C(s)$ is any curve in $H$ such that $C(s)\big|_{s=0}=1_H$ and $dC(s)/ds\big|_{s=0}=X$. 
Likewise, with the group morphism $m:G\rightarrow \Aut(H)$ there
is associated the Lie algebra morphism $\widehat{m}:\mathfrak{g}\rightarrow \mathfrak{der}(\mathfrak{h})$ 
as follows. 
For $a\in G$, let $\dot m(a):\mathfrak{h}\rightarrow \mathfrak{h}$ be the 
vector space morphism given by 
\begin{equation}
\dot m(a)(X)=\frac{dm(a)(C(s))}{ds}\Big|_{s=0}, 
\label{st2grst2lie2}
\end{equation} 
where $C(s)$ is any curve in $H$ such that $C(s)\big|_{s=0}=1_H$ and $dC(s)/ds\big|_{s=0}=X$. 
Then, viewing $\widehat{m}:\mathfrak{g}\otimes\mathfrak{h}\rightarrow \mathfrak{h}$, we have  
\begin{equation}
\widehat{m}(x)(X)=\frac{d\dot m(c(u)))(X)}{du}\Big|_{u=0},
\label{st2grst2lie3}
\end{equation} 
where $c(u)$ is any curve in $G$ such that $c(u)\big|_{u=0}=1_G$ and $dc(u)/du\big|_{u=0}=x$.

We can now attach to a Lie crossed module $(G,H,t,m)$ canonically a  differential Lie crossed module 
$(\mathfrak{g},\mathfrak{h},\tau,\mu)$ as follows. 
\begin{enumerate}

\item $\mathfrak{g}=\Lie G$; 

\item $\mathfrak{h}=\Lie H$.

\item $\tau=\dot t$;

\item $\mu=\widehat{m}$.

\end{enumerate}

The resulting correspondence can be phrased in the language of strict $2$--groups 
and strict $2$--term $L_\infty$ algebras using the results of subsects.
\ref{sec:difliecr}, \ref{sec:twogr}.


\subsection{\normalsize \textcolor{blue}{The strict Lie $2$--group of $2$--term $L_\infty$ 
algebra automorphisms}}\label{sec:linftyauto} 

\hspace{.5cm} The notion of $2$--term $L_\infty$ algebra automorphism is central 
in the theory of $2$--term $L_\infty$ algebras. 
In this section, we shall show that the automorphisms of
a $2$--term $L_\infty$ algebra $\mathfrak{v}$ form a strict $2$--group 
$\Aut(\mathfrak{v})$. See again \cite{Baez:2003fs}.

A {\it $1$--automorphism} of $\mathfrak{v}$ is $2$--term $L_\infty$ algebra $1$--morphism 
$\phi:\mathfrak{v}\rightarrow\mathfrak{v}$ such that 
$\phi_0:\mathfrak{v}_0\rightarrow\mathfrak{v}_0$
and $\phi_1:\mathfrak{v}_1\rightarrow\mathfrak{v}_1$ 
(cf. subsect. \ref{sec:linftymorph}).
We shall denote the set of all $1$--automorphisms of $\mathfrak{v}$ 
by $\Aut_1(\mathfrak{v})$.

For any two $1$--automorphisms $\phi,\psi$, a {\it $2$--automorphism}
from $\phi$ to $\psi$ is just a $2$--term $L_\infty$ algebra $2$--morphism $\varPhi:\phi\Rightarrow\psi$
(cf. subsect. \ref{sec:linftymorph}).
We shall denote the set of all $2$--automorphisms 
$\varPhi:\phi\Rightarrow\psi$ by $\Aut_2(\mathfrak{v})(\phi,\psi)$
and the set of all $2$--automorphisms $\varPhi$ by $\Aut_2(\mathfrak{v})$. 

A $1$--automorphism $\phi\in\Aut_1(\mathfrak{v})$ is invertible as a $1$--morphism
$\phi:\mathfrak{v}\rightarrow\mathfrak{v}$, that is there exists a $1$--morphism 
$\phi^{-1_\circ}:\mathfrak{v}\rightarrow\mathfrak{v}$ such that $\phi^{-1_\circ}\circ\phi=\phi\circ\phi^{-1_\circ}=
\id_{\mathfrak{v}}$. 
Explicitly, writing $\phi^{-1_\circ}=(\phi^{-1_\circ}{}_0,\phi^{-1_\circ}{}_1,\phi^{-1_\circ}{}_2)$, we have 
\begin{subequations}
\label{mor3/2tlinalg}
\begin{align}
&\phi^{-1_\circ}{}_0(x)=\phi_0{}^{-1}(x), \hspace{4.05cm}
\vphantom{\Big]}
\label{mor3/2tlinalgd}
\\
&\phi^{-1_\circ}{}_1(X)=\phi_1{}^{-1}(X), 
\vphantom{\Big]}
\label{mor3/2tlinalge}
\\
&\phi^{-1_\circ}{}_2(x,y)=-\phi_1{}^{-1}\phi_2(\phi_0{}^{-1}(x),\phi_0{}^{-1}(y)).
\vphantom{\Big]}
\label{mor3/2tlinalgf}
\end{align}
\end{subequations}

A $2$--automorphism $\varPhi\in\Aut_2(\mathfrak{v})$ is both horizontally and vertically invertible as 
a $2$--morphism $\varPhi:\phi\Rightarrow\psi$, that is there are a $2$--morphism 
$\varPhi^{-1_\circ}:\phi^{-1_\circ}\Rightarrow\psi^{-1_\circ}$ such that 
$\varPhi^{-1_\circ}\circ\varPhi=\varPhi\circ\varPhi^{-1_\circ}=\mathrm{Id}_{\id}$,
and a $2$--morphism $\varPi^{-1_\bfdot}:\psi\Rightarrow\phi$ such that 
$\varPhi^{-1_\bfdot}\bfdot\varPhi=\mathrm{Id}_{\phi}$, $\varPhi\bfdot\varPhi^{-1_\bfdot}=\mathrm{Id}_{\psi}$,
Explicitly, we have 
\begin{align}
&\varPhi^{-1_\circ}(x)=-\phi_1{}^{-1}\varPhi\psi_0{}^{-1}(x)=-\psi_1{}^{-1}\varPhi\phi_0{}^{-1}(x),
\vphantom{\Big]}
\label{mor4/1tlinalgb}
\\
&\varPhi^{-1_\bfdot}(x)=-\varPhi(x).
\vphantom{\Big]}
\label{mor4/1tlinalgd}
\end{align}

$\Aut_1(\mathfrak{v})$ and $\Aut_2(\mathfrak{v})$ are subsets of the set 
$2$--term $L_\infty$ algebra $1$--morphism and $2$--morphisms, respectively. 
$\Aut_1(\mathfrak{v})$ is so endowed with a composition and an inversion law 
and a unit.  $\Aut_2(\mathfrak{v})$ is similarly endowed with  horizontal and vertical composition and inversion laws
and units. It is straightforward though lengthy to check that 
these composition, inversion and unit structures satisfy the axioms \eqref{twogr1}
rendering $(\Aut_1(\mathfrak{v})$, $\Aut_2(\mathfrak{v}))$  
a strict $2$--group, as announced. 

$\Aut(\mathfrak{v})$ is actually a strict Lie $2$--group. 
Its associated strict $2$--term $L_\infty$ Lie algebra $\mathfrak{aut}(\mathfrak{v})$ is 
is described as follows. 

An element of $
\mathfrak{aut}_0(\mathfrak{v})$ consists of three mappings.
\begin{enumerate}

\item a vector space morphism $\alpha_0:\mathfrak{v}_0\rightarrow\mathfrak{v}_0$;

\item a vector space morphism $\alpha_1:\mathfrak{v}_1\rightarrow\mathfrak{v}_1$;

\item a vector space morphism $\alpha_2:\mathfrak{v}_0\wedge\mathfrak{v}_0\rightarrow\mathfrak{v}_1$.

\end{enumerate} 
These must satisfy the following relations:
\begin{subequations}
\label{mor5tlinalg}
\begin{align}
&\alpha_0(\partial X)-\partial\alpha_1(X)=0,
\vphantom{\Big]}
\label{mor5tlinalga}
\\
&\alpha_0([x,y])-[\alpha_0(x),y]-[x,\alpha_0(y)]-\partial\alpha_2(x,y)=0,
\hspace{1cm}
\vphantom{\Big]}
\label{mor5tlinalgb}
\\
&\alpha_1([x,X])-[\alpha_0(x),X]-[x,\alpha_1(X)]-\alpha_2(x,\partial X)=0,
\vphantom{\Big]}
\label{mor5tlinalgc}
\\
&[x,\alpha_2(y,z)]+[y,\alpha_2(z,x)]+[z,\alpha_2(x,y)]
+\alpha_2(x,[y,z])
\vphantom{\Big]}
\label{mor5tlinalgd}
\end{align}
\begin{align} 
&+\alpha_2(y,[z,x])+\alpha_2(z,[x,y])-\alpha_1([x,y,z])+
[x,y,\alpha_0(z)]
\vphantom{\Big]}
\nonumber
\\
&+[y,z,\alpha_0(x)]+[z,x,\alpha_0(y)]=0.
\vphantom{\Big]}
\nonumber
\end{align}
\end{subequations}

An element of $
\mathfrak{aut}_1(\mathfrak{v})$ consists of a single mapping mapping.
\begin{enumerate}

\item a vector space morphism $\varGamma:\mathfrak{v}_0\rightarrow\mathfrak{v}_1$.

\end{enumerate} 
No restrictions are imposed on it. 

The boundary map and the brackets of $\mathfrak{aut}(\mathfrak{v})$ 
are given by the expressions 
\begin{subequations}
\label{mor7tlinalg}
\begin{align}
&\partial_{\mathrm{aut}}\varGamma_0(x)=-\partial\varGamma(x), \hspace{6.1cm}
\vphantom{\Big]}
\label{mor7tlinalgx}
\\
&\partial_{\mathrm{aut}}\varGamma_1(X)=-\varGamma(\partial X),
\vphantom{\Big]}
\label{mor7tlinalgy}
\\
&\partial_{\mathrm{aut}}\varGamma_2(x,y)=[x,\varGamma(y)]-[y,\varGamma(x)]-\varGamma([x,y]),
\vphantom{\Big]}
\label{mor7tlinalgz}
\\
&[\alpha,\beta]_{\mathrm{aut} 0}(x)=\alpha_0\beta_0(x)-\beta_0\alpha_0(x), 
\vphantom{\Big]}
\label{mor7tlinalga}
\\
&[\alpha,\beta]_{\mathrm{aut} 1}(X)=\alpha_1\beta_1(X)-\beta_1\alpha_1(X),
\vphantom{\Big]}
\label{mor7tlinalgb}
\\
&[\alpha,\beta]_{\mathrm{aut} 2}(x,y)=\alpha_1\beta_2(x,y)+\alpha_2(\beta_0(x),y)+\alpha_2(x,\beta_0(y))
\vphantom{\Big]}
\label{mor7tlinalgc}
\\
&\qquad\qquad\qquad\qquad\qquad -\beta_1\alpha_2(x,y)-\beta_2(\alpha_0(x),y)-\beta_2(x,\alpha_0(y)),
\vphantom{\Big]}
\nonumber
\\
&[\alpha,\varGamma]_{\mathrm{aut}}(x)=\alpha_1\varGamma(x)-\varGamma\alpha_0(x),
\vphantom{\Big]}
\label{mor7tlinalgv}
\\
&[\alpha,\beta,\gamma]_{\mathrm{aut}}(x)=0.
\vphantom{\Big]}
\label{mor7tlinalgw}
\end{align}
\end{subequations}
Relations  \eqref{mor5tlinalg} ensure that the basic relations \eqref{2tlinalg}
are satisfied.

For a Lie group $G$ with Lie algebra $\mathfrak{g}$, there exists 
is a canonical group morphism of $G$ into $\Aut(\mathfrak{g})$ 
stemming from the Lie group structure of $G$, 
defining the adjoint representation of $G$. We are now going to see that 
this property generalizes to a strict Lie $2$--group $V$ with strict Lie $2$--algebra $\mathfrak{v}$ 
by constructing a canonical $2$--group morphism of $V$ into $\Aut(\mathfrak{v})$. 
Though we have not defined the notion of representation of a strict $2$--group, we can  
consider rightfully this morphism to be the strict $2$--group generalization of the adjoint 
representation of an ordinary Lie group. 
To this end, we view $V$ as a Lie crossed module 
$(G,H)$ and $\mathfrak{v}$ as the corresponding differential Lie crossed module
$(\mathfrak{g},\mathfrak{h})$. With any $a\in V_1$, we associate
a $1$--automorphism of $\mathfrak{v}$ defined by 
\begin{subequations}
\label{linftymorph1}
\begin{align}
&\phi_{a0}(x)=axa^{-1},
\vphantom{\Big]}
\label{linftymorph1a}
\\
&\phi_{a1}(X)=\dot m(a)(X),
\vphantom{\Big]}
\label{linftymorph1b}
\\
&\phi_{a2}(x,y)=0.
\vphantom{\Big]}
\label{linftymorph1c}
\end{align}
\end{subequations}
Further, with any $(a,A)\in V_2(a,b)$ with $b=t(A)a$, we associate a 
$2$--automorphism $\varPhi_{a,A}:\phi_a\Rightarrow \phi_b$ of $\mathfrak{v}$ defined by 
\begin{equation}
\varPhi_{a,A}(x)=Q(axa^{-1},A), 
\label{linftymorph2}
\end{equation}
where, for $A\in H$, $Q(\cdot,A):\mathfrak{g}\rightarrow\mathfrak{h}$ 
is the vector space morphism defined by  
\begin{equation}
Q(x,A)=\frac{d}{du}m(c(u))(A)A^{-1}\Big|_{u=0}, 
\label{linftymorph3}
\end{equation}
with $c(u)$ is a curve in $G$ such that $c(0)=1_G$ and $dc(u)/du\big|_{u=0}=x$ 
\footnote{ $\vphantom{\bigg]}$ $Q$ has the following properties, which turn out to be relevant,
\vspace{-.1cm}
\begin{subequations}
\label{linftymorph4}
\begin{align}
&Q([x,y],A)+[Q(x,A),Q(y,A)]-[x,Q(y,A)]+[y,Q(x,A)]=0,
\vphantom{\Big]}
\label{linftymorph4a}
\\
&Q(x,AB)=Q(x,A)+AQ(x,B)A^{-1},
\vphantom{\Big]}
\label{linftymorph4b}
\\
&Q(axa^{-1},A)=\dot m(a)(Q(x,m(a^{-1})(A))).
\vphantom{\Big]}
\label{linftymorph4c}
\end{align}
\end{subequations}
By relation \eqref{linftymorph4a}, for fixed $A$, the mapping $x\rightarrow x-Q(x,A)$ is a Lie algebra 
morphism of $\mathfrak{g}$ into the semidirect sum $\mathfrak{g}\subset\!\!\!\!\!\! +\,\mathfrak{h}$.
By relation \eqref{linftymorph4b}, for fixed $x$, $A\rightarrow Q(x,A)$ is a $\mathfrak{h}$--valued 
$H$ $1$--cocycle. Relation \eqref{linftymorph4c} implies that 
$x\rightarrow Q(x,1_H)$ is $G$--equivariant}.
Now, it is straightforward to verify that the mappings $a\rightarrow \phi_a$ and 
$(a,A)\rightarrow \varPhi_{a,A}$ define a strict $2$--group $1$--morphism from $V$
to $\Aut(\mathfrak{v})$ as desired.


\subsection{\normalsize \textcolor{blue}{Examples of Lie $2$--groups}}\label{sec:2grex}

\hspace{.5cm} Below, we shall illustrate some simple but important examples of 
$2$--groups.$\vphantom{\ul{\ul{x}}}$


\vspace{.25cm}

{\it 1. Lie groups}

Every Lie group $L$ can be regarded as a strict Lie $2$--group,
denoted by the same symbol. As a Lie crossed module, $L$ is 
defined by the data $(G,H,t,m)$, where 
$G=L$, $H=1$,  $t:H\rightarrow G$ vanishes and 
$m:G\rightarrow \Aut(H)$ is trivial. 
The Lie $2$--algebra of $L$ as a Lie $2$--group 
is the Lie algebra $\mathfrak{l}$ of $L$ as a Lie group 
regarded as Lie $2$--algebra (cf. sect. \ref{sec:linftyex}). 

{\it 2. Inner automorphism Lie $2$--groups} 

With any Lie group $L$, there is associated canonically a strict Lie $2$--group $\Inn(L)$ 
defined as follows. As a Lie crossed module, $\Inn(L)$ is the quadruple of data 
$(G,H,t,m)$, where $G=L$, $H=L$, $t:H\rightarrow G$ is the identity $\id_{L}$
and $m:G\rightarrow \Aut(H)$ is the adjoint action $\Ad_{L}$
of $L$ on itself. $\Inn(L)$ is called {\it inner automorphism Lie $2$--group of $L$}.
The Lie $2$--algebra of $\Inn(L)$ is the 
inner derivation Lie $2$--algebra $\mathfrak{inn(l)}$ of the Lie algebra $\mathfrak{l}$ of $L$
(cf. sect. \ref{sec:linftyex}). 

{\it 3. Automorphism Lie $2$--groups} 

The automorphisms of a Lie  group $L$, $\Aut(L)$ 
form a Lie group and thus also a strict $2$--group, by example 1. 
However, $\Aut(L)$ has a second strict $2$--group structure defined as follows. 
Viewed again as a Lie crossed module, $\Aut(L)$ is specified by the data $(G,H, t,m)$, 
where $G=\Aut(L)$, $H=L$, $t:H\rightarrow G$ is the adjoint Lie group morphism $\Ad_{L}$
and $m:G\rightarrow \Aut(H)$ is the identity $\id_{\Aut(L)}$. 
$\Aut(L)$ is called {\it automorphism Lie $2$--group of $L$}.
The Lie $2$--algebra of $\Aut(L)$ is the derivation Lie $2$--algebra $\mathfrak{der(l)}$ of 
the Lie algebra $\mathfrak{l}$ of $L$ (cf. sect. \ref{sec:linftyex}). 

{\it 4. Central extension Lie $2$--groups}

Consider a central extension of a Lie group $L$ by an Abelian Lie
group $A$, that is a third Lie group $E$
fitting in  a short exact sequence of Lie groups 
\begin{equation}
\xymatrix{1\ar[r]&A\ar[r]&E\ar[r]&L\ar[r]&1,}
\label{2grex1}
\end{equation}
with the image of $A$ contained in the center of $E$.
With the extension, there is associated a canonical Lie crossed module
$(G,H, t,m)$, hence 
a strict $2$--group, as follows. $G=L$, $H=E$.
$t:H\rightarrow G$ is the third morphism in the sequence 
\eqref{2grex1}. $m:G\rightarrow \Aut(H)$ is defined by choosing a linear mapping
$s:L\rightarrow E$ such that $t\circ s=\id_{L}$
and setting $m(a)(A)=s(a)As(a)^{-1}$. As $s$ is defined mod $\ker t$ which is 
contained in the center of $E$, $m$ is well--defined. 
The resulting strict Lie $2$--group
$C_E$ is called the {\it central extension Lie $2$--group of $E$}. 
The Lie algebra $\mathfrak{e}$ of $E$ is a central extension of the Lie algebra 
$\mathfrak{l}$ of $L$ by the Abelian Lie algebra $\mathfrak{a}$ of $A$.
The Lie $2$--algebra of $C_E$ is the central extension Lie $2$--algebra 
$\mathfrak{c}_{\mathfrak{e}}$ of $\mathfrak{e}$ (cf. sect. \ref{sec:linftyex}, ex. 4).

Next, we consider a few non strict examples.

{\it 5. Non associative structures and Moufang loops}

In sect. \ref{sec:linftyex}, we have introduced the Jacobiator Lie $2$--algebra $\mathfrak{j}_{\mathfrak{l}}$
associated to a Lie prealgebra $\mathfrak{l}$, which is generally non strict. It is natural to wonder whether 
there are higher group like structures which have such Lie $2$--algebras as infinitesimal counterparts. 
When $\mathfrak{l}$ is a Lie algebra, $\mathfrak{j}_{\mathfrak{l}}=\mathfrak{inn(l)}$, which is the Lie 
$2$--algebra of the inner automorphism $2$--group $\Inn(L)$, where $L$ is a Lie group integrating $L$. 
When $\mathfrak{l}$ is not a Lie algebra, things are not so clear. To the best of our knowledge, 
not much is known about this matter. Consider for instance the Jacobiator Lie $2$--algebra of the imaginary octonions 
$\Ima\mathbb{O}$. $\Ima\mathbb{O}$ is the tangent space at $1$ 
of the $7$--sphere of the unit norm octonions $\mathrm{U}(\mathbb{O})$. As octonions are not associative,
$\mathrm{U}(\mathbb{O})$ is neither a group nor a strict $2$--group under octonionic multiplication. 
But $\mathrm{U}(\mathbb{O})$ does not constitute even a coherent $2$--group. 
It has rather the structure of a {\it Moufang loop} \cite{Moufang:1935}. 
A loop is a set $S$ equipped with a generally non associative 
binary operation $(x,y)\rightarrow xy$ with the following properties:
\begin{enumerate}

\item for any $x,y\in S$ there exists $u,v\in S$ satisfying the relations $ux=y$, $xv=y$;

\item there distinguished element $1\in S$ such that $x1=1x=x$ for $x\in S$.

\end{enumerate}
A Moufang Loop is a loop where the Moufang identity $(zx)(yz) = (z(xy))z=z((xy)z)$ 
is satisfied identically for $x,y,z\in S$. 
The stronger condition $w(x(yz)) = ((wx)y)z$ does not hold, as no associator is available in
a Moufang loop. 

{\it 6. The string $2$--group}

Let $G$ be a simply-connected, connected, compact simple Lie group $G$. 
The loop group $\Omega G$ of $G$ is the infinite dimensional Lie group of the 
smooth loops of $G$ based at $1_G$ equipped with pointwise multiplication. 
In ref. \cite{PressSeg}, Pressley and Segal showed that, for each integer $k$, 
$\Omega G$ has a central extension 
\begin{equation}
\xymatrix{1\ar[r]&U(1)\ar[r]&\widehat{\Omega}_kG\ar[r]&\Omega G\ar[r]&1.}
\end{equation}
$k$ is called the level of the extension. The extensions of different levels are inequivalent.

Proceeding as explained in ex. 4, one builds the infinite dimensional central extension Lie $2$--group
of $\widehat{\Omega}_kG$, the {\it level $k$ loop $2$--group} $L_k(G)$ of $G$. In ref. \cite{BCSS},  
Baez et al. showed that $L_k(G)$ fits into an exact sequence
\begin{equation}
\xymatrix{1\ar[r]&L_k(G)\ar[r]&\mathrm{String}_k(G)\ar[r]&G\ar[r]&1,}
\end{equation}
of strict Lie $2$--groups. The middle term of the sequence is an infinite dimensional
strict Lie $2$--group, the {\it level $k$ string $2$--group} $\mathrm{String}_k(G)$  of $G$.
The infinite-dimensional strict Lie $2$--algebra of $\mathrm{String}_k(G)$ is equivalent 
to the  string Lie $2$--algebra $\mathfrak{string}_k(\mathfrak{g})$.
Note that $\mathrm{String}_k(G)$ is defined only for integer $k$ while
$\mathfrak{string}_k(\mathfrak{g})$ is for any real $k$.
The relationship just described between $\mathrm{String}_k(G)$ and 
$\mathfrak{string}_k(\mathfrak{g})$, so, holds only when
$k$ is integer.

In \cite{BCSS} Baez et al. did not integrate the string Lie 
$2$--algebra $\mathfrak{string}_k(\mathfrak{g})$. They only constructed
$\mathrm{String}_k(G)$ and showed that its infinite dimensional 
Lie 2-algebra is equivalent to $\mathfrak{string}_k(\mathfrak{g})$. 
In \cite{Hen}, Henriques worked out a procedure to integrate
$\mathfrak{string}_1(\mathfrak{g})$. The resulting model of $\mathrm{String}_1(G)$
is again an infinite dimensional Lie $2$--group. 

In ref. \cite{SchPr}, Schommer-Pries managed to produce a finite dimensional 
model of the string $2$--group $\mathrm{String}_1(G)$. The price paid for this is the need for
a notion of Lie $2$--group weaker than the customary one of coherent Lie $2$--group. 
Schommer-Pries' smooth $2$--groups are $2$--group objects in a bicategory 
of Lie groupoids, left-principal bibundles, and bibundle maps
rather than the $2$--category of Lie groupoids, smooth functors and smooth natural transformations.

The above discussion shows once more that the relation between 
coherent Lie $2$--groups and  Lie $2$--algebras is rather subtle.
While any strict Lie $2$--algebra always integrates to a strict Lie $2$--group, 
not so  Lie $2$--algebras. 

\vfill\eject

\section{\normalsize \textcolor{blue}{Semistrict higher gauge theory}}\label{sec:linfgauge}

\vfil
\hspace{.5cm} We can now start illustrating our formulation of 
{\it semistrict algebra gauge theory}. 
For an alternative approaches to higher gauge theory see ref. \cite{Baez:2010ya}.

\vfil
As we recalled in the introduction, 
the basic geometric structure of ordinary gauge theory is a principal bundle 
$P(M,G)$ on a manifold $M$ with structure Lie group $G$. The Lie algebra of $G$, 
$\mathfrak{g}$, is a derived secondary object. 
A formulation on the same lines of semistrict higher gauge theory 
would require a principal $2$--bundle $P(M,V)$ on a $2$--manifold 
$M$ with a structure Lie $2$--group $V$ of the appropriate type along the lines of \cite{
Baez:2004in,Baez:2005qu}. Again, the 
Lie $2$--algebra of $V$, $\mathfrak{v}$, would be a derived secondary object. 
This type of approach, while the most powerful in theory, 
is likely to be very difficult to implement in practice. 
The relation between Lie $2$--groups and Lie $2$-algebras
is a sticky matter beyond the strict case. If we demand as we do that $\mathfrak{v}$
be a semistrict Lie $2$--algebra then $V$ may be something more general than a mere 
coherent Lie $2$--group, as the case of the string Lie $2$--algebra 
shows. In particular, it may be infinite dimensional. 
Though general techniques to cope with these difficulties have been developed 
recently \cite{Fiorenza2011,Schreiber2011}, their abstractness
makes it difficult their application to detailed calculations of the type presented in the second half 
of this paper. For this reason, we have opted for a more conventional and conservative approach, which 
exploits as much as possible the computational effectiveness of $2$--term $L_\infty$ algebras, as we 
illustrate next.

\vfil
In ordinary gauge theory, the fields are $\mathfrak{g}$--valued and their global properties are 
controlled by a set of \v Cech gluing data hinging on an $\Aut(\mathfrak{g})$--valued cocycle
acting by gauge transformations and obeying certain coherence conditions.  
The theory, so, can be formulated to a substantial extent relying on the structure Lie algebra 
$\mathfrak{g}$ only and its topological features are encoded in those data. 
In the same way, in our formulation of semistrict higher gauge theory, the fields are $\mathfrak{v}$--valued 
and their global properties are controlled by a set of \v Cech gluing data 
rooted in a higher $\Aut(\mathfrak{v})$--valued cocycle acting by higher gauge transformations and 
obeying higher coherence conditions. 
The theory, then, is formulated in terms of $\mathfrak{v}$ only, conveniently seen as a 
$2$--term $L_\infty$ algebra, and its topological features are implicitly contained in the data.
This is the line of thought followed below. As we shall see in due course, this way of proceeding 
works quite well for perturbative Lagrangian field theory, but it is of little us at the non 
perturbative level. 

In this section, we shall first analyze the local aspects of $2$--term $L_\infty$ algebra gauge 
theory, neglecting global issues altogether. Later, we shall tackle the problem of assembling 
locally defined gauge theoretic data in a globally consistent manner by means of suitable 
gluing data. We shall also point out the strengths and weaknesses of our approach and 
endeavour to relate our formulation with others which have appeared in the 
literature.

\subsection{\normalsize \textcolor{blue}{$2$--term $L_\infty$ algebra cohomology}}\label{subsec:linfweil}

\hspace{.5cm} 
The Chevalley--Eilenberg complex of a Lie algebra $\mathfrak{g}$ encodes the structure
of $\mathfrak{g}$ \cite{Chevalley1}. It also abstracts the formal algebraic properties of the flat connections 
of a principal $G$--bundle, where $G$ is a Lie group integrating $\mathfrak{g}$. 
The Weil complex of $\mathfrak{g}$ extends the Chevalley--Eilenberg complex in that it encapsulates
the algebraic properties of the connections and curvatures thereof of a $G$--principal bundle
and constitute the basic framework of the Chern--Weil theory of characteristic classes
\cite{Weil1,HCartan2,HCartan1} (see also \cite{Ehresmann1}). 
These well--known classical facts 
generalize to Lie
$2$--algebras and principal $2$--bundle and beyond as worked out in refs.
\cite{Fiorenza2011,Schreiber2011}. We review these matters in this subsection
and the next one, but we shall not go through the Chern--Weil theory which, albeit very important on its own,  
lies outside the scope of this paper.

Let $\mathfrak{g}$ be a Lie algebra. The {\it Chevalley--Eilenberg algebra} 
$\mathrm{CE}(\mathfrak{g})$ of $\mathfrak{g}$ is the graded commutative algebra 
$S(\mathfrak{g}^\vee[1])\simeq \bigwedge^*\mathfrak{g}^\vee$ 
generated by $\mathfrak{g}^\vee[1]$, the $1$ step degree shifted dual of $\mathfrak{g}$
\footnote{$\vphantom{\bigg[}$ 
For any vector space $X$, $X[q]$ is 
$X$ itself with Grassmann 
degree shifted of $q$ units. If $q$ is odd, $S(X[q])$ is isomorphic to the exterior algebra 
$\bigwedge^*X$ of $X$ with $X$ in degree $q$.  If $q$ is even, $S(X[q])$ is isomorphic to the 
customary symmetric algebra $\bigvee^* X$ of $X$ with $X$ in degree $q$.}.
The {\it Chevalley--Eilenberg differential} $\mathcal{Q}_{\mathrm{CE}(\mathfrak{g})}$ 
is the degree $1$ differential defined as follows 
\footnote{$\vphantom{\bigg[}$ A degree $p$ differential $d$ on a graded commutative algebra $A$ is
a vector endomorphism $d:A\rightarrow A$ such that 
$d(ab)=dab+(-1)^{p\deg b}adb$.}. Let $\{e_a\}$ be a basis of $\mathfrak{g}$ and let $\{\pi^a\}$ be the basis 
of $\mathfrak{g}^\vee[1]$ dual to $\{e_a\}$. Set  \hphantom{xxxxxxxxxxxxxxxxxxx}
\begin{equation}
\pi=\pi^a\otimes e_a,
\label{xgammaCdef}
\end{equation}
Then, $\mathcal{Q}_{\mathrm{CE}(\mathfrak{g})}$ is given compactly by 
\begin{equation}
\mathcal{Q}_{\mathrm{CE}(\mathfrak{g})}\pi=-\frac{1}{2}[\pi,\pi].
\label{x2tlinalgQ}
\end{equation}
It is immediately verified that $\mathcal{Q}_{\mathrm{CE}(\mathfrak{g})}$ is nilpotent, 
\begin{equation}
\mathcal{Q}_{\mathrm{CE}(\mathfrak{g})}{}^2=0,
\label{xQ2=0}
\end{equation}
$(\mathrm{CE}(\mathfrak{g}),\mathcal{Q}_{\mathrm{CE}(\mathfrak{g})})$ is a so cochain complex. 
Its cohomology $H_{CE}^*(\mathfrak{g})$ is the {\it Chevalley--Eilenberg cohomology}, 
also known as {\it Lie algebra cohomology}, of $\mathfrak{g}$. 

The nilpotence of $\mathcal{Q}_{\mathrm{CE}(\mathfrak{g})}$ is equivalent to the bracket
$[\cdot,\cdot]$ satisfying the Jacobi identity, as is readily verified. Indeed, 
there is a one--to--one correspondence between Lie algebra structures on a vector space 
$\mathfrak{g}$ and nilpotent degree $1$ differentials $\mathcal{Q}$ on $S(\mathfrak{g}^\vee[1])$. 

As it is apparent from \eqref{xgammaCdef}, \eqref{x2tlinalgQ}, the Chevalley--Eilenberg complex 
$\mathrm{CE}(\mathfrak{g})$ formalizes the algebraic properties of the 
flat connections of a principle $G$--bundle, where $\Lie G=\mathfrak{g}$. (The precise meaning of this 
statement will be explained in the subsection.)
The structure which does the same job for all connections of the $G$--bundle
is the Weil complex of $\mathfrak{g}$, which we define next. $\vphantom{\ul{\ul{x}}}$

The {\it Weil algebra} $\mathrm{W}(\mathfrak{g})$ of $\mathfrak{g}$ 
is the graded commutative algebra 
$S(\mathfrak{g}^\vee[1]\oplus\mathfrak{g}^\vee[2])\simeq \bigwedge^*(\mathfrak{g}^\vee\oplus\mathfrak{g}^\vee[1])$
generated by $\mathfrak{g}^\vee[1]\oplus\mathfrak{g}^\vee[2]$. 
The {\it Weil differential} $\mathcal{Q}_{\mathrm{W}(\mathfrak{g})}$ is defined as follows. 
Let again $\{e_a\}$ be a basis of $\mathfrak{g}$ and let $\{\pi^a\}$, $\{\gamma^a\}$ be the bases of 
$\mathfrak{g}^\vee[1]$, $\mathfrak{g}^\vee[2]$ dual to $\{e_a\}$, respectively. 
We define $\pi$ as in \eqref{xgammaCdef} and set 
\begin{equation}
\gamma=\gamma^a\otimes e_a.
\label{linfweil1}
\end{equation}
Then, $\mathcal{Q}_{\mathrm{W}(\mathfrak{g})}$ is given by 
\begin{subequations}
\label{linfweil2}
\begin{align}
\mathcal{Q}_{\mathrm{W}(\mathfrak{g})}\pi&=-\frac{1}{2}[\pi,\pi]+\gamma,
\vphantom{\Big]}
\label{linfweil2a}
\\
\mathcal{Q}_{\mathrm{W}(\mathfrak{g})}\gamma&=-[\pi,\gamma].
\vphantom{\Big]}
\label{linfweil2b}
\end{align}
\end{subequations}
It is readily checked that \hphantom{xxxxxxxxxxxxxxxxxxx}
\begin{equation}
\mathcal{Q}_{\mathrm{W}(\mathfrak{g})}{}^2=0.
\label{linfweil3}
\end{equation}
$(\mathrm{W}(\mathfrak{g}),\mathcal{Q}_{\mathrm{W}(\mathfrak{g})})$ is a so cochain complex. 
Its cohomology $H_{W}^*(\mathfrak{g})$ turns out to be trivial
in positive degree
\footnote{$\vphantom{\bigg[}$ Adding contractions $I_x$ 
along the Lie algebra elements $x\in \mathfrak{g}$ to $\mathcal{Q}_{\mathrm{W}(\mathfrak{g})}$,
it is possible to define the basic cohomology $H_{W\mathrm{bas}}^*(\mathfrak{g})$  
of $\mathrm{W}(\mathfrak{g})$, which
is isomorphic to $S(\mathfrak{g}^\vee[2])_{\mathrm{inv}}$ and so generally non trivial. 
This plays a pivotal role in Chern--Weil theory.}. 

Requiring that $\mathcal{Q}_{\mathrm{W}(\mathfrak{g})}\big|_{\mathfrak{g}^\vee[1]}=\mathcal{Q}_{\mathrm{CE}(\mathfrak{g})}+\sigma$, 
where $\sigma:\mathfrak{g}^\vee[1]\rightarrow\mathfrak{g}^\vee[2]$ is the degree shift vector morphism, and 
that $\mathcal{Q}_{\mathrm{W}(\mathfrak{g})}$ is nilpotent fully determines the Weil differential $\mathcal{Q}_{\mathrm{W}(\mathfrak{g})}$.
Further, the projection $\mathfrak{g}^\vee[1]\oplus\mathfrak{g}^\vee[2]\rightarrow\mathfrak{g}^\vee[1]$
induces a canoni\-cal differential graded commutative algebra epimorphism
$\mathrm{W}(\mathfrak{g})\rightarrow\mathrm{CE}(\mathfrak{g})$.

From \eqref{xgammaCdef}, \eqref{linfweil1}, \eqref{linfweil2}, it is apparent that 
the Weil complex $\mathrm{W}(\mathfrak{g})$ formalizes the algebraic properties of the 
connections of a principle $G$--bundle, \eqref{linfweil2a}, \eqref{linfweil2b}
corresponding to the definition of the curvature of a connection and to
the Bianchi identity this obeys, respectively. 

The above superalgebraic construction generalizes straightforwardly to Lie $2$--algebras. 
Let $\mathfrak{v}$ be such an algebra seen as a $2$--term $L_\infty$ algebra.$\vphantom{\ul{\ul{x}}}$
Similarly to ordinary Lie algebras, the {\it Chevalley--Eilenberg algebra} 
$\mathrm{CE}(\mathfrak{v})$ of $\mathfrak{v}$ is the graded commutative algebra 
$S(\mathfrak{v}^\vee[1])\simeq \bigwedge^*\mathfrak{v}^\vee$ 
generated by $\mathfrak{v}^\vee[1]$.
The {\it Chevalley--Eilenberg differential} $\mathcal{Q}_{\mathrm{CE}(\mathfrak{v})}$ 
is the degree $1$ differential defined as follows.
Let $\{e_a\}$, $\{E_\alpha\}$ be bases of $\hat{\mathfrak{v}}_0$, $\hat{\mathfrak{v}}_1$
\footnote{$\vphantom{\dot{\dot{\dot{\dot{x}}}}}$ \label{foot:hat} Here and below, 
for a graded vector space $X$, $\hat X$ is $X$ with its grading set to $0$.}
and let $\{\pi^a\}$, $\{\varPi^\alpha\}$ be the bases of $\mathfrak{v}_0{}^\vee[1]$, 
$\mathfrak{v}_1{}^\vee[1]$ dual to $\{e_a\}$, $\{E_\alpha\}$, respectively. Define now
\begin{subequations}
\label{gammaCdef}
\begin{align}
\pi&=\pi^a\otimes e_a,
\vphantom{\Big]}
\label{gammadef}
\\
\varPi&=\varPi^\alpha\otimes E_\alpha,
\vphantom{\Big]}
\label{Cdef}
\end{align}
\end{subequations}
in analogy to \eqref{xgammaCdef}. Then, $\mathcal{Q}_{\mathrm{CE}(\mathfrak{v})}$ is given succinctly by 
\begin{subequations}
\label{2tlinalgQ}
\begin{align}
\mathcal{Q}_{\mathrm{CE}(\mathfrak{v})}\pi&=-\frac{1}{2}[\pi,\pi]+\partial \varPi,
\vphantom{\Big]}
\label{2tlinalgQa}
\\
\mathcal{Q}_{\mathrm{CE}(\mathfrak{v})}\varPi&=-[\pi,\varPi]+\frac{1}{6}[\pi,\pi,\pi].
\vphantom{\Big]}
\label{2tlinalgQb}
\end{align}
\end{subequations}
The form of $\mathcal{Q}_{\mathrm{CE}(\mathfrak{v})}$ is determined by the requirement that it is nilpotent,
\begin{equation}
\mathcal{Q}_{\mathrm{CE}(\mathfrak{v})}{}^2=0,
\label{wQ2=0}
\end{equation}
$(\mathrm{CE}(\mathfrak{v}),\mathcal{Q}_{\mathrm{CE}(\mathfrak{v})})$ is a so cochain complex. 
The associated Chevalley--Eilenberg cohomology
$H_{CE}^*(\mathfrak{v})$ is the Lie $2$--algebra cohomology of $\mathfrak{v}$
generalizing ordinary Lie algebra cohomology. 

The nilpotence of $\mathcal{Q}_{\mathrm{CE}(\mathfrak{v})}$ is equivalent to the brackets
$\partial,[\cdot,\cdot],[\cdot,\cdot,\cdot]$ satisfying the relations \eqref{2tlinalg}
characterizing a $2$--term $L_\infty$ algebra
\footnote{$\vphantom{\dot{\dot{\dot{\dot{x}}}}}$ 
The condition $\mathcal{Q}_{\mathrm{CE}(\mathfrak{v})}{}^2=0$ 
translates into the following relations equivalent to \eqref{2tlinalg}
\begin{subequations}
\label{2tlincoh}
\begin{align}
&[\pi,\partial \varPi]-\partial[\pi,\varPi]=0,
\vphantom{\Big]}
\vphantom{\dot{\dot{\dot{\dot{x}}}}}
\label{2tlincoha}
\\
&[\partial \varPi,\varPi]=0,
\vphantom{\Big]}
\label{2tlincohb}
\\
&3[\pi,[\pi,\pi]]-\partial[\pi,\pi,\pi]=0,
\vphantom{\Big]}
\label{2tlincohc}
\\
&2[\pi,[\pi,\varPi]]-[[\pi,\pi],\varPi]-[\pi,\pi,\partial \varPi]=0,
\vphantom{\Big]}
\label{2tlincohd}
\\
&4[\pi,[\pi,\pi,\pi]]-6[\pi,\pi,[\pi,\pi]]=0.
\vphantom{\Big]}
\label{2tlincohe}
\end{align}
\end{subequations}
\vspace{-9truemm}
}. 
Indeed, $\vphantom{\ul{\ul{x}}}$again as for ordinary Lie algebras, 
there is a one--to--one correspondence between $2$--term $L_\infty$ algebra structures on a graded vector space 
$\mathfrak{v}=\mathfrak{v}_0\oplus\mathfrak{v}_1$ 
and nilpotent degree $1$ differentials $\mathcal{Q}$ on $S(\mathfrak{v}^\vee[1])$. 

Generalizing the Lie algebraic case, we can assume that the Chevalley--Eilen\-berg complex 
$\mathrm{CE}(\mathfrak{v})$ formalizes the algebraic properties of the 
flat $2$--connections of a principle $V$--$2$--bundle, where $V$ is the appropriate kind 
of $2$--group having $\mathfrak{v}$ as infinitesimal counterpart,  
\eqref{gammaCdef}, \eqref{2tlinalgQ} defining the flatness conditions. 

The {\it Weil algebra} $\mathrm{W}(\mathfrak{v})$ of $\mathfrak{v}$ 
is the graded commutative algebra 
$S(\mathfrak{v}^\vee[1]\oplus\mathfrak{v}^\vee[2])\simeq \bigwedge^*(\mathfrak{v}^\vee\oplus\mathfrak{v}^\vee[1])$
generated by $\mathfrak{v}^\vee[1]\oplus\mathfrak{v}^\vee[2]$. 
The {\it Weil differential} $\mathcal{Q}_{\mathrm{W}(\mathfrak{v})}$ is defined as follows. 
Let again $\{e_a\}$, $\{E_\alpha\}$ be bases of $\hat{\mathfrak{v}}_0$, $\hat{\mathfrak{v}}_1$
and let $\{\pi^a\}$, $\{\gamma^a\}$, $\{\varPi^\alpha\}$, $\{\varGamma^\alpha\}$
 be the bases of $\mathfrak{v}_0{}^\vee[1]$, $\mathfrak{v}_0{}^\vee[2]$, 
$\mathfrak{v}_1{}^\vee[1]$, $\mathfrak{v}_1{}^\vee[2]$ dual to $\{e_a\}$, $\{E_\alpha\}$, respectively.  
We define $\pi$, $\varPi$ as in \eqref{gammaCdef} and set 
\begin{subequations}
\label{sgammaCdef}
\begin{align}
\gamma&=\gamma^a\otimes e_a,
\vphantom{\Big]}
\label{sgammadef}
\\
\varGamma&=\varGamma^\alpha\otimes E_\alpha,
\vphantom{\Big]}
\label{sCdef}
\end{align}
\end{subequations}
Then, $\mathcal{Q}_{\mathrm{W}(\mathfrak{v})}$ is given by 
\begin{subequations}
\label{slinfweil2}
\begin{align}
\mathcal{Q}_{\mathrm{W}(\mathfrak{v})}\pi&=-\frac{1}{2}[\pi,\pi]+\partial\varPi+\gamma,
\vphantom{\Big]}
\label{slinfweil2a}
\\
\mathcal{Q}_{\mathrm{W}(\mathfrak{v})}\varPi&=-[\pi,\varPi]+\frac{1}{6}[\pi,\pi,\pi]+\varGamma,
\vphantom{\Big]}
\label{slinfweil2b}
\\
\mathcal{Q}_{\mathrm{W}(\mathfrak{v})}\gamma&=-[\pi,\gamma]-\partial\varGamma,
\vphantom{\Big]}
\label{slinfweil2c}
\\
\mathcal{Q}_{\mathrm{W}(\mathfrak{v})}\varGamma&=-[\pi,\varGamma]+[\gamma,\varPi]-\frac{1}{2}[\pi,\pi,\gamma].
\vphantom{\Big]}
\label{slinfweil2d}
\end{align}
\end{subequations}
Again, it is checked that \hphantom{xxxxxxxxxxxxxxxxxxx}
\begin{equation}
\mathcal{Q}_{\mathrm{W}(\mathfrak{v})}{}^2=0.
\label{slinfweil3}
\end{equation}
$(\mathrm{W}(\mathfrak{v}),\mathcal{Q}_{\mathrm{W}(\mathfrak{v})})$ is a so cochain complex. 
Its cohomology $H_{W}^*(\mathfrak{v})$ turns out to be trivial
in positive degree as for ordinary Lie algebras.

Again, requiring that $\mathcal{Q}_{\mathrm{W}(\mathfrak{v})}\big|_{\mathfrak{v}^\vee[1]}
=\mathcal{Q}_{\mathrm{CE}(\mathfrak{v})}+\sigma$, 
\pagebreak 
where $\sigma:\mathfrak{v}^\vee[1]\rightarrow\mathfrak{v}^\vee[2]$ is the degree shift vector morphism, and 
that $\mathcal{Q}_{\mathrm{W}(\mathfrak{v})}$ is nilpotent fully determines the Weil differential $\mathcal{Q}_{\mathrm{W}(\mathfrak{v})}$.
Further, the projection $\mathfrak{v}^\vee[1]\oplus\mathfrak{v}^\vee[2]\rightarrow\mathfrak{v}^\vee[1]$
induces a canoni\-cal differential graded commutative algebra epimorphism
$\mathrm{W}(\mathfrak{v})\rightarrow\mathrm{CE}(\mathfrak{v})$.

Extending the Lie algebraic framework once more, we can think of
the Weil complex $\mathrm{W}(\mathfrak{v})$ as an algebraic model describing  the $2$--con\-nections of 
a principle $V$--$2$--bundle, \eqref{slinfweil2a}, \eqref{slinfweil2b} corresponding 
to the definition of the curvature
components and \eqref{slinfweil2c}, \eqref{slinfweil2d} expressing 
the Bianchi identities which these obey.

\subsection{\normalsize \textcolor{blue}{$2$--term $L_\infty$ algebra gauge theory, 
local aspects}}\label{subsec:linfdoub}

\hspace{.5cm} In ordinary as well as higher gauge theory, fields propagate on a fixed $d$--fold $M$.
Each field is characterized by its {\it form degree} $m$ and {\it ghost number degree} 
$n$ for some integers $m\geq 0$ and $n$. In that case, the field is said to have 
{\it bidegree} $(m,n)$. 

To study the local aspects of the theory, we first assume that
$M$ is diffeomorphic to $\mathbb{R}^d$. On such an $M$, 
a field of bidegree $(m,n)$ is then  an element of the space 
$\Omega^m(M,E[n])$ of $m$--forms on $M$ with values in $E[n]$, where $E$ is
some vector space. 
In an ordinary gauge theory with structure Lie algebra $\mathfrak{g}$, fields are generally 
drawn from the spaces
$\Omega^m(M,\mathfrak{g}[n])$ and $\Omega^m(M,\mathfrak{g}^\vee[n])$. 
In the first case, they are called bidegree $(m,n)$ fields, in the second, bidegree $(m,n)$ dual fields.

The main field of the gauge theory is the {\it connection} $\omega$, which is a bidegree $(1,0)$ field.
$\omega$ is characterized by its {\it curvature} $f$, 
the bidegree $(2,0)$ field given by 
\begin{equation}
f=d\omega+\frac{1}{2}[\omega,\omega].
\vphantom{\Big]}
\label{xfcurv}
\end{equation}
$f$ satisfies the standard {\it Bianchi identity}
\begin{equation}
df+[\omega,f]=0.
\label{xfBianchi}
\end{equation}

It is a classic result \pagebreak 
that the assignment of a connection $\omega$ is equivalent 
to that of a differential graded commutative algebra morphism 
$\mathrm{W}(\mathfrak{g})\rightarrow \Omega^*(M)$ 
from the Weil algebra of $\mathfrak{g}$ (cf. subsect. \ref{subsec:linfweil})
to the differential forms of $M$. The morphism is the one mapping the generators 
$\pi$, $\gamma$ of $\mathrm{W}(\mathfrak{g})$ respectively in $\omega$, $f$
and its being differential is evident from comparing eqs. \eqref{linfweil2} with eqs. \eqref{xfcurv},
\eqref{xfBianchi}. 

The connection $\omega$ is {\it flat} if the curvature $f=0$. This  happens precisely 
when the associated 
morphism $\mathrm{W}(\mathfrak{g})\rightarrow \Omega^*(M)$
factors as $\mathrm{W}(\mathfrak{g})\rightarrow \mathrm{CE}(\mathfrak{g})\rightarrow \Omega^*(M)$,
where $\mathrm{W}(\mathfrak{g})\rightarrow \mathrm{CE}(\mathfrak{g})$ is the canonical 
morphism of the Weil onto the Chevalley--Eilenberg algebra of $\mathfrak{g}$ 
(cf. subsect. \ref{subsec:linfweil}), as follows from 
\eqref{x2tlinalgQ} and \eqref{xfcurv}. 

The {\it covariant derivative} of a field $\phi$ is given by the well--known expression 
\begin{equation}
D\phi=d\phi+[\omega,\phi]
\vphantom{\Big]}
\label{xfcovder}
\end{equation}
and 
satisfies the standard Ricci identity 
\begin{equation}
DD\phi=[f,\phi].
\label{xfRicci}
\end{equation}
The covariant derivative of a dual field $\upsilon$ is given similarly by
\begin{equation}
D\upsilon=d\upsilon+[\omega,\upsilon]^\vee,
\label{xdufcovder}
\end{equation}
the Ricci identity being \hphantom{xxxxxxxxxxxxxxxxxxx}
\begin{equation}
DD\upsilon=[f,\upsilon]^\vee
\label{xdufRicci}
\end{equation}
\footnote{$\vphantom{\bigg[}$
Using the canonical duality pairing $\langle\cdot,\cdot\rangle$ of 
$\mathfrak{g}{}^\vee,\mathfrak{g}$, we define the dual brackets in $\mathfrak{g}^\vee$ by
\begin{equation}
\langle[x,\xi]^\vee,z\rangle=-\langle\xi,[x, z]\rangle.
\label{xco2tlinalgb}
\end{equation}
Similarly, we can associate with any
automorphism  $\phi$ of $\mathfrak{g}$ its dual automorphism  $\phi^\vee$ of $\mathfrak{g}^\vee$ 
by 
\begin{equation}
\langle\phi^\vee(\xi),x\rangle=\langle\xi,\phi^{-1}(x)\rangle,
\vphantom{\Big]}
\label{xco3tlinalga}
\end{equation}
\vspace{-1cm}
}.
The covariant derivative preserves the canonical pairing of fields and dual fields
\begin{equation}
d\langle\upsilon,\phi\rangle
=\langle D\upsilon,\phi\rangle+(-1)^{r+s}\langle \upsilon,D\phi\rangle,
\vphantom{\ul{\ul{\ul{\ul{\ul{\ul{\ul{x}}}}}}}}
\vphantom{\bigg]}
\label{xdbltparts}
\end{equation}
if $\upsilon$ has bidegree $(r,s)$.

The Bianchi identity \eqref{xfBianchi} 
obeyed $f$ can be written compactly as 
\begin{equation}
Df=0. 
\label{xcpfBianchi}
\end{equation}
The Bianchi identity, so,  contains information sufficient to recover 
the form of the covariant differentiation operator $D$ 
on fields. Imposing that \eqref{xdbltparts} holds determines the form of $D$ on dual fields. 

The familiar properties of connections recalled above provide us with important clues 
about the definition of $2$--connection appropriate for semistrict higher gauge theory
and suggest how to construct the covariant derivative operator in such context. 
This, we shall do next. Our treatment is actually a particular case of the general formulation of 
\cite{Fiorenza2011,Schreiber2011}. (See subsect. \ref{subsec:schreib} for a further discussion.)

In $2$--term $L_\infty$ algebra gauge theory, as a rule, fields organize in {\it field doublets} 
$(\phi,\varPhi)\in\Omega^m(M,\hat{\mathfrak{v}}_0[n])\times \Omega^{m+1}(M,\hat{\mathfrak{v}}_1[n])$ and 
{\it dual field doublets} 
$(\varUpsilon,\upsilon)\in\Omega^m(M,\hat{\mathfrak{v}}_1{}^\vee[n])\times 
\Omega^{m+1}(M,\hat{\mathfrak{v}}_0{}^\vee[n])$, 
where $-1\leq m\leq d$ (see fn. \ref{foot:hat} for the definition of the hat notation).
If $m=-1$, the first component of the doublet vanishes.
If $m=d$, the second component does. The doublets of this form are said to have bidegree
$(m,n)$.   

There is a distinguished field doublet in the theory, the {\it connection doublet} $(\omega,\varOmega)$
of bidegree $(1,0)$. Associated with it is the {\it curvature doublet} $(f,F)$ of 
bidegree $(2,0)$ defined by the expressions
\begin{subequations}
\label{fFcurv}
\begin{align}
&f=d\omega+\frac{1}{2}[\omega,\omega]-\partial\varOmega,
\vphantom{\Big]}
\label{fcurv}
\\
&F=d\varOmega+[\omega,\varOmega]-\frac{1}{6}[\omega,\omega,\omega].
\vphantom{\Big]}
\label{Fcurv}
\end{align}
\end{subequations}
From \eqref{fFcurv}, it is readily 
verified that $(f,F)$ satisfies 
the {\it Bianchi identities}
\begin{subequations}
\label{fFBianchi}
\begin{align}
&df+[\omega,f]+\partial F=0,
\vphantom{\Big]}
\label{fBianchi}
\\
&dF+[\omega,F]-[f,\varOmega]+\frac{1}{2}[\omega,\omega,f]=0
\vphantom{\Big]}
\label{FBianchi}
\end{align}
\end{subequations}
analogous to  the Bianchi identity \eqref{xfBianchi} of ordinary gauge theory. 

The above definition is justified by the request that the assignment of a 
connection doublet be equivalent to that of a differential graded commutative 
algebra morphism $\mathrm{W}(\mathfrak{v})\rightarrow \Omega^*(M)$ 
from the Weil algebra of $\mathfrak{v}$ (cf. subsect. \ref{subsec:linfweil}) 
to the differential forms of $M$, generalizing the corresponding property of 
connections in ordinary gauge theory. The morphism is the one mapping 
the generators $\pi$, $\varPi$ $\gamma$, $\varGamma$ of $\mathrm{W}(\mathfrak{v})$ respectively 
to $\omega$, $\varOmega$, $f$, $F$ and its differential property is evident from 
the comparison of eqs. \eqref{slinfweil2} with eqs. \eqref{fFcurv}, \eqref{fFBianchi}.

The connection doublet $(\omega,\varOmega)$ is said {\it flat} if the curvature doublet 
$(f,F)=(0,0)$, with an obvious naming.  $(\omega,\varOmega)$ is flat precisely 
when the associated 
morphism $\mathrm{W}(\mathfrak{v})\rightarrow \Omega^*(M)$
factors as $\mathrm{W}(\mathfrak{v})\rightarrow \mathrm{CE}(\mathfrak{v})\rightarrow \Omega^*(M)$,
where $\mathrm{W}(\mathfrak{v})\rightarrow \mathrm{CE}(\mathfrak{v})$ is the canonical 
morphism of the Weil onto the Chevalley--Eilenberg algebra of $\mathfrak{v}$ 
(cf. subsect. \ref{subsec:linfweil}), generalizing again the corresponding 
property of ordinary connections, as it is apparent from inspecting 
eqs. \eqref{2tlinalgQ} and \eqref{fFcurv}. 

Let $(\phi,\varPhi)$ be a field doublet of bidegree $(p,q)$. 
The {\it covariant derivative doublet} of $(\phi,\varPhi)$ is
the field doublet $(D\phi,D\varPhi)$ of bidegree $(p+1,q)$ defined by
\begin{subequations}
\label{fFcovder}
\begin{align}
&D\phi=d\phi+[\omega,\phi]+(-1)^{p+q}\partial\varPhi,
\vphantom{\Big]}
\label{fcovder}
\\
&D\varPhi=d\varPhi+[\omega,\varPhi]-(-1)^{p+q}[\phi,\varOmega]+\frac{(-1)^{p+q}}{2}[\omega,\omega,\phi].
\vphantom{\Big]}
\label{Fcovder}
\end{align}
\end{subequations}
The sign  $(-1)^{p+q}$ is conventional, since the relative sign of $\phi$, $\varPhi$
cannot be fixed in any natural manner. From \eqref{fFcovder}, we deduce easily the
appropriate version of the Ricci identities, 
\begin{subequations}
\label{fFRicci}
\begin{align}
&DD\phi=[f,\phi],
\vphantom{\Big]}
\label{fRicci}
\\
&DD\varPhi=[f,\varPhi]-[\phi,F]-[\phi,\omega,f].
\vphantom{\ul{\ul{\ul{\ul{\ul{x}}}}}}
\vphantom{\Big]}
\label{FRicci}
\end{align}
\end{subequations}
The explicit apparence of the connection component $\omega$ in the right hand side of \eqref{FRicci}
is a consequence of the presence of a term quadratic in $\omega$ in \eqref{Fcovder}.

Let $(\varUpsilon,\upsilon)$ be a dual field doublet of bidegree $(r,s)$. The {\it covariant derivative
dual doublet}
of $(\varUpsilon,\upsilon)$ is the dual field doublet $(D\varUpsilon,D\upsilon)$ of bidegree $(r+1,s)$ 
defined by
\begin{subequations}
\label{dufFcovder}
\begin{align}
&D\varUpsilon=d\varUpsilon+[\omega,\varUpsilon]^\vee-(-1)^{r+s}\partial^\vee\upsilon,
\vphantom{\Big]}
\label{dufcovder}
\\
&D\upsilon=d\upsilon+[\omega,\upsilon]^\vee-(-1)^{r+s}[\varOmega,\varUpsilon]^\vee
-\frac{(-1)^{r+s}}{2}[\omega,\omega,\varUpsilon]^\vee,
\vphantom{\Big]}
\label{duFcovder}
\end{align}
\end{subequations}
analogously to \eqref{fFcovder}
\footnote{$\vphantom{\bigg[}$
Using the canonical duality pairing $\langle\cdot,\cdot\rangle$ of 
$\hat{\mathfrak{v}}_0{}^\vee,\hat{\mathfrak{v}}_0$ and $\hat{\mathfrak{v}}_1{}^\vee,\hat{\mathfrak{v}}_1$, 
we obtain a canonical {\it $2$--term $L_\infty$ algebra costructure}.
This consists of the linear maps 
$\partial^\vee:\hat{\mathfrak{v}}_0{}^\vee\rightarrow\hat{\mathfrak{v}}_1{}^\vee$,
$[\cdot,\cdot]^\vee:\hat{\mathfrak{v}}_0\otimes \hat{\mathfrak{v}}_0{}^\vee
\rightarrow \hat{\mathfrak{v}}_0{}^\vee$, \
$[\cdot,\cdot]^\vee:\hat{\mathfrak{v}}_0\otimes \hat{\mathfrak{v}}_1{}^\vee
\rightarrow \hat{\mathfrak{v}}_1{}^\vee$, \
$[\cdot,\cdot]^\vee:\hat{\mathfrak{v}}_1\otimes \hat{\mathfrak{v}}_1{}^\vee
\rightarrow \hat{\mathfrak{v}}_0{}^\vee$, \
$[\cdot,\cdot,\cdot]^\vee:(\hat{\mathfrak{v}}_0\wedge \hat{\mathfrak{v}}_0)\otimes \hat{\mathfrak{v}}_1{}^\vee
\rightarrow \hat{\mathfrak{v}}_0{}^\vee$ $\vphantom{\Big[}$ 
defined by 
\begin{subequations}
\label{co2tlinalg}
\begin{align}
&\langle\partial^\vee\xi,X\rangle=\langle\xi,\partial X\rangle, 
\vphantom{\Big]}
\label{co2tlinalga}
\\
&\langle[x,\xi]^\vee,z\rangle=-\langle\xi,[x, z]\rangle,
\vphantom{\Big]}
\label{co2tlinalgb}
\\
&\langle[x,\varXi]^\vee,Z\rangle=-\langle\varXi,[x,Z]\rangle,
\vphantom{\Big]}
\label{co2tlinalgc}
\\
&\langle[X,\varXi]^\vee,z\rangle=+\langle\varXi,[z,X]\rangle, 
\vphantom{\Big]}
\label{co2tlinalgd}
\\
&\langle[x,y,\varXi]^\vee,z\rangle=-\langle\varXi,[x,y,z]\rangle,
\vphantom{\Big]}
\label{co2tlinalge}
\end{align}
\end{subequations}
where 
$\xi\in\hat{\mathfrak{v}}_0{}^\vee$, $\varXi\in\hat{\mathfrak{v}}_1{}^\vee$ 
Again, we use the notation $[\cdot,\cdot]^\vee$ for all 
$2$--argument cobrackets. 

Using the duality pairing again, we can associate with any
automorphism  $\phi=(\phi_0,\phi_1,\phi_2)$ of $\mathfrak{v}$ its
{\it dual automorphism} of $\phi$. This consists of the linear maps 
$\phi^\vee{}_0 :\hat{\mathfrak{v}}_0{}^\vee\rightarrow\hat{\mathfrak{v}}_0{}^\vee$,
$\phi^\vee{}_1:\hat{\mathfrak{v}}_1{}^\vee\rightarrow\hat{\mathfrak{v}}_1{}^\vee$,
$\phi^\vee{}_2:\hat{\mathfrak{v}}_0\otimes \hat{\mathfrak{v}}_1{}^\vee\rightarrow\hat{\mathfrak{v}}_0{}^\vee$ 
defined by the relations 
\begin{subequations}
\label{co3tlinalg}
\begin{align}
&\langle\phi^\vee{}_0(\xi),x\rangle=\langle\xi,\phi^{-1}{}_0(x)\rangle,
\vphantom{\Big]}
\label{co3tlinalga}
\\
&\langle\phi^\vee{}_1(\Xi),X\rangle=\langle\Xi,\phi^{-1}{}_1(X)\rangle,
\vphantom{\Big]}
\label{co3tlinalgb}
\\
&\langle\phi^\vee{}_2(x,\Xi),y\rangle=\langle\Xi,\phi^{-1}{}_2(x,y)\rangle.
\vphantom{\Big]}
\label{co3tlinalgc}
\end{align}
\end{subequations}
We shall denote the dual of $\phi$ by 
$\phi^\vee$ or $(\phi^\vee{}_0,\phi^\vee{}_1,\phi^\vee{}_2)$.}.  
The Ricci identities then read as
\begin{subequations}
\label{dufFRicci}
\begin{align}
&DD\varUpsilon=[f,\varUpsilon]^\vee,
\vphantom{\Big]}
\label{dufRicci}
\\
&DD\upsilon=[f,\upsilon]^\vee-(-1)^{r+s}[F,\varUpsilon]^\vee-(-1)^{r+s}[f,\omega,\varUpsilon]^\vee,
\vphantom{\ul{\ul{\ul{\ul{\ul{\ul{x}}}}}}}
\vphantom{\Big]}
\label{duFRicci}
\end{align}
\end{subequations} 
analogously to \eqref{fFRicci}. 

There exists a natural pairing of field and$\vphantom{\ul{\ul{x}}}$  dual field doublets.
The pairing of a field doublet
$(\phi,\varPhi)$ of bidegree $(p,q)$ and a dual field doublet$\vphantom{\ul{\ul{x}}}$  
$(\varUpsilon,\upsilon)$ of bidegree
$(r,s)$ is a the scalar valued field of bidegree $(p+r+1,q+s)$ given by 
\begin{equation}
\langle (\varUpsilon,\upsilon),(\phi,\varPhi)\rangle=\langle\upsilon,\phi\rangle
-(-1)^{p+q}\langle\varUpsilon,\varPhi\rangle.
\label{dbltpair}
\end{equation}
The basic property of the pairing is that
\begin{equation}
d\langle (\varUpsilon,\upsilon),(\phi,\varPhi)\rangle
=\langle (D\varUpsilon,D\upsilon),(\phi,\varPhi)\rangle
-(-1)^{r+s}\langle (\varUpsilon,\upsilon),(D\phi,D\varPhi)\rangle.
\vphantom{\bigg]}
\label{dbltparts}
\end{equation}
By the Stokes' theorem, upon integration on $M$, this relation yields an integration by parts
formula for the covariant derivative of (dual) field doublets. 

The above definition of covariant differentiation is yielded 
by the request that the Bianchi identities \eqref{fFBianchi} 
be expressed as the vanishing of the covariant derivative
doublet $(Df,DF)$ of the curvature doublet $(f,F)$ 
\begin{subequations}
\label{cpfFBianchi}
\begin{align}
&Df=0,
\vphantom{\Big]}
\label{cpfBianchi}
\\
&DF=0
\vphantom{\Big]}
\label{cpFBianchi}
\end{align}
\end{subequations} 
as it is the case for the Bianchi identity of ordinary gauge theory, 
eq. \eqref{xcpfBianchi}.
Imposing that \eqref{dbltparts} holds determines the the action of $D$ on dual fields.

\subsection{\normalsize \textcolor{blue}{The $2$--group of $2$--term $L_\infty$ algebra gauge transformations}}
\label{subsec:linfgautrsf}

\hspace{.5cm} One expects that there is a notion of 
semistrict higher gauge transformation generalizing the corresponding notion of ordinary gauge theory
and that this plays an important role in 
semistrict higher gauge theory. This is indeed so. 

In fact, $2$--term $L_\infty$ algebra gauge transformations can be meaningfully defined. 
Further, they organize in an infinite dimensional strict $2$--group described below
(cf. subsect. \ref{sec:twogr}). We assume again that $M$ is diffeomorphic 
to $\mathbb{R}^d$. 

The definition of $2$--term $L_\infty$ algebra $1$--gauge transformation
given below is not straightforward and needs to be justified. 
To this end, we begin by considering an ordinary gauge theory
with structure Lie algebra $\mathfrak{g}$.  
A gauge transformation is a map $g\in\Map(M,\Aut(\mathfrak{g}))$
of a special form: its range consists of inner automorphisms of 
$\mathfrak{g}$. So, letting $G$ be a Lie group integrating $\mathfrak{g}$, 
there is a map $\gamma\in \Map(M,G)$ 
such that $g(x)=\Ad\gamma(x)=\gamma x\gamma^{-1}$. 
When we try to define a gauge transformation in a $2$--term $L_\infty$ algebra gauge theory with  
structure algebra $\mathfrak{v}$ following the same line, we soon run into trouble,
as there is no natural notion of inner automorphism of $\mathfrak{v}$. 

We circumvent this difficulty as follows. We note that $\sigma_g=\gamma^{-1}d\gamma$
is a flat connection such that $dg(x)=g([\sigma_g,x])$. 
Thus, we may extend the notion of gauge transformation by defining it as a pair of
\begin{enumerate}

\item a map $g\in\Map(M,\Aut(\mathfrak{g}))$,

\item a flat connection $\sigma_g$, \hphantom{xxxxxxxxxxxxxxxxxx}
\begin{equation}
d\sigma_g+\frac{1}{2}[\sigma_g,\sigma_g]=0,
\label{x1linfdgloba}
\end{equation}

\item related to $g$ through the condition \hphantom{xxxxxxxxxxxxxxxxxx}
\begin{equation}
g^{-1}dg(x)-[\sigma_g,x]=0.
\label{x1linfdglobb}
\end{equation}
\end{enumerate}
We shall denote the gauge transformation by $(g,\sigma_g)$ or simply by $g$, 
having in mind that now $\sigma_g$ is not determined by $g$ but participates with $g$
in the transformation. Further, we shall denote by $\Gau(M,\mathfrak{g})$ the set of 
all such extended gauge transformations.

Albeit not all $g\in\Gau(M,\mathfrak{g})$ correspond to conventional gauge transformations
\footnote{$\vphantom{\bigg[}$ 
If $g\in\Map(M,\Aut(\mathfrak{g}))$ has the property that there is a flat connection
$\sigma_g$ such that $g^{-1}dg(x)=[\sigma_g,x]$ for $x\in \mathfrak{g}$, 
$\vphantom{\dot{\dot{\dot{\dot{x}}}}}$then there is 
a map $\gamma\in \Map(M,G)$ and a constant $g_0\in\Aut(\mathfrak{g})$ such that 
$g=g_0\Ad\gamma$. In general, $g_0\not=1_{\mathfrak{g}}$. 
Thus, the range of $g$ does not necessarily consist of 
inner automorphisms of $\mathfrak{g}$.}$\vphantom{\ul{\ul{x}}}$, 
the above observation
provides clues which indicate the direction along which to construct
the generalization of the notion of  gauge transformation appropriate for 
$2$--term $L_\infty$ algebra gauge theory. 

A {\it $2$--term $L_\infty$ algebra $1$--gauge transformation} consists of the following set of data.
\begin{enumerate}

\item a map $g\in\Map(M,\Aut_1(\mathfrak{v}))$ (cf. sect. \ref{sec:linftyauto});

\item a flat connection doublet $(\sigma_g,\varSigma_g)$, 
\begin{subequations}
\label{1linfdglob}
\begin{align}
&d\sigma_g+\frac{1}{2}[\sigma_g,\sigma_g]-\partial\varSigma_g=0,
\vphantom{\Big]}
\label{1linfdgloba}
\\
&d\varSigma_g+[\sigma_g,\varSigma_g]-\frac{1}{6}[\sigma_g,\sigma_g,\sigma_g]=0;
\vphantom{\Big]}
\label{1linfdglobb}
\end{align}
\end{subequations}

\item an element $\tau_g$ of $\Omega^1(M,\Hom(\hat{\mathfrak{v}}_0,\hat{\mathfrak{v}}_1))$  
satisfying 
\begin{align}
&d\tau_g(x)+[\sigma_g,\tau_g(x)]-[x,\varSigma_g]+\frac{1}{2}[\sigma_g,\sigma_g,x]
\vphantom{\Big]}
\label{2linfdglob}
\\
&\qquad\qquad\qquad\qquad\qquad\qquad
+\tau_g([\sigma_g,x]+\partial\tau_g(x))=0. \vphantom{\ul{\ul{\ul{\ul{\ul{\ul{x}}}}}}}
\vphantom{\Big]}
\nonumber
\end{align}
\end{enumerate}
These data are required to satisfy the following relations. 
If $g=(g_0,g_1,g_2)$ (cf. sect. \ref{sec:linftymorph}), then one has $\vphantom{\ul{\ul{x}}}$ 
\begin{subequations}
\label{3linfdglob}
\begin{align}
&g_0{}^{-1}dg_0(x)-[\sigma_g,x]-\partial\tau_g(x)=0,
\vphantom{\Big]}
\label{3linfdgloba}
\\
&g_1{}^{-1}dg_1(X)-[\sigma_g,X]-\tau_g(\partial X)=0,
\vphantom{\Big]} 
\label{3linfdglobb}
\\
&g_1{}^{-1}(dg_2(x,y)-g_2(g_0{}^{-1}dg_0(x),y)-g_2(x,g_0{}^{-1}dg_0(y)))
\vphantom{\Big]}
\label{3linfdglobc}
\\
&\qquad\qquad -[\sigma_g,x,y]-\tau_g([x,y])+[x,\tau_g(y)]-[y,\tau_g(x)]=0
\vphantom{\Big]}
\nonumber
\end{align}
\end{subequations}
hold. 
In the following, we are going to denote a $2$--term $L_\infty$ algebra $1$--gauge transformation
such as the above as $(g,\sigma_g,\varSigma_g,\tau_g)$ or simply as $g$.
We remark that, in so doing, we are not implying that $\sigma_g$, $\varSigma_g$, $\tau_g$
are determined by $g$, but only that they are the partners of $g$ in the 
gauge transformation. We shall denote the set of all $2$--term $L_\infty$ algebra $1$--gauge 
transformations by $\Gau_1(M,\mathfrak{v})$. 

The remarks made at the beginning of this subsection already justify to a considerable extent the 
definition of $2$--term $L_\infty$ algebra $1$--gauge transformation given above. When the Lie algebra $\mathfrak{g}$ 
gets replaced by a more general $2$--term $L_\infty$ algebra $\mathfrak{v}$, the flat
connection $\sigma_g$ gets promoted to a flat connection doublet $(\sigma_g,\varSigma_g)$ as is appropriate. 
We obtain in this way eqs. \eqref{1linfdglob}. The point is that this is not sufficient to fully explain the form
of relations \eqref{3linfdglob}, for reasons explained next. 

For the ordinary gauge transformation considered above, in order the Maurer--Cartan equation
$d(g^{-1}dg)+g^{-1}dgg^{-1}dg=0$ to be satisfied, it is sufficient that $\sigma_g$ 
is flat. The proof of this requires crucially the use of the Jacobi identity of the Lie algebra 
$\mathfrak{g}$. When $\mathfrak{g}$ is replaced by a 
$2$--term $L_\infty$ algebra $\mathfrak{v}$, that identity is no longer available.
This forces one to introduce another object, namely $\tau_g$, and modify the naive 
relations $g_0^{-1}dg_0(x)=[\sigma_g,x]$, $g_1^{-1}dg_1(X)=[\sigma_g,X]$,  
as shown in \eqref{3linfdgloba}, \eqref{3linfdglobb}.
If $\tau_g$ vanished, for the the Maurer--Cartan equations
$d(g_0{}^{-1}dg_0)+g_0{}^{-1}dg_0g_0{}^{-1}dg_0=0$,
$d(g_1{}^{-1}dg_1)+g_1{}^{-1}dg_1g_1{}^{-1}dg_1=0$ 
to be satisfied, the flatness relations \eqref{1linfdglob} would not be sufficiently by themselves:
one would need an extra condition, namely
$-[x,\varSigma_g]+\frac{1}{2}[\sigma_g,\sigma_g,x]=0$. 
This latter, a purely algebraic requirement on the flat connection doublet $(\sigma_g,\varSigma_g)$, 
does not fit into our higher gauge theoretic set-up in any  natural way, 
and, so, it is hardly acceptable. 
Once we allow for $\tau_g$, however, this condition takes the natural form of
a differential consistency relation satisfied by $\tau_g$, viz \eqref{2linfdglob}. The reasoning just 
expounded justifies calling \eqref{2linfdglob} ``$2$--Maurer--Cartan equation''. 
As to relation \eqref{3linfdglobc}, it is just a natural coherence condition ensuring the 
compatibility of \eqref{3linfdgloba}, \eqref{3linfdglobb} and 
\eqref{mor2tlinalg}. 

For any two 
$1$--gauge transformations $g,h\in\Gau_1(M,\mathfrak{v})$,
a {\it $2$--term $L_\infty$ algebra $2$--gauge transformation from $g$ to $h$} consists of the following data.
\begin{enumerate}

\item a map $F\in\Map(M,\Aut_2(\mathfrak{v}))(g,h)$, where $\Map(M,\Aut_2(\mathfrak{v}))(g,h)$ is the space of 
sections of the fiber bundle $\bigcup_{m\in M}\Aut_2(\mathfrak{v})(g(m),h(m))\rightarrow M$ 
(cf. sect. \ref{sec:linftyauto});

\item an element $A_F\in\Omega^1(M,\hat{\mathfrak{v}}_1)$.

\end{enumerate}
They are required to satisfy the following relations, 
\begin{subequations}
\label{0linfdglob}
\begin{align}
&\sigma_g-\sigma_h=\partial A_F, 
\vphantom{\Big]}
\label{0linfdgloba}
\\
&\varSigma_g-\varSigma_h=dA_F+[\sigma_h,A_F]+\frac{1}{2}[\partial A_F,A_F], \hspace{2.6cm}
\vphantom{\Big]}
\label{0linfdglobb}
\\
&\tau_g(x)-\tau_h(x)=[x,A_F]+g_1{}^{-1}\big(dF(x)-F([\sigma_h,x]+\partial\tau_h(x))\big).
\vphantom{\Big]}
\label{0linfdglobc}
\end{align}
\end{subequations}
In the following, we are going to denote a $2$--term $L_\infty$ algebra $2$--gauge transformation
such as the above as $(F,A_F)$ or simply as $F$. Again, in so doing, we are not implying 
that $A_F$ is determined by $F$, but only that it is the partner of $F$ in the 
gauge transformation. We shall also write $F:g\Rightarrow h$ to emphasize its source and target. 
We shall denote the set of all $2$--term $L_\infty$ algebra $2$--gauge transformations 
$F:g\Rightarrow h$  by $\Gau_2(M,\mathfrak{v})(g,h)$ 
and the set of all $2$--gauge transformations $F$  by $\Gau_2(M,\mathfrak{v})$.

To justify the above definition of $2$--term $L_\infty$ algebra $2$--gauge transformation, 
the following remarks are in order.
Suppose that $(g,\sigma_g,\varSigma_g,\tau_g)$ is a 
$2$--term $L_\infty$ algebra $1$--gauge transformation. Let us ask what the most natural 
class of deformations of $(g,\sigma_g,\varSigma_g,\tau_g)$ which preserve 
its being a $1$--gauge transformation and which can be  parametrized in terms of elementary fields is. 
As $g,h\in\Map(M,\Aut_1(\mathfrak{v}))$, it is natural to demand that $g,h$ are the source and the target
of some $F\in\Map(M,\Aut_2(\mathfrak{v}))(g,h)$. Once this is done, the only remaining
deformational degree of freedom is an element $A\in\Omega^1(M,\hat{\mathfrak{v}}_1)$ turning
$\sigma_g$ into $\sigma_h=\sigma_g-\partial A$. We require $A$ to be $\hat{\mathfrak{v}}_1$--valued
in order it to be utilizable to deform $\varSigma_g$ into 
$\varSigma_h=\varSigma_g-dA+\frac{1}{2}[\partial A,A]+\cdots$
and $\tau_g(x)$ into $\tau_h(x)=\tau_g(x)-[x,A]+\cdots$. 
Requiring that $(h,\sigma_h,\varSigma_h,\tau_h)$ is a $1$--gauge transformation 
fixes the form of the terms not shown. 

In ordinary gauge theory, gauge transformations form a group, the guage group of the gauge theory. 
This remains true also for the more general gauge transformations, 
which we have defined at the beginning of this subsection
(cf. eqs. \eqref{x1linfdgloba}, \eqref{x1linfdglobb}). Define
\begin{subequations}
\label{x4linfdglob}
\begin{align}
&h\diamond g=hg, 
\vphantom{\Big]}
\label{x4linfdglobz}
\\
&\sigma_{h\,\diamond \,g}
=\sigma_g+ g^{-1}(\sigma_h),
\vphantom{\Big]}
\label{x4linfdgloba}
\\
&g^{-1_\diamond}=g^{-1},
\vphantom{\Big]}
\label{x6linfdglobz}
\\
&\sigma_{g^{-1_\diamond}}=-g(\sigma_g),
\vphantom{\Big]}
\label{x6linfdgloba}
\\
&i=\id_{\mathfrak{g}},
\vphantom{\Big]}
\label{x5linfdglobz}
\\
&\sigma_i=0,
\vphantom{\Big]}
\label{x5linfdgloba}
\end{align}
\end{subequations}
where $g,h\in\Gau(M,\mathfrak{g})$ and, in \eqref{x4linfdglobz}, \eqref{x6linfdglobz}, \eqref{x5linfdglobz},
the composition, inversion and unit in the right hand side are those of $\Aut(\mathfrak{g})$
thought of as holding pointwise on $M$. Then, as it is immediately checked, $\Gau(M,\mathfrak{g})$ is an ordinary 
group, the extended gauge group of the theory. 
Inspection of \eqref{x4linfdglob} shows that
$\Gau(M,\mathfrak{g})$ is a subgroup of the semidirect product $\Omega^1(M,\mathfrak{g})\rtimes \Map(M,\Aut(\mathfrak{g}))$
associated with the right action of $\Map(M,\Aut(\mathfrak{g}))$ on $\Omega^1(M,\mathfrak{g})$
induced by the right action of $\Aut(\mathfrak{g})$ on $\mathfrak{g}$. 
$\Gau(M,\mathfrak{g})$ is a proper subgroup, when $M$ is not a point, 
because its elements satisfy the additional differential relations 
\eqref{x1linfdgloba}, \eqref{x1linfdglobb}.

The property which we have 
found to hold in ordinary gauge theory generalizes
to $2$--term $L_\infty$ algebra gauge theory. 
Indeed, as we show below in detail, it is possible to define a composition and an inversion law 
and a unit in $\Gau_1(M,\mathfrak{v})$  and horizontal and vertical composition and inversion laws
and units in $\Gau_2(M,\mathfrak{v})$, making $\Gau(M,\mathfrak{v})
=(\Gau_1(M,\mathfrak{v}),\Gau_2(M,\mathfrak{v}))$  
a strict $2$--group (cf. subsect. \ref{sec:twogr}). 

The composition and inversion laws and the unit of  $1$--gauge transformation are defined by 
the relations 
\begin{subequations}
\label{4linfdglob}
\begin{align}
&h\diamond g=h\circ g, 
\vphantom{\Big]}
\label{4linfdglobz}
\\
&\sigma_{h\,\diamond \,g}
=\sigma_g+ g_0{}^{-1}(\sigma_h),
\vphantom{\Big]}
\label{4linfdgloba}
\\
&\varSigma_{h\,\diamond \,g}
=\varSigma_g+ g_1{}^{-1}\Big(\varSigma_h
+\frac{1}{2} g_2(g_0{}^{-1}(\sigma_h),g_0{}^{-1}(\sigma_h))\Big)-\tau_g(g_0{}^{-1}(\sigma_h)),
\vphantom{\Big]}
\label{4linfdglobb}
\\
&\tau_{h\,\diamond \,g}(x)
=\tau_g(x)+ g_1{}^{-1}\big(\tau_h(g_0(x))-g_2(g_0{}^{-1}(\sigma_h),x)\big),
\vphantom{\Big]}
\label{4linfdglobc}
\\
&g^{-1_\diamond}=g^{-1_\circ},
\vphantom{\Big]}
\label{6linfdglobz}
\\
&\sigma_{g^{-1_\diamond}}=-g_0(\sigma_g),
\vphantom{\Big]}
\label{6linfdgloba}
\\
&\varSigma_{g^{-1_\diamond}}=- g_1(\varSigma_g+\tau_g(\sigma_g))-\frac{1}{2} g_2(\sigma_g,\sigma_g),
\label{6linfdglobb}
\\
&\tau_{g^{-1_\diamond}}(x)=- g_1(\tau_g( g_0{}^{-1}(x)))- g_2(\sigma_g, g_0{}^{-1}(x)),
\vphantom{\Big]}
\label{6linfdglobc}
\\
&i=\id_{\mathfrak{v}},
\vphantom{\Big]}
\label{5linfdglobz}
\\
&\sigma_i=0,
\vphantom{\Big]}
\label{5linfdgloba}
\\
&\varSigma_i=0,
\vphantom{\Big]}
\label{5linfdglobb}
\\
&\tau_i(x)=0, \vphantom{\ul{\ul{\ul{\ul{\ul{\ul{x}}}}}}}
\vphantom{\Big]}
\label{5linfdglobc}
\end{align}
\end{subequations}
where $g,h\in\Gau_1(M,\mathfrak{v})$. 
In \eqref{4linfdglobz}, \eqref{6linfdglobz}, \eqref{5linfdglobz},
the composition, inversion and unit in the right hand side are those of $\Aut_1(\mathfrak{v})$
thought of as holding pointwise on $M$ (cf. eqs. \eqref{mor3tlinalg}, \eqref{mor3/1tlinalg},
\eqref{mor3/2tlinalg}). 

The horizontal and vertical composition and inversion laws and the units of $2$--gauge
transformations are defined by the relations 
\begin{subequations}
\label{50linfdglob}
\begin{align}
&G\diamond F(x)=G\circ F(x),   
\vphantom{\Big]}
\label{50linfdgloba}
\\
&A_{G\,\diamond\, F}=A_F+h^{-1}{}_1(A_G)-g_1{}^{-1}Fh_0{}^{-1}(\sigma_k),
\vphantom{\Big]}
\label{50linfdglobb}
\end{align}
\begin{align}
&F^{-1_\diamond}(x)=F^{-1_\circ}(x),
\vphantom{\dot{\dot{\dot{\dot{\dot{\dot{x}}}}}}}
\vphantom{\Big]}
\label{50linfdglobc}
\\
&A_{F^{-1_\diamond }}=-g_1(A_F)-F(\sigma_h),
\vphantom{\Big]}
\label{50linfdglobd}
\\
&K\bullet H(x)=K\bfdot H(x),
\vphantom{\Big]}
\label{50linfdglobe}
\\
&A_{K\,\bullet\, H}=A_H+A_K, \hspace{3.5cm}
\vphantom{\Big]}
\label{50linfdglobf}
\\
&H^{-1_\bullet}(x)=H^{-1_\bfdot}(x),
\vphantom{\Big]}
\label{50linfdglobg}
\\
&A_{H^{-1_\bullet}}=-A_H,
\vphantom{\Big]}
\label{50linfdglobh}
\\
&I_g(x)=\mathrm{Id}_g(x), 
\vphantom{\Big]}
\label{50linfdglobi}
\\
&A_{I_g}=0,
\vphantom{\Big]}
\label{50linfdglobj}
\end{align}
\end{subequations}  
where $g,h,k,l\in\Gau_1(M,\mathfrak{v})$ and $F,G,H,K\in\Gau_2(M,\mathfrak{v})$, 
with $F:g\Rightarrow h$, $G:k\Rightarrow l$ and $H,K$ composible. 
In \eqref{50linfdgloba}, \eqref{50linfdglobc}, \eqref{50linfdglobe}, \eqref{50linfdglobg},
\eqref{50linfdglobi},
the horizontal and vertical composition and inversion and the units in the right hand side 
are those of $\Aut_2(\mathfrak{v})$ thought of as holding pointwise on $M$ (cf. eqs. 
\eqref{mor4tlinalga}--\eqref{mor4tlinalgc}, \eqref{mor4/1tlinalgb}, \eqref{mor4/1tlinalgd}).
The expressions can be written in several other equivalent ways
using repeatedly relations \eqref{mor0tlinalg} with $\phi,\psi,\varPhi$ replaced by
$g,h,F$ or $k,l,G$. 

The composition, inversion and unit structures
just defined satisfy the axioms \eqref{twogr1}
rendering $(\Gau_1(M,\mathfrak{v})$, $\Gau_2(M,\mathfrak{v}))$  
a strict $2$--group, as announced.
From \eqref{4linfdglob}, it appears that the $1$--cell group
$\Gau_1(M,\mathfrak{v})$ is a subgroup of the semidirect product $\Omega^1(M,\hat{\mathfrak{v}}_0)\oplus
\Omega^2(M,\hat{\mathfrak{v}}_1)\oplus \Omega^1(M,\Hom(\hat{\mathfrak{v}}_0,\hat{\mathfrak{v}}_1))
\rtimes \Map(M,\Aut_1(\mathfrak{v}))$ associated with 
a certain right action of $\Map(M,\Aut_1(\mathfrak{v}))$ on $\Omega^1(M,\hat{\mathfrak{v}}_0)\oplus
\Omega^2(M,\hat{\mathfrak{v}}_1)\oplus \Omega^1(M,\Hom(\hat{\mathfrak{v}}_0,\hat{\mathfrak{v}}_1))$. 
$\Gau_1(M,\mathfrak{v})$ is a proper subgroup, when $M$ is not a point, 
as its elements satisfy the additional differential relations 
\eqref{1linfdglob}--\eqref{3linfdglob}. These findings suggest that
the full $2$--group $\Gau(M,\mathfrak{v})$ may be a $2$--subgroup of 
a conjectural semidirect product $2$--group defined by the relations \eqref{4linfdglob}, \eqref{50linfdglob}.
We shall not elaborate on this point any further.


\subsection{\normalsize \textcolor{blue}{The gauge transformation action}}
\label{subsec:gauact}

\hspace{.5cm} 
An important test of the viability of the definition of 
semistrict higher gauge 
transformation we have worked out in subsect. \ref{subsec:linfgautrsf} is the existence of a suitable 
gauge transformation action on fields. In principle, several prescriptions are possible 
and a full inspection of all options is out of question. 
To select the appropriate definition, we proceed once more from standard gauge
theory. Here, we assume again that $M$ is diffeomorphic to $\mathbb{R}^d$. 

Consider an ordinary gauge theory with structure Lie algebra $\mathfrak{g}$.
Conventionally, a gauge transformation is a mapping $\gamma\in\Map(M,G)$,
where $G$ is a Lie group integrating $\mathfrak{g}$, acting on a connection 
$\omega$ by ${}^\gamma\omega=\gamma\omega\gamma^{-1}-d\gamma\gamma^{-1}$ and 
on its curvature $f$ by ${}^\gamma f=\gamma f\gamma^{-1}$. As argued in subsect.
\ref{subsec:linfgautrsf}, when aiming to construct higher generalizations,
it is useful to extend the range of gauge transformations to all 
$g\in\Gau(M,\mathfrak{g})$. 
Noticing that, 
in the usual case just considered, 
$(g,\sigma_g)=(\Ad\gamma,\gamma^{-1}d\gamma)$, we realize immediately that   
the gauge transform by $g$ of the connection $\omega$ and its curvature $f$ must have the form 
\begin{equation}
{}^g\omega=g(\omega-\sigma_g)
\label{xgauact1}
\end{equation}
and \hphantom{xxxxxxxxxxxxxxxxxxxxxxxxxxxxxxxxxxxxxxxxxxxxxxxxxxxxx}
\begin{equation}
{}^gf=g(f).
\label{xgauact2}
\end{equation}

If $\phi$ is a field, the gauge transform of $\phi$ under a standard gauge transformation $\gamma\in\Map(M,G)$
is given by ${}^\gamma\phi=\gamma\phi\gamma^{-1}$ and that of its covariant derivative $D\phi$
by ${}^\gamma D\phi=\gamma D\phi\gamma^{-1}$, as is well--known. These relations generalize immediately to 
any gauge transformation $g\in\Gau(M,\mathfrak{g})$, yielding
\begin{equation}
{}^g\phi=g(\phi)
\label{xgauact3}
\end{equation}
and likewise \hphantom{xxxxxxxxxxxxxxxxxxx}
\begin{equation}
{}^gD\phi=g(D\phi)
\label{xgauact4}
\end{equation}
\footnote{$\vphantom{\bigg[}$ For any field $\mathcal{F}$,  
${}^gD\mathcal{F}\equiv{}^g(D\mathcal{F})$ is defined by 
replacing each occurrence of each field $\mathcal{G}$ in the expression of 
$D\mathcal{F}$ by ${}^g\mathcal{G}$.\label{foot:gDf}}.
For a dual field $\upsilon$, we have similarly \hphantom{xxxxxxxxxxxxxxxxxxx}
\begin{equation}
{}^g\upsilon=g^\vee(\upsilon) 
\vphantom{\dot{\dot{\dot{\dot{\dot{x}}}}}}
\label{xgauact5}
\end{equation}
and \hphantom{xxxxxxxxxxxxxxxxxxxxxxxxxxxxxxxxxxxxxxxxxxxxxxx}
\begin{equation}
{}^gD\upsilon=g^\vee(D\upsilon). 
\label{xgauact6}
\end{equation}

Gauge transformation of connections and (dual) fields constitutes a left action of the 
gauge transformation group $\Gau(M,\mathfrak{g})$ on fields, that is, 
for any two gauge transformations $g,h\in \Gau(M,\mathfrak{g})$, 
${}^{g\diamond h}\mathcal{F}={}^g({}^h\mathcal{F})$,
where $\mathcal{F}=\omega$, $\phi$, $\upsilon$, $f$, $D\phi$, $D\upsilon$ 
is anyone of the fields considered above.

Next, we shift to $2$--term $L_\infty$ algebra gauge theory 
and look for a sensible definition of gauge transformation action 
in this higher context extending without trivializing the ordinary
gauge transformation action as formulated above. 
But, before doing that, a preliminary issue must be settled. 
$2$--term $L_\infty$ gauge 
transformations form a strict $2$--group $\Gau(M,\mathfrak{v})$ 
comprising $1$-- and $2$--gauge transformations (cf. subsect. \ref{subsec:linfgautrsf}). 
The natural question arises  
about whether one or both types of gauge transformations act on fields. 
From our analysis of $\Gau(M,\mathfrak{v})$, it emerges that 
it is the $1$--gauge transformations which answer to 
the gauge transformations of ordinary gauge theory, while $2$--gauge transformations 
constitute what may be called gauge 
for gauge transformations. 
It is thus natural to assume that only $1$--gauge transformations 
act effectively on fields. So, we shall restrict to the $1$--cell set 
$\Gau_1(M,\mathfrak{v})$ of $\Gau(M,\mathfrak{v})$, 
which, we recall, is an ordinary group. 
The role of the $2$--cell set $\Gau_2(M,\mathfrak{v})$ of $\Gau(M,\mathfrak{v})$
will become clear in the analysis of the global properties of the theory. 

In ordinary gauge theory, covariant differentiation is gauge covariant. 
This is often stated by saying$\vphantom{\ul{\ul{x}}}$  that for any gauge
transformation $g$ and any field $\mathcal{F}$,  
${}^gD\mathcal{F}$ has the same form as ${}^g\mathcal{F}$.
An equivalent way of telling this 
is that ${}^gD\mathcal{F}$ is given by formal covariant differentiation
of ${}^g\mathcal{F}$ treating $g$ and $\sigma_g$ as if they were formally covariantly constant,
a property ultimately guaranteed by \eqref{x1linfdgloba}, \eqref{x1linfdglobb}. 
It is reasonable to require that covariant differentiation
in $2$--term $L_\infty$ algebra gauge theory has 
the same basic feature. So, the condition determining the form of
${}^g\mathcal{F}$  in $2$--term $L_\infty$ algebra gauge theory
is that, for any gauge transformation $g$, 
${}^gD\mathcal{F}$ is given by formal covariant differentiation of ${}^g\mathcal{F}$ treating 
the components $g_0$, $g_1$, $g_2$ of $g$ as well as 
$\sigma_g$, $\varSigma_g$, $\tau_g$ as if they were formally covariantly constant.
Eventually, 
this property will ensue from the basic identities  \eqref{1linfdglob}--\eqref{3linfdglob}. 
Proceeding in this way, we obtain the expressions of the gauge transforms 
of the various types of fields reported below. 
As it turns out, the fact that  
covariant differentiation mixes 
fields of the same doublet 
and depends on an assigned connection doublet makes 
${}^g\mathcal{F}$ depend in general 
on $\mathcal{F}$ and its doublet partner as well as  
the connection doublet, a property which has no analog in ordinary gauge theory.

Let $g\in\Gau_1(M,\mathfrak{v})$ be a $1$--gauge transformation.

We consider first a connection doublet $(\omega,\varOmega)$. 
The gauge transformed connection doublet $({}^g\omega,{}^g\varOmega)$ is defined to be
\begin{subequations}
\label{7linfdglob}
\begin{align}
&{}^g\omega=g_0(\omega-\sigma_g), 
\vphantom{\Big]}
\label{7linfdgloba}
\\
&{}^g\varOmega=g_1(\varOmega-\varSigma_g+\tau_g(\omega-\sigma_g))-\frac{1}{2}g_2(\omega-\sigma_g,\omega-\sigma_g).
\vphantom{\Big]}
\label{7linfdglobb}
\end{align}
\end{subequations}
We verify that  this prescription satisfies the requirements
established above, by computing the gauge transformed curvature  
doublet $({}^gf,{}^gF)$ and checking that $({}^gf,{}^gF)$ is given by
formal covariant differentiation of $({}^g\omega,{}^g\varOmega)$ with 
$g$ treated as covariantly constant \footnote{$\vphantom{\bigg[}$
Here, 
the covariant derivative of a connection is taken conventionally 
to be its curvature.}. 
Indeed,  substituting \eqref{7linfdglob} into \eqref{fFcurv}, we find 
\begin{subequations}
\label{8linfdglob}
\begin{align}
&{}^gf=g_0(f), 
\vphantom{\ul{\ul{\ul{\ul{\ul{\ul{\ul{x}}}}}}}}
\vphantom{\Big]}
\label{8linfdgloba}
\end{align}
\begin{align}
&{}^gF=g_1(F-\tau_g(f))+g_2(\omega-\sigma_g,f).
\vphantom{\ul{\ul{\ul{\ul{\ul{\ul{\ul{x}}}}}}}}
\vphantom{\Big]}
\label{8linfdglobb}
\end{align}
\end{subequations}  

Let a connection doublet $(\omega,\varOmega)$ be fixed. 
A bidegree $(p,q)$ field doublet $(\phi,\varPhi)$ is said {\it canonical},
if the gauge transformed field doublet $({}^g\phi,{}^g\varPhi)$ is given by
\begin{subequations}
\label{9linfdglob}
\begin{align}
&{}^g\phi=g_0(\phi),
\vphantom{\Big]}
\label{9linfdgloba}
\\
&{}^g\varPhi=g_1(\varPhi-(-1)^{p+q}\tau_g(\phi))+(-1)^{p+q}g_2(\omega-\sigma_g,\phi).
\vphantom{\Big]}
\label{9linfdglobb}
\end{align}
\end{subequations}
Again, to see that this prescription has the properties required above, we compute 
the gauge transformed covariant derivative field doublet $({}^gD\phi,{}^gD\varPhi)$
and check that $({}^gD\phi,{}^gD\varPhi)$ is given by formal covariant
differentiation of $({}^g\phi,{}^g\varPhi)$ with $g$ assumed covariantly constant. 
Substituting \eqref{7linfdglob}, \eqref{9linfdglob}
into \eqref{fFcovder} and using \eqref{fFcurv}, we find indeed
\begin{subequations}
\label{10linfdglob}
\begin{align}
&{}^gD\phi=g_0(D\phi), 
\vphantom{\Big]}
\label{10linfdgloba}
\\
&{}^gD\varPhi=g_1(D\varPhi+(-1)^{p+q}\tau_g(D\phi))
\vphantom{\Big]}
\label{10linfdglobb}
\\
&\qquad\qquad\qquad-(-1)^{p+q}g_2(\omega-\sigma_g,D\phi)
+(-1)^{p+q}g_2(f,\phi).
\vphantom{\Big]}
\nonumber
\end{align}
\end{subequations}
Note that the gauge transformation 
$\vphantom{\ul{\ul{x}}}$action depends explicitly on $\omega$, 
as predicted earlier. 
Note also that the field doublet $(D\phi,D\varPhi)$ is not canonical:
\eqref{10linfdglobb} cannot be recovered from \eqref{9linfdglobb} just
by replacing $(\phi,\varPhi)$ with $(D\phi,D\varPhi)$ and shifting $p$ into $p+1$
because of the extra $f$ dependent term in the right hand side. 
This is an unavoidable consequence of explicit $\omega$ dependence.
Finally, we observe 
that the curvature doublet $(f,F)$ is a bidegree $(2,0)$
canonical field doublet. $\vphantom{\ul{\ul{x}}}$

Similarly, 
a bidegree $(r,s)$ dual field doublet $(\varUpsilon,\upsilon)$ is said canonical,
if the gauge transformed dual field doublet $({}^g\varUpsilon,{}^g\upsilon)$ is given by
\begin{subequations}
\label{11linfdglob}
\begin{align}
&{}^g\varUpsilon=g^\vee{}_1(\varUpsilon),
\vphantom{\Big]}
\label{11linfdgloba}
\\
&{}^g\upsilon=g^\vee{}_0(\upsilon-(-1)^{r+s}\tau_g{}^\vee(\varUpsilon))-
(-1)^{r+s}g^\vee{}_2(g_0(\omega-\sigma_g),\varUpsilon),
\vphantom{\Big]}
\label{11linfdglobb}
\end{align}
\end{subequations}
where $\tau_g{}^\vee$ is defined by the relation \pagebreak 
$\langle \Xi,\tau_g(x)\rangle=\langle\tau_g{}^\vee(\Xi),x\rangle$.
The gauge transformed covariant derivative dual field doublet $({}^gD\varUpsilon,{}^gD\upsilon)$
can be obtained readily by substituting \eqref{7linfdglob}, \eqref{11linfdglob}
into \eqref{dufFcovder} and using \eqref{fFcurv}, 
\begin{subequations}
\label{12linfdglob}
\begin{align}
&{}^gD\varUpsilon=g^\vee{}_1(D\varUpsilon),
\vphantom{\Big]}
\label{12linfdgloba}
\\
&{}^gD\upsilon=g^\vee{}_0(D\upsilon+(-1)^{r+s}\tau_g{}^\vee(D\varUpsilon))
\vphantom{\Big]}
\label{12linfdglobb}
\\
&\qquad\qquad\qquad+(-1)^{r+s}g^\vee{}_2(g_0(\omega-\sigma_g),D\varUpsilon)
-(-1)^{r+s}g^\vee{}_2(g_0(f),\varUpsilon)
\vphantom{\Big]}
\nonumber
\end{align}
\end{subequations} 
and, as before, has the required properties. 
Again, the gauge transformation action is explicitly $\omega$ dependent 
and $(D\varUpsilon,D\upsilon)$ is not canonical.

Gauge transformation preserves the field/dual field doublet pairing:
\begin{equation}
\langle ({}^g\varUpsilon,{}^g\upsilon),({}^g\phi,{}^g\varPhi)\rangle
=\langle (\varUpsilon,\upsilon),(\phi,\varPhi)\rangle,
\label{13linfdglob}
\end{equation}
a simple consequence of  \eqref{dbltpair}, \eqref{9linfdglob}, \eqref{11linfdglob}.

By \eqref{8linfdgloba}, the $2$--form curvature $f$ has this remarkable property:
$f=0$ implies that ${}^gf=0$ for all $1$--gauge transformations $g\in\Gau_1(M,\mathfrak{v})$.
Thus, the vanishing $2$--form curvature condition 
\begin{equation}
f=0
\label{14linfdglob}
\end{equation}
can be imposed consistently with gauge covariance. Indeed, it is rather natural 
to do so, as is immediate to see. By \eqref{8linfdglobb}, when \eqref{14linfdglob} holds, 
the $3$--form curvature gauge transforms very simply as   
\begin{equation}
{}^gF=g_1(F).
\label{15linfdglob}
\end{equation}
Further, by \eqref{10linfdglob}, if $(\phi,\varPhi)$ is a canonical field doublet, then also 
$(D\phi,D\varPhi)$ is and similarly, by \eqref{12linfdglob},  for a canonical 
dual field doublet $(\varUpsilon,\upsilon)$. In fact, condition \eqref{14linfdglob}
is closely related to the so called ``vanishing fake curvature condition'' 
arising in other formulations of higher gauge theory with strict structure $2$--group $V$
\cite{Breen:2001ie,Girelli:2003ev,Baez:2004in,Baez:2005qu}.
Such condition guarantees that $2$--parallel transport is a $2$--functor from the path $2$--groupoid 
$P_2(M)$ of $M$ to the delooping $2$--groupoid $BV$ of $V$ generalizing the analogous property 
of parallel transport in standard gauge theory. 
To know whether such condition arises naturally also in our formulation of 
semistrict higher gauge theory, one would need a suitable definition of parallel transport.
But, as we explained at the beginning of this section, our approach,  
relying on the automorphism $2$--group $\Aut(\mathfrak{v})$ of a structure Lie $2$--algebra $\mathfrak{v}$
rather than a structure $2$--group $V$, is apparently unsuitable  for the treatment of parallel transport. 
The issue is thus still open. More on this in subsect. \ref{subsec:schreib}.

Gauge transformation of connection and canonical field/dual field 
doublets constitutes a left action of the $1$--gauge transformation group
$\Gau_1(M,\mathfrak{v})$ on fields.
Indeed, for any two $1$--gauge transformations $g,h\in \Gau_1(M,\mathfrak{v})$, 
${}^{g\diamond h}\mathcal{F}={}^g({}^h\mathcal{F})$,
where $\mathcal{F}=\omega$, $\varOmega$, $ \phi$, $\varPhi$, $\varUpsilon$, $\upsilon$, $f$, $F$, $D\phi$, $D\varPhi$, 
$D\varUpsilon$, $D\upsilon$ 
is anyone of the fields considered above, 
as can be verified 
straightforwardly from \eqref{7linfdglob}
--\eqref{12linfdglob}
using \eqref{4linfdglobz}--\eqref{4linfdglobc} systematically. 
%
This is a very basic property and its holding indicates that
our definition of gauge transformation is sound.

Later, we shall encounter other more complicated forms of gauge transformations action
involving combinations of a large number of fields. The one presented above is however
canonical in many ways.


\subsection{\normalsize \textcolor{blue}{$2$--term $L_\infty$ algebra gauge rectifier}}
\label{subsec:rectf}

\hspace{.5cm} In the previous subsection, we have found that gauge transformation
acts linearly on the components of canonical field doublets (cf. eqs. \eqref{9linfdglob}).
The transformation however mixes the components and further depends on
a preassigned connection doublet. The gauge transformation action on canonical 
dual field doublets has analogous features (cf. eqs. \eqref{11linfdglob}).
The complicated way these doublets behave under
gauge transformation makes
it difficult to control gauge covariance and turns out to be a major obstacle
to consistently carrying out gauge fixing in 
semistrict higher gauge theory.
Fortunately, using gauge rectifiers, it is possible to perform 
field redefinitions
turning canonical (dual) field doublets into rectified ones 
transforming linearly, with no mixing and independently from any connection doublet
under the gauge transformation action. 

A field doublet $(\phi,\varPhi)$ is said {\it rectified}, if, under any 
$1$--gauge transformation $g\in\Gau_1(M,\mathfrak{v})$, it transforms as 
\begin{subequations}
\label{rectf6}
\begin{align}
&{}^g\phi=g_0(\phi),
\vphantom{\Big]}
\label{rectf6a}
\\
&{}^g\varPhi=g_1(\varPhi).
\vphantom{\Big]}
\label{rectf6b}
\end{align}
\end{subequations}
Similarly, a dual field doublet $(\varUpsilon,\upsilon)$ is rectified, if, under 
$g$, it transforms as 
\begin{subequations}
\label{rectf7}
\begin{align}
&{}^g\varUpsilon=g^\vee{}_1(\varUpsilon),
\vphantom{\Big]}
\label{rectf7a}
\\
&{}^g\upsilon=g^\vee{}_0(\upsilon).
\vphantom{\Big]}
\label{rectf7b}
\end{align}
\end{subequations}
By definition, then, the gauge transformation action on rectified (dual) field doublets 
is linear, free of  component mixing and independent from any given connection doublet.
These makes rectified doublets very convenient to handle
in field theoretic applications. 

Comparing eqs. \eqref{rectf6}, \eqref{9linfdglob}, it appears that a 
canonical field doublet $(\phi,\varPhi)$ is not rectified. 
Similarly, from inspecting \eqref{rectf7}, \eqref{11linfdglob}, we find that 
a canonical dual field doublet $(\varUpsilon,\upsilon)$ is not rectified either.
Gauge rectifiers remedy this defect, as we now show. 

We assume again that $M$ is diffeomorphic to $\mathbb{R}^d$. 
A pair $(\lambda,\mu)$ with $\lambda\in\Omega^0(M,\Hom(\hat{\mathfrak{v}}_0\wedge\hat{\mathfrak{v}}_0,
\hat{\mathfrak{v}}_1))$,
$\mu\in\Omega^1(M,\Hom(\hat{\mathfrak{v}}_0,\hat{\mathfrak{v}}_1))$,  is a {\it $2$--term $L_\infty$ algebra 
gauge rectifier} if,  under any $1$--gauge transformation $g\in\Gau_1(M,\mathfrak{v})$, it transforms as
follows 
\begin{subequations}
\label{rectf1}
\begin{align}
&{}^g\lambda(x,y)=g_1(\lambda(g^{-1}{}_0(x),g^{-1}{}_0(y))+g^{-1}{}_2(x,y)), \hspace{2cm}
\vphantom{\Big]}
\label{rectf1a}
\\
&{}^g\mu(x)=g_1(\mu(g^{-1}{}_0(x))-\tau_g(g^{-1}{}_0(x))-\lambda(\sigma_g,g^{-1}{}_0(x))).
\vphantom{\Big]}
\label{rectf1b}
\end{align}
\end{subequations}
Gauge rectifiers span an affine space and, \pagebreak in this respect, they are akin to 
ordinary gauge connections. 
It is immediately checked using \eqref{mor3tlinalgc}, \eqref{4linfdgloba}, \eqref{4linfdglobc}
that, for any two $1$--gauge transformations $g,h\in \Gau_1(M,\mathfrak{v})$, 
${}^{g\diamond h}\mathcal{F}={}^g({}^h\mathcal{F})$, where $\mathcal{F}=\lambda,\mu$.
Therefore, $1$--gauge transformation of gauge rectifiers constitutes a left action of 
the group $\Gau_1(M,\mathfrak{v})$ on their space, as for fields. 
We have not been able to ascertain whether gauge rectifiers exist in general. 
We assume that they do in what follows. 

Given $\lambda\in \Hom(\hat{\mathfrak{v}}_0\wedge\hat{\mathfrak{v}}_0,\hat{\mathfrak{v}}_1)$, let \vspace{-1mm}
\begin{subequations}
\label{rectf2}
\begin{align}
&[x,y]_\lambda=[x,y]-\partial\lambda(x,y),
\vphantom{\Big]}
\label{rectf2a}
\\
&[x,X]_\lambda=[x,X]-\lambda(x,\partial X),
\vphantom{\Big]}
\label{rectf2b}
\\
&[x,y,z]_\lambda=[x,y,z]-[x,\lambda(y,z)]]-[y,\lambda(z,x)]]-[z,\lambda(x,y)]
\vphantom{\Big]}
\label{rectf2c}
\\
&\qquad\qquad\qquad
-\lambda(x,[y,z])-\lambda(y,[z,x])-\lambda(z,[x,y])
\vphantom{\Big]}
\nonumber
\\
&\qquad\qquad\qquad
+\lambda(x,\partial\lambda(y,z))+\lambda(y,\partial\lambda(z,x))+\lambda(z,\partial\lambda(x,y)).
\vphantom{\Big]}
\nonumber
\end{align}
\end{subequations}
Then, as is easily verified, $\mathfrak{v}_\lambda=
(\mathfrak{v}_0,\mathfrak{v}_1,\partial,[\cdot,\cdot]_\lambda,[\cdot,\cdot,\cdot]_\lambda)$
is a $2$--term $L_\infty$ algebra, the {\it $\lambda$--deformation of $\mathfrak{v}$}.
Thus, once a gauge rectifier $(\lambda,\mu)$ is assigned, we have a 
$\lambda$--deformation $\mathfrak{v}_\lambda$ of $\mathfrak{v}$ pointwise on $M$. 

The $\lambda$--deformed brackets transform covariantly under $1$--gauge transformation, that is, 
for $g\in\Gau_1(M,\mathfrak{v})$, 
\begin{subequations}
\label{3rectf}
\begin{align}
&[x,y]_{{}^g\lambda}=g_0([g^{-1}{}_0(x),g^{-1}{}_0(y)]_\lambda), 
\vphantom{\Big]}
\label{3rectfa}
\\
&[x,X]_{{}^g\lambda}=g_1([g^{-1}{}_0(x),g^{-1}{}_1(X)]_\lambda), 
\vphantom{\Big]}
\label{3rectfb}
\\
&[x,y,z]_{{}^g\lambda}=g_1([g^{-1}{}_0(x),g^{-1}{}_0(y),g^{-1}{}_0(z)]_\lambda).
\vphantom{\Big]}
\label{3rectfc}
\end{align}
\end{subequations}
This can be shown straightforwardly by combining \eqref{mor2tlinalgb}--\eqref{mor2tlinalgd} with 
\eqref{rectf2}. By this remarkable property, the deformed brackets are the truly natural ones under 
$1$--gauge transformation. 

With a gauge rectifier $(\lambda,\mu)$, there is associated a {\it derived rectifier}
$(v_{\lambda,\mu},w_{\lambda,\mu})$, where 
$v_{\lambda,\mu}\in\Omega^1(M,\Hom(\hat{\mathfrak{v}}_0\wedge\hat{\mathfrak{v}}_0,\hat{\mathfrak{v}}_1))$, 
$w_{\lambda,\mu}\in\Omega^2(M,\Hom(\hat{\mathfrak{v}}_0,\hat{\mathfrak{v}}_1))$ and 
\begin{subequations}
\label{rectf5}
\begin{align}
&v_{\lambda,\mu}(x,y)=d\lambda(x,y)-\mu([x,y]_\lambda)+[x,\mu(y)]_\lambda-[y,\mu(x)]_\lambda,
\vphantom{\Big]}
\label{rectf5a}
\\
&w_{\lambda,\mu}(x)=d\mu(x)+\mu(\partial\mu(x)). \hspace{3.8truecm}
\vphantom{\Big]}
\label{rectf5b}
\end{align}
\end{subequations}
Under any $1$--gauge transformation $g\in\Gau_1(M,\mathfrak{v})$, we have
\begin{subequations}
\label{rectf4}
\begin{align}
&v_{{}^g\lambda,{}^g\mu}(x,y)=g_1(v_{\lambda,\mu}(g^{-1}{}_0(x),g^{-1}{}_0(y))
-[\sigma_g,g^{-1}{}_0(x),g^{-1}{}_0(y)]_\lambda),
\vphantom{\Big]}
\label{rectf4a}
\\
&w_{{}^g\lambda,{}^g\mu}(x)=g_1\Big(w_{\lambda,\mu}(g^{-1}{}_0(x))-v_{\lambda,\mu}(\sigma_g,g^{-1}{}_0(x))
\vphantom{\Big]}
\label{rectf4b}
\\
&\qquad\qquad
-[g^{-1}{}_0(x),\varSigma_g-\frac{1}{2}\lambda(\sigma_g,\sigma_g)+\mu(\sigma_g)]_\lambda
+\frac{1}{2}[\sigma_g,\sigma_g,g^{-1}{}_0(x)]_\lambda\Big).
\vphantom{\Big]}
\nonumber
\end{align}
\end{subequations}
These relations follow from writing $g_2$ and $\tau_g$ in terms 
of $\lambda$, $\mu$ via \eqref{rectf1} and substituting the resulting expressions
in \eqref{3linfdglobc} and \eqref{2linfdglob}.
Though this is not immediately evident, $v_{\lambda,\mu}$, $w_{\lambda,\mu}$
are natural differential expressions in $\lambda$, $\mu$ which 
appear repeatedly as building blocks of $1$--gauge covariant expressions. 

We now show how one can turn canonical (dual) field doublets into rectified ones
using a chosen $2$--term $L_\infty$ algebra gauge rectifier $(\lambda,\mu)$. 
Suppose that a connection doublet $(\omega,\varOmega)$ is given. 
Let $(\phi,\varPhi)$ be a bidegree $(p,q)$ canonical field doublet. 
Then, naturally associated with $(\phi,\varPhi)$, there is a rectified 
field doublet $(\phi_{\lambda,\mu},\varPhi_{\lambda,\mu})$, where 
$\phi_{\lambda,\mu}=\phi$ and 
\begin{equation}
\varPhi_{\lambda,\mu}=\varPhi+(-1)^{p+q}\lambda(\omega,\phi)-(-1)^{p+q}\mu(\phi).
\label{rectf8}
\end{equation}
Similarly, a bidegree $(r,s)$ canonical dual 
field doublet $(\varUpsilon,\upsilon)$ yields naturally a rectified dual 
field doublet $(\varUpsilon_{\lambda,\mu},\upsilon_{\lambda,\mu})$, where 
$\varUpsilon_{\lambda,\mu}=\varUpsilon$ and 
\begin{equation}
\upsilon_{\lambda,\mu}=\upsilon-(-1)^{r+s}\lambda^\vee(\omega,\varUpsilon)+(-1)^{r+s}\mu^\vee(\varUpsilon)
\label{rectf9}
\end{equation}
and the gauge corectifier $(\lambda^\vee,\mu^\vee)$ is defined by 
$\langle \lambda^\vee(x,\varXi),y\rangle=-\langle\varXi, \lambda(x,y)\rangle$, 
$\langle \mu^\vee(\varXi),x\rangle=-\langle\varXi,\mu(x)\rangle$. 

Given a connection doublet $(\omega,\varOmega)$ 
and a gauge rectifier $(\lambda,\mu)$, 
it is possible to define a rectified covariant derivative $D_{\lambda,\mu}$
mapping rectified (dual) field doublets into rectified ones. 
For a bidegree $(p,q)$ rectified field doublet $(\phi,\varPhi)$, 
the bidegree $(p+1,q)$ rectified covariant derivative field doublet  $(D_{\lambda,\mu}\phi,D_{\lambda,\mu}\varPhi)$
is given by  
\begin{subequations}
\label{rectf10}
\begin{align}
&D_{\lambda,\mu}\phi=d\phi+[\omega,\phi]_\lambda+\partial\mu(\phi),
\vphantom{\Big]}
\label{rectf10a}
\\
&D_{\lambda,\mu}\varPhi=d\varPhi+[\omega,\varPhi]_\lambda+\mu(\partial\varPhi).
\vphantom{\Big]}
\label{rectf10b}
\end{align}
\end{subequations}
Similarly, for a bidegree $(r,s)$ rectified dual field doublet $(\varUpsilon,\upsilon)$, 
the bidegree $(r+1,s)$ rectified covariant derivative dual 
field doublet $(D_{\lambda,\mu}\varUpsilon,D_{\lambda,\mu}\upsilon)$ is 
\begin{subequations}
\label{rectf11}
\begin{align}
&D_{\lambda,\mu}\varUpsilon=d\varUpsilon+[\omega,\varUpsilon]_\lambda{}^\vee+\partial^\vee\mu^\vee(\varUpsilon),
\vphantom{\Big]}
\label{rectf11a}
\\
&D_{\lambda,\mu}\upsilon=d\upsilon+[\omega,\upsilon]_\lambda{}^\vee+\mu^\vee(\partial^\vee\upsilon),
\vphantom{\Big]}
\label{rectf11b}
\end{align}
\end{subequations} 
where the $\lambda$--deformed cobracket $[\cdot,\cdot]_\lambda{}^\vee$ is defined in the same way
as the undeformed one (cf. eqs. \eqref{co2tlinalgb}, \eqref{co2tlinalgc}).
Rectified covariant differentiation intertwines naturally with the duality pairing,
\begin{subequations}
\label{rectf12}
\begin{align}
&d\langle\upsilon,\phi\rangle=\langle D_{\lambda,\mu}\upsilon,\phi\rangle
-(-1)^{r+s}\langle\upsilon,D_{\lambda,\mu}\phi\rangle,
\vphantom{\Big]}
\label{rectf12a}
\\
&d\langle\varUpsilon,\varPhi\rangle=\langle D_{\lambda,\mu}\varUpsilon,\varPhi\rangle
+(-1)^{r+s}\langle\varUpsilon,D_{\lambda,\mu}\varPhi\rangle.
\vphantom{\Big]}
\label{rectf12b}
\end{align}
\end{subequations}
Compare with \eqref{dbltparts}.


\vspace{.0cm}

\subsection{\normalsize \textcolor{blue}{The $2$--group of crossed module gauge transformations}}
\label{subsec:crmdgautrsf}

\hspace{.5cm} 
We are going to define a notion of gauge transformation 
naturally hinged on an arbitrary Lie crossed module $(G,H,t,m)$.  
We than shall show that these crossed module gauge transformations
form naturally a strict $2$--group. 
We consider again the case where $M$ is diffeomorphic to $\mathbb{R}^d$.

A {\it crossed module $1$--gauge transformation} consists of the following data:
\begin{enumerate}

\item a map $\gamma\in\Map(M,G)$;

\item an element $\chi_\gamma\in \Omega^1(M,\mathfrak{h})$. 

\end{enumerate}
In the following,  we are going to denote a crossed module $1$--gauge transformation
such as the above as $(\gamma,\chi_\gamma)$ or simply as $\gamma$.
Again, as before, in so doing, we are not implying that $\chi_\gamma$ 
is determined by $\gamma$, but only that its the partner of $\gamma$ in the 
gauge transformation. We shall denote the set of crossed $1$--gauge 
transformation by $\Gau_1(M,G,H)$. The notion of gauge transformation defined here 
coincides with that given in refs. \cite{Baez:2004in,Baez:2005qu}.

For any two crossed module $1$--gauge transformations $\gamma,\beta\in\Gau_1(M,G,H)$,
a {\it crossed module $2$--gauge transformation} from $\gamma$ to $\beta$ consists of the following data:
\begin{enumerate}

\item a map $\varTheta\in\Map(M,H)$;

\item an element $B_\varTheta\in\Omega^1(M,\mathfrak{h})$.

\end{enumerate}
They are required to satisfy the following relations,
\begin{subequations}
\label{crmdgautrsf1}
\begin{align}
&\beta=t(\varTheta)\gamma,
\vphantom{\Big]}
\label{crmdgautrsf1a}
\\
&\chi_\gamma-\chi_\beta=B_\varTheta.
\vphantom{\Big]}
\label{crmdgautrsf1b}
\end{align}
\end{subequations}
We shall denote a crossed module $2$--gauge transformation
like this one as $(\varTheta,B_\varTheta)$ or simply as $\varTheta$, meaning as usual
in the former case that $B_\varTheta$ is the partner of $\varTheta$.
We shall also write 
$\varTheta:\gamma\Rightarrow \beta$ to emphasize its source and target. 
We shall denote the set 
of all crossed  module $2$--gauge transformations 
$\varTheta:\gamma\Rightarrow \beta$  by $\Gau_2(M,G,H)(\gamma,\beta)$
and the set of all $2$--gauge transformations $\varTheta$ by $\Gau_2(M,G,H)$.
Note that, by \eqref{crmdgautrsf1b}, the datum $B_\varTheta$ is determined $\gamma,\beta$
and, so, is essentially redundant.

Next, we shall show that it is possible to define a composition and an inversion law 
and a unit in $\Gau_1(M,G,H)$  and horizontal and vertical composition and inversion laws
and units in $\Gau_2(M,G,H)$, making $(\Gau_1(M,G,H),\Gau_2(M,G$, $H))$  
a strict $2$--group (cf. subsect. \ref{sec:twogr}). 

\vspace{1mm}\pagebreak

The composition and inversion laws and the unit of $1$--gauge transformation are defined by 
the relations \hphantom{xxxxxxxxxxxxxxxxxxxxxxxxxxxx} 
\begin{subequations}
\label{crmdgautrsf3}
\begin{align}
&\beta\diamond\gamma=\beta\gamma,
\vphantom{\Big]}
\label{crmdgautrsf3a}
\\
&\chi_{\beta\,\diamond\,\gamma}=\chi_\beta+\dot m(\beta)(\chi_\gamma),
\vphantom{\Big]}
\label{crmdgautrsf3b}
\\
&\gamma^{-1_\diamond}=\gamma^{-1},
\vphantom{\Big]}
\label{crmdgautrsf3c}
\\
&\chi_{\gamma^{-1_\diamond}}=-\dot m(\gamma^{-1})(\chi_\gamma),
\vphantom{\Big]}
\label{crmdgautrsf3d}
\\
&\iota=1_G,
\vphantom{\Big]}
\label{crmdgautrsf3e}
\\
&\chi_\iota=0.
\vphantom{\Big]}
\label{crmdgautrsf3f}
\end{align}
\end{subequations}
where $\gamma,\beta\in\Gau_1(M,G,H)$ and $\dot m$ is defined in \eqref{st2grst2lie2}.
In \eqref{crmdgautrsf3a}, \eqref{crmdgautrsf3c}, \eqref{crmdgautrsf3e}, 
the composition, inver\-sion and unit in the right hand side are those of 
$G$ thought of as holding pointwise on $M$.

The horizontal and vertical composition and inversion laws and the units of $2$--gauge
transformations are defined by the relations
\begin{subequations}
\label{crmdgautrsf4}
\begin{align}
&\varLambda\diamond\varTheta=\varLambda m(\zeta)(\varTheta),
\vphantom{\Big]}
\label{crmdgautrsf4a}
\\
&B_{\varLambda\,\diamond\,\varTheta}=B_\varLambda+\dot m(\zeta)(B_\varTheta)+Q(\zeta \dot t(\chi_\beta)\zeta^{-1},\varLambda),
\hspace{1.6cm}
\vphantom{\Big]}
\label{crmdgautrsf4b}
\\
&\varTheta^{-1_\diamond}=m(\gamma^{-1})(\varTheta^{-1}),
\vphantom{\Big]}
\label{crmdgautrsf4c}
\\
&B_{\varTheta^{-1_\diamond}}=-\dot m(\gamma^{-1})(B_\varTheta)
+\dot m(\gamma^{-1})(Q(t(\varTheta)^{-1}\dot t(\chi_\beta)t(\varTheta),\varTheta)),
\vphantom{\Big]}
\label{crmdgautrsf4d}
\\
&\varPi\bullet\varXi=\varPi\varXi,
\vphantom{\Big]}
\label{crmdgautrsf4e}
\\
&B_{\varPi\,\bullet\,\varXi}=B_\varXi+B_\varPi,
\vphantom{\Big]}
\label{crmdgautrsf4f}
\\
&\varXi^{-1_\bullet}=\varXi^{-1},
\vphantom{\Big]}
\label{crmdgautrsf4g}
\\
&B_{\varXi^{-1_\bullet}}=-B_\varXi,
\vphantom{\Big]}
\label{crmdgautrsf4h}
\\
&I_\gamma=1_H,
\vphantom{\Big]}
\label{crmdgautrsf4i}
\\
&B_{I_\gamma}=0, 
\vphantom{\Big]}
\label{crmdgautrsf4j}
\end{align}
\end{subequations}
where $\gamma,\beta,\zeta,\eta\in\Gau_1(M,G,H)$ and $\varTheta,\varLambda,\varXi,\varPi
\in\Gau_2(M,G,H)$, 
with $\varTheta:\gamma\Rightarrow \beta$, $\varLambda:\zeta\Rightarrow \eta$ and 
$\varXi,\varPi$ composible  and $\dot t$ and $Q$ are defined in \eqref{st2grst2lie1} and 
\eqref{linftymorph3}. In \eqref{crmdgautrsf4a}, \eqref{crmdgautrsf4c}, \eqref{crmdgautrsf4e}, 
\eqref{crmdgautrsf4g}, \eqref{crmdgautrsf4i}, 
the composition, inversion and unit in the right hand side are those of 
$H$ holding pointwise on $M$ and similarly for the $G$--action $m$. 

It is straightforward to verify that  the composition, inversion and unit structures
just defined satisfy the axioms \eqref{twogr1}, so that $(\Gau_1(M,G,H),\Gau_2(M$, $G,H))$ is indeed 
a strict $2$--group, as claimed.

The pair $(\Map(M,G),\Map(M,H))$ has a structure of infinite dimensional Lie crossed module 
induced by that of $(G,H)$ pointwise on $M$. $(\Map(M,G),\Map(M$, $H))$ in turn can be viewed as 
an infinite dimensional strict $2$--group $\Map(M,V)$ using the conversion prescriptions listed in 
subsect. \ref{sec:twogr}. Eqs. \eqref{crmdgautrsf3a}, \eqref{crmdgautrsf3c}, \eqref{crmdgautrsf3e}, 
\eqref{crmdgautrsf4a}, \eqref{crmdgautrsf4c}, \eqref{crmdgautrsf4e}, 
\eqref{crmdgautrsf4g}, \eqref{crmdgautrsf4i} define precisely the $2$--group operations of
$\Map(M,V)$ expressed in terms of the crossed module structure of $(\Map(M,G),\Map(M,H))$.

Once the above is realized, inspection of \eqref{crmdgautrsf3} reveals that
the $1$--cell group $\Gau_1(M,G,H)$ is the semidirect product 
$\Omega^1(M,\mathfrak{h})\rtimes \Map(M,V_1)$ associated with 
a certain right action of $\Map(M,V_1)$ on $\Omega^1(M,\mathfrak{h})$. Unlike 
the group of $2$--term $L_\infty$ algebra $1$--gauge transformation $\Gau_1(M,\mathfrak{v})$ 
in subsect. \ref{subsec:linfgautrsf}, $\Gau_1(M,G,H)$ is not a proper subgroup of 
$\Omega^1(M,\mathfrak{h})\rtimes \Map(M,V_1)$, since there are no differential relations obeyed 
by its elements analogous to \eqref{1linfdglob}--\eqref{3linfdglob}. 
Again, this leads us to conjecture that the full $2$--group $\Gau(M,G,H)$ may be described 
as a semidirect product $2$--group defined by the relations \eqref{crmdgautrsf3},
\eqref{crmdgautrsf4}, for a suitable definition of the latter notion.

There exists a natural strict $2$--group $1$--morphism (cf. sect. \ref{sec:twogr}, eq. \eqref{twogr2})
from the crossed module gauge transformation $2$--group $\Gau(M,G,H)$ to 
the gauge transformation $2$--group $\Gau(M,\mathfrak{v})$, where $\mathfrak{v}$
is the strict $2$--term $L_\infty$ algebra corresponding to 
the differential Lie crossed module $(\mathfrak{g},\mathfrak{h})$. 
The morphism is defined by following expressions 
\begin{subequations}
\label{crmdgautrsf5}
\begin{align}
&g_\gamma=\phi_\gamma, 
\vphantom{\Big]}
\label{crmdgautrsf5a}
\\
&\sigma_{g_\gamma}=\gamma^{-1}d\gamma+\gamma^{-1}\dot t(\chi_\gamma)\gamma,
\vphantom{\Big]}
\label{crmdgautrsf5b}
\\
&\varSigma_{g_\gamma}=\dot m(\gamma^{-1})\Big(d\chi_\gamma+\frac{1}{2}[\chi_\gamma,\chi_\gamma]\Big),
\vphantom{\Big]}
\label{crmdgautrsf5c}
\\
&\tau_{g_\gamma}(x)=\widehat{m}(x)(\dot m(\gamma^{-1})(\chi_\gamma)),
\vphantom{\Big]}
\label{crmdgautrsf5d}
\\
&F_\varLambda(x)=\varPhi_{\zeta,\varLambda}(x), 
\vphantom{\Big]}
\label{crmdgautrsf5e}
\\
&A_{F_\varLambda}=\dot m(\zeta^{-1})(-\varLambda^{-1}d\varLambda+\chi_\zeta+\varLambda^{-1}(B_\varLambda-\chi_\zeta)\varLambda),
\vphantom{\Big]}
\label{crmdgautrsf5f}
\end{align}
\end{subequations}
for $\gamma,\zeta,\eta\in\Gau_1(M,G,H)$ and $\varLambda\in\Gau_2(M,G,H)$ 
with $\varLambda:\zeta\Rightarrow \eta$, 
where the right hand sides of \eqref{crmdgautrsf5a}
and \eqref{crmdgautrsf5e} are defined by \eqref{linftymorph1} and \eqref{linftymorph2} 
pointwise on $M$ and $\widehat{m}$ is defined in \eqref{st2grst2lie3}.

The mappings $\gamma\rightarrow \phi_\gamma$ and 
$(\zeta,\varLambda)\rightarrow \varPhi_{\zeta,\varLambda}$ define a strict $2$--group $1$--morphim from $\Map(M,V)$
to $\Map(M,\Aut(\mathfrak{v}))$ which is the pointwise version of the ``adjoint'' $2$--group $1$--morphim from $V$
to $\Aut(\mathfrak{v})$ defined in sect. \ref{sec:linftyauto}.
\eqref{crmdgautrsf5} above extend such morphism to one from $\Gau(M,G,H)$ to $\Gau(M,\mathfrak{v})$.

Let $(\omega,\varOmega)$ be a connection doublet.
Inserting eqs. \eqref{crmdgautrsf5b}--\eqref{crmdgautrsf5d}
into the relations \eqref{7linfdglob}, we obtain
\begin{subequations}%
\label{gliecross6}
\begin{align}
&{}^{g_\gamma}\omega=\gamma\omega\gamma^{-1}-d\gamma\gamma^{-1}-\dot t(\chi_\gamma), 
\vphantom{\Big]}
\label{gliecross6a}
\\
&{}^{g_\gamma}\varOmega=\dot m(\gamma)(\varOmega)
-\widehat{m}(\gamma\omega\gamma^{-1}-d\gamma\gamma^{-1}-\dot t(\chi_\gamma))(\chi_\gamma)
-d\chi_\gamma-\frac{1}{2}[\chi_\gamma,\chi_\gamma].
\vphantom{\Big]}
\label{gliecross6b}
\end{align}
\end{subequations} 
Inserting eqs. \eqref{crmdgautrsf5b}--\eqref{crmdgautrsf5d} into \eqref{8linfdglob},
we find further
\begin{subequations}%
\label{gliecross7}
\begin{align}
&{}^{g_\gamma}f=\gamma f\gamma^{-1}, 
\vphantom{\Big]}
\label{gliecross7a}
\\
&{}^{g_\gamma}F=\dot m(\gamma)(F)
-\widehat{m}(\gamma f\gamma^{-1})(\chi_\gamma).
\vphantom{\Big]}
\label{gliecross7b}
\end{align}
\end{subequations} 
Remarkably, these expressions are identical to those obtained originally in refs.
\cite{Baez:2004in,Baez:2005qu}. 
This shows that \eqref{st2grst2lie3}, \eqref{8linfdglob} provide the appropriate generalization 
of the gauge transformation action for a $2$--term $L_\infty$ gauge theory. 


\subsection{\normalsize \textcolor{blue}{Review of principal $2$--bundle theory}}
\label{subsec:twobund}

\vspace{1mm}
\hspace{.5cm} 
The analysis of the global aspects in gauge theory consists in determining how locally defined data
glue in a globally consistent manner. 
In the same way as the global properties of ordinary gauge theory are described by the theory of principal 
bundles, those of higher gauge theory are expressed by the theory of principal $2$--bundles.
So, it is appropriate at this point to review these topics,
recalling well--known basic facts of differential topology of principal bundles and then showing how these 
generalize to principal $2$--bundles. Our presentation has no pretense of completeness or mathematical rigour 
and serves only the purpose of setting the terminology and the notation used later.

The definition of local data requires the choice 
of an {\it open covering} $U=\{U_i\}$ of $M$, that is a collection of open subsets $U_i\subset M$
such that
\begin{equation}
M=\bigcup\nolimits_iU_i.
\label{cech1}
\end{equation}
The covering $U$ is characterized by its {\it nerve}, which is the the collection of all
non empty intersections $U_{i_0\ldots i_n}=U_{i_0}\cap\ldots\cap U_{i_n}\not=\emptyset$ with $n\geq 0$.

Let $G$ be a Lie group. We define a groupoid $\check P(U,G)$ as follows.
\begin{enumerate}

\item A $0$--cell of $\check P(U,G)$ is collection $g=\{g_{ij}\}$ 
of smooth maps $g_{ij}\in\Map(U_{ij},G)$ satisfying the condition 
\begin{equation}
g_{ik}=g_{ij}g_{jk}, \qquad \text{on $U_{ijk}$}.
\label{cech2}
\end{equation}

\item For any two $0$--cell $g$, $g'$ of $\check P(U,G)$, a $1$--cell $g\rightarrow g'$
is a collection $\{h_i\}$ of smooth maps $h_i\in\Map(U_i,G)$ such that
\begin{equation}
g_{ij} h_j=h_ig'{}_{ij}, \qquad \text{on $U_{ij}$}.
\label{cech3}
\end{equation}

\item For any $0$--cell $g$, the identity $1$--cell $\id_g$ of $g$ 
is the collection $\{1_{Gi}\}$ of constant maps of $\Map(U_i,G)$ with value $1_G$.

\item For any $1$--cell $h:g\rightarrow g'$, the inverse $1$--cell 
$h^{-1}:g'\rightarrow g$ is defined by
\begin{equation}
h^{-1}=\{h_i{}^{-1}\}.
\label{cech4}
\end{equation}

\item For any two $1$--cells $h:g\rightarrow g'$, $h':g'\rightarrow g''$, the composition  $1$--cell 
$h'h:g\rightarrow g''$ is defined by \hphantom{xxxxxxxxxxxxxxxxxxxxxx}
\begin{equation}
h'h=\{h_ih'{}_i\}.
\label{cech5}
\end{equation}
\end{enumerate}
It is straightforward to check that $\check P(U,G)$ is indeed a groupoid as anticipated.  

The set of isomorphism classes of $0$--cells $g$ is nothing but the 1st \v Cech cohomology
$\check H^1(U,G)$ of the covering $U$ with coefficients in $G$. 
The dependence on $U$ can be eliminated by switching to 
1st \v Cech cohomology $\check H^1(M,G)$ of $M$ with coefficients in $G$, which is defined as the 
direct limit under covering refinement of the cohomology $\check H^1(U,G)$, 
\hphantom{xxxxxxxxxxxxxxxxxxxxxxxxx}
\begin{equation}
\check H^1(M,G)=\lim_{\overrightarrow{U}}\check H^1(U,G). 
\label{cech6}
\end{equation}
In differential topology,  $\check H^1(M,G)$ has a well-known interpretation: 
it is the set of isomorphism classes of smooth principal $G$--bundles $P$. 

For any two $0$--cells $g$, $g'$, consider the set $\check H^2(U,G;g,g')$ of 
$1$--cells $g\rightarrow g'$. If it is non empty, $\check H^2(U,G;g,g')$  
depends only to the common isomorphism class of $g$ and $g'$ in $\check H^1(U,G)$ up to bijection,
so that we can set $g'=g$ right away. The dependence on $U$ is eliminated by switching  
2nd \v Cech cohomology $\check H^2(P)$
\begin{equation}
\check H^2(P)=\lim_{\overrightarrow{U}}\check H^2(U,G;g,g), 
\label{cech7}
\end{equation}
where $P$ is a principal $G$--bundles in the isomorphism class 
associated with $g$. 
In differential topology,  $\check H^2(P)$ has also a well-known interpretation. 
If $P$ is represented by a $0$--cell $g$, $\check H^2(P)$ is represented by the group
of $1$--cells $g\rightarrow g$. $\check H^2(P)$ is thus the group of automorphisms of $P$,
i. e.  the gauge group $\Gau(P)$.

Let $V$ be a strict Lie $2$--group. 
Isomorphism classes of principal $V$--$2$--bundle $P$, the gauge group of 
one such bundle $P$ and other appended structures can be characterized 
in a way which is a direct generalization of that of ordinary principal bundle theory
illustrated above. We shall now go through this more explicitly following loosely the 
treatment of $2$--bundles of refs. \cite{Baez:2004in,Baez:2005qu}, 
to which we refer the reader. 
(See also \cite{Waldorf1} for a comparison of different approaches.)

We recall first the definition of strict $2$--groupoid. 
A {\it strict $2$--groupoid} consists of the following set of data: 
\begin{enumerate}

\item a set of $0$-cells $V_0$;

\item for each pair of $0$--cells $x,y$, a set of $1$-cells $V_1(x,y)$;

\item for each triple of $0$--cells $x,y,z$, a composition law of $1$--cells 
$\circ: V_1(x,y)\times V_1(y,z)\rightarrow V_1(x,z)$;

\item for each pair of $0$--cells $x,y$, a inversion law of $1$--cells ${}^{-1_\circ}: V_1(x,y)\rightarrow V_1(y,z)$;

\item for each $0$--cell $x$, a distinguished unit $1$--cell $1_x\in V_1(x,x)$;

\item for each pair of $0$--cells $x,y$
and for each pair of $1$--cells $a,b\in V_1(x,y)$, a set of $2$--cells $V_2(a,b)$;

\item for each triple of $0$--cells $x,y,z$ and for each pair of $1$--cells $a,c\in V_1(x,y)$ and 
for each pair of $1$--cells $b,d\in V_1(y,z)$, a horizontal composition law of $2$--cells
$\circ:V_2(a,c)\times V_2(b,d)\rightarrow V_2(b\circ a,d\circ c)$;

\item for each pair of $0$--cells $x,y$ and for each pair of $1$--cells $a,b\in V_1(x,y)$,
a horizontal inversion law of $2$--cells ${}^{-1_\circ}: V_2(a,b)\rightarrow V_2(a^{-1_\circ},b^{-1_\circ})$;

\item for each pair of $0$--cells $x,y$ and for each triple of $1$--cells $a,b,c\in V_1(x,y)$, 
a vertical composition law of $2$--cells $\bfdot:V_2(a,b)\times V_2(b,c)\rightarrow V_2(a,c)$;

\item for each pair of $0$--cells $x,y$ and for each pair of $1$--cells $a,b\in V_1(x,y)$, 
a vertical inversion law of $2$--cells ${}^{-1_\bfdot}\!\!: V_2(a,b)\rightarrow V_2(b,a)$;

\item for each pair of $0$--cells $x,y$ and for each $1$--cell $a\in V_1(x,y)$, a distinguished 
unit $2$--cell $1_a\in V_2(a,a)$.
\end{enumerate}
 These are required to satisfy the following axioms. 
\begin{subequations}
\label{twogroid1}
\begin{align}
&(c\circ b)\circ a=c\circ(b\circ a),
\vphantom{\Big]}
\label{twogroid1a}
\\
&a^{-1_\circ}\circ a=1_x, \qquad a\circ a^{-1_\circ}=1_y,
\vphantom{\Big]}
\label{twogroid1b}
\\
&a\circ 1_x=1_y\circ a=a,
\vphantom{\Big]}
\label{twogroid1c}
\\
&(C\circ B)\circ A=C\circ(B\circ A),
\vphantom{\Big]}
\label{twogroid1d}
\\
&A^{-1_\circ}\circ A=1_{1_x}\qquad A\circ A^{-1_\circ}=1_{1_y},
\vphantom{\Big]}
\label{twogroid1e}
\\
&A\circ 1_{1_x}=1_{1_y}\circ A=A,
\vphantom{\Big]}
\label{twogroid1f}
\\
&(C\bfdot B)\bfdot A=C\bfdot(B\bfdot A),
\vphantom{\Big]}
\label{twogroid1g}
\\
&A^{-1_\bfdot}\!\bfdot A=1_a,\qquad A\bfdot A^{-1_\bfdot}=1_b,
\vphantom{\Big]}
\label{twogroid1h}
\\
&A\bfdot 1_a=1_b\bfdot A=A,
\vphantom{\Big]}
\label{twogroid1i}
\\
&(D\bfdot C)\circ(B\bfdot A)=(D\circ B)\bfdot(C\circ A).
\vphantom{\Big]}
\label{twogroid1j}
\end{align}
\end{subequations}%
Here and in the following, $x,y,z,\dots\in V_0$, $a,b,c,\dots\in V_1$, $A,B,C,\dots\in V_2$, where 
$V_1$ and $V_2$ denote the set of all $1$-- and $2$--cells, respectively. For clarity, we often denote 
$a\in V_1(x,y)$ as $a:x\rightarrow y$ and $A\in V_2(a,b)$
as $A:a\Rightarrow b$. 
All identities involving the horizontal and vertical composition and inversion hold whenever defined.  
Relation \eqref{twogroid1j} is called again interchange law. 
In the following, we shall denote a $2$--groupoid such as the above as $V$ or $(V_0,V_1,V_2)$
or $(V_0,V_1,V_2,\circ,{}^{-1_\circ},\bfdot,{}^{-1_\bfdot},1_-)$ to emphasize the underlying structure.

If $(V_0,V_1,V_2,\circ,{}^{-1_\circ},\bfdot,{}^{-1_\bfdot},1_-)$ is a strict $2$--groupoid, then 
$(V_0,V_1,\circ,{}^{-1_\circ},1_-)$ and $(V_1,V_2,\bfdot,{}^{-1_\bfdot},1_-)$ are both groupoids. 
$V$ can also be viewed as a strict 
$2$--category in which all $1$--morphisms are invertible  and all $2$--morphisms
are both horizontal and vertical invertible. 
A strict $2$--group $(V_1,V_2)$ is just a strict $2$--groupoid $(V_0,V_1,V_2)$
such that $V_0$ is the singleton set (cf. subsect. \ref{sec:twogr}). 

Let $(G,H)$ be a Lie crossed module associated with the strict $2$--group $V$. 
We define a strict $2$--groupoid $\check P_2(U,G,H)$ as follows.
\begin{enumerate}

\item A $0$--cell of $\check P_2(U,G,H)$ is a collection $(g,W)=\{g_{ij},W_{ijk}\}$
of smooth maps $g_{ij}\in\Map(U_{ij},G)$, $W_{ijk}\in\Map(U_{ijk},H)$ 
satisfying the relations 
\begin{subequations}
\begin{align}
&g_{ij}g_{jk}=t(W_{ijk})g_{ik}, \qquad \text{on $U_{ijk}$},
\vphantom{\Big]}
\label{cech10a}
\\
&m(g_{ij})(W_{jkl})W_{ijl}=W_{ijk}W_{ikl}, \qquad \text{on $U_{ijkl}$}.
\vphantom{\Big]}
\label{cech10b}
\end{align}
\label{cech10}
\end{subequations}
\vspace{-1.1cm}

\item For any two $0$--cells $(g,W)$, $(g',W')$, a $1$--cell $(g,W)\rightarrow (g',W')$ of $\check P_2(U,G,H)$
is a collection $(h,J)=\{h_i,J_{ij}\}$ of smooth maps $h_i\in\Map(U_i,G)$, $J_{ij}\in\Map(U_{ij},H)$ satisfying
the relations
\begin{subequations}
\begin{align}
&h_ig'_{ij}=t(J_{ij})g_{ij}h_j, \qquad \text{on $U_{ij}$},
\vphantom{\Big]}
\label{cech11a}
\\
&J_{ij}m(g_{ij})(J_{jk})W_{ijk}=m(h_i)(W'{}_{ijk})J_{ik}, \qquad \text{on $U_{ijk}$}.
\vphantom{\Big]}
\label{cech11b}
\end{align}
\label{cech11}
\end{subequations}
\vspace{-1.1cm}

\item For any two $1$--cells $(h,J), (h',J'):(g,W)\rightarrow (g',W')$, a $2$--cell 
$(h,J)\Rightarrow (h',J')$  of $\check P_2(U,G,H)$ is a collection  $K=\{K_i\}$
of smooth maps $K\in\Map(U_i,H)$ such that
\begin{subequations}
\begin{align}
&h'{}_i=t(K_i)h_i,
\vphantom{\Big]}
\label{cech12a}
\\
&J'{}_{ij}m(g_{ij})(K_j)=K_iJ_{ij}.
\vphantom{\Big]}
\label{cech12b}
\end{align}
\label{cech12}
\end{subequations}
\vspace{-1.1cm}

\item For any $0$--cell $(g,W)$, 
the identity $1$--cell $\id_{(g,W)}$ of $(g,W)$ 
is the collection $\{1_{Gi},1_{Hij}\}$ constant maps of $\Map(U_i,G)$ with value $1_G$
and $\Map(U_{ij},H)$ with value $1_H$.

\item For any $1$--cell $(h,J):(g,W)\rightarrow (g',W')$, \pagebreak
the inverse $1$--cell 
$(h,J)^{-1_\circ}:(g',W')$ $\rightarrow (g,W)$ is defined by
\begin{equation}
(h,J)^{-1_\circ}=\{h_i{}^{-1},m(h_i{}^{-1})(J_{ij}{}^{-1})\}.
\label{cech13}
\end{equation}

\item For any two $1$--cells $(h,J):(g,W)\rightarrow (g',W')$, $(h',J'):(g',W')\rightarrow (g'',W'')$, 
the composition $1$--cell 
$(h',J')\circ (h,J):(g,W)\rightarrow (g'',W'')$ is defined by 
\begin{equation}
(h',J')\circ (h,J)=\{h_ih'{}_i,m(h_i)(J'{}_{ij})J_{ij}\}.
\label{cech14}
\end{equation}

\item For any $1$--cell $(h,J)$, the identity $2$--cell $\id_{(h,J)}$ of $(h,J)$ 
is the collection $\{1_{Hi}\}$ of constant maps $U_i\rightarrow H$ with value $1_H$.

\item For any $2$--cell $K:(h,J)\Rightarrow (h',J')$, the vertical inverse $2$--cell
$K^{-1_\bfdot}:(h',J')\Rightarrow (h,J)$ is defined by  
\begin{equation}
K^{-1_\bfdot}=\{K_i{}^{-1}\}.
\label{cech15}
\end{equation}

\item For any two $2$--cells $K:(h,J)\Rightarrow (h',J')$, $K':(h',J')\Rightarrow (h'',J'')$,
the vertical composition $2$--cell $K'\bfdot K:(h,J)\Rightarrow (h'',J'')$ is defined by 
\begin{equation}
K'\bfdot K=\{K'{}_iK_i\}.
\label{cech16}
\end{equation}

\item For any $2$--cell $K:(h,J)\Rightarrow (h',J')$,  the horizontal inverse $2$--cell
$K^{-1_\circ}:(h,J)^{-1_\circ}\Rightarrow (h',J')^{-1_\circ}$ is defined by  
\begin{equation}
K^{-1_\circ}=\{m(h_i{}^{-1})(K_i{}^{-1})\}
\label{cech17}
\end{equation}

\item For any two $2$--cells $K:(h,J)\Rightarrow (h'',J'')$, $K':(h',J')\Rightarrow (h''',J''')$
such that the composition $1$--cells $(h',J')\circ (h,J)$, $(h''',J''')\circ (h'',J'')$ are defined 
the horizontal composition $2$--cell $K'\circ K:(h',J')\circ (h,J)\Rightarrow (h''',J''')\circ (h'',J'')$ 
is defined by \hphantom{xxxxxxxxxxxxxxxxxxxxxxxx}
\begin{equation}
K'\circ K=\{K_im(h_i)(K'{}_i)\}.
\label{cech18}
\end{equation}

\end{enumerate}
It is straightforward to check that $\check P_2(U,G,H)$ is indeed a strict  $2$--groupoid. 

The set $1$--isomorphisms classes of $0$--cells $(g,W)$ is the 1st \v Cech cohomology
$\check H^1(U,G,H)$ of the covering $U$ with coefficients in $(G,H)$. 
Again, the dependence on $U$ can be eliminated by switching to 
1st \v Cech cohomology $\check H^1(M,G,H)$ of $M$ with coefficients in $(G,H)$, the 
direct limit under  covering refinement of the cohomology $\check H^1(U,G,H)$, 
\begin{equation}
\check H^1(M,G,H)=\lim_{\overrightarrow{U}}\check H^1(U,G,H). 
\label{cech19}
\end{equation}
By analogy to the theory of ordinary principal bundles,  $\check H^1(M,G,H)$ is regarded 
as the set of isomorphism classes of smooth principal $(G,H)$--$2$--bundles $P$. 

For any two $0$--cells $(g,W)$, $(g',W')$, let $\check H^2(U,G,H;(g,W),(g',W'))$ be the set of
vertical $2$--isomorphism classes of $1$--cells $(g,W)\rightarrow (g',W')$. In case that it is non empty, 
$\check H^2(U,G,H;(g,W),(g',W'))$ depends 
only to the common $1$--isomorphism class of $(g,W)$ and $(g',W')$ in $\check H^1(U,G,H)$ up to bijection
so that we can set $(g',W')=(g,W)$ without loss of generality. 
Again, the dependence on $U$ can be eliminated by switching to 
2nd \v Cech cohomology $\check H^2(P)$ defined by 
\begin{equation}
\check H^2(P)=\lim_{\overrightarrow{U}}\check H^2(U,G,H;(g,W),(g,W)), 
\label{cech20}
\end{equation}
where $P$ is a principal $(G,H)$--$2$--bundles in the isomorphism class 
corresponding to $(g,W)$.
If $P$ is represented by a $0$--cell $(g,W)$, then $\check H^2(P)$ is represented by the group 
of vertical $2$--isomorphism classes of $1$--cells $(g,W)\rightarrow (g,W)$. $\check H^2(P)$ 
is so the group of isomorphism
classes of $1$--automorphisms of $P$, which we can identify with the gauge group $\Gau(P)$
again by analogy with the theory of ordinary principal bundles. 

The process does not stop here. 
For any two $1$--cells $(h,J),(h',J'):(g,W)\rightarrow(g,W)$, consider the set $\check H^3(U,G,H;(h,J),(h',J'))$ of 
$2$--cells $(h,J)\Rightarrow (h',J')$. If it is non empty, $\check H^3(U,G,H;(h,J),(h',J'))$  
depends only to the common isomorphism class of $(h,J)$, $(h',J')$ in $\check H^2(U,G,H;(g,W),(g,W))$ up to bijection,
so that we can set $(h',J')=(h,J)$. Once more, the dependence on $U$ is eliminated by switching  
3rd \v Cech cohomology $\check H^2(\Gamma)$
\begin{equation}
\check H^3(P,\varGamma)=\lim_{\overrightarrow{U}} \check H^3(U,G,H;(h,J),(h,J)), 
\label{cech21}
\end{equation}
where $P$ is a principal $(G,H)$--$2$--bundles in the isomorphism class 
corresponding to $(g,W)$ and $\varGamma$ is an $1$--automorphisms of $P$ in the isomorphism class associated with 
$(h,J)$. If $P$ is represented by the $0$--cell $(g,W)$ and $\varGamma$ is represented by a $1$--cell $(h,J)$, 
then $\check H^3(P,\varGamma)$ is represented by the vertical group of 
$2$--cells $(h,J)\Rightarrow (h,J)$. $\check H^3(P,\varGamma)$ is thus the group of $2$--automorphisms of $\varGamma$, 
which we may view as the gauge for gauge group $\Gau(P,\varGamma)$.  

The description of principal $2$--bundle we have formulated above looks more transparent though
a bit more abstract when rephrased directly in terms of the strict Lie $2$--group $V$ underlying
the Lie crossed module $(G,H)$. Let us elaborate on this point. 

Let $G$ be a Lie group. As we have reviewed above, isomorphism classes of ordinary principal $G$--bundles 
are characterized in terms of gluing data $g=\{g_{ij}\}$ satisfying the condition 
\eqref{cech2} and determined up to an equivalence defined in terms of intertwiner data $h=\{h_i\}$ 
satisfying \eqref{cech3}. The intuitive idea underlying the definition of principal $2$--bundles 
is that of extending the Lie group $G$ to a strict Lie $2$--group $V=(V_1,V_2)$
with $G=V_1$ and ``weakening'' \eqref{cech2}, \eqref{cech3} so that they hold only up to 
$2$--cell errors $W=\{W_{ijk}\}$ and $J=\{J_{ij}\}$ drawn from $V_2$ and satisfying natural 
coherence conditions. Correspondingly, the groupoid $\check P(U,G)$ extends to a strict 
$2$--groupoid $\check P_2(U,V)$. In Lie crossed module theoretic terms, $\check P_2(U,V)$ is just 
the $2$--groupoid $\check P_2(U,G,H)$ we have studied above. 

Explicitly, the content of $\check P_2(U,V)$ can be described as follows.
\begin{enumerate}

\item A $0$--cell of $\check P_2(U,V)$ is a collection $(g,W)=\{g_{ij},W_{ijk}\}$
of smooth maps $g_{ij}\in\Map(U_{ij},V_1)$, $W_{ijk}\in\Map(U_{ijk},V_2)$ such that 
\hphantom{xxxxxxxxxxxxxxxxx}
\begin{subequations}
\begin{equation}
W_{ijk}:g_{ik}\Rightarrow g_{ij}\circ g_{jk}, \qquad \text{on $U_{ijk}$},
\label{cech22a}
\end{equation}
where $\circ$ denotes pointwise multiplication in $V_1$, satisfying 
in addition the coherence condition
\begin{equation}
(1_{g_{ij}}\circ W_{jkl})\bfdot W_{ijl}=(W_{ijk}\circ 1_{g_{kl}})\bfdot W_{ikl},
\qquad \text{on $U_{ijkl}$},
\label{cech22b}
\end{equation}
\label{cech22}
\end{subequations}
\!\!where $\circ$ and $\bfdot$ denote pointwise horizontal and vertical multiplication in $V_2$,
respectively. \eqref{cech22b} follows from equating the two pointwise $2$--cells 
$g_{il}\Rightarrow g_{ij}\circ g_{jk}\circ g_{kl}$ that can be built using \eqref{cech22a}.

\item  For any two $0$--cells $(g,W)$, $(g',W')$, a $1$--cell $(g,W)\rightarrow (g',W')$ of $\check P_2(U,V)$
is a collection $(h,J)=\{h_i,J_{ij}\}$ of smooth maps $h_i\in\Map(U_i,V_1)$,
$J_{ij}\in\Map(U_{ij},V_2)$ such that 
\begin{subequations}
\begin{equation}
J_{ij}:g_{ij}\circ h_j\Rightarrow h_i\circ g'{}_{ij}, \qquad \text{on $U_{ij}$},
\label{cech23a}
\end{equation}
satisfying the coherence condition
\begin{equation}
(J_{ij}\circ 1_{g'{}_{jk}})\bfdot (1_{g_{ij}}\circ J_{jk})\bfdot (W_{ijk}\circ 1_{h_k})
=(1_{h_i}\circ W'{}_{ijk})\bfdot J_{ik},\qquad \text{on $U_{ijk}$}
\label{cech23b}
\end{equation}
\label{cech23}
\end{subequations}
\!\!stemming from imposing that the two pointwise $2$--cells 
$g_{ik}\circ h_k\Rightarrow h_i\circ g'{}_{ij}\circ g'{}_{jk}$ which can be built using
\eqref{cech22a}, \eqref{cech23a} are equal. 

\item For any two $1$--cells $(h,J), (h',J'):(g,W)\rightarrow (g',W')$, a $2$--cell 
$(h,J)\Rightarrow (h',J')$  of $\check P_2(U,V)$ is a collection  $K=\{K_i\}$ of smooth maps 
$K_i\in\Map(U_i$, $V_2)$ satisfying \hphantom{xxxxxxxxxxxxxxxxxxxxxx}
\begin{subequations}
\begin{equation}
K_i:h_i\Rightarrow h'{}_i, \qquad \text{on $U_i$},
\label{cech24a}
\end{equation}
and the coherence condition 
\begin{equation}
J'{}_{ij}\bfdot(1_{g_{ij}}\circ K_j)=(K_i\circ 1_{g'{}_{ij}})\bfdot J_{ij}, \qquad \text{on $U_{ij}$},
\label{cech24b}
\end{equation}
\label{cech24}
\end{subequations}
\!\!equating the two pointwise $2$--cells $g_{ij}\circ h_j\Rightarrow h'{}_i\circ g'{}_{ij}$ 
built using \eqref{cech23a}, \eqref{cech24a}. 

\end{enumerate}
It is straightforward to check that, \pagebreak when written in $(G,H)$ terms, \eqref{cech22}--\eqref{cech24} 
precisely reproduce \eqref{cech10}--\eqref{cech12}, respectively.  

The description of principal bundles in terms of gluing data and 
their equivalence can be approached from alternative point of view 
advocated by Schreiber \cite{Schreiber:2005ff,Schreiber2011}, developing upon the results of refs.
\cite{Baez:2004in,Baez:2005qu}. We shall  describe how the construction works for principal bundles
and then we shall show how it generalizes to principal $2$--bundles.

We begin by introducing the {\it \v Cech groupoid} $\check C(U)$ of the covering $U=\{U_i\}$.
$\check C(U)$ is defined as follows. 
\begin{enumerate}

\item A $0$--cell of $\check C(U)$ is a pair $(m,i)$ with $m\in U_i$.

\item A $1$--cell of $\check C(U)$ is a triple $(m,i,j)$ with
$m\in U_{ij}$, constituting $1$--arrows $(m,i)\rightarrow (m,j)$.

\item The identity $1_{(m,i)}$ of the $0$--cell $(m,i)$ is the $1$--cell $(m,i,i)$. 

\item The composition $(m',j,k)\circ(m,i,j)$ of two $1$--cells $(m,i,j)$, $(m',j,k)$ with 
$m=m'$ is the $1$--cell $(m,i,k)$.

\item The inverse $(m,i,j)^{-1_\circ}$ of a $1$--cells $(m,i,j)$ is the $1$--cell $(m,j,i)$.

\end{enumerate}
The sets of $0$-- and $1$--cells of $\check C(U)$ are the disjoint unions $\coprod_iU_i$ and $\coprod_{ij}U_{ij}$, 
respectively, and so have a smooth structure induced by that of $M$, providing a smooth structure 
to $\check C(U)$. 

Let $G$ be a Lie group. The delooping groupoid $BG$ of $G$ is just $G$ seen as the groupoid such that 
$BG_0=\{*\}$, the singleton set, and $BG_1=G$ with the smooth structure induced by that of $G$. 

A $0$--cell $g$ of the groupoid $\check P(U,G)$ is equivalent to a smooth groupoid morphism 
$g:\check C(U)\rightarrow BG$
from the \v Cech groupoid $\check C(U)$ of $U$ to the delooping groupoid $BG$ of $G$. 
Indeed, a set of smooth gluing data $g=\{g_{ij}\}$ satisfying \eqref{cech2}
defines a smooth functor from $\check C(U)$ to $BG$ 
mapping the $1$--cell $(m,i,j)$ to the $1$--cell $g_{ji}(m)$. 
A $1$--cell $h:g\rightarrow g'$ of the groupoid $\check P(U,G)$ is equivalent to a smooth natural
isomorphism of the corresponding groupoid morphisms $g, g':\check C(U)\rightarrow BG$.
Indeed, a set of smooth intertwining data $h=\{h_i\}$ satisfying \eqref{cech3}
defines an invertible smooth natural transformation of the functors $g,g'$ 
mapping the $0$--cell $(m,i)$ to the $1$--cell $h_i(m)$. In this way,
a bijection is established between the isomorphism classes of 
smooth principal $G$--bundles $P$ and the natural isomorphism classes
of smooth functors $\check C(U)\rightarrow BG$. The theory of such principal bundle classes 
is therefore fully encoded the functor category $[\check C(U),BG]$.

The \v Cech $2$--groupoid $\check C_2(U)$ of the covering $U$ 
is the smooth strict $2$--groupoid obtained by promoting 
the \v Cech groupoid $\check C(U)$ to a $2$--groupoid by adding identity $2$--cells.

Let $V$ be a Lie $2$--group. The delooping $2$--groupoid $BV$ of $V$ is just $V$ seen as the strict $2$--groupoid 
such that $BV_0=\{*\}$, the singleton set, $BV_1=V_1$ and $BV_2=V_2$ with the smooth structure induced by 
that of $V$. 

A $0$--cell $(g,W)$ of the groupoid $\check P_2(U,V)$ is equivalent to a smooth $2$--groupoid pseudomorphism 
$(g,W):\check C_2(U)\rightarrow BV$ from the \v Cech $2$--groupoid $\check C_2(U)$ of $U$ to the delooping $2$--groupoid 
$BV$ of $V$. A $2$--groupoid pseudomorphism is a pseudofunctor, that is an arrow $\varPhi:A\rightarrow B$ of 
$2$--categories associating to each $0$-- and each $1$--cell of $A$ a $0$-- and $1$--cell of $B$, respectively,  
as a functor does, but preserving identity $1$--cells and $1$--cell compositions only up to $2$--cells satisfying 
certain coherence conditions. A set of smooth data $\{g_{ij},W_{ijk}\}$ satisfying \eqref{cech22} 
precisely defines a smooth pseudofunctor from $\check C_2(U)$ to $BV$ 
mapping each $1$--cell $(m,i,j)$ to the $1$--cell $g_{ji}(m)$
preserving the identities $1$--cells $(m,i,i)$ only up to the $2$--cells 
$1_{g_{ii}{}^{-1}}\circ W_{iii}:1\Rightarrow g_{ii}$
and the $1$--cell compositions $(m,i,k)=(m,j,k)\circ(m,i,j)$ only up the $2$--cells 
$W_{kji}:g_{ki}\Rightarrow g_{kj}\circ g_{ji}$. 

A $1$--cell $(h,J):(g,W)\rightarrow (g',W')$ of the groupoid $\check P_2(U,V)$ is equivalent to a smooth 
pseudonatural isomorphism of the corresponding $2$--groupoid pseudomorphisms 
$(g,W), (g',W'):\check C_2(U)\rightarrow BV$. Given two arrows $\varPhi,\varPsi:A\rightarrow B$ of 
$2$--categories, a pseudonatural isomorphism is a $2$--arrow $\alpha: \varPhi\Rightarrow \varPsi$ 
associating to each object $a$ of $A$ an arrow $\alpha_a:\varPhi(a)\rightarrow \varPsi(a)$ of $B$
in such a way that the well--known conditions defining a natural transformation of functors is fulfilled 
only up to a $2$--cells of $B$ again satisfying certain coherence conditions. 
A set of smooth data $\{h_i,J_{ij}\}$ satisfying \eqref{cech23} precisely 
defines a smooth pseudonatural isomorphism of the pseudofunctors $(g,W)$, $(g',W')$
mapping the each $0$--cell $(m,i)$ to the $1$--cell $h_i(m)$
intertwining the $1$--cells $g_{ji}(m)$, $g'{}_{ji}(m)$ corresponding to the $(m,i,j)$ 
only up to the $2$--cells $J_{ij}:g_{ij}\circ h_j\Rightarrow h_i\circ g'{}_{ij}$.

In this way,
a bijection is established between the isomorphism classes pf 
smooth principal $V$--$2$--bundles $P$ and the pseudonatural isomorphism classes
of pseudofunctors $\check C_2(U)\rightarrow BV$. The theory of such principal $2$--bundle classes 
is therefore completely encoded in the pseudofunctor category $[\check C_2(U),BV]$.


\subsection{\normalsize \textcolor{blue}{$2$--term $L_\infty$ algebra gauge theory, global  aspects}}
\label{subsec:linfdglob}

\hspace{.5cm} Now, we have all the elements necessary for the analysis of 
the global aspects of 
semistrict higher gauge theory. 

Let $M$ be a smooth $d$--fold. Though $M$ is not necessarily diffeomorphic to 
$\mathbb{R}^d$, it admits an open covering $\{U_i\}$ such that the all sets $U_i$ 
as well as all their non empty finite intersections are. 

To understand fully  how things work in higher gauge theory, we begin again 
with considering what happens in an ordinary gauge theory with structure 
Lie algebra $\mathfrak{g}$. 
A generic field $\mathcal{F}$ on $M$ is not in general a vector valued function globally 
defined on $M$, but instead is given as a collection $\{\mathcal{F}_i\}$, where 
$\mathcal{F}_i$ is a vector valued function defined locally on $U_i$. 
$\mathcal{F}_i$ can be viewed as the representation of
$\mathcal{F}$ with respect to a local vector frame on $U_i$.
$\mathcal{F}_i$, $\mathcal{F}_j$ are thus related by a frame change 
on every $U_{ij}$. The $\mathcal{F}_i$ are typically $\mathfrak{g}$ valued fields 
and are so acted upon by the gauge transformation group $\Gau(U_i,\mathfrak{g})$
(cf. subsect. \ref{subsec:gauact}). 
It is natural to require that the frame change occurring on each 
$U_{ij}$ is given by a gauge transformation $g_{ij}\in\Gau(U_{ij},\mathfrak{g})$, 
so that we have 
\begin{equation}
\mathcal{F}_i={}^{g_{ij}}\mathcal{F}_j,\qquad\text{on $U_{ij}$}. 
\label{plinfdglob0}
\end{equation}
The gluing data $g_{ij}$  are defined up to a certain equivalence relation
amd must satisfy a coherence condition analogous to those of principal bundle theory.
To study the associated class of topological and geometrical structures, 
we introduce a groupoid $\check{\mathcal{P}}(U,\mathfrak{g})$
defined analogously to the groupoid $\check P(U,G)$ of 
subsect. \ref{subsec:twobund} by replacing the mapping group $\Map(\cdot,G)$ by the gauge transformation 
group $\Gau(\cdot,\mathfrak{g})$ throughout (cf. subsect. \ref{subsec:linfgautrsf}). 
$\check{\mathcal{P}}(U,\mathfrak{g})$ can thus be described in the following terms.
\begin{enumerate}

\item A $0$--cell of $\check{\mathcal{P}}(U,\mathfrak{g})$ is collection $g=\{g_{ij}\}$ 
of gauge transformations $g_{ij}\in\Gau(U_{ij},\mathfrak{g})$ satisfying the condition 
\begin{equation}
g_{ij}\diamond g_{jk}=g_{ik}, \qquad \text{on $U_{ijk}$}.
\label{pcech2}
\end{equation}

\item For any two $0$--cells $g$, $g'$ of $\check{\mathcal{P}}(U,\mathfrak{g})$, a $1$--cell $g\rightarrow g'$
is a collection $\{h_i\}$ of gauge transformations $h_i\in\Gau(U_i,\mathfrak{g})$ such that
\begin{equation}
g_{ij}\diamond h_j=h_i\diamond  g'{}_{ij}, \qquad \text{on $U_{ij}$}.
\label{pcech3}
\end{equation}

\end{enumerate}
Further, the groupoid operations are defined formally in the same way as those of  
$\check P(U,G)$.

Concretely, a $0$--cell $g$ is equivalent to a collection of data
$\{g_{ij},\sigma_{ij}\}$ with $g_{ij}\in\Map(U_{ij}$, $\Aut(\mathfrak{g}))$
and $\sigma_{ij}$ a connection on $U_{ij}$ such that
\begin{subequations}
\label{px1linfdglob}
\begin{align}
&d\sigma_{ij}+\frac{1}{2}[\sigma_{ij},\sigma_{ij}]=0,
\vphantom{\Big]}
\label{px1linfdgloba}
\\
&g_{ij}{}^{-1}dg_{ij}(x)-[\sigma_{ij},x]=0
\vphantom{\Big]}
\label{px1linfdglobb}
\end{align}
\end{subequations}
\!\!satisfying the coherence conditions 
\vskip-.78truecm
\begin{subequations}
\label{px4linfdglob}
\begin{align}
&g_{ij}g_{jk}=g_{ik},
\vphantom{\Big]}
\label{px4linfdglobz}
\\
&\sigma_{ik}-\sigma_{jk}-g_{jk}{}^{-1}(\sigma_{ij})=0.
\vphantom{\Big]}
\label{px4linfdgloba}
\end{align}
\end{subequations}
For any two $0$--cell $g$, $g'$ of $\check{\mathcal{P}}(U,\mathfrak{g})$, a $1$--cell 
$h:g\rightarrow g'$ is equivalent to a collection of data $\{h_i,\pi_i\}$ with 
$h_i\in\Map(U_i,\Aut(\mathfrak{g}))$ and $\pi_i$ a connection on $U_i$ with 
\begin{subequations}
\label{qx1linfdglo}
\begin{align}
&d\pi_i+\frac{1}{2}[\pi_i,\pi_i]=0,
\vphantom{\Big]}
\label{qx1linfdgloba}
\\
&h_i{}^{-1}dh_i(x)-[\pi_i,x]=0
\vphantom{\Big]}
\label{qx1linfdglobb}
\end{align}
\end{subequations}
\!\!satisfying the coherence conditions
\begin{subequations}
\label{qx4linfdglob}
\begin{align}
&g_{ij}h_j=h_ig'{}_{ij},
\vphantom{\Big]}
\label{qx4linfdglobz}
\\
&\sigma'{}_{ij}-h_j{}^{-1}(\sigma_{ij})-\pi_j+g'{}_{ij}{}^{-1}(\pi_i)=0.
\vphantom{\Big]}
\label{qx4linfdgloba}
\end{align}
\end{subequations}
By construction, 
the above topological set up is formally analogous to that of principal bundle theory. 
Yet it is not possible 
to frame it in principal bundle theoretic terms.
Any attempt at casting the groupoid $\check{\mathcal{P}}(U,\mathfrak{g})$
as a groupoid of the form $\check P(U,\widehat{G})$ for some Lie group
$\widehat{G}$ and interpret $1$--cell isomorphism classes of $0$--cells
of $\check{\mathcal{P}}(U,\mathfrak{g})$ as ones of $\check P(U,\widehat{G})$, hence  
as isomorphisms classes of principal 
$\widehat{G}$--bundles, will fail because of the differential conditions
\eqref{px1linfdglob}, \eqref{qx1linfdglo} obeyed by the $1$--form data. 
We note however that a $1$--cell isomorphism class of $0$--cells
in $\check{\mathcal{P}}(U,\mathfrak{g})$ yields a $1$--cell isomorphism class of $0$--cells
in $\check P(U,\Aut(\mathfrak{g}))$ obtained by the forgetful map that 
keeps the data $g_{ij}$ and $h_i$ but drops the data $\sigma_{ij}$ and $\pi_i$, 
thus an isomorphism class of principal $\Aut(\mathfrak{g})$--bundles. 
Thus, a principal $\Aut(\mathfrak{g})$--bundle is part of the
structure defined by the $0$--cell, but it does exhaust it. 
Similar remarks holds for the counterpart of principal bundle 
automorphisms.

It is important to emphasize the points where the above gauge theoretic framework 
differs from the standard one. Consider a conventional gauge theory with structure 
group $G$. The topological background of the theory is then a principal $G$--bundle $P$ 
codified in a $1$--cell isomorphism class of $0$--cells $\gamma=\{\gamma_{ij}\}$
of $\check P(U,G)$. Since in gauge theory all fields are in the adjoint of $G$, 
the effective structure group is the adjoint group $\Ad G=G/Z(G)$ rather than $G$.  
Can we, then, replace $G$ by $\Ad G$? 
The crucial point is whether from the knowledge of the data $g_{ij}=\Ad\gamma_{ij}$ 
it is possible to reconstruct the data $\sigma_{ij}=\gamma_{ij}{}^{-1}d\gamma_{ij}$ 
which control the global matching of local 
connections. It depends on the structure group $G$. If $G$ is semisimple, e. g. $G=\mathrm{SU}(n)$, it 
can indeed be done. If $G$ is not semisimple, e. g. $G=\mathrm{U}(1)$, then it is no longer 
possible. However, we can still work with $\Ad G$ rather than $G$, if we give up the condition 
that the $\sigma_{ij}$ be determined by the $g_{ij}$ and regard the former as data (partially) 
independent from the latter, switching from $\check P(U,G)$ to $\check{\mathcal{P}}(U,\mathfrak{g})$ 
and controlling the global definedness of the fields using $1$--cell isomorphism class of 
$0$--cells $g=\{g_{ij},\sigma_{ij}\}$. 
It is clear that this can only work perturbatively, as important 
non perturbative effects are attached to the center $Z(G)$ of $G$. 
As $\Ad G \simeq\Inn(\mathfrak{g})\subset \Aut(\mathfrak{g})$, however, by 
working with $\check{\mathcal{P}}(U,\mathfrak{g})$ we are generalizing gauge theory 
including the case where the data $g_{ij}$ take values in the non inner automorphisms
of $\mathfrak{g}$. 

Let us now assume that the global properties of the fields of a gauge theory 
with structure Lie algebra $\mathfrak{g}$ of the generalized sort described in the previous paragraph
are defined by a $1$--cell isomorphism class of $0$--cells $g=\{g_{ij},\sigma_{ij}\}$ 
of $\check{\mathcal{P}}(U,\mathfrak{g})$. 

A connection $\omega$ on $M$ is a collection $\{\omega_i\}$ 
of connections $\omega_i$ on the sets $U_i$ satisfying the matching relation  
\eqref{plinfdglob0} with $\mathcal{F}=\omega$, the right hand side being given 
by \eqref{xgauact1}. 
The curvature $f$ of $\omega$ on $M$ is the collection $\{f_i\}$ of the curvatures 
$f_i$ of the $\omega_i$ and satisfies \eqref{plinfdglob0} 
with $\mathcal{F}=f$ and the right hand side given by \eqref{xgauact2}. 

A field $\phi$ on $M$ is a collection $\{\phi_i\}$ of fields
$\phi_i$ of the same type on the sets $U_i$ 
satisfying matching relations of the form \eqref{plinfdglob0} 
with $\mathcal{F}=\phi$,
the right hand side being given by \eqref{xgauact3}. 
The covariant derivative field  $D\phi$ of $\phi$ on $M$ 
is the collection $\{D\phi_i\}$ of the covariant derivative fields
$D\phi_i$ of the $\phi_i$ and satisfies also \eqref{plinfdglob0} 
with $\mathcal{F}=D\phi$ and the right hand side given by \eqref{xgauact4}. 

Similarly, a dual field $\upsilon$ on $M$ is a collection $\{\upsilon_i\}$ 
of dual fields $\upsilon_i$ of the same type on the sets $U_i$ such that \eqref{plinfdglob0}  
holds with $\mathcal{F}=\upsilon$ and the right hand side given by \eqref{xgauact5}. 
The covariant derivative dual field $D\upsilon$ on $M$ 
is the collection $\{D\upsilon_i\}$ of the covariant derivative dual fields $D\upsilon_i$ 
of the $\upsilon_i$ and satisfies matching relations of the form 
\eqref{plinfdglob0} with $\mathcal{F}=D\upsilon$ and
the right hand side given by \eqref{xgauact6}. 

In a $2$--term $L_\infty$ algebra gauge theory with structure algebra $\mathfrak{v}$, things proceed
much in the same way. A generic field $\mathcal{F}$ on $M$ is again a collection 
$\{\mathcal{F}_i\}$ of local fields $\mathcal{F}_i$. Here, the $\mathcal{F}_i$
are fields of one of the types considered in subsect. \ref{subsec:linfdoub}
and so are acted upon by the $1$--gauge transformation group $\Gau_1(U_i,\mathfrak{v})$ 
(cf. subsect. \ref{subsec:gauact}). 
It is natural to require that the relationship of the local fields $\mathcal{F}_i$, $\mathcal{F}_j$
on $U_{ij}$ is given by a $1$--gauge transformation $g_{ij}\in\Gau_1(U_{ij},\mathfrak{v})$, 
\begin{equation}
\mathcal{F}_i={}^{g_{ij}}\mathcal{F}_j,\qquad\text{on $U_{ij}$}. \vphantom{\Bigg]}
\label{2plinfdglob0}
\end{equation}
in analogy to \eqref{plinfdglob0}.
Again, as in principal bundle theory, 
the gluing data $g_{ij}$  are defined up to a certain equivalence relation
and must satisfy a compatibility condition. 
This latter,  however, 
cannot be expressed in terms of the $g_{ij}$ only but requires the introduction of 
further gluing data not entering \eqref{2plinfdglob0}, 
$2$--gauge transformations $W_{ijk}\in\Gau_2(U_{ijk},\mathfrak{v})$,
satisfying a coherence condition.
In analogy with ordinary gauge theory, 
to study the associated class of topological and geometrical structures, 
we introduce a strict $2$--groupoid $\check{\mathcal{P}}_2(U,\mathfrak{v})$
defined analogously to the groupoid $\check P_2(U,V)$ studied in subsect. 
\ref{subsec:twobund} by replacing the mapping strict $2$--group $\Map(\cdot,V)$ by the $2$--term $L_\infty$ algebra
gauge transformation $2$--group $\Gau(\cdot,\mathfrak{v})$ throughout 
(cf. subsect. \ref{subsec:linfgautrsf}). 
$\check{\mathcal{P}}_2(U,\mathfrak{v})$ can be described in the following terms. 
\begin{enumerate}

\item A $0$--cell of $\check{\mathcal{P}}_2(U,\mathfrak{v})$ 
is a collection $(g,W)=\{g_{ij},W_{ijk}\}$
of $1$--gauge transformations $g_{ij}\in\Gau_1(U_{ij},\mathfrak{v})$ 
and $2$--gauge transformations $W_{ijk}\in\Gau_2(U_{ijk},\mathfrak{v})$ such that 
one has \hphantom{xxxxxxxxxxxxxxxxx}
\begin{subequations}
\begin{equation}
W_{ijk}:g_{ik}\Rightarrow g_{ij}\diamond g_{jk}, \qquad \text{on $U_{ijk}$},
\label{pcech22a}
\end{equation}
where $\diamond$ denotes the composition of $1$--cells in $\Gau_1(\cdot 
,\mathfrak{v})$, 
and satisfying the coherence condition
\begin{equation}
(I_{g_{ij}}\diamond W_{jkl})\bullet W_{ijl}=(W_{ijk}\diamond I_{g_{kl}})\bullet W_{ikl}.
\qquad \text{on $U_{ijkl}$},
\label{pcech22b}
\end{equation}
\label{pcech22}
\end{subequations}
\!\!where $\diamond$ and $\bullet$ denote 
the horizontal and vertical composition of $2$--cells in $\Gau_2(\cdot 
,\mathfrak{v})$,
respectively (cf. subsect. \ref{subsec:linfgautrsf}). 

\item  For any two $0$--cells $(g,W)$, $(g',W')$, a $1$--cell $(g,W)\rightarrow (g',W')$ of 
$\check{\mathcal{P}}_2(U,\mathfrak{v})$
is a collection $(h,J)=\{h_i,J_{ij}\}$ of $1$--gauge transformations $h_i\in\Gau_1(U_i,\mathfrak{v})$
and $2$--gauge transformations $J_{ij}\in\Gau_2(U_{ij},\mathfrak{v})$ such that 
\begin{subequations}
\begin{equation}
J_{ij}:g_{ij}\diamond h_j\Rightarrow h_i\diamond g'{}_{ij}, \qquad \text{on $U_{ij}$},
\label{pcech23a}
\end{equation}
satisfying the coherence condition
\begin{equation}
(J_{ij}\diamond I_{g'{}_{jk}})\bullet (I_{g_{ij}}\diamond J_{jk})\bullet (W_{ijk}\diamond I_{h_k})
=(I_{h_i}\diamond W'{}_{ijk})\bullet J_{ik}, \quad \text{on $U_{ijk}$}.
\label{pcech23b}
\end{equation}
\label{pcech23}
\end{subequations}
\vspace{-.5truecm}

\item For any two $1$--cells $(h,J), (h',J'):(g,W)\rightarrow (g',W')$, a $2$--cell 
$(h,J)\Rightarrow (h',J')$  of $\check{\mathcal{P}}_2(U,\mathfrak{v})$ is a collection  $K=\{K_i\}$ 
of $2$--gauge transformations 
$K_i\in\Gau_2(U_i$, $\mathfrak{v})$ satisfying 
\begin{subequations}
\begin{equation}
K_i:h_i\Rightarrow h'{}_i, \qquad \text{on $U_i$},
\label{pcech24a}
\end{equation}
and the coherence condition 
\begin{equation}
J'{}_{ij}\bullet(I_{g_{ij}}\diamond K_j)=(K_i\diamond I_{g'{}_{ij}})\bullet J_{ij}, \qquad \text{on $U_{ij}$}.
\label{pcech24b}
\end{equation}
\label{pcech24}
\end{subequations}
\end{enumerate}
\vspace{-.5truecm}
\eject\noindent
The $2$--groupoid operations are defined formally in the same way as in $\check P_2(U,V)$. 

Concretely, a $0$--cell $(g,W)$ is equivalent to a collection of data 
$\{g_{ij},\sigma_{ij},\varSigma_{ij}$, $\tau_{ij},W_{ijk},A_{ijk}\}$ 
with $g_{ij}\in\Map(U_{ij},\Aut_1(\mathfrak{v}))$, 
$(\sigma_{ij},\varSigma_{ij})$ a connection doublet on $U_{ij}$, 
$\tau_{ij}\in\Omega^1(U_{ij},\Hom(\hat{\mathfrak{v}}_0,\hat{\mathfrak{v}}_1))$,  
$W_{ijk}\in\Map(U_{ijk},\Aut_2(\mathfrak{v}))$ and 
$A_{ijk}\in\Omega^1(U_{ijk},\hat{\mathfrak{v}}_1)$  
satisfying the relations 
\begin{subequations}
\label{p1linfdglob}
\begin{align}
&d\sigma_{ij}+\frac{1}{2}[\sigma_{ij},\sigma_{ij}]-\partial\varSigma_{ij}=0,
\vphantom{\Big]}
\label{p1linfdgloba}
\\
&d\varSigma_{ij}+[\sigma_{ij},\varSigma_{ij}]-\frac{1}{6}[\sigma_{ij},\sigma_{ij},\sigma_{ij}]=0,
\vphantom{\Big]}
\label{p1linfdglobb}
\\
&d\tau_{ij}(x)+[\sigma_{ij},\tau_{ij}(x)]-[x,\varSigma_{ij}]+\frac{1}{2}[\sigma_{ij},\sigma_{ij},x]
\vphantom{\Big]}
\label{p2linfdglob}
\\
&\qquad\qquad\qquad\qquad\qquad\qquad
+\tau_{ij}([\sigma_{ij},x]+\partial\tau_{ij}(x))=0, \hspace{3cm}
\vphantom{\Big]}
\nonumber
\\
&g_{ij0}{}^{-1}dg_{ij0}(x)-[\sigma_{ij},x]-\partial\tau_{ij}(x)=0,
\vphantom{\Big]}
\label{p3linfdgloba}
\\
&g_{ij1}{}^{-1}dg_{ij1}(X)-[\sigma_{ij},X]-\tau_{ij}(\partial X)=0,
\vphantom{\Big]} 
\label{p3linfdglobb}
\\
&g_{ij1}{}^{-1}(dg_{ij2}(x,y)-g_{ij2}(g_{ij0}{}^{-1}dg_{ij0}(x),y)-g_{ij2}(x,g_{ij0}{}^{-1}dg_{ij0}(y)))
\vphantom{\Big]}
\label{p3linfdglobc}
\\
&\qquad\qquad\qquad -[\sigma_{ij},x,y]-\tau_{ij}([x,y])+[x,\tau_{ij}(y)]-[y,\tau_{ij}(x)]=0,
\vphantom{\Big]}
\nonumber
\\
&W_{ijk}:g_{ik}\Rightarrow g_{ij}\circ g_{jk},
\vphantom{\Big]}
\label{ppcech22a}
\\
&\sigma_{jk}-\sigma_{ik}+g_{jk0}{}^{-1}(\sigma_{ij})+\partial A_{ijk}=0,
\vphantom{\Big]}
\label{p0linfdgloba}
\\
&\varSigma_{jk}
-\varSigma_{ik}+g_{jk1}{}^{-1}(\varSigma_{ij})
+\frac{1}{2} g_{jk1}{}^{-1}g_{jk2}(g_{jk0}{}^{-1}(\sigma_{ij}),g_{jk0}{}^{-1}(\sigma_{ij}))
\vphantom{\Big]}
\label{p0linfdglobb}
\\
&\qquad\qquad-\tau_{jk}(g_{jk0}{}^{-1}(\sigma_{ij}))
+dA_{ijk}+[\sigma_{ik},A_{ijk}]-\frac{1}{2}[\partial A_{ijk},A_{ijk}]=0,
\vphantom{\Big]}
\nonumber
\\
&\tau_{jk}(x)-\tau_{ik}(x)+g_{jk1}{}^{-1}(\tau_{ij}(g_{jk0}(x)))
-g_{jk1}{}^{-1}g_{jk2}(g_{jk0}{}^{-1}(\sigma_{ij}),x)
\vphantom{\Big]}
\label{p0linfdglobc}
\\
&\qquad\qquad+[x,A_{ijk}]+(g_{ij1}g_{jk1})^{-1}\big(dW_{ijk}(x)-W_{ijk}([\sigma_{ik},x]+\partial\tau_{ik}(x))\big)=0.
\vphantom{\Big]}
\nonumber
\end{align}
\end{subequations}
\!\!and satisfying the coherence conditions 
\begin{subequations}
\begin{align}
&(1_{g_{ij}}\circ W_{jkl})\bfdot W_{ijl}=(W_{ijk}\circ 1_{g_{kl}})\bfdot W_{ikl},
\vphantom{\Big]}
\label{pppcech22a}
\\
&A_{jkl}-A_{ikl}+A_{ijl}-g_{kl1}{}^{-1}(A_{ijk})-g_{jl1}{}^{-1}W_{jkl}(g_{jk0}g_{kl0})^{-1}(\sigma_{ij})=0.
\vphantom{\Big]}
\label{pppcech22b}
\end{align}
\label{ppcech22}
\end{subequations}
\!\!For any two $0$--cells $(g,W)$, $(g',W')$, a $1$--cell $(g,W)\rightarrow (g',W')$ \pagebreak 
is equivalent to a collection of data 
$\{h_i,\pi_i,\varPi_i,\rho_i,J_{ij},D_{ij}\}$ 
with $h_i\in\Map(U_i,\Aut_1(\mathfrak{v}))$, 
$(\pi_i,\varPi_i)$ a connection doublet on $U_i$, 
$\rho_i\in\Omega^1(U_i,\Hom(\hat{\mathfrak{v}}_0,\hat{\mathfrak{v}}_1))$,  
$J_{ij}\in\Map(U_{ij},\Aut_2(\mathfrak{v}))$ and 
$D_{ij}\in\Omega^1(U_{ij},\hat{\mathfrak{v}}_1)$  
satisfying the relations 
\begin{subequations}
\label{q1linfdglob}
\begin{align}
&d\pi_i+\frac{1}{2}[\pi_i,\pi_i]-\partial\varPi_i=0,
\vphantom{\Big]}
\label{q1linfdgloba}
\\
&d\varPi_i+[\pi_i,\varPi_i]-\frac{1}{6}[\pi_i,\pi_i,\pi_i]=0,
\vphantom{\Big]}
\label{q1linfdglobb}
\\
&d\rho_i(x)+[\pi_i,\rho_i(x)]-[x,\varPi_i]+\frac{1}{2}[\pi_i,\pi_i,x]
\vphantom{\Big]}
\label{q2linfdglob}
\\
&\qquad\qquad\qquad\qquad\qquad\qquad
+\rho_i([\pi_i,x]+\partial\rho_i(x))=0,
\vphantom{\Big]}
\nonumber
\\
&h_{i0}{}^{-1}dh_{i0}(x)-[\pi_i,x]-\partial\rho_i(x)=0,
\vphantom{\Big]}
\label{q3linfdgloba}
\\
&h_{i1}{}^{-1}dh_{i1}(X)-[\pi_i,X]-\rho_i(\partial X)=0,
\vphantom{\Big]} 
\label{q3linfdglobb}
\\
&h_{i1}{}^{-1}(dh_{i2}(x,y)-h_{i2}(h_{i0}{}^{-1}dh_{i0}(x),y)-h_{i2}(x,h_{i0}{}^{-1}dh_{i0}(y)))
\vphantom{\Big]}
\label{q3linfdglobc}
\\
&\qquad\qquad -[\pi_i,x,y]-\rho_i([x,y])+[x,\rho_i(y)]-[y,\rho_i(x)]=0,
\vphantom{\Big]}
\nonumber
\\
&J_{ij}:g_{ij}\circ h_j\Rightarrow h_i\circ g'{}_{ij}, 
\vphantom{\Big]}
\label{qqcech23a}
\\
&\sigma'{}_{ij}-h_{j0}{}^{-1}(\sigma_{ij})-\pi_j+g'{}_{ij0}{}^{-1}(\pi_i)+\partial D_{ij}=0, 
\vphantom{\Big]}
\label{q0linfdgloba}
\\
&\varSigma'{}_{ij}-h_{j1}{}^{-1}(\varSigma_{ij})
-\frac{1}{2} h_{j1}{}^{-1}h_{j2}(h_{j0}{}^{-1}(\sigma_{ij}),h_{j0}{}^{-1}(\sigma_{ij}))
\vphantom{\Big]}
\label{q0linfdglobb}
\\
&+\rho_j(h_{j0}{}^{-1}(\sigma_{ij}))-\varPi_j+g'{}_{ij1}{}^{-1}(\varPi_i)
+\frac{1}{2} g'{}_{ij1}{}^{-1}g'{}_{ij2}(g'{}_{ij0}{}^{-1}(\pi_i),g'{}_{ij0}{}^{-1}(\pi_i))
\vphantom{\Big]}
\nonumber
\\
&-\tau'{}_{ij}(g'{}_{ij0}{}^{-1}(\pi_i))+dD_{ij}+[\sigma'{}_{ij}+g'{}_{ij0}{}^{-1}(\pi_i),D_{ij}]+\frac{1}{2}[\partial D_{ij},D_{ij}]=0, 
\vphantom{\Big]}
\nonumber
\\
&\tau'{}_{ij}(x)-h_{j1}{}^{-1}(\tau_{ij}(h_{j0}(x)))+h_{j1}{}^{-1}h_{j2}(h_{j0}{}^{-1}(\sigma_{ij}),x)
\vphantom{\Big]}
\label{q0linfdglobc}
\\
&-\rho_j(x)+g'{}_{ij1}{}^{-1}(\rho_i(g'{}_{ij0}(x)))
-g'{}_{ij1}{}^{-1}g'{}_{ij2}(g'{}_{ij0}{}^{-1}(\pi_i),x)
\vphantom{\Big]}
\nonumber
\\
&+[x,D_{ij}]+(g_{ij1}h_{j1})^{-1}\big(dJ_{ij}(x)-J_{ij}\big([\sigma'{}_{ij}+g'{}_{ij0}{}^{-1}(\pi_i),x]
\vphantom{\Big]}
\nonumber
\\
&+\partial(\tau'{}_{ij}(x)+g'{}_{ij1}{}^{-1}(\rho_i(g'{}_{ij0}(x)))
-g'{}_{ij1}{}^{-1}g'{}_{ij2}(g'{}_{ij0}{}^{-1}(\pi_i),x))\big)\big)=0
\vphantom{\Big]}
\nonumber
\end{align}
\end{subequations}
\!\!and satisfying the coherence conditions 
\begin{subequations}
\begin{align}
&(J_{ij}\circ 1_{g'{}_{jk}})\bfdot (1_{g_{ij}}\circ J_{jk})\bfdot (W_{ijk}\circ 1_{h_k})
=(1_{h_i}\circ W'{}_{ijk})\bfdot J_{ik},
\vphantom{\Big]}
\label{qqcech23b}
\end{align}
\begin{align}
&D_{jk}-D_{ik}+g'{}_{jk1}{}^{-1}(D_{ij})-(g_{jk1}h_{k1})^{-1}J_{jk}(h_{j0}g'{}_{jk0})^{-1}(\sigma_{ij})
\vphantom{\Big]}
\label{qqqcech23b}
\\
&+g'{}_{ik1}{}^{-1}W'{}_{ijk}(g'{}_{ij0}g'{}_{jk0})^{-1}(\pi_i)
-A'{}_{ijk}+h_{k1}{}^{-1}(A_{ijk})=0
\vphantom{\Big]}
\nonumber
\end{align}
\label{qqcech23}
\end{subequations}
\!\!For any two $1$--cells $(h,J), (h',J'):(g,W)\rightarrow (g',W')$, a $2$--cell 
$(h,J)\Rightarrow (h',J')$ is equivalent to a collection  of data $\{K_i,C_i\}$ 
with $K_i\in\Map(U_i,\Aut_2(\mathfrak{v}))$ and $C_i\in\Omega^1(U_i,\hat{\mathfrak{v}}_1)$  
satisfying the relations 
\begin{subequations}
\begin{align}
&K_i:h_i\Rightarrow h'{}_i
\vphantom{\Big]}
\label{qqcech24a}
\\
&\pi_i-\pi'{}_i=\partial C_i, 
\vphantom{\Big]}
\label{qqcech24b}
\\
&\varPi_i-\varPi'{}_i=dC_i+[\pi'{}_i,C_i]+\frac{1}{2}[\partial C_i,C_i], 
\vphantom{\Big]}
\label{qqcech24c}
\\
&\rho_i(x)-\rho'{}_i(x)=[x,C_i]+h_{i1}{}^{-1}(dK_i(x)-K_i([\pi'{}_i,x]+\partial\rho'{}_i(x))).
\vphantom{\Big]}
\label{}
\end{align}
\label{qqcech24}
\end{subequations}
\!\!and satisfying the coherence condition 
\begin{equation}
D'{}_{ij}-D_{ij}+C_j-g'{}_{ij1}{}^{-1}(C_i)-h_{j1}{}^{-1}K_jh'{}_{j0}{}^{-1}(\sigma_{ij})=0.
\label{qqcech25}
\end{equation}

By construction, the above topological set up is formally analogous to that of
principal $2$--bundle theory of subsect. \ref{subsec:twobund}, but, 
similarly to ordinary gauge theory,  it is not possible to frame it in principal $2$--bundle theoretic terms.
The $2$--groupoid $\check{\mathcal{P}}_2(U,\mathfrak{v})$ cannot be cast 
as a $2$--groupoid of the form $\check P_2(U,\widehat{V})$ for some strict Lie $2$--group
$\widehat{V}$, $1$--cell isomorphism classes of $0$--cells
of $\check{\mathcal{P}}_2(U,\mathfrak{v})$ cannot be interpreted as ones of $\check P_2(U,\widehat{V})$, hence  
as isomorphism classes of principal $\widehat{V}$--$2$--bundles, because of the differential conditions
\eqref{p1linfdglob}, \eqref{q1linfdglob} obeyed by the $1$--and $2$ form data. 
Similarly to ordinary gauge theory again, a $1$--cell isomorphism class of $0$--cells
in $\check{\mathcal{P}}_2(U,\mathfrak{v})$ yields a $1$--cell isomorphism class of $0$--cells
in $\check P_2(U,\Aut(\mathfrak{v}))$ obtained by the forgetful map that 
keeps the data $g_{ij}$, $W_{ijk}$ and $h_i$, $J_{ij}$ but drops the data $\sigma_{ij}$, $\varSigma_{ij}$, $\tau_{ij}$,
$A_{ijk}$ and $\pi_i$, $\varPi_i$, $\rho_i$, $D_{ij}$, 
thus an isomorphism class of principal $\Aut(\mathfrak{v})$--$2$--bundles. 
In this way, a principal $\Aut(\mathfrak{v})$--$2$--bundle is part of the set up without exhausting it. 
Similar remarks hold for the counterpart of principal $2$--bundle 
automorphisms and automorphism for automorphisms.

It is important to relate the above gauge theoretic framework with others which have appeared previously 
in the literature, in particular \cite{Baez:2004in,Baez:2005qu} (see also \cite{Aschieri:2003mw}). 
Consider a higher gauge theory with a strict structure $2$--group $V$. The topological background of the theory 
is then a principal $V$--$2$--bundle $P$ codified in a $1$--cell isomorphism class of $0$--cells 
$(\gamma,\varTheta)=\{\gamma_{ij},\varTheta_{ijk}\}$ of $\check P_2(U,G,H)$, where $(G,H)$ is the Lie crossed 
module corresponding to $V$ (cf. subsect. \ref{subsec:twobund}). %
Unlike standard gauge theory, the data $\gamma_{ij}$, $\varTheta_{ijk}$ are not sufficient by themselves
to control the global matching of local connections. Further independent data $\chi_{ij}\in\Omega^1(U_{ij},\mathfrak{h})$ 
are required (cf. subsect. \ref{subsec:crmdgautrsf}). To frame all these data in a coherent whole
and to study the associated class of topological and geometrical structures, 
we introduce a strict $2$--groupoid $\check{\mathcal{P}}_2(U,G,H)$
defined analogously to the groupoid $\check P_2(U,G,H)$  
by replacing the mapping strict $2$--group $\Map(\cdot,G,H)$ by the crossed module 
gauge transformation strict $2$--group $\Gau(\cdot,G,H)$ studied in subsect. \ref{subsec:crmdgautrsf}
throughout. $\check{\mathcal{P}}_2(U,G,H)$ can be described as follows. 
\begin{enumerate}

\item A $0$--cell of $\check{\mathcal{P}}_2(U,G,H)$ 
is a collection $(\gamma,\varTheta)=\{\gamma_{ij},\varTheta_{ijk}\}$
of $1$--gauge transformations $\gamma_{ij}\in\Gau_1(U_{ij},G,H)$ 
and $2$--gauge transformations $\varTheta_{ijk}\in\Gau_2(U_{ijk},G,H)$ such that 
\begin{subequations}
\begin{equation}
\varTheta_{ijk}:\gamma_{ik}\Rightarrow \gamma_{ij}\diamond \gamma_{jk}, \qquad \text{on $U_{ijk}$},
\label{ypcech22a}
\end{equation}
where $\diamond$ denotes the composition of $1$--cells in $\Gau_1(\cdot 
,G,H)$, 
and satisfying the coherence condition
\begin{equation}
(I_{\gamma_{ij}}\diamond \varTheta_{jkl})\bullet \varTheta_{ijl}=(\varTheta_{ijk}\diamond I_{\gamma_{kl}})
\bullet \varTheta_{ikl}.
\qquad \text{on $U_{ijkl}$},
\label{ypcech22b}
\end{equation}
\label{ypcech22}
\end{subequations}
\!\!where $\diamond$ and $\bullet$ denote 
the horizontal and vertical composition of $2$--cells in $\Gau_2(\cdot,G,H)$,
respectively (cf. subsect. \ref{subsec:crmdgautrsf}). 

\item  For any two $0$--cells $(\gamma,\varTheta)$, $(\gamma',\varTheta')$, 
a $1$--cell $(\gamma,\varTheta)\rightarrow (\gamma',\varTheta')$ of 
$\check{\mathcal{P}}_2(U,G,H)$
is a collection $(\eta,\varUpsilon)=\{\eta_i,\varUpsilon_{ij}\}$ of $1$--gauge transformations 
$\eta_i\in\Gau_1(U_i,$ $G,H)$ and $2$--gauge transformations $\varUpsilon_{ij}\in\Gau_2(U_{ij},G,H)$ such that 
\begin{subequations}
\begin{equation}
\varUpsilon_{ij}:\gamma_{ij}\diamond \eta_j\Rightarrow \eta_i\diamond \gamma'{}_{ij}, \qquad \text{on $U_{ij}$},
\label{ypcech23a}
\end{equation}
satisfying the coherence condition
\begin{equation}
(\varUpsilon_{ij}\diamond I_{\gamma'{}_{jk}})\bullet (I_{\gamma_{ij}}\diamond \varUpsilon_{jk})
\bullet (\varTheta_{ijk}\diamond I_{\eta_k})
=(I_{\eta_i}\diamond \varTheta'{}_{ijk})\bullet \varUpsilon_{ik}, \quad \text{on $U_{ijk}$}.
\label{ypcech23b}
\end{equation}
\label{ypcech23}
\end{subequations}
\vspace{-.95truecm}

\item For any two $1$--cells $(\eta,\varUpsilon), (\eta',\varUpsilon'):(\gamma,\varTheta)
\rightarrow (\gamma',\varTheta')$, a $2$--cell 
$(\eta,\varUpsilon)\Rightarrow (\eta',\varUpsilon')$  of $\check{\mathcal{P}}_2(U,G,H)$ 
is a collection  $\varLambda=\{\varLambda_i\}$ 
of $2$--gauge transformations $\varLambda_i\in\Gau_2(U_i$, $G,H)$ satisfying 
\begin{subequations}
\begin{equation}
\varLambda_i:\eta_i\Rightarrow \eta'{}_i, \qquad \text{on $U_i$},
\label{ypcech24a}
\end{equation}
and the coherence condition 
\begin{equation}
\varUpsilon'{}_{ij}\bullet(I_{\gamma_{ij}}\diamond \varLambda_j)=(\varLambda_i\diamond 
I_{\gamma'{}_{ij}})\bullet \varUpsilon_{ij}, \qquad \text{on $U_{ij}$}.
\label{ypcech24b}
\end{equation}
\label{ypcech24}
\end{subequations}
\end{enumerate}
\vspace{-.95truecm}
The $2$--groupoid operations are defined formally in the same way as $\check P_2(U,G,H)$. 

Concretely, a $0$--cell $(\gamma,\varTheta)$ is equivalent to a collection of data
$\{\gamma_{ij},\chi_{ij},\varTheta_{ijk}$, $B_{ijk}\}$ with $\gamma_{ij}\in\Map(U_{ij},G)$
$\chi_{ij}\in\Omega^1(U_{ij},\mathfrak{h})$, $\varTheta_{ijk}\in\Map(U_{ijk},H)$,
$B_{ijk}\in\Omega^1(U_{ijk},\mathfrak{h})$ satisfying the relations
\begin{subequations}
\begin{align}
&\gamma_{ij}\gamma_{jk}=t(\varTheta_{ijk})\gamma_{ik},
\vphantom{\Big]}
\label{fpppcech21a}
\\
&\dot m(\gamma_{ij})(\chi_{jk})-\chi_{ik}+\chi_{ij}+B_{ijk}=0
\vphantom{\Big]}
\label{fpppcech21b}
\end{align}
\label{fppcech21}
\end{subequations}
\!\!and the coherence conditions
\begin{subequations}
\begin{align}
&m(\gamma_{ij})(\varTheta_{jkl})\varTheta_{ijl}=\varTheta_{ijk}\varTheta_{ikl},
\vphantom{\Big]}
\label{fpppcech22a}
\\
&\dot m(\gamma_{ij})(B_{jkl})-B_{ikl}+B_{ijl}-B_{ijk}-Q(\gamma_{ik}\dot t(\chi_{kl})\gamma_{ik}{}^{-1},\varTheta_{ijk})=0.
\vphantom{\Big]}
\label{fpppcech22b}
\end{align}
\label{fppcech22}
\end{subequations}
\!\!For any two $0$--cells $(\gamma,\varTheta)$, $(\gamma',\varTheta')$, \pagebreak
a $1$--cell $(\eta,\varUpsilon):(\gamma,\varTheta)\rightarrow (\gamma',\varTheta')$ 
is the same as a collection of data $\{\eta_i,\lambda_i,\varUpsilon_{ij},E_{ij}\}$ with 
$\eta_i\in\Map(U_i,G)$, 
$\lambda_i\in\Omega^1(U_i,\mathfrak{h})$, $\varUpsilon_{ij}\in\Map(U_{ij},H)$,
$E_{ij}\in\Omega^1(U_{ij},\mathfrak{h})$ such that 
\begin{subequations}
\begin{align}
&\eta_i\gamma'{}_{ij}=t(\varUpsilon_{ij})\gamma_{ij}\eta_j,
\vphantom{\Big]}
\label{fpppcech23a}
\\
&\dot m(\eta_i)(\chi'{}_{ij})-\chi_{ij}-\dot m(\gamma_{ij})(\lambda_j)+\lambda_i+E_{ij}=0,
\vphantom{\Big]}
\label{fpppcech23b}
\end{align}
\label{fppcech23}
\end{subequations}
\!\!and the coherence conditions
\begin{subequations}
\begin{align}
&\varUpsilon_{ij}m(\gamma_{ij})(\varUpsilon_{jk})\varTheta_{ijk}=m(\eta_i)(\varTheta'{}_{ijk})\varUpsilon_{ik},
\vphantom{\Big]}
\label{fpppcech24a}
\\
&\dot m(\eta_i)(B'{}_{ijk})-B_{ijk}-\dot m(\gamma_{ij})(E_{jk})+E_{ik}-E_{ij}
\vphantom{\Big]}
\label{fpppcech24b}
\\
&\hspace{.7cm}-Q(\gamma_{ij}\eta_j\dot t(\chi_{kl})\eta_j{}^{-1}\gamma_{ij}{}^{-1},\varUpsilon_{ij})
-Q(\gamma_{ik}\dot t(\lambda_k)\gamma_{ik}{}^{-1},\varTheta_{ijk})=0.
\vphantom{\Big]}
\nonumber
\end{align}
\label{fppcech24}
\end{subequations}
\!\!Finally, for any two $1$--cells $(\eta,\varUpsilon), (\eta',\varUpsilon'):(\gamma,\varTheta)
\rightarrow (\gamma',\varTheta')$, a $2$--cell $\varLambda:(\eta,\varUpsilon)\Rightarrow (\eta',\varUpsilon')$ 
is equivalent to a collection of data $\{\varLambda_i,L_i\}$ with $\varLambda_i\in\Map(U_i,H)$,
$L_i\in\Omega^1(U_i,\mathfrak{h})$ satisfying the relations 
\begin{subequations}
\begin{align}
&\eta'{}_i=t(\varLambda_i)\eta_i,
\vphantom{\Big]}
\label{fpppcech25a}
\\
&\lambda_i-\lambda'{}_i=L_i,
\vphantom{\Big]}
\label{fpppcech25b}
\end{align}
\label{fppcech25}
\end{subequations}
\!\!and the coherence conditions
\begin{subequations}
\begin{align}
&\varUpsilon'{}_{ij}m(\gamma_{ij})(\varLambda_j)=\varLambda_i\varUpsilon_{ij},
\vphantom{\Big]}
\label{fpppcech26a}
\\
&E'{}_{ij}-E_{ij}-\dot m(\gamma'{}_{ij})(L_j)+L_i
-Q(\eta_i\dot t(\chi'{}_{ij})\eta_i{}^{-1},\varLambda_i)=0.
\vphantom{\Big]}
\label{fpppcech26b}
\end{align}
\label{fppcech26}
\end{subequations}
\!\!Here, unlike the cases considered before, there are no differential conditions which the $1$--form data must obey. 
It should then be possible to cast $2$--groupoid $\check{\mathcal{P}}_2(U,G,H)$ 
as a $2$--groupoid of the form $\check P_2(U,\widehat{G},\widehat{H})$ for some  Lie crossed module
$(\widehat{G},\widehat{H})$ and $1$--cell isomorphism classes of $0$--cells
of $\check{\mathcal{P}}_2(U,G,H)$ may be interpreted as ones of $\check P_2(U,\widehat{G},\widehat{H})$, hence  
as isomorphism classes of principal $(\widehat{G},\widehat{H})$--$2$--bundles.
We shall not attempt to describe  $(\widehat{G},\widehat{H})$. We shall limit ourselves to note that
a $1$--cell isomorphism class of $0$--cells in $\check{\mathcal{P}}_2(U,G,H)$ \pagebreak yields 
a $1$--cell isomorphism class of $0$--cells in $\check P_2(U,G,H)$ 
obtained by the forgetful map that 
keeps the data $\gamma_{ij}$, $\varTheta_{ijk}$ and $\eta_i$, $\varUpsilon_{ij}$ dropping the data 
$\chi_{ij}$, $B_{ijk}$, $\lambda_i$, $E_{ij}$, the isomorphism class of principal $(G,H)$--$2$--bundles
which we started with. 

In subsect. \ref{subsec:crmdgautrsf}, we have shown that there is
a natural strict $2$--group $1$--morphism from the crossed module gauge transformation $2$--group 
$\Gau(M,G,H)$ to the gauge transformation $2$--group $\Gau(M,\mathfrak{v})$, where $\mathfrak{v}$
is the strict $2$--term $L_\infty$ algebra corresponding to the differential Lie crossed module $(\mathfrak{g},\mathfrak{h})$.
Now, the operations of the $2$--groupoids $\check{\mathcal{P}}_2(U,G,H)$ and $\check{\mathcal{P}}_2(U,\mathfrak{v})$ 
are defined completely in terms of those of the $2$--groups $\Gau(M,G,H)$ and $\Gau(M,\mathfrak{v})$, 
respectively. A natural strict $2$--groupoid $1$--morphism from 
$\check{\mathcal{P}}_2(U,G,H)$ to $\check{\mathcal{P}}_2(U,\mathfrak{v})$ 
is thus induced, furnishing us with a dictionary translating the 
formulation of higher gauge of refs. \cite{Baez:2004in,Baez:2005qu} extended in the way we have indicated 
into the one worked out in this paper. This parallels what happens in ordinary gauge theory, 
though in a rather non trivial way.
As in ordinary gauge theory, working with $\check{\mathcal{P}}_2(U,\mathfrak{v})$ 
involves a loss of central information, which may relevant beyond the perturbative level. 
Our approach has however the virtue of working for an arbitrary $2$--term $L_\infty$ algebra 
$\mathfrak{v}$ not necessarily arising from a differential Lie crossed module $(\mathfrak{g},\mathfrak{h})$.

Let us now assume that the global properties of the fields of a $2$--term $L_\infty$ algebra gauge theory 
with structure Lie algebra $\mathfrak{v}$ 
are defined by a $1$--cell isomorphism class of $0$--cells $g=\{g_{ij},\sigma_{ij},\varSigma_{ij},\tau_{ij}\}$ 
of $\check{\mathcal{P}}_2(U,\mathfrak{v})$.

The fact that the gluing data $g_{ij}$ do not satisfy the standard $1$--cocycle 
relation analogous to \eqref{px4linfdglobz} but the weaker condition \eqref{ppcech22a} is in general 
incompatible with the global single valuedness of the fields, by a well--known argument. 
As observed by Baez and Schreiber in \cite{Baez:2004in,Baez:2005qu}, 
single valuedness is recovered imposing certain conditions 
involving simultaneosly the gluing data and the fields and 
ensuring that the relations 
\begin{equation}
{}^{g_{ij}\diamond g_{jk}}\mathcal{F}_k={}^{g_{ik}}\mathcal{F}_k,\qquad \text{on $U_{ijk}$}.
\vphantom{\ul{\ul{\ul{\ul{\ul{\ul{\ul{x}}}}}}}}
\label{bash1}
\end{equation}
hold. Using crossed module notation, in which the $2$--cell $W_{ijk}:g_{ik}\Rightarrow g_{ij}\diamond g_{jk}$
is represented as a pair $(g_{ik},W_{ijk})$ with $W_{ijk}\in\Gau_2(U_{ijk},\mathfrak{v})$ such that 
$s(W_{ijk})=i$ and $t(W_{ijk})=g_{ij}\diamond g_{jk}\diamond g_{ik}{}^{-1}$,
\eqref{bash1} can be cast more compactly as
\begin{equation}
\mathcal{F}_i={}^{t(W_{ijk})}\mathcal{F}_i,\qquad \text{on $U_{ijk}$}.
\label{bash2}
\end{equation}
In general, the conditions do not take directly the form \eqref{bash1} or \eqref{bash2}
but are conditions sufficient for these to hold  
emerging naturally in a higher categorical formulation of the 
theory. 

Relation \eqref{2plinfdglob0} describing the global matching of the  local representations
$\mathcal{F}_i$ of a field $\mathcal{F}$ is schematic and must be made more precise. 
In subsect. \ref{subsec:linfdoub}, we saw that,
when $M$ is diffeomorphic to $\mathbb{R}^d$, the fields of $2$--term 
$L_\infty$ algebra gauge theory organize in (dual) field doublets. In subsects.
\ref{subsec:gauact}, 
we found further that the gauge transformation 
group acts naturally on doublets rather than their individual components. 
These properties should continue to hold in the appropriate form 
when $M$ is a general $d$--fold. 
We are thus led to define a (dual) field doublet 
$(\mathcal{F}^{(1)},\mathcal{F}^{(2)})$ on $M$ to be a collection 
$\{(\mathcal{F}^{(1)}{}_{i},\mathcal{F}^{(2)}{}_{i})\}$ 
of doublets $(\mathcal{F}^{(1)}{}_{i},\mathcal{F}^{(2)}{}_{i})$ of the same type 
on the open sets $U_i$ of the covering such that
\begin{subequations}
\label{linfdglob3/1}
\begin{align}
&\mathcal{F}^{(1)}{}_{i}={}^{g_{ij}}\mathcal{F}^{(1)}{}_j,
\vphantom{\Big]}
\label{linfdglob3/1a}
\\
&\mathcal{F}^{(2)}{}_{i}={}^{g_{ij}}\mathcal{F}^{(2)}{}_j,
\qquad\text{on $U_{ij}$}.
\vphantom{\Big]}
\label{linfdglob3/1b}
\end{align}
\end{subequations}
In the concrete cases we have studied, this works out as follows. 

A connection doublet $(\omega,\Omega)$ on $M$ is a collection $\{(\omega_i,\Omega_i)\}$ 
of connection doublets $(\omega_i,\Omega_i)$ on the sets $U_i$ such that 
\eqref{linfdglob3/1} holds with $(\mathcal{F}^{(1)},\mathcal{F}^{(2)})=(\omega,\Omega)$, 
the right hand side being given by \eqref{7linfdglob}.
Associated with the connection doublet, there is a curvature doublet
$(f,F)$ on $M$ given locally as the collection $\{(f_i,F_i)\}$ of the curvature doublets 
$(f_i,F_i)$ on the $U_i$  and satisfying the matching relation \eqref{linfdglob3/1}
with $(\mathcal{F}^{(1)},\mathcal{F}^{(2)})=(f,F)$ and the right hand side given by  
\eqref{8linfdglob}.

Let a connection doublet \pagebreak $(\omega,\Omega)$ on $M$ be fixed. 
A bidegree $(p,q)$ canonical field doublet $(\phi,\Phi)$ on $M$ is a 
collection $\{(\phi_i,\Phi_i)\}$ of bidegree $(p,q)$ canonical field doublets 
$(\phi_i,\Phi_i)$ on the sets $U_i$ such that 
\eqref{linfdglob3/1} holds with $(\mathcal{F}^{(1)},\mathcal{F}^{(2)})=(\phi,\Phi)$,
the right hand side being given by \eqref{9linfdglob}. 
In addition to the field doublet, we have  
a covariant derivative field doublet $(D\phi,D\Phi)$ on $M$ 
given locally as the collection $\{(D\phi_i,D\Phi_i)\}$ 
of the covariant derivative field doublets
$(D\phi_i,D\Phi_i)$ on the $U_i$ 
and satisfying matching relations of the form \eqref{linfdglob3/1}
with $(\mathcal{F}^{(1)},\mathcal{F}^{(2)})=(D\phi,D\Phi)$ 
and the right hand side given by \eqref{10linfdglob}.

Similarly, a bidegree $(r,s)$ dual field doublet $(\Upsilon,\upsilon)$ on $M$ 
is a collection $\{(\Upsilon_i,\upsilon_i)\}$ 
of bidegree $(r,s)$ canonical dual field doublets $(\Upsilon_i,\upsilon_i)$ 
on the sets $U_i$ such that \eqref{linfdglob3/1} holds with 
$(\mathcal{F}^{(1)},\mathcal{F}^{(2)})=(\Upsilon,\upsilon)$ and
the right hand side given by \eqref{11linfdglob}. 
Further, we have a covariant derivative dual field doublet $(D\Upsilon,D\upsilon)$ on $M$ 
given locally as the collection $\{(D\Upsilon_i,D\upsilon_i)\}$ 
of the covariant derivative dual field doublets $(D\Upsilon_i,D\upsilon_i)$ 
on the $U_i$ and satisfying matching relations of the form 
\eqref{linfdglob3/1} with $(\mathcal{F}^{(1)},\mathcal{F}^{(2)})=(D\Upsilon,D\upsilon)$ and
the right hand side given by \eqref{12linfdglob}. 



Next, let us examine whether the above results can be adapted to the rectified set--up described 
in  subsect. \ref{subsec:rectf}. Rectification requires gauge rectifiers. On a non trivial manifold
$M$, a rectifier must be assigned on each open set of the chosen covering and appropriate matching 
relations must be satisfied for consistency. We find then that a $2$--term $L_\infty$ algebra gauge 
rectifier $(\lambda,\mu)$ on $M$ is to be defined as a collection $\{(\lambda_i,\mu_i)\}$ 
of gauge rectifiers $(\lambda_i,\mu_i)$ on the sets $U_i$ such that 
\begin{subequations}
\label{linfdglob3/2}
\begin{align}
&\lambda_{i}(x,y)={}^{g_{ij}}\lambda_j(x,y),
\vphantom{\Big]}
\label{linfdglob3/2a}
\\
&\mu_{i}(x)={}^{g_{ij}}\mu_j(x),
\qquad\qquad\text{on $U_{ij}$},
\vphantom{\Big]}
\label{linfdglob3/2b}
\end{align}
\end{subequations}
the right hand side being given by \eqref{rectf1}.    
The compatibility of \eqref{ppcech22a} with the single valuedness
of the rectifier requires that 
\begin{subequations}
\label{linfdglob3/3}
\begin{align}
&\lambda_{i}(x,y)={}^{t(W_{ijk})}\lambda_i(x,y),
\vphantom{\Big]}
\label{linfdglob3/3a}
\\
&\mu_{i}(x)={}^{t(W_{ijk})}\mu_i(x),
\qquad\qquad\text{on $U_{ijk}$},
\vphantom{\Big]}
\label{linfdglob3/3b}
\end{align}
\end{subequations}
where crossed module notation is used, analogously to \eqref{bash2}.
It is likely that strong topologocal conditions must be satisfied in order
for gauge rectifiers to exit. We leave the solution of this problem to future work.

With a gauge rectifier at one's disposal, it is possible to rectify
canonical (dual) field doublets as well as define
rectified covariant derivatives using 
eqs. \eqref{rectf8}, \eqref{rectf9} and \eqref{rectf10}, \eqref{rectf11}, respectively. 
The analysis we have carried out for canonical (dual) field doublets
and their covariant derivatives can be repeated almost {\it verbatim} also for 
their rectified counterparts. 
The compatibility of \eqref{ppcech22a} with the single valuedness of the rectified fields
requires that \eqref{bash2} is still satisfied. Further, the  matching relations have 
still the form \eqref{linfdglob3/1},
but the gauge transformations of the fields occurring in the right hand side are now
given by \eqref{rectf6}, \eqref{rectf7} for rectified doublets and similarly 
for their rectified covariant derivatives. 
Unlike in the non rectified case, all these matching relations 
are independent of any preassigned connection doublet $(\omega,\Omega)$.


The above analysis can be generalized to more complicated situations, in which the fields
do not group in doublets and are instead subject to more complex forms 
of gauge transformation involving several fields.
It is only required that gauge transformation group action 
is left. 

The results of this subsection provide us with the theoretical tools necessary 
for assessing whether a $2$--term $L_\infty$ algebra gauge theory defined on 
a $d$--fold $M$ diffeomorphic to $\mathbb{R}^d$ can be defined globally also
on a generic $d$--fold $M$: it is sufficient to check its gauge covariance. 
This requires the prior specification of the gauge transformation prescription of the fields,
without which no statement about gauge covariance can be made.

Before concluding this subsection, an 
important remark is in order. Gauge covariance must not be confused with gauge symmetry.
Gauge symmetry represents an objective property of a gauge theory, the invariance of the action 
and the observables under gauge symmetry variations of the fields.
Gauge covariance, instead, reflects the independence of a gauge theory from subjective 
frame choices and is manifest in the covariance of the basic equations under gauge transformation
of the fields.
Gauge symmetry variation is {\it active}: it does change the fields.
Gauge transformation is {\it passive}: it does not, it simply  
governs the way the representations of the fields with respect to frames 
transform when the frames are changed. 
In many cases, gauge symmetry and gauge covariance are intimately related ed essentially 
equivalent (e. g. diffeomorphism symmetry and general covariance in general relativity), 
but this is not always so. These remarks should be kept in mind by the reader when studying the 
global properties of the $2$--term $L_\infty$ algebra gauge theories constructed in the next section.

\vspace{1mm}

\subsection{\normalsize \textcolor{blue}{Relation with other formulations}}\label{subsec:schreib}

We conclude this subsection with the following remarks. In ordinary gauge theory, 
a connection on a principal $G$--bundle $P\rightarrow M$ is defined as a Lie algebra valued 
form $\omega\in\Omega^1(P)\otimes\mathfrak{g}$ satisfying the two Ehresmann conditions
\cite{Ehresmann1}. The first of these requires that $\omega$ equals the left invariant Maurer--Cartan 
form along the fibers of $P$, the second imposes that $\omega$ is $G$--equivariant.
In an equivalent definition more suitable to our purposes,  
a connection $\omega$ is a differential 
graded commutative algebra morphism $\mathrm{W}(\mathfrak{g})\rightarrow\Omega^*(P)$,
whose vertical projection $\mathrm{W}(\mathfrak{g})\rightarrow\Omega_{\mathrm{vert}}{}^*(P)$
along the fibers of $P$ is flat, and so it factors through $\mathrm{CE}(\mathfrak{g})$, 
and whose restriction to the invariant subalgebra
$\ker d_{\mathrm{W}(\mathfrak{g})}\big|_{S(\mathfrak{g}^\vee[2])}\rightarrow\Omega^*(P)$ is basic and so 
it factors trough $\Omega^*(M)$ (cf. subsect. \ref{subsec:linfweil}). 
When $M$ is diffeomorphic to $\mathbb{R}^d$, $P\simeq M\times G$ 
and the second definition of connection boils down to the one we have given 
in subsect. \ref{subsec:linfdoub} as a morphism 
$\mathrm{W}(\mathfrak{g})\rightarrow\Omega^*(M)$. When $M$ is not, one has to pick a cover 
$U=\{U_i\}$ of $M$ and define locally a connection as a morphism 
$\mathrm{W}(\mathfrak{g})\rightarrow\Omega^*(U_i)$ for each $i$. The resulting local data must then 
be assembled in a globally consistent way. The way of doing so is dictated by the topology of the bundle
$P$ and codified in a $G$--valued cocycle. The problem with this classic approach is that it cannot be 
straightforwardly extended as it is to semistrict higher gauge theory, because there is no notion of total 
space of a principal $2$--bundle that can be handled with the same ease. 
The authors of refs. \cite{Fiorenza2011,Schreiber2011} tackle this problem
reformulating ordinary gauge theory in a way that it can be directly generalized to the higher 
case. (See also the recent papers \cite{Nikolaus1,Nikolaus2}).

Given a Lie algebra $\mathfrak{g}$ and a Cartesian space $S=\mathbb{R}^d$, they consider the 
simplicial set of differential graded commutative algebra morphisms 
$\mathrm{W}(\mathfrak{g})\rightarrow\Omega^*(S\times\Delta^k)$, where $\Delta^k$ is the standard 
geometric $k$--simplex with $k\geq 0$, whose vertical projection along $\Delta^k$ factors through
$\mathrm{CE}(\mathfrak{g})$ and whose restriction to the invariant subalgebra 
$\ker d_{\mathrm{W}(\mathfrak{g})}\big|_{S(\mathfrak{g}^\vee[2])}$ factors through $\Omega^*(S)$.
For each $k$, a morphism $\mathrm{W}(\mathfrak{g})\rightarrow\Omega^*(S\times\Delta^k)$
is equivalent to a connection $1$--form $\omega$ on the trivial bundle $G\times(S\times\Delta^k)$, whose curvature
$2$--form $f$ has components only along $S$.
For $k=0$, $\Delta^0$ is the singleton $0$ and a morphism $\mathrm{W}(\mathfrak{g})\rightarrow\Omega^*(S\times\Delta^0)$
as above reduces to an assignment of a connection $1$--form $\omega$ on $S$, as we have already seen.
For $k=1$, $\Delta^1$ is the $1$--simplex $0\rightarrow 1$ (an interval)
and a morphism $\mathrm{W}(\mathfrak{g})\rightarrow\Omega^*(S\times\Delta^1)$
as above encodes a differential equation, which, once integrated, yields a gauge transformation $g$ on $S$
connecting the components of the connection $1$--form $\omega$ along $S$ at the extremes $0,1$ of 
$0\rightarrow 1$. For $k=2$, $\Delta^2$ is the $2$--simplex
$0\rightarrow 1\rightarrow 2$
(a triangle) and a morphism 
$\mathrm{W}(\mathfrak{g})\rightarrow\Omega^*(S\times\Delta^2)$ identifies the 
gauge transformation $g_{21}g_{10}$ acting along the edges $0\rightarrow 1$, $1\rightarrow 2$
with that $g_{20}$ acting along $0\rightarrow 2$. On a non trivial manifold $M$ equipped with
a covering $U=\{U_i\}$, one considers morphisms with $S=U_i$, $S=U_{ij}$ and $S=U_{ijk}$ 
for $k=0,~1~2$, respectively, which are compatible in  the following sense:
the restrictions of the resulting local data associated with the inclusions
$U_{ijk}\subseteq U_{ij}\subseteq U_i$ coincide with the
restrictions associated with the face inclusions $\Delta^0\subset\Delta^1\subset\Delta^2$. 
Then, those local data reduce to the familiar one defining a principal $G$--bundle bundle with connection:
the Lie valued connection $1$--forms $\omega_i$ and the group valued transition functions $g_{ij}$ satisfying the usual
cocycle condition and relating $\omega_i$, $\omega_j$ through gauge transformation.
The advantage of this approach is that it generalizes directly to the higher case.
One simply replaces the Lie algebra $\mathfrak{g}$ with a general 
$L_\infty$ algebra $\mathfrak{v}$ (or even an $L_\infty$ algebroid $\mathfrak{a}$) and goes through analogous steps.

A simplicial presheaf $\mathcal{G}$ is a presheaf over the category of Cartesian spaces $\mathrm{CartSp}$ 
such that for any Cartesian space $S\in \mathrm{CartSp}$, $\mathcal{G}(S)=\{\mathcal{G}_k(S)\}$ is a 
simplicial set. A simplicial presheaf is just (a presentation of) a smooth $\infty$--groupoid, 
an $\infty$ category in which $k$--morphisms are equivalences for all $k$ 
with an appropriate notion of smoothness. For an $L_\infty$ algebra $\mathfrak{v}$, 
the mapping $S\mapsto\{\mathrm{W}(\mathfrak{v})\rightarrow\Omega^*(S\times\Delta^k)\}$, 
introduced in the previous paragraph, defines in the form of a simplicial presheaf a smooth $\infty$--groupoid
$\exp(\mathfrak{v})_{\mathrm{conn}}$. 
The induced mapping $S\mapsto\{\mathrm{CE}(\mathfrak{v})\rightarrow\Omega_{\mathrm{vert}}{}^*(S\times\Delta^k)\}$
defines a second smooth $\infty$--groupoid $\exp(\mathfrak{v})$, which may be viewed as
the one that integrates $\mathfrak{v}$ in the sense of Lie theory. 
Replacing the Cartesian spaces $S$ with the open sets of the nerve of a \v Cech covering $U=\{U_i\}$ 
of a manifold $M$ defines an $\infty$ groupoid morphism from $M$, seen as an $\infty$ groupoid, 
to $\exp(\mathfrak{v})_{\mathrm{conn}}$ or $\exp(\mathfrak{v})$. 
The morphism encodes the set of local data defining an $\infty$ differential \v Cech cocycle describing an 
$\infty$ principal bundle $P$ over $M$ equipped with an $\infty$ connection $\omega$,
if $\exp(\mathfrak{v})_{\mathrm{conn}}$ is used, or the subset of data describing $P$ alone, if 
$\exp(\mathfrak{v})$ is used instead. Note that we may view $\exp(\mathfrak{v})$
as the ``structure'' $\infty$ groupoid of $P$. 

For a $2$--term $L_\infty$ algebra $\mathfrak{v}$, the method of refs.\cite{Fiorenza2011,Schreiber2011} 
yields the same definition of $2$--connection on a trivial $M$ we gave in subsect. \ref{subsec:linfdoub} 
as a differential graded commutative algebra morphism $\mathrm{W}(\mathfrak{v})\rightarrow\Omega^*(M)$.
The approach to semistrict higher gauge theory, which we have described at length in this section, however 
differs from that of \cite{Fiorenza2011,Schreiber2011} in that it is ``effective'' in the following sense. 
We bypassed the difficulty of dealing with the $\infty$ groupoid $\exp(\mathfrak{v})$, 
which in its present abstract formulation does not lend itself easily to detailed calculations, and relied 
instead on the automorphism $2$--group $\Aut(\mathfrak{v})$. In this way we have avoided 
operating with gauge transformations as they are at their most basic level, roughly $\exp(\mathfrak{v})$--valued 
maps, and we have reduced ourselves to work with objects which somehow encapsulate the action 
of gauge transformations on fields in its most concrete form. We did this in subsects. 
\ref{subsec:linfgautrsf}, \ref{subsec:gauact}, where we defined a $2$--term $L_\infty$ algebra gauge transformation 
as an $\Aut_1(\mathfrak{v})$--valued map $g$ plus a set of appended
$\hat{\mathfrak{v}}_0$-- and $\hat{\mathfrak{v}}_1$-- and $\Hom(\hat{\mathfrak{v}}_0,\hat{\mathfrak{v}}_1)$--valued 
forms, $\sigma_g$, $\varSigma_g$ and $\tau_g(\cdot)$ respectively, satisfying certain relations
and expressing the action of gauge transformation on fields in terms of these. Some guesswork was involved 
in this. We tested our approach in subsect. \ref{subsec:crmdgautrsf}, where we checked that it reproduces 
the usual notions of gauge transformation when $\mathfrak{v}$ is strict. We were however unable to verify
this property beyond the strict case. At any rate, we expect that the effective objects $g$, $\sigma_g$, $\varSigma_g$ 
and $\tau_g(\cdot)$ are expressible in terms of the more basic gauge transformations of the theory of 
\cite{Fiorenza2011,Schreiber2011}, with the relations, which $g$, $\sigma_g$, $\varSigma_g$ and $\tau_g(\cdot)$ 
obey and which we treated as axioms, emerging as theorems. Our approach, though admittedly non fundamental, 
turns out to be quite efficient in the analysis of gauge covariance in field theoretic applications, as we shall 
show in the next section.

\vfill\eject

\section{\normalsize \textcolor{blue}{$2$--term $L_\infty$ algebra BF gauge theory}}
\label{sec:bvlinf}

\hspace{.5cm} 
In this section, we shall construct and analyze the semistrict Lie $2$--algebra 
analog of the standard BF theory \cite{Schwarz:1978,Horowitz:1989ng} 
providing in this way a simple but non trivial example of semistrict higher 
gauge theory. 
We shall first study the classical theory and, then, using an AKSZ approach
\cite{AKSZ}, we shall work out the quantum theory and obtain by suitable gauge fixing
a topological field theory.


Below, $\mathfrak{v}$ is the $2$--term $L_\infty$ structure algebra of the model 
and $M$ is the oriented manifold on which fields propagate. $M$ is taken $3$--dimensional, 
because this is the simplest situation in which a $3$--form curvature does not 
vanish identically.

\subsection{\normalsize \textcolor{blue}{Classical $\mathfrak{v}$ BF gauge theory}}
\label{subsec:clbf2tlinf}

\hspace{.5cm} We consider first the case where $M$ is diffeomorphic to
$\mathbb{R}^3$ avoiding in this way the problems related to the global definedness 
of the theory. 

The fields of classical $\mathfrak{v}$ BF gauge theory organize in a bidegree $(1,0)$ 
connection doublet $(\omega,\varOmega)$ and a further bidegree $(0,0)$ dual field doublet 
$(B^+,b^+)$ (cf. sect. \ref{subsec:linfdoub}).
Fields are supposed to fall off rapidly at the boundary of $M$, so that integration on $M$ 
is convergent and integration by parts can be carried out without generating boundary 
contributions. The classical action 
of the theory is a rather straightforward generalization of that
of standard BF gauge theory,
\begin{equation}
S_{\mathrm{cl}}=\int_M\Big[\langle b^+,f\rangle-\langle B^+,F\rangle\Big],
\label{clbfact}
\end{equation}
where the curvature doublet $(f,F)$ is given by \eqref {fFcurv}.
The field equations are 
\begin{subequations}
\label{clbfeqs}
\begin{align}
&f=0,
\vphantom{\Big]}
\label{clbfeqsa}
\\
&F=0,
\vphantom{\Big]}
\label{clbfeqsb}
\\
&DB^+=0,
\vphantom{\Big]}
\label{clbfeqsc}
\\
&Db^+=0,
\vphantom{\Big]}
\label{clbfeqsd}
\end{align}
\end{subequations}
\vskip-.5truecm\eject\noindent
where the covariant derivation $D$ is defined in subsect. \ref{subsec:linfdoub}.
They imply in part\-icular that the connection doublet $(\omega,\varOmega)$ is flat, as
in ordinary BF theory.

Classical $\mathfrak{v}$ BF gauge theory enjoys a high amount of gauge symmetry.
The gauge symmetry variations of the fields are expressed in terms of ghost fields
organized in a bidegree $(0,1)$ field doublet $(c,C)$ and a bidegree $(-1,1)$ dual 
field doublet $(0,\beta^+)$, 
\begin{subequations}
\label{clbffvar}
\begin{align}
&\delta_{\mathrm{cl}}\omega=-Dc,
\vphantom{\Big]}
\label{clbffvara}
\\
&\delta_{\mathrm{cl}}\varOmega=-DC,
\vphantom{\Big]}
\label{clbffvarb}
\\
&\delta_{\mathrm{cl}}B^+=-[c,B^+]^\vee+\partial^\vee\beta^+,
\vphantom{\Big]}
\label{clbffvarc}
\\
&\delta_{\mathrm{cl}}b^+=-[c,b^+]^\vee+[C,B^+]^\vee+[\omega,c,B^+]^\vee-D\beta^+.
\vphantom{\Big]}
\label{clbffvard}
\end{align}
\end{subequations}
The action 
is invariant under the symmetry, 
\begin{equation}
\delta_{\mathrm{cl}}S_{\mathrm{cl}}=0,
\label{dlcclbfact}
\end{equation}
as is easily verified. 
The expressions of the variations $\delta_{\mathrm{cl}}\omega$, $\delta_{\mathrm{cl}}\varOmega$ 
could be guessed on the basis of a formal similarity to that of the gauge field in standard gauge 
theory. Taking these for granted, the expressions of $\delta_{\mathrm{cl}}B^+$, $\delta_{\mathrm{cl}}b^+$ 
follow then from the requirement of invariance of the action. 

For consistency, it should be possible to define 
gauge symmetry variations of the ghost fields 
rendering the gauge field variation operator $\delta_{\mathrm{cl}}$ nilpotent at least on--shell. 
These variations depend on 
the ghost field doublets $(c,C)$, $(0,\beta^+)$
and a further ghost for ghost field seen as a bidegree $(-1,2)$ field doublet $(0,\varGamma)$, 
\begin{subequations}
\label{ghclbffvar}
\begin{align}
&\delta_{\mathrm{cl}}c=-\frac{1}{2}[c,c]+\partial\varGamma,
\vphantom{\Big]}
\label{ghclbffvara}
\\
&\delta_{\mathrm{cl}}C=-[c,C]+\frac{1}{2}[\omega,c,c]-D\varGamma,
\vphantom{\Big]}
\label{ghclbffvarb}
\\
&\delta_{\mathrm{cl}}\beta^+=-[c,\beta^+]^\vee+\frac{1}{2}[c,c,B^+]^\vee+[\varGamma,B^+]^\vee,
\vphantom{\Big]}
\label{ghclbffvarc}
\\
&\delta_{\mathrm{cl}}\varGamma=-[c,\varGamma]+\frac{1}{6}[c,c,c].
\vphantom{\Big]}
\label{ghclbffvard}
\end{align}
\end{subequations}
The expressions of the variations $\delta_{\mathrm{cl}}c$, $\delta_{\mathrm{cl}}C$, 
$\delta_{\mathrm{cl}}\beta^+$ with $\varGamma$ formally set to $0$ may be 
inferred viewing $c$ as akin to the standard 
gauge theory ghost. This suggests the first term in the right hand side
of each of them. The remaining contributions are determined 
by the requirement of the on--shell nilpotency of $\delta_{\mathrm{cl}}$ upon 
taking into account that the bracket $[\cdot,\cdot]$ has a generally non trivial 
Jacobiator $[\cdot,\cdot,\cdot]$. The terms depending on $\varGamma$ reflect the existence
of a gauge for gauge symmetry. The expression of $\delta_{\mathrm{cl}}\varGamma$ 
follows from a similar reasoning. 

Using \eqref{clbffvar}, \eqref{ghclbffvar}, we find that $\delta_{\mathrm{cl}}{}^2\mathcal{F}=0$
for all fields and ghost fields $\mathcal{F}$ except for $\varOmega$, $b^+$, in which case one has
\begin{subequations}
\label{sqclbffvar}
\begin{align}
&\delta_{\mathrm{cl}}{}^2\varOmega=\frac{1}{2}[f,c,c]-[f,\varGamma],
\vphantom{\Big]}
\label{sqclbffvarb}
\\
&\delta_{\mathrm{cl}}{}^2b^+=\frac{1}{2}[c,c,DB^+]^\vee+[\varGamma,DB^+]^\vee.
\vphantom{\Big]}
\label{sqclbffvard}
\end{align}
\end{subequations}
Thus, on account of \eqref{clbfeqs}, $\delta_{\mathrm{cl}}$ 
is indeed nilpotent on shell, as required. We observe 
that the inclusion of the ghost for ghost $\varGamma$ 
is crucial for the nilpotency of $\delta_{\mathrm{cl}}$
in the ghost sector. Had $\varGamma$ not been there, 
$\delta_{\mathrm{cl}}{}^2$ would have vanished only up to a ghost field dependent 
gauge for gauge symmetry variation 
\footnote{$\vphantom{\bigg[}$ If we had omitted $\varGamma$, we would have had in fact
\vspace{-.2cm}
\begin{subequations}
\label{ghsqclbffvar}
\begin{align}
&\delta_{\mathrm{cl}}{}^2c=-\frac{1}{6}\partial([c,c,c]),
\vphantom{\Big]}
\label{ghsqclbffvara}
\\
&\delta_{\mathrm{cl}}{}^2C=-\frac{1}{6}D([c,c,c]),
\vphantom{\Big]}
\label{ghsqclbffvarb}
\\
&\delta_{\mathrm{cl}}{}^2\beta^+=-\frac{1}{6}[[c,c,c],B^+]^\vee,
\vphantom{\Big]}
\label{ghsqclbffvarc}
\end{align}
\end{subequations}
where $(0,[c,c,c])$ is treated as a bidegree $(-1,3)$ field doublet.
The right hand side of each of these relations is a gauge for gauge symmetry variation
of the relevant ghost field with degree $3$ parameter $-(1/6)[c,c,c]$.
In gauge theory, indeed, one can expect $\delta_{\mathrm{cl}}$ to be nilpotent
on--shell only up to gauge symmetry variations with ghost field dependent parameters.}.

Since the gauge field variation operator $\delta_{\mathrm{cl}}$ is not nilpotent off--shell,
the gauge symmetry algebra of the theory is open.
Further, as the gauge symmetry admits a gauge for gauge symmetry, 
the gauge symmetry algebra is (at least) first stage
\vspace{.15cm} \eject\noindent
reducible.
These diseases will be cured by a suitable AKSZ reformulation of the model in the next subsection. 

Let us now see whether $\mathfrak{v}$ BF gauge theory can be consistently formulated on a
general closed $3$--fold $M$. For reasons explained at the end of subsect. \ref{subsec:linfdglob}
\footnote{$\vphantom{\bigg[}$ The reader is invited to keep in mind
the remarks at the close of subsect. \ref{subsec:linfdglob} below.
},
it is sufficient to check the gauge covariance of the model when $M$ is diffeomorphic 
to $\mathbb{R}^3$. Recall that this requires that hypotheses on the the gauge transformation prescriptions 
of the fields and ghost fields be made.

Let $g\in\Gau_1(M,\mathfrak{v})$ be any $1$--gauge transformation (cf. subsect. \ref{subsec:linfgautrsf}).
We assume that the connection doublet $(\omega,\varOmega)$ transforms as in \eqref{7linfdglob} 
and the dual field doublet $(B^+,b^+)$ is canonical and, so, does as in \eqref{11linfdglob}. 
We then find that the classical Lagrangian $\mathcal{L}_{\mathrm{cl}}$ 
(the integrand of $S_{\mathrm{cl}}$ in \eqref{clbfact}) 
is gauge invariant, ${}^g\mathcal{L}_{\mathrm{cl}}=\mathcal{L}_{\mathrm{cl}}$. Hence, 
the action $S_{\mathrm{cl}}$ can be defined also on a general $3$--fold $M$.

Gauge covariance of the gauge symmetry field variations requires that 
${}^g\delta_{\mathrm{cl}}\mathcal{F}=\delta_{\mathrm{cl}}{}^g\mathcal{F}$
for all fields and ghost fields $\mathcal{F}$
\footnote{$\vphantom{\big[}$ ${}^g\delta_{\mathrm{cl}}\mathcal{F}$ is defined by replacing each occurrence
of each field $\mathcal{G}$ in $\delta_{\mathrm{cl}}\mathcal{F}$ by ${}^g\mathcal{G}$.
\label{foot:gdf}}.
We assume that the ghost field doublet $(c,C)$ is canonical and, so, transforms as
in \eqref{9linfdglob}. Then, if we insist that the above property be satisfied in the ghost sector, 
the ghost dual field doublet $(0,\beta^+)$ and ghost for ghost field doublet $(0,\varGamma)$ cannot be 
canonical, but, instead, they must transform in a more complicated way, viz
\begin{subequations}
\label{vbfth1}
\begin{align}
&{}^g\beta^+=g^\vee{}_0(\beta^+)-g^\vee{}_2(g_0(c),B^+),
\vphantom{\Big]}
\label{vbfth1a}
\\
&{}^g\varGamma=g_1(\varGamma)-\frac{1}{2}g_2(c,c).
\vphantom{\Big]}
\label{vbfth1b}
\end{align}
\end{subequations}
We find then that ${}^g\delta_{\mathrm{cl}}\mathcal{F}=\delta_{\mathrm{cl}}{}^g\mathcal{F}$
for all fields and ghost fields $\mathcal{F}$ but $\varOmega$, $b^+$, 
\begin{subequations}
\label{vbfth2}
\begin{align}
&{}^g\delta_{\mathrm{cl}}\varOmega=\delta_{\mathrm{cl}}{}^g\varOmega-g_2(c,f), \hspace{2.4cm}
\vphantom{\Big]}
\label{vbfth2b}
\\
&{}^g\delta_{\mathrm{cl}}b^+=\delta_{\mathrm{cl}}{}^gb^+-g^\vee{}_2(g_0(c),DB^+).
\vphantom{\Big]}
\label{vbfth2d}
\end{align}
\end{subequations}
\vfil\eject\noindent
Thus, the gauge field variation operator $\delta_{\mathrm{cl}}$ is covariant only on shell.
The gauge symmetry field variations, so, as given in \eqref{clbffvar}, \eqref{ghclbffvar},
cannot be defined on a general $3$--fold $M$.

The failure of gauge symmetry field variation to be gauge covariant 
spoils the gauge symmetry invariance of the action $S_{\mathrm{cl}}$ when $M$ is not
diffeomorphic to $\mathbb{R}^3$.
The verification of \eqref{dlcclbfact} requires the use of Stokes theorem to eliminate a term
of the form $\int_Md\langle\beta^+,f\rangle$. On a non trivial $M$, this is legitimate only if 
$\langle\beta^+,f\rangle$ is gauge invariant. Unfortunately, this is 
not the case. Thus, as expected, \eqref{dlcclbfact} fails to hold for a general $3$--fold $M$. 
This negative result can be traced back to the lack of gauge covariance of the 
gauge symmetry field variations. Indeed, it is easily verified that the combination 
$\langle\beta^+,f\rangle+\langle B^+,\delta_{\mathrm{cl}}\varOmega\rangle-\langle b^+,
\delta_{\mathrm{cl}}\omega\rangle$ is gauge invariant. 
Therefore, the offending term causing the break down of the proof of \eqref{dlcclbfact} 
may be substituted by 
$-\int_Md(\langle B^+,\delta_{\mathrm{cl}}\varOmega\rangle-\langle b^+,\delta_{\mathrm{cl}}\omega\rangle)$, 
indicating that the origin of the problem is the gauge covariance failure of the variation 
$\delta_{\mathrm{cl}}\varOmega$. Again, this disease will be cured by the AKSZ reformulation 
of the model worked out in the next subsection.

\subsection{\normalsize \textcolor{blue}{AKSZ reformulation of  $\mathfrak{v}$ BF gauge theory}}
\label{subsec:linfvmod} 

\hspace{.5cm} AKSZ theory \cite{AKSZ} is a method of constructing field theories satisfying 
the requirements of the BV quantization scheme \cite{BV1,BV2}  using solely the  
geometric data at hand. Following the AKSZ approach ensures that the resulting 
field theory can be consistently quantized on one hand and renders the whole 
construction in a way canonical on the other. 

The basic steps of the AKSZ approach are the following. 

1. The definition of a graded field space $\matheul{F}$, the {\it BV field space}.
 
\noindent

2.  The assignment of a degree $-1$ symplectic form
$\varOmega_{\mathrm{BV}}$ on $\matheul{F}$, the {\it BV form}.  

\noindent
Canonically associated with $\varOmega_{\mathrm{BV}}$ is 
a degree $1$ Gerstenhaber bracket $(\cdot,\cdot)_{\mathrm{BV}}$
on the functional algebra $\Fun(\matheul{F})$ of $\matheul{F}$, the {\it BV bracket}.

3. The construction of the appropriate degree $0$ field functional  
$S_{\mathrm{BV}}$, the {\it BV master action}, satisfying the 
{\it classical BV master equation} 
\begin{equation}
(S_{\mathrm{BV}},S_{\mathrm{BV}})_{\mathrm{BV}}=0.
\label{SWSW=0}
\end{equation}

\noindent
Then, according to BV theory, the field theory governed by $S_{\mathrm{BV}}$
is consistent and suitable for quantization. 

Associated with $S_{\mathrm{BV}}$ is the {\it BV field variation operator}
$\delta_{\mathrm{BV}}=(S_{\mathrm{BV}},\cdot)_{\mathrm{BV}}$ on $\Fun(\matheul{F})$.  
From \eqref{SWSW=0}, it follows that $\delta_{\mathrm{BV}}$ is nilpotent 
\begin{equation}
\delta_{\mathrm{BV}}{}^2=0.
\label{dBV2=0}
\end{equation}
The cohomology of $\delta_{\mathrm{BV}}$, $H_{BV}^*$, is the theory's {\it BV cohomology}. 
Again from \eqref{SWSW=0}, 
\begin{equation}
\delta_{\mathrm{BV}}S_{\mathrm{BV}}=0.
\label{dBVsBV=0}
\end{equation}
The BV action $S_{\mathrm{BV}}$ is so BV invariant.

As anticipated in the previous subsection, the problems of $\mathfrak{v}$ BF gauge theory can be fixed 
through an AKSZ reformulation of the model. Next, we shall describe this in detail.
Using the AKSZ method, we shall work out a field theory canonically associated to 
$\mathfrak{v}$, {\it AKSZ $\mathfrak{v}$ BF gauge theory}, 
and show that this is indeed the BV extension of $\mathfrak{v}$ 
BF gauge theory curing this latter's diseases. 

The AKSZ construction is best implemented by using the  
{\it superfield formalism}, which we now briefly recall. 
For any vector space $V$, the space of $V[p]$--valued superfields is 
$\varGamma(T[1]M,V[p])$, where $T[1]M$ is the parity $1$--shifted tangent bundle  
of $M$. In $3$ dimensions, 
an element $\mathbfs{\varphi}\in\varGamma(T[1]M,V[p])$ has a component expansion 
$\mathbfs{\varphi}=\varphi^{0,p}+\varphi^{1,p-1}+\varphi^{2,p-2}+\varphi^{3,p-3}$, where
$\varphi^{r,s}$ is a $V$--valued form degree $r$ and ghost number degree $s$ field.
Superfields can be integrated on $T[1]M$. If $\mathbfs{\varphi}$ is as above, then,
by definition, $\int_{T[1]M}\varrho\,\mathbfs{\varphi}=\int_M\varphi^{3,p-3}$, where 
$\varrho$ is the standard supermeasure of $T[1]M$. 

As usual, we first consider the case where $M$ is diffeomorphic
to $\mathbb{R}^3$. The field content of AKSZ $\mathfrak{v}$ BF gauge theory   
consists of 
superfields $\mathbfs{p}\in\varGamma(T[1]M,\hat{\mathfrak{v}}_0{}^\vee[1])$, 
$\mathbfs{q}\in\varGamma(T[1]M,\hat{\mathfrak{v}}_0[1])$,
$\mathbfs{P}\in\varGamma(T[1]M,\hat{\mathfrak{v}}_1{}^\vee[0])$, $\mathbfs{Q}\in\varGamma(T[1]M,\hat{\mathfrak{v}}_1[2])$.
The quadruple $(\mathbfs{p},\mathbfs{q},\mathbfs{P},\mathbfs{Q})$ 
can be packaged into a superfield $\mathbfs{\mathcal{A}}\in\varGamma(T[1]M,T^*[2](\hat{\mathfrak{v}}_0[1]
\oplus\hat{\mathfrak{v}}_1[2]))$. Recall that $\hat{\mathfrak{v}}_0$, $\hat{\mathfrak{v}}_1$ are assumed 
conventionally to have ghost number degree $0$. 

The BV symplectic form 
of AKSZ $\mathfrak{v}$ BF gauge theory has the canonical form
\begin{equation}
\varOmega_{\mathrm{BV}}=\int_{T[1]M}\varrho\Big[\langle\delta\mathbfs{p},\delta\mathbfs{q}\rangle
+\langle\delta\mathbfs{P},\delta\mathbfs{Q}\rangle\Big].
\label{OmegaW}
\end{equation}
$\varOmega_{\mathrm{BV}}$ is just the pull--back by $\mathbfs{\mathcal{A}}$ of the canonical ghost 
number degree $2$ symplectic form of $T^*[2](\hat{\mathfrak{v}}_0[1]\oplus\hat{\mathfrak{v}}_1[2])$.
From BV theory, 
associated with $\varOmega_{\mathrm{BV}}$ is the BV bracket $(\cdot,\cdot)_{\mathrm{BV}}$.

The BV action 
of AKSZ $\mathfrak{v}$ BF gauge theory is given by
\begin{align}
S_{\mathrm{BV}}&=\int_{T[1]M}\varrho\Big[-\langle\mathbfs{p},\mathbfs{d}\mathbfs{q}
-\frac{1}{2}[\mathbfs{q},\mathbfs{q}]+\partial\mathbfs{Q}\rangle
\vphantom{\Big[}
\label{SW}
\\
&\hspace{5cm}+\langle\mathbfs{P}, \mathbfs{d}\mathbfs{Q}-[\mathbfs{q},\mathbfs{Q}]
+\frac{1}{6}[\mathbfs{q},\mathbfs{q},\mathbfs{q}]\rangle\Big].
\vphantom{\Big[}
\nonumber
\end{align}
$S_{\mathrm{BV}}$ satisfies the classical BV master equation \eqref{SWSW=0}
and, so, according to BV theory, the model is consistent and quantizable.
The action $S_{\mathrm{BV}}$ is canonical\-ly associated with $\mathfrak{v}$, since 
the master equation is satisfied if and only if 
$\mathfrak{v}=(\mathfrak{v}_0,\mathfrak{v}_1,\partial,[\cdot,\cdot],[\cdot,\cdot,\cdot])$
is a $2$--term $L_\infty$ algebra (cf. subsect. \ref{sec:linfty}). 

The BV variations 
of the AKSZ $\mathfrak{v}$ BF gauge theory superfields are 
\begin{subequations}
\label{dWsuperfields}
\begin{align}
&\delta_{\mathrm{BV}}\mathbfs{p}
=\mathbfs{d}\mathbfs{p}-[\mathbfs{q},\mathbfs{p}]^\vee+[\mathbfs{Q},\mathbfs{P}]^\vee
+\frac{1}{2}[\mathbfs{q},\mathbfs{q},\mathbfs{P}]^\vee,
\vphantom{\Big]}
\label{dWbibfs}
\\
&\delta_{\mathrm{BV}}\mathbfs{q}
=\mathbfs{d}\mathbfs{q}-\frac{1}{2}[\mathbfs{q},\mathbfs{q}]+\partial\mathbfs{Q},
\vphantom{\Big]}
\label{dWcibfs}
\\
&\delta_{\mathrm{BV}}\mathbfs{P}
=\mathbfs{d}\mathbfs{P}-[\mathbfs{q},\mathbfs{P}]^\vee+\partial^\vee\mathbfs{p},
\vphantom{\Big]}
\label{dWBibfs}
\\
&\delta_{\mathrm{BV}}\mathbfs{Q}
=\mathbfs{d}\mathbfs{Q}-[\mathbfs{q},\mathbfs{Q}]+\frac{1}{6}[\mathbfs{q},\mathbfs{q},\mathbfs{q}].
\vphantom{\Big]}
\label{dWCibfs}
\end{align}
\end{subequations}
Since $S_{\mathrm{BV}}$ 
solves the BV master equation \eqref{SWSW=0}, 
the BV superfield variation operator $\delta_{\mathrm{BV}}$ satisfies \eqref{dBV2=0}, 
and so is nilpotent,
as can also be directly verified from \eqref{dWsuperfields}.
For the same reason, the BV action 
$S_{\mathrm{BV}}$ satisfies \eqref{dBVsBV=0} and so is BV invariant.

Let us now see whether our AKSZ $\mathfrak{v}$ BF gauge theory can be consistently formulated 
on a general closed $3$--fold $M$. Again, 
as explained in subsect. \ref{subsec:linfdglob},
it is enough to find the appropriate gauge transformation rules 
of the basic super\-fields and then check the model's gauge covariance for $M$ diffeomorphic 
to $\mathbb{R}^3$.

For any $1$--gauge transformation $g\in\Gau_1(M,\mathfrak{v})$, 
the appropriate expression of the gauge transformed superfields turns out to be 
\begin{subequations}
\label{gsuperfields}
\begin{align}
&{}^g\mathbfs{p}=g^\vee{}_0(\mathbfs{p}+\mathbfs{\tau}_g{}^\vee(\mathbfs{P}))
-g^\vee{}_2(g_0(\mathbfs{q}+\mathbfs{\sigma}_g),\mathbfs{P}),
\vphantom{\Big]}
\label{gbibfs}
\\
&{}^g\mathbfs{q}=g_0(\mathbfs{q}+\mathbfs{\sigma}_g),
\vphantom{\Big]}
\label{gcibfs}
\\
&{}^g\mathbfs{P}=g^\vee{}_1(\mathbfs{P}),
\vphantom{\Big]}
\label{gBibfs}
\\
&{}^g\mathbfs{Q}
=g_1(\mathbfs{Q}-\mathbfs{\varSigma}_g-\mathbfs{\tau}_g(\mathbfs{q}+\mathbfs{\sigma}_g))
-\frac{1}{2}g_2(\mathbfs{q}+\mathbfs{\sigma}_g,\mathbfs{q}+\mathbfs{\sigma}_g),
\vphantom{\Big]}
\label{gCibfs}
\end{align}
\end{subequations}  
where $\mathbfs{\sigma}_g$, $\mathbfs{\varSigma}_g$, $\mathbfs{\tau}_g$ are just 
$\sigma_g$, $\varSigma_g$, $\tau_g$ seen as one--component superfields
(cf. subsect. \ref{subsec:linfgautrsf}). 
\eqref{gsuperfields} can be justified as follows. %
The 
usual connection doublet
$(\omega,\varOmega)$ must appear in the component expansion of the superfield pair
$(-\mathbfs{q},\mathbfs{Q})$, as is evident by matching of their form and ghost number degrees, 
the minus sign being conventional. 
So, the pair 
$(-\mathbfs{q},\mathbfs{Q})$ must transform as if it were a connection doublet 
(cf. eqs. \eqref{7linfdglob}). This explains the form of \eqref{gcibfs}, \eqref{gCibfs}.
Similarly, to have gauge covariance, 
$(-\mathbfs{P},\mathbfs{p})$ must transform 
as a bidegree $(0,0)$ canonical dual 
field doublet (cf. eqs. \eqref{11linfdglob}), yielding \eqref{gbibfs}, \eqref{gBibfs}. 

Let $\mathbfs{\varLambda}_{\mathrm{BV}}$ be the integrand superfield 
of $\varOmega_{\mathrm{BV}}$ in \eqref{OmegaW}. It is easy to check that 
$\mathbfs{\varLambda}_{\mathrm{BV}}$ is gauge invariant, 
${}^g\mathbfs{\varLambda}_{\mathrm{BV}}=\mathbfs{\varLambda}_{\mathrm{BV}}$.
Hence, the BV form $\varOmega_{\mathrm{BV}}$ 
can be defined also on a general $3$--fold $M$.  

Let $\mathbfs{\mathcal{L}}_{\mathrm{BV}}$ be the BV Lagrangian, the integrand superfield 
of $S_{\mathrm{BV}}$ in \eqref{SW}. A simple calculation shows that $\mathbfs{\mathcal{L}}_{\mathrm{BV}}$
is gauge invariant, ${}^g\mathbfs{\mathcal{L}}_{\mathrm{BV}}=\mathbfs{\mathcal{L}}_{\mathrm{BV}}$.
Hence, the BV action $S_{\mathrm{BV}}$, also,  
can be defined on a general $3$--fold $M$.  

In the verification of the BV master equation \eqref{SWSW=0}, one uses 
Stokes' theorem to eliminate a term of the form $2\int_{T[1]M}\mathbfs{d}\mathbfs{\mathcal{L}}_{\mathrm{BV}}$.  
This is legitimate also on a non trivial $M$, as 
${}^g\mathbfs{\mathcal{L}}_{\mathrm{BV}}=\mathbfs{\mathcal{L}}_{\mathrm{BV}}$.
So, the master equation holds on any $3$--fold $M$.

Gauge covariance of the BV field variations \eqref{dWsuperfields} requires that 
${}^g\delta_{\mathrm{BV}}\mathbfs{\mathcal{F}}=\delta_{\mathrm{BV}}{}^g\mathbfs{\mathcal{F}}$
for all superfields $\mathbfs{\mathcal{F}}$
(cf. fn. \ref{foot:gdf}). This is expected by the gauge covariance 
of the BV master equation \eqref{SWSW=0} and it can be directly verified
using \eqref{dWsuperfields}, \eqref{gsuperfields}. 
So, the field variations \eqref{dWsuperfields} are globally defined on a general 
$3$--fold $M$.

We can now show that AKSZ $\mathfrak{v}$ BF gauge theory is the appropriate BV extension
of classical $\mathfrak{v}$ BF gauge theory curing all the diseases of this latter. To do so,
we expand the basic superfields $\mathbfs{p}$, $\mathbfs{q}$, $\mathbfs{P}$, $\mathbfs{Q}$
in components 
\begin{subequations}
\label{bcBCi}
\begin{align}
&\mathbfs{p}=-\beta^++b^+-\omega^++c^+,
\vphantom{\Big]}
\label{bi}
\\
&\mathbfs{q}=c-\omega+b-\beta,
\vphantom{\Big]}
\label{ci}
\\
&\mathbfs{P}=-B^++\varOmega^+-C^++\varGamma^+,
\vphantom{\Big]}
\label{Bi}
\\
&\mathbfs{Q}=\varGamma-C+\varOmega-B,
\vphantom{\Big]}
\label{Ci}
\end{align}
\end{subequations}
where the terms in the right hand side are written down in increasing order of 
form degree and decreasing order of ghost number degree and 
the choice of the signs of the component fields is conventional
\footnote{$\vphantom{\bigg[}$ Here, $\phi^+$ is the antifield of $\phi$.
In order to have the component fields organized naturally
in doublets as in subsect. \ref{subsec:linfdoub}, 
we have not followed the convention, common in BV theory, of requiring the antifields
to have negative ghost number degree. In AKSZ theory, this is allowed since 
the field/antifield splitting is simply a conventional choice of local Darboux coordinates 
for the BV form in BV field space. A redefinition of such separation is just 
a BV field symplectomorphism leaving the BV form invariant. }. 
We then write down the BV action 
and the BV superfield variations
in terms of the components by substituting the expansions \eqref{bcBCi}
into \eqref{SW} and \eqref{dWsuperfields}, respectively. 
The resulting expressions are 
rather messy and are collected in app. \ref{app:vakszcomp}  for the interested reader.
However, it is not difficult to see that the truncation of the BV action 
\eqref{SW} to the ghost 
\vspace{.1cm}\eject\noindent
number degree $0$ fields $\omega$, $\varOmega$, $B^+$, $b^+$    
reproduces precisely the classical action 
of $\mathfrak{v}$ BF gauge theory given in eq. \eqref{clbfact}. 
Further, the truncation of the BV variations of the ghost number degree $0$ fields 
$\omega$, $\varOmega$, $B^+$, $b^+$ to the ghost number degree $0$, 
$1$ fields 
$\omega$, $\varOmega$, $B^+$, $b^+$, $c$, $C$, $\beta^+$ reproduces 
precisely the corresponding classical gauge symmetry variations 
given in eqs. \eqref{clbffvar}.
Similarly, we obtain  the classical gauge symmetry variations of the ghost number 
degree $1$, $2$ fields $c$, $C$, $\beta^+$, $\varGamma$ given in eqs. \eqref{ghclbffvar} by truncating the 
BV variations of those fields to ghost number degree $0$, $1$, $2$ fields
$\omega$, $\varOmega$, $B^+$, $b^+$, $c$, $C$, $\beta^+$, $\varGamma$. 

AKSZ $\mathfrak{v}$ BF gauge theory, however, is not affected by the problems 
plaguing classical $\mathfrak{v}$ BF gauge theory: unlike its classical counterpart
$\delta_{\mathrm{cl}}$, the BV field variation operator 
$\delta_{\mathrm{BV}}$ is nilpotent off--shell and gauge covariant. 
There is nevertheless a cost for this gain. The component fields 
$\omega$, $\varOmega$ no longer behave under gauge transformation as
the components of a connection doublet (cf. eqs. \eqref{7linfdglob}), but mix with the 
component fields $b$, $c$. This renders the geometrical interpretation of the 
component fields less evident.

\subsection{\normalsize \textcolor{blue}{Rectified AKSZ $\mathfrak{v}$ BF gauge theory}}
\label{subsec:linfgaufix}

\hspace{.5cm} The non linear nature of the superfield gauge transformations 
\eqref{gsuperfields} 
in AKSZ $\mathfrak{v}$ BF gauge theory makes it difficult to control gauge covariance 
and carry out gauge fixing. It is possible to reformulate the theory in an equivalent way 
that is covariant under a completely linear rectified gauge transformation action.
We call the resulting model {\it rectified AKSZ $\mathfrak{v}$ BF gauge theory}.
The price  for this 
is that 
the BV action and field variations are  definitely 
more complicated. 

The field content of rectified AKSZ $\mathfrak{v}$ BF gauge theory
consists of four superfields 
$\mathbfs{p}$, $\mathbfs{q}$, $\mathbfs{P}$, $\mathbfs{Q}$ of the same type as that
of plain AKSZ $\mathfrak{v}$ BF gauge theory.
The BV symplectic form 
is given again by \eqref{OmegaW}. 

The BV action 
of rectified AKSZ $\mathfrak{v}$ BF gauge theory
is given instead by an  expression different from 
\eqref{SWSW=0} involving a $2$--term $L_\infty$ algebra gauge rectifier $(\lambda,\mu)$ 
and a background connection doublet $(\bar\omega,\bar\varOmega)$ with curvature doublet 
$(\bar f,\bar F)$ (cf. subsects. \ref{subsec:linfdoub}, \ref{subsec:rectf}). Explicitly, 
the action is given by 
\begin{align}
S_{\mathrm{BV}}=\int_{T[1]M}\varrho\Big[&-\langle\mathbfs{p},-\bar{\mathbfs{f}}_{\lambda,\mu}
+\bar{\mathbfs{D}}_{\lambda,\mu}\mathbfs{q}
-\frac{1}{2}[\mathbfs{q},\mathbfs{q}]_\lambda+\partial\mathbfs{Q}\rangle
\vphantom{\Big[}
\label{rSW}
\\
&+\langle\mathbfs{P}, \bar{\mathbfs{F}}_{\lambda,\mu}+\bar{\mathbfs{D}}_{\lambda,\mu}\mathbfs{Q}
-[\mathbfs{q},\mathbfs{Q}+\bar{\mathbfs{\varOmega}}
-\frac{1}{2}\mathbfs{\lambda}(\bar{\mathbfs{\omega}},\bar{\mathbfs{\omega}})
+\mathbfs{\mu}(\bar{\mathbfs{\omega}})]_\lambda
\nonumber
\\
&+\frac{1}{6}[\mathbfs{q}-\bar{\mathbfs{\omega}},
\mathbfs{q}-\bar{\mathbfs{\omega}},\mathbfs{q}-\bar{\mathbfs{\omega}}]_\lambda
+\frac{1}{6}[\bar{\mathbfs{\omega}},\bar{\mathbfs{\omega}},\bar{\mathbfs{\omega}}]_\lambda
\vphantom{\Big[}
\nonumber
\\
&+\frac{1}{2}\mathbfs{v}_{\lambda,\mu}(\mathbfs{q}-\bar{\mathbfs{\omega}},\mathbfs{q}-\bar{\mathbfs{\omega}})
-\frac{1}{2}\mathbfs{v}_{\lambda,\mu}(\bar{\mathbfs{\omega}},\bar{\mathbfs{\omega}})
+\mathbfs{w}_{\lambda,\mu}(\mathbfs{q})\rangle\Big],
\nonumber
\end{align}
where $\mathbfs{\lambda}$, $\mathbfs{\mu}$, $\mathbfs{v}_{\lambda,\mu}$, 
$\mathbfs{w}_{\lambda,\mu}$, $\bar{\mathbfs{\omega}}$, $\bar{\mathbfs{\varOmega}}$,
$\bar{\mathbfs{f}}_{\lambda,\mu}$, $\bar{\mathbfs{F}}_{\lambda,\mu}$, $\bar{\mathbfs{D}}_{\lambda,\mu}$ 
are just $\lambda$, $\mu$, $v_{\lambda,\mu}$, $w_{\lambda,\mu}$,
$\bar\omega$, $\bar\varOmega$, $\bar f_{\lambda,\mu}$, $\bar F_{\lambda,\mu}$,
$D_{\lambda,\mu}$ seen as one--component superfields.
Above, the deformed brackets $[\cdot,\cdot]_\lambda$, $[\cdot,\cdot,\cdot]_\lambda$ are defined in
\eqref{rectf2}. The derived rectifiers $v_{\lambda,\mu}$, $w_{\lambda,\mu}$  are defined in \eqref{rectf5}. 
$\bar f_{\lambda,\mu}$, $\bar F_{\lambda,\mu}$ are the rectified counterpart of the curvature components
$\bar f$, $\bar F$ and are defined according to \eqref{rectf8}, viz
$\bar f_{\lambda,\mu}=\bar f$, $\bar F_{\lambda,\mu}=\bar F+\lambda(\bar \omega,\bar f)
-\mu(\bar f)$. 
The rectified covariant derivative $D_{\lambda,\mu}$ is defined according to 
\eqref{rectf10}, \eqref{rectf11} with $\omega$ replaced by $\bar\omega$. Again, 
$S_{\mathrm{BV}}$ satisfies the classical BV master equation \eqref{SWSW=0}
and, so,  the rectified model is consistent and quantizable. 

The BV variations of the rectified $\mathfrak{v}$ 
AKSZ BF gauge theory superfields are 
\begin{subequations}
\label{rdWsuperfields}
\begin{align}
&\delta_{\mathrm{BV}}\mathbfs{p}
=\bar{\mathbfs{D}}_{\lambda,\mu}\mathbfs{p}-[\mathbfs{q},\mathbfs{p}]_\lambda{}^\vee
+[\mathbfs{Q}+\bar{\mathbfs{\varOmega}}
-\frac{1}{2}\mathbfs{\lambda}(\bar{\mathbfs{\omega}},\bar{\mathbfs{\omega}})
+\mathbfs{\mu}(\bar{\mathbfs{\omega}}),\mathbfs{P}]_\lambda{}^\vee
\label{rdWbibfs}
\\
&\hspace{1.5cm}
+\frac{1}{2}[\mathbfs{q}-\bar{\mathbfs{\omega}},\mathbfs{q}-\bar{\mathbfs{\omega}},\mathbfs{P}]_\lambda{}^\vee
+\mathbfs{v}_{\lambda,\mu}{}^\vee(\mathbfs{q}-\bar{\mathbfs{\omega}},\mathbfs{P})
+\mathbfs{w}_{\lambda,\mu}{}^\vee(\mathbfs{P}),
\vphantom{\Big]}
\nonumber
\\
&\delta_{\mathrm{BV}}\mathbfs{q}
=-\bar{\mathbfs{f}}_{\lambda,\mu}
+\bar{\mathbfs{D}}_{\lambda,\mu}\mathbfs{q}
-\frac{1}{2}[\mathbfs{q},\mathbfs{q}]_\lambda+\partial\mathbfs{Q},
\vphantom{\Big]}
\label{rdWcibfs}
\\
&\delta_{\mathrm{BV}}\mathbfs{P}
=\bar{\mathbfs{D}}_{\lambda,\mu}\mathbfs{P}-[\mathbfs{q},\mathbfs{P}]_\lambda{}^\vee+\partial^\vee\mathbfs{p},
\vphantom{\Big]}
\label{rdWBibfs}
\\
&\delta_{\mathrm{BV}}\mathbfs{Q}=
 \bar{\mathbfs{F}}_{\lambda,\mu}+\bar{\mathbfs{D}}_{\lambda,\mu}\mathbfs{Q}
-[\mathbfs{q},\mathbfs{Q}+\bar{\mathbfs{\varOmega}}
-\frac{1}{2}\mathbfs{\lambda}(\bar{\mathbfs{\omega}},\bar{\mathbfs{\omega}})
+\mathbfs{\mu}(\bar{\mathbfs{\omega}})]_\lambda
\label{rdWCibfs}
\\
&\hspace{1.5cm}+\frac{1}{6}[\mathbfs{q}-\bar{\mathbfs{\omega}},
\mathbfs{q}-\bar{\mathbfs{\omega}},\mathbfs{q}-\bar{\mathbfs{\omega}}]_\lambda
+\frac{1}{6}[\bar{\mathbfs{\omega}},\bar{\mathbfs{\omega}},\bar{\mathbfs{\omega}}]_\lambda
\vphantom{\Big[}
\nonumber
\\
&\hspace{1.5cm}+\frac{1}{2}\mathbfs{v}_{\lambda,\mu}(\mathbfs{q}-\bar{\mathbfs{\omega}},
\mathbfs{q}-\bar{\mathbfs{\omega}})
-\frac{1}{2}\mathbfs{v}_{\lambda,\mu}(\bar{\mathbfs{\omega}},\bar{\mathbfs{\omega}})
+\mathbfs{w}_{\lambda,\mu}(\mathbfs{q}).
\vphantom{\Big]}
\nonumber
\end{align}
\end{subequations}
Above, the deformed cobrackets $[\cdot,\cdot]_\lambda{}^\vee$, $[\cdot,\cdot,\cdot]_\lambda{}^\vee$
are defined according to the same prescription as their undeformed counterparts
(cf. eqs. \eqref{co2tlinalgb}--\eqref{co2tlinalge}). 
The derived gauge corectifier $(v_{\lambda,\mu}{}^\vee, w_{\lambda,\mu}{}^\vee)$
are defined by the relations 
$\langle\Xi,v_{\lambda,\mu}(x,y)\rangle=-\langle v_{\lambda,\mu}{}^\vee(x,\Xi),y\rangle$, 
$\langle\Xi,w_{\lambda,\mu}(x)\rangle=-\langle w_{\lambda,\mu}{}^\vee(\Xi),y\rangle$. 
As before, $\delta_{\mathrm{BV}}$ satisfies \eqref{dBV2=0}, and is thus nilpotent, and 
$S_{\mathrm{BV}}$ satisfies \eqref{dBVsBV=0}, and is thus BV invariant.

In  rectified AKSZ $\mathfrak{v}$ BF gauge theory, the gauge transformation action
is no longer given by \eqref{gsuperfields} but, instead, takes the fully linear rectified 
form
\begin{subequations}
\label{grsuperfields}
\begin{align}
&{}^g\mathbfs{p}=g^\vee{}_0(\mathbfs{p})
\vphantom{\Big]}
\label{grbibfs}
\\
&{}^g\mathbfs{q}=g_0(\mathbfs{q}),
\vphantom{\Big]}
\label{grcibfs}
\\
&{}^g\mathbfs{P}=g^\vee{}_1(\mathbfs{P}),
\vphantom{\Big]}
\label{grBibfs}
\\
&{}^g\mathbfs{Q}=g_1(\mathbfs{Q}),
\vphantom{\Big]}
\label{grCibfs}
\end{align}
\end{subequations}  
for $g\in\Gau_1(M,\mathfrak{v})$. 

It can be verified that the integrand superfields 
in the rectified analog of \eqref{OmegaW} and \eqref{rSW} are gauge invariant. 
So, for reasons explained in subsect. \ref{subsec:linfdglob},
the rectified BV form $\varOmega_{\mathrm{BV}}$ and action $S_{\mathrm{BV}}$ 
can be defined also on a general $3$--fold $M$. 
One also finds that 
the rectified BV superfield variations \eqref{rdWsuperfields}
are gauge covariant, ensuring that they also can be defined globally on a general 
$3$--fold $M$. 

Non rectified and rectified AKSZ $\mathfrak{v}$ BF gauge theories are related by an invertible  
superfield redefinition
depending on the gauge rectifier $(\lambda,\mu)$ mapping the fields of the former into those of the latter,
\begin{subequations}
\label{rectsfields}
\begin{align}
&\mathbfs{p}_{\mathrm{r}}=\mathbfs{p}_{\mathrm{nr}}
-\mathbfs{\lambda}^\vee(\mathbfs{q}_{\mathrm{nr}},\mathbfs{P}_{\mathrm{nr}})
-\mathbfs{\mu}^\vee(\mathbfs{P}_{\mathrm{nr}}),
\vphantom{\Big]}
\label{rectbibfs}
\\
&\mathbfs{q}_{\mathrm{r}}=\mathbfs{q}_{\mathrm{nr}}+\bar{\mathbfs{\omega}},
\vphantom{\Big]}
\label{rectcibfs}
\\
&\mathbfs{P}_{\mathrm{r}}=\mathbfs{P}_{\mathrm{nr}},
\vphantom{\Big]}
\label{rectBibfs}
\\
&\mathbfs{Q}_{\mathrm{r}}=\mathbfs{Q}_{\mathrm{nr}}
-\frac{1}{2}\mathbfs{\lambda}(\mathbfs{q}_{\mathrm{nr}},\mathbfs{q}_{\mathrm{nr}})
-\mathbfs{\mu}(\mathbfs{q}_{\mathrm{nr}})-
\bar{\mathbfs{\varOmega}}+\frac{1}{2}\mathbfs{\lambda}(\bar{\mathbfs{\omega}},\bar{\mathbfs{\omega}})
-\mathbfs{\mu}(\bar{\mathbfs{\omega}}).
\vphantom{\Big]}
\label{rectCibfs}
\end{align}
\end{subequations} 
Above,  the subscript $\mathrm{r}$ and $\mathrm{nr}$ mark the rectified and non rectified 
version of the 
\eject\noindent
superfields, respectively. The gauge corectifier 
$(\lambda^\vee,\mu^\vee)$ is defined below eq. \eqref{rectf9}.
Remarkably, \eqref{rectsfields} is BV canonical field map: it transforms the BV symplectic form 
$\varOmega_{\mathrm{BV}}$
of the non rectified theory into that of the rectified one
(cf. eq. \eqref{OmegaW} and its rectified counterpart). It also maps the BV action 
$S_{\mathrm{BV}}$ and the BV superfield variations 
of the non rectified 
theory (cf. eqs. \eqref{SW}, \eqref{dWsuperfields})
into those of the rectified one (cf. eqs. \eqref{rSW}, \eqref{rdWsuperfields}). 
Last but not least, it maps the non rectified superfields gauge transforming according to
\eqref{gsuperfields} into their rectified counterparts gauge transforming instead according to
\eqref{grsuperfields}. 
The non rectified and the rectified theories are therefore fully equivalent. So, 
since the former is independent from the gauge rectifier $(\lambda,\mu)$ 
and the background connection doublet $(\bar\omega,\bar\varOmega)$, the latter also is, in
spite of appearances. 

Rectified AKSZ $\mathfrak{v}$ BF gauge theory can be written in terms 
of superfield components. The component expansions of the superfields are 
again given by eqs. \eqref{bcBCi}, but now the components organize in rectified (dual) 
field doublets. The component expressions of the BV action and field variations are 
reported in app. \ref{app:rvakszcomp} for the interested reader.

The advantage of using the rectified version of AKSZ $\mathfrak{v}$ BF gauge theory
will become clear in the procedure of 
gauge fixing of the theory, which is the topic of the next subsection. 

\subsection{\normalsize \textcolor{blue}{Gauge fixing of  AKSZ $\mathfrak{v}$ BF gauge theory}}
\label{subsec:2tligfix}
 
\hspace{.5cm} 
In BV field theory, gauge fixing is implemented by restricting the theory's
field configurations to lie on a suitable Lagrangian submanifold $\matheul{L}$ 
of BV field space $\matheul{F}$ \cite{BV1,BV2}. The generating functional $\varPsi$ of
$\matheul{L}$ is called {\it gauge fermion}. By definition, $\matheul{L}$ is determined by $\varPsi$ 
as the locus in $\matheul{F}$ where $\phi_i{}^+=\delta_r\varPsi/\delta\phi_i$ holds, where the $\phi_i$, 
$\phi_i{}^+$ are the appropriate set of fields/antifields and $\delta_r/\delta\phi_i$ denotes 
right functional differentiation. 

In general, for a given BV field theory, the construction of $\varPsi$ is not possible without 
suitably modifying the theory. Extra fields and their antifields must be added to $\matheul{F}$. 
These in turn introduce further contributions to the BV odd symplectic form $\varOmega_{\mathrm{BV}}$
and master action $S_{\mathrm{BV}}$. In the extended BV field theory so obtained, $\varPsi$ can be built and 
the gauge fixing can be carried out by restricting to the associated field space 
Lagrangian submanifold $\matheul{L}$ as indicated above. 

The extra fields and antifields required by the construction of the gauge fermion 
$\varPsi$ mentioned in the previous paragraph organize in trivial pairs. 
For each gauge fixing condition, a trivial pair of fields and 
the trivial pair of their antifields is needed. 
The nature of the trivial pairs
is determined by the form of gauge fixing conditions and the requirement that
$\varPsi$ has ghost number degree $-1$.

In more detail, $\varPsi$ is constructed through the following procedure.

Let $V$ be a vector space. A bidegree $(m,n)$ $V$ {\it trivial pair} is a
pair of fields $(\phi,\psi)$ with $\phi\in\Omega^m(M,V^\vee[n])$, $\psi\in\Omega^m(M,V^\vee[n+1])$.
The antifields of the fields of the pair $(\phi,\psi)$ 
form a bidegree $(d-m,-n-2)$ $V^\vee$ trivial pair $(\psi^+,\phi^+)$, the {\it antipair} of $(\phi,\psi)$,
where $d=\dim M$ (in our case $d=3$). 

The BV odd symplectic form $\varOmega_{\mathrm{tBV}}$ of the trivial pair $(\phi,\psi)$, $(\psi^+,\phi^+)$ 
system has the canonical form \hphantom{xxxxxxxxxxxxxxxxxxx}
\begin{equation}
\varOmega_{\mathrm{tBV}}=\int_M\Big[\langle\delta\phi^+,\delta\phi\rangle_{V^\vee}
+\langle\delta\psi^+,\delta\psi\rangle_{V^\vee}\Big].
\label{2tligfix1}
\end{equation}
The BV action $S_{\mathrm{tBV}}$ of the $(\phi,\psi)$, $(\psi^+,\phi^+)$ system is 
\begin{equation}
S_{\mathrm{tBV}}=\int_M\langle\psi,\phi^+\rangle_V
\label{2tligfix2}
\end{equation}
and is easily seen to verify the BV master equation \eqref{SWSW=0}.
The BV field variations of the pair fields are given by
\begin{subequations}
\label{2tligfix4}
\begin{align}
&\delta_{\mathrm{tBV}}\phi=(-1)^{m+n}\psi, \qquad\,\delta_{\mathrm{tBV}}\psi=0,
\vphantom{\Big]}
\label{2tligfix4a}
\\
&\delta_{\mathrm{tBV}}\psi^+=-\phi^+, \qquad\qquad\delta_{\mathrm{tBV}}\phi^+=0.
\vphantom{\Big]}
\label{2tligfix4c}
\end{align}
\end{subequations}  
As usual, $\delta_{\mathrm{tBV}}$ satisfies \eqref{dBV2=0}, 
and is thus nilpotent, and 
$S_{\mathrm{tBV}}$ satisfies \eqref{dBVsBV=0}, 
and is thus BV invariant.
When several trivial pairs and their antipairs are involved,
the BV form $\varOmega_{\mathrm{tBV}}$ is again of the form \eqref{2tligfix1}
with a contribution from 
each pair/antipair and similarly for the BV 
action $S_{\mathrm{tBV}}$ in eq. \eqref{2tligfix2}.

If a gauge fixing condition is of the form $\varpi=0$, where $\varpi$ is a  bidegree $(p,q)$
$V$--valued field expression, one needs a bidegree $(d-p,-q-1)$ $V$
trivial pair $(\phi,\psi)$ together with its bidegree $(p,q-1)$ $V^\vee$
trivial antipair $(\psi^+,\phi^+)$. The corresponding contribution to the 
gauge fermion $\varPsi$ is then given by 
\begin{equation}
\varPsi_\varpi=\int_M\langle\phi,\varpi\rangle_V. \vphantom{\Bigg]}
\label{2tligfix8}
\end{equation}
There is one such contribution for each gauge fixing condition $\varpi$.

The reason why, instead of relying on the plain 
AKSZ 
gauge theory
of subsect. \ref{subsec:linfvmod}, we endeavoured to formulate its rectified 
version in subsect. \ref{subsec:linfgaufix} is that gauge fixing is carried out in a manifestly 
gauge covariant manner far more easily in the latter than the former. 
In fact, the gauge fixing conditions involve the single components of the superfields
and this makes controlling gauge covariance problematic in the non rectified theory
where the components mix under gauge transformation, while it poses no problem in the
rectified one, where they do not (cf. subsects. \ref{subsec:linfvmod}, \ref{subsec:linfgaufix}).

In AKSZ $\mathfrak{v}$ BF gauge theory, a gauge fixing condition is required for 
every field of positive form degree acted upon by the exterior differential $d$. 
From \eqref{rSW} and \eqref{ci}, \eqref{Ci}, we need then a condition for each of the components
$\omega$, $b$, $\beta$, $C$, $\varOmega$, $B$, which we choose to be of the form \hphantom{xxxxxxxxxxx} 
\begin{subequations}
\label{2tligfix7}
\begin{align}
&\bar D_{\lambda,\mu}*\omega=0,
\vphantom{\Big]}
\label{2tligfix7a}
\\
&\bar D_{\lambda,\mu}*b=0,
\vphantom{\Big]}
\label{2tligfix7b}
\\
&*\beta=0,
\vphantom{\Big]}
\label{2tligfix7c}
\\
&\bar D_{\lambda,\mu}*C=0,
\vphantom{\Big]}
\label{2tligfix7d}
\end{align}
\vskip-.3truecm
\eject\noindent
\begin{align}
&\bar D_{\lambda,\mu}*\varOmega=0,
\vphantom{\Big]}
\label{2tligfix7e}
\\
&*B=0,
\vphantom{\Big]}
\label{2tligfix7f}
\end{align}
\end{subequations}
where $*$ is the Hodge star operator associated with some metric $h$
on $M$. \eqref{2tligfix7a}, \eqref{2tligfix7b}, \eqref{2tligfix7d}, \eqref{2tligfix7e}
are standard Lorenz gauge fixing conditions.  
\eqref{2tligfix7c}, \eqref{2tligfix7f} are background gauge fixing conditions.
The reason why the gauge fixing prescriptions of $\beta$, $B$ are stronger than those
of $\omega$, $b$, $C$, $\varOmega$ is that $d\beta=0$, $dB=0$ identically in $d=3$ dimensions. 
Proceeding as explained above and taking into account the form of the 
conditions \eqref{2tligfix7}, 
we introduce three $\hat{\mathfrak{v}}_0$ trivial pairs
$(\tilde c,\tilde \theta)$, $(\tilde\omega,\tilde\xi)$, $(\tilde q,\tilde \chi)$ 
of bidegree $(0,-1)$, $(1,0)$, $(3,1)$ 
together with their $\hat{\mathfrak{v}}_0{}^\vee$ trivial antipairs
$(\tilde\theta^+,\tilde c^+)$, $(\tilde\xi^+,\tilde\omega^+)$, $(\tilde \chi^+,\tilde q^+)$ 
of bidegree $(3,-1)$, $(2,-2)$, $(0,-3)$ for the 
conditions \eqref{2tligfix7a}--\eqref{2tligfix7c} and three $\hat{\mathfrak{v}}_1$ trivial pairs
$(\tilde\varGamma, \tilde H)$, $(\tilde C,\tilde\varTheta)$, $(\tilde F, \tilde\varSigma)$
of bidegree $(0,-2)$, $(1,-1)$, $(3,0)$ together with their 
$\hat{\mathfrak{v}}_1{}^\vee$ trivial antipairs
$(\tilde H^+,\tilde\varGamma^+)$, $(\tilde\varTheta^+,\tilde C^+)$, $(\tilde\varSigma^+,\tilde F^+)$
of bidegree $(3,0)$, $(2,-1)$, $(0,-2)$ for the 
conditions \eqref{2tligfix7d}--\eqref{2tligfix7f}, respectively. The above gauge fixing 
is however not sufficient. The resulting gauge fermion $\varPsi$ contains a portion 
$\int_M[\langle\tilde \omega,\bar D_{\lambda,\mu}*b\rangle
+\langle\tilde C,\bar D_{\lambda,\mu}*\varOmega\rangle]$ whereby 
two of the constrains which define the field space Lagrangian submanifold $\matheul{L}$ 
take the form $b^+=*D_{\lambda,\mu}\tilde \omega$ and $\varOmega^+=*D_{\lambda,\mu}\tilde C$. 
As $\tilde \omega$, $\tilde C$ have positive form degree and are acted upon by $d$, 
two more gauge fixing conditions are required, 
\begin{subequations}
\label{2tligfix9}
\begin{align}
&\bar D_{\lambda,\mu}*\tilde\omega=0,
\vphantom{\Big]}
\label{2tligfix9a}
\\
&\bar D_{\lambda,\mu}*\tilde C=0.
\vphantom{\Big]}
\label{2tligfix9b}
\end{align}
\end{subequations} 
We thus introduce one $\hat{\mathfrak{v}}_0{}^\vee$ trivial pair
$(a,\theta)$ of bidegree $(0,-1)$ together with its 
$\hat{\mathfrak{v}}_0$ trivial antipair $(\theta^+,a^+)$ of bidegree $(3,-1)$ 
for the gauge fixing conditions \eqref{2tligfix9a} and one $\hat{\mathfrak{v}}_1{}^\vee$ 
trivial pair $(\varPhi, H)$ of bidegree $(0,0)$ together with its 
$\hat{\mathfrak{v}}_0$ trivial antipair $(H^+,\varPhi^+)$ of bidegree $(3,-2)$ 
for the gauge fixing conditions \eqref{2tligfix9b}, respectively.
The trivial pair BV action and field variations can now be written down easily 
using \eqref{2tligfix2} and \eqref{2tligfix4}.

From \eqref{2tligfix8}, the gauge fermion associated with the above gauge fixing is
\begin{align}
&\varPsi=\int_M\Big[\langle\tilde c,\bar D_{\lambda,\mu}*\omega\rangle
+\langle\tilde \omega,\bar D_{\lambda,\mu}*b\rangle
+\langle\tilde q,*\beta\rangle+\langle \tilde\varGamma, \bar D_{\lambda,\mu}*C\rangle
\vphantom{\Big]}
\label{2tligfix10}
\\
&\hspace{2.cm}+\langle\tilde C,\bar D_{\lambda,\mu}*\varOmega\rangle
+\langle\tilde F, *B\rangle+\langle a,\bar D_{\lambda,\mu}*\tilde\omega\rangle
+\langle\varPhi,\bar D_{\lambda,\mu}*\tilde C\rangle\Big].
\vphantom{\Big]}
\nonumber
\end{align}
The constraints defining the associated BV field space Lagrangian submanifold $\matheul{L}$ are
written down explicitly in app. \ref{app:psilagr}.

The gauge fixed action is now given by $I=(S_{\mathrm{BV}}+S_{\mathrm{tBV}})|_{\matheul{L}}$.
Explicitly 
\begin{align}
I&=\int_M\Big[\langle *\bar D_{\lambda,\mu}\tilde\omega,
\bar f_{\lambda,\mu}+\bar D_{\lambda,\mu}\omega
+\frac{1}{2}[\omega,\omega]_\lambda
+[c,b]_\lambda-\partial\varOmega\rangle
\vphantom{\Big]}
\label{2tligfix12}
\\
&\hspace{1.4cm}
-\langle *\tilde F,\bar F_{\lambda,\mu}+\bar D_{\lambda,\mu}\varOmega
+[\omega,\varOmega+\bar\varOmega-\frac{1}{2}\lambda(\bar\omega,\bar\omega)+\mu(\bar\omega)]_\lambda
\vphantom{\Big]}
\nonumber
\\
&\hspace{1.4cm}+[c,B]_\lambda+[b,C]_\lambda+[\beta,\varGamma]_\lambda
-[\omega+\bar\omega,c,b]_\lambda-\frac{1}{2}[c,c,\beta]_\lambda
\vphantom{\Big]}
\nonumber
\\
&\hspace{1.4cm}-\frac{1}{6}[\omega+\bar\omega,\omega+\bar\omega,\omega+\bar\omega]_\lambda
+\frac{1}{6}[\bar\omega,\bar\omega,\bar\omega]_\lambda+v_{\lambda,\mu}(c,b)
\vphantom{\Big]}
\nonumber
\\
&\hspace{1.4cm}+\frac{1}{2}v_{\lambda,\mu}(\omega+\bar\omega,\omega+\bar\omega)
-\frac{1}{2}v_{\lambda,\mu}(\bar\omega,\bar\omega)-w_{\lambda,\mu}(\omega)
\rangle
\vphantom{\Big]}
\nonumber
\\
&\hspace{1.4cm}
-\langle*\bar D_{\lambda,\mu}\tilde c,\bar D_{\lambda,\mu}c+[\omega,c]_\lambda-\partial C\rangle
-\langle*\bar D_{\lambda,\mu}\tilde C,\bar D_{\lambda,\mu}C+[\omega,C]_\lambda
\vphantom{\Big]}
\nonumber
\\
&\hspace{1.4cm}
+[c,\varOmega+\bar\varOmega-\frac{1}{2}\lambda(\bar\omega,\bar\omega)+\mu(\bar\omega)]_\lambda
+[b,\varGamma]_\lambda-\frac{1}{2}[c,c,b]_\lambda
\vphantom{\Big]}
\nonumber
\\
&\hspace{1.4cm}
-\frac{1}{2}[\omega+\bar\omega,\omega+\bar\omega,c]_\lambda+v_{\lambda,\mu}(\omega+\bar\omega,c)
-w_{\lambda,\mu}(c)\rangle+
\vphantom{\Big]}
\nonumber
\\
&\hspace{1.4cm}
+\langle *\bar D_{\lambda,\mu}\tilde\varGamma,\bar D_{\lambda,\mu}\varGamma+[\omega,\varGamma]_\lambda
+[c,C]_\lambda-\frac{1}{2}[\omega+\bar\omega,c,c]_\lambda
\vphantom{\Big]}
\nonumber
\\
&\hspace{1.4cm}+\frac{1}{2}v_{\lambda,\mu}(c,c)\rangle
-\langle*\tilde q,\bar D_{\lambda,\mu}b+[\omega,b]_\lambda+[c,\beta]_\lambda
-\partial B\rangle
\vphantom{\Big]}
\nonumber
\\
&\hspace{1.4cm}-\langle\tilde \theta,\bar D_{\lambda,\mu}*\omega\rangle
-\langle\tilde \xi,\bar D_{\lambda,\mu}*b+*\bar D_{\lambda,\mu}a\rangle
+\langle\tilde \chi,*\beta\rangle
\vphantom{\Big]}
\nonumber
\\
&\hspace{1.4cm}+\langle\tilde H,\bar D_{\lambda,\mu}*C\rangle
+\langle\tilde\varTheta,\bar D_{\lambda,\mu}*\varOmega-*\bar D_{\lambda,\mu}\varPhi\rangle
-\langle\tilde \varSigma,*B\rangle
\vphantom{\Big]}
\nonumber
\\
&\hspace{1.4cm}-\langle \theta,\bar D_{\lambda,\mu}*\tilde\omega\rangle
+\langle H,\bar D_{\lambda,\mu}*\tilde C\rangle\Big].
\nonumber
\end{align}
The BRST field variation operator is 
$s=(\delta_{\mathrm{BV}}+\delta_{\mathrm{tBV}})|_{\matheul{L}}$. One finds so
\begin{subequations}
\label{2tligfix13}
\begin{align}
&sc
=-\frac{1}{2}[c,c]_\lambda+\partial\varGamma, \hspace{8.1cm}
\vphantom{\Big]}
\label{2tligfix131}
\\
&s\omega
=-\bar D_{\lambda,\mu}c-[\omega,c]_\lambda+\partial C,
\vphantom{\Big]}
\label{2tligfix132}
\end{align}
\begin{align}
&sb
=-\bar f_{\lambda,\mu}-\bar D_{\lambda,\mu}\omega-\frac{1}{2}[\omega,\omega]_\lambda-[c,b]_\lambda
+\partial\varOmega, \hspace{3.5cm}
\vphantom{\Big]}
\label{2tligfix133}
\\
&s\beta
=-\bar D_{\lambda,\mu}b-[\omega,b]_\lambda-[c,\beta]_\lambda+\partial B,
\vphantom{\Big]}
\label{2tligfix134}
\\
&s\varGamma
=-[c,\varGamma]_\lambda+\frac{1}{6}[c,c,c]_\lambda,
\vphantom{\Big]}
\label{2tligfix135}
\\
&sC
=-\bar D_{\lambda,\mu}\varGamma-[\omega,\varGamma]_\lambda
-[c,C]_\lambda+\frac{1}{2}[\omega+\bar\omega,c,c]_\lambda-\frac{1}{2}v_{\lambda,\mu}(c,c),
\vphantom{\Big]}
\label{2tligfix136}
\\
&s\varOmega
=-\bar D_{\lambda,\mu}C-[\omega,C]_\lambda
-[c,\varOmega+\bar\varOmega-\frac{1}{2}\lambda(\bar\omega,\bar\omega)+\mu(\bar\omega)]_\lambda
\vphantom{\Big]}
\label{2tligfix137}
\\
&\hspace{2cm}
-[b,\varGamma]_\lambda+\frac{1}{2}[c,c,b]_\lambda
+\frac{1}{2}[\omega+\bar\omega,\omega+\bar\omega,c]_\lambda
\vphantom{\Big]}
\nonumber
\\
&\hspace{2cm}-v_{\lambda,\mu}(\omega+\bar\omega,c)+w_{\lambda,\mu}(c),
\vphantom{\Big]}
\nonumber
\\
&sB
=-\bar F_{\lambda,\mu}-\bar D_{\lambda,\mu}\varOmega
-[\omega,\varOmega+\bar\varOmega-\frac{1}{2}\lambda(\bar\omega,\bar\omega)+\mu(\bar\omega)]_\lambda
\vphantom{\Big]}
\label{2tligfix138}
\\
&\hspace{2cm}-[c,B]_\lambda-[b,C]_\lambda-[\beta,\varGamma]_\lambda
+[\omega+\bar\omega,c,b]_\lambda+\frac{1}{2}[c,c,\beta]_\lambda
\vphantom{\Big]}
\nonumber
\\
&\hspace{2cm}+\frac{1}{6}[\omega+\bar\omega,\omega+\bar\omega,\omega+\bar\omega]_\lambda
-\frac{1}{6}[\bar\omega,\bar\omega,\bar\omega]_\lambda-v_{\lambda,\mu}(c,b)
\hspace{2cm}
\vphantom{\Big]}
\nonumber
\\
&\hspace{2cm}-\frac{1}{2}v_{\lambda,\mu}(\omega+\bar\omega,\omega+\bar\omega)
+\frac{1}{2}v_{\lambda,\mu}(\bar\omega,\bar\omega)+w_{\lambda,\mu}(\omega),
\nonumber
\vphantom{\Big]}
\\
&s\tilde c=-\tilde \theta, 
\vphantom{\Big]}
\label{2tligfix139}
\\
&s\tilde \theta=0,
\vphantom{\Big]}
\label{2tligfix1310}
\\
&s\tilde C=\tilde \varTheta, 
\vphantom{\Big]}
\label{2tligfix1311}
\\
&s\tilde \varTheta=0,
\vphantom{\Big]}
\label{2tligfix1312}
\\
&s\varPhi=H,
\vphantom{\Big]}
\label{2tligfix1313}
\pagebreak[4]
\\
&sH=0,
\vphantom{\Big]}
\label{2tligfix1314}
\\
&s\tilde \varGamma=\tilde H,
\vphantom{\Big]}
\label{2tligfix1315}
\\
&s\tilde H=0,
\vphantom{\Big]}
\label{2tligfix1316}
\\
&s\tilde\omega=-\tilde\xi,
\vphantom{\Big]}
\label{2tligfix1317}
\\
&s\tilde\xi=0,
\vphantom{\Big]}
\label{2tligfix1318}
\\
&sa=-\theta,
\vphantom{\Big]}
\label{2tligfix1319}
\end{align}
\vfil\eject\noindent
\begin{align}
&s\theta=0,
\vphantom{\Big]}
\label{2tligfix1320}
\\
&s\tilde F=-\tilde\varSigma, 
\vphantom{\tilde{\Big]}}
\label{2tligfix1321}
\\
&s\tilde\varSigma=0, \hspace{10.3cm}
\vphantom{\Big]}
\label{2tligfix1322}
\\
&s\tilde q=\tilde\chi,
\vphantom{\Big]}
\label{2tligfix1323}
\\
&s\tilde\chi=0.
\vphantom{\Big]}
\label{2tligfix1324}
\end{align}
\end{subequations} 

The BRST field variation operator $s$ is nilpotent off--shell
\begin{equation}
s^2=0.
\label{2tligfix15}
\end{equation}
This property of $s$ is fundamental. There is no need to directly verify it:
its holding is obvious a priori. From the component expansions \eqref{bcBCi} and 
the expressions of the BV field variations \eqref{rdWsuperfields}, \eqref{2tligfix4a}, 
it is apparent that fields are closed under the action of the BV field variation 
operators $\delta_{\mathrm{BV}}$, $\delta_{\mathrm{tBV}}$, unlike antifields. 
Thus, $s$ reduces simply to the restriction of 
$\delta_{\mathrm{BV}}$, $\delta_{\mathrm{tBV}}$ to the field sector of the extended BV 
field space. The nilpotence of $s$ is therefore an immediate consequence of that of 
$\delta_{\mathrm{BV}}$, $\delta_{\mathrm{tBV}}$.  

The BRST invariant action $I$ is simply related to the gauge fermion $\varPsi$, 
\begin{equation}
I=s\varPsi.
\label{2tligfix14}
\end{equation}
From \eqref{2tligfix15}, \eqref{2tligfix14}, it is then immediate that
\begin{equation}
sI=0.
\label{2tligfix16}
\end{equation}
The action $I$ is thus BRST invariant as required.

Let us now pose to discuss the results just obtained.
The use of rectified fields avoids field mixing under gauge transformation and 
renders all the gauge fixing conditions manifestly gauge covariant. 
It also makes 
the trivial pair BV action $S_{\mathrm{tBV}}$ 
and the gauge fermion $\varPsi$ and, 
consequently, also the gauge fixed action $I$ and the BRST field variation operator $s$
manifestly gauge invariant. In this way, 
\eject\noindent
on a non trivial $3$--fold $M$, 
rectified fields behave much as sections of ordinary vector bundles. 
Further, $S_{\mathrm{tBV}}$, $\varPsi$, $I$ and $s$, once expressed in terms of rectified fields,
are all manifestly globally defined.
In the non rectified theory, the same could also be achieved of course, but at the price 
of dealing with fields mixing under gauge transformation in a complicated way
and, so, loosing manifest gauge covariance. 
On a non trivial $3$--fold $M$, non rectified fields belong to a rather non standard non linear 
geometry in which the fields' global properties are not describable in the familiar terms of ordinary vector bundle
geometry (cf. subsect. \ref{subsec:linfdglob}). 
Further, $S_{\mathrm{tBV}}$, $\varPsi$, $I$ and $s$, when expressed in terms of non rectified fields, 
are not manifestly globally defined. 
On the other hand, certain expressions are simpler in non rectified form.
Derectification can be achieved, if one wishes so, by substituting 
the expressions \eqref{rectsfields} in component form in the above relations. 

\subsection{\normalsize \textcolor{blue}{Topological $\mathfrak{v}$ BF gauge theory}}
\label{subsec:2tligtop}

\hspace{.5cm} By \eqref{2tligfix14}, the gauge fixed AKSZ $\mathfrak{v}$ BF gauge theory is topological.
We call this theory {\it topological $\mathfrak{v}$ BF gauge theory}
because it stems from classical $\mathfrak{v}$ BF gauge theory introduced 
in subsect. \ref{subsec:clbf2tlinf}.
 
In topological field theory, topological operators are of particular salience, 
because their correlators are independent from
all background fields which enter only in the gauge fermion
and so compute topological invariants of the background geometry. 
The topological operators are the BRST cohomology classes. 
A thorough discussion of the BRST cohomology of 
topological $\mathfrak{v}$ BF gauge theory is therefore in order.
This requires the prior study of the BV cohomology of
AKSZ $\mathfrak{v}$ BF gauge theory, beginning with the non rectified version of the model
and then proceeding to the rectified one. 

In subsect. \ref{subsec:linfweil}, we have seen that the $2$--term $L_\infty$ algebra 
$\mathfrak{v}$ is characterized by the cohomology $H_{CE}^*(\mathfrak{v})$
of the Chevalley--Eilenberg cochain complex $(\mathrm{CE}(\mathfrak{v}),\mathcal{Q}_{\mathrm{CE}(\mathfrak{v})})$, where 
$\mathrm{CE}(\mathfrak{v})=S(\mathfrak{v}^\vee[1])$ and 
$\mathcal{Q}_{\mathrm{CE}(\mathfrak{v})}$ is defined in \eqref{2tlinalgQ}. 

Assume first that $M$ is diffeomorphic to $\mathbb{R}^3$.
The algebra $\Omega^*(M,\mathrm{CE}(\mathfrak{v}))$ of $\mathrm{CE}(\mathfrak{v})$--valued forms 
is graded by the total form plus $\mathrm{CE}(\mathfrak{v})$ degree.
Further, $\Omega^*(M,\mathrm{CE}(\mathfrak{v}))$ admits a natural coboundary operator 
given by 
\begin{equation}
d_{\mathrm{CE}(\mathfrak{v})}=d-\mathcal{Q}_{\mathrm{CE}(\mathfrak{v})},
\label{2tligtop1}
\end{equation}
where $d$ is the exterior differential. So, 
$(\Omega^*(M,\mathrm{CE}(\mathfrak{v})),d_{\mathrm{CE}(\mathfrak{v})})$ is a cochain complex, with which there is associated 
a cohomology $H_{dCE}^*(M,\mathrm{CE}(\mathfrak{v}))$. 

Let $\varkappa\in \Omega^n(M,\mathrm{CE}(\mathfrak{v}))$. Then, $\varkappa$ has an expansion
\begin{equation}
\varkappa=\sum_{p,q\geq 0}\frac{1}{p!q!}\varkappa_{p,q}(\pi,\ldots,\pi;\varPi,\ldots,\varPi),
\label{2tligtop2}
\end{equation}
where $\varkappa_{p,q}$ is a $n-p-2q$ form with values in $\bigwedge^p\hat{\mathfrak{v}}_0{}^\vee\otimes
\bigvee^q\hat{\mathfrak{v}}_1{}^\vee$ (see subsect. \ref{sec:linfty}). 
$\varkappa_{p,q}$ vanishes unless $n-d,0\leq p+2q\leq n$, where $d=3$ in our case. 
In non rectified AKSZ $\mathfrak{v}$ BF gauge theory, we associate with $\varkappa$ the 
degree $n$ superfield
\begin{equation}
\mathbfs{\mathcal{O}}_\varkappa
=\sum_{p,q\geq 0}\frac{1}{p!q!}\mathbfs{\varkappa}_{p,q}(\mathbfs{p},\ldots,\mathbfs{p};
\mathbfs{P},\ldots,\mathbfs{P}), 
\label{2tligtop3}
\end{equation}
where $\mathbfs{\varkappa}_{p,q}$ is $\varkappa_{p,q}$ seen as a superfield. 
A simple calculation shows that 
\begin{equation}
\delta_{\mathrm{BV}}\mathbfs{\mathcal{O}}_\varkappa=\mathbfs{d}\mathbfs{\mathcal{O}}_\varkappa
-\mathbfs{\mathcal{O}}_{d_{\mathrm{CE}(\mathfrak{v})}\varkappa}. \vphantom{\bigg[}
\label{2tligtop4}
\end{equation}
Thus, if $\varkappa$ is a $d_{\mathrm{CE}(\mathfrak{v})}$--cocycle, $d_{\mathrm{CE}(\mathfrak{v})}\varkappa=0$, then 
\begin{equation}
\delta_{\mathrm{BV}}\mathbfs{\mathcal{O}}_\varkappa=\mathbfs{d}\mathbfs{\mathcal{O}}_\varkappa.
\vphantom{\bigg[}
\label{2tligtop5}
\end{equation}
$\mathbfs{\mathcal{O}}_\varkappa$ is then $\delta_{\mathrm{BV}}$--closed mod $\mathbfs{d}$.
Further, when $\varkappa$ is a $d_{\mathrm{CE}(\mathfrak{v})}$--coboundary, 
$\varkappa=d_{\mathrm{CE}(\mathfrak{v})}\vartheta$ for some $\vartheta\in 
\Omega^{n-1}(M,\mathrm{CE}(\mathfrak{v}))$, then 
\begin{equation}
\mathbfs{\mathcal{O}}_\varkappa=\mathbfs{d}\mathbfs{\mathcal{O}}_\vartheta
-\delta_{\mathrm{BV}}\mathbfs{\mathcal{O}}_\vartheta.
\vphantom{\bigg[}
\label{2tligtop6}
\end{equation}
$\mathbfs{\mathcal{O}}_\varkappa$ is then $\delta_{\mathrm{BV}}$--exact mod $\mathbfs{d}$.
The algorithm followed here, so, produces only mod $\mathbfs{d}$ BV cohomology classes, 
while we want to obtain true BV cohomology classes. 
The way of achieving this is well--established. 

\vfill\eject

A supercycle (superboundary) $\mathbfs{\gamma}$ is a formal sum of ordinary cycles (boundaries),
$\mathbfs{\gamma}=\gamma_0+\gamma_1+\gamma_2+\gamma_3$. Superfields can be integrated on supercycles.
If $\mathbfs{\varphi}=\varphi^0+\varphi^1+\varphi^2+\varphi^3$ is a superfield expanded in its components 
and $\mathbfs{\gamma}$ is a supercycle, then we have $\int_{\mathbfs{\gamma}}\mathbfs{\varphi}=
\sum_{r=1}^3\int_{\gamma_r}\phi^r$. By Stokes theorem, we have $\int_{\mathbfs{\gamma}}\mathbfs{d\varphi}=0$. 
For any supercycle $\mathbfs{\gamma}$, we define
\begin{equation}
\langle\mathbfs{\gamma},\mathbfs{\mathcal{O}}_\varkappa\rangle
=\int_{\mathbfs{\gamma}}\mathbfs{\mathcal{O}}_\varkappa.
\label{2tligtop7}
\end{equation}
From \eqref{2tligtop5}, when $\varkappa$ is a cocycle, 
\begin{equation}
\delta_{\mathrm{BV}}\langle\mathbfs{\gamma},\mathbfs{\mathcal{O}}_\varkappa\rangle=0.
\label{2tligtop8}
\end{equation}
Further, from \eqref{2tligtop6}, when $\varkappa$ is a coboundary, 
\begin{equation}
\langle\mathbfs{\gamma},\mathbfs{\mathcal{O}}_\varkappa\rangle
=-\delta_{\mathrm{BV}}\langle\mathbfs{\gamma},\mathbfs{\mathcal{O}}_\vartheta\rangle.
\label{2tligtop9}
\end{equation}
It follows that, for a fixed  supercycle $\mathbfs{\gamma}$, 
the mapping $\varkappa\mapsto \langle\mathbfs{\gamma},\mathbfs{\mathcal{O}}_\varkappa\rangle$
is a homomorphism of the cohomology $H_{dCE}^*(M,\mathrm{CE}(\mathfrak{v}))$ into the model's 
BV cohomology $H_{BV}^*$.

We require the gauge invariance of the superfields $\mathbfs{\mathcal{O}}_\varkappa$. 
An action of $\Gau_1(M,\mathfrak{v})$ on $\Omega^*(M,\mathrm{CE}(\mathfrak{v}))$
must be defined such that, for $g\in\Gau_1(M,\mathfrak{v})$,  
\begin{equation}
\mathbfs{\mathcal{O}}_{{}^g\varkappa}({}^g\mathbfs{p};{}^g\mathbfs{P})
=\mathbfs{\mathcal{O}}_\varkappa(\mathbfs{p};\mathbfs{P}), 
\label{2tligtop10}
\end{equation}
where we have explicitly indicated the dependence of $\mathbfs{\mathcal{O}}_\varkappa$ on the basic
superfields $\mathbfs{p}$, $\mathbfs{P}$ for clarity and the action of $\Gau_1(M,\mathfrak{v})$ on 
$\mathbfs{p}$, $\mathbfs{P}$ is given by \eqref{gcibfs}, \eqref{gCibfs}. 
Given the non linear nature of the gauge transformation action on superfields, 
a simple explicit expression of ${}^g\varkappa$ cannot be written. 
However, it is easy to see that ${}^g\varkappa$ is linear in $\varkappa$
\footnote{$\vphantom{\Big[}$ Indeed, by \eqref{2tligtop2} and \eqref{2tligtop10}, one has 
$\mathbfs{\mathcal{O}}_{{}^g(a_1\varkappa_1+a_2\varkappa_2)}({}^g\mathbfs{p};{}^g\mathbfs{P})
=\mathbfs{\mathcal{O}}_{a_1\varkappa_1+a_2\varkappa_2}(\mathbfs{p};\mathbfs{P})
=a_1\mathbfs{\mathcal{O}}_{\varkappa_1}(\mathbfs{p};\mathbfs{P})
+a_2\mathbfs{\mathcal{O}}_{\varkappa_2}(\mathbfs{p};\mathbfs{P})
=a_1\mathbfs{\mathcal{O}}_{{}^g\varkappa_1}({}^g\mathbfs{p};{}^g\mathbfs{P})
+a_2\mathbfs{\mathcal{O}}_{{}^g\varkappa_2}({}^g\mathbfs{p};{}^g\mathbfs{P})
=\mathbfs{\mathcal{O}}_{a_1{}^g\varkappa_1+a_2{}^g\varkappa_2}({}^g\mathbfs{p};{}^g\mathbfs{P})$,
which implies that ${}^g(a_1\varkappa_1+a_2\varkappa_2)=a_1{}^g\varkappa_1+a_2{}^g\varkappa_2$}.

When $M$ is a non trivial $3$--fold, \pagebreak the above analysis must be modified as follows.
In accordance with the general discussion of subsect. \ref{subsec:linfdglob}, 
fields are defined only locally in $M$ and the matching of the local representation of the fields 
on the sets of an open covering $\{U_i\}$ of $M$ is governed
by suitable 
matching data $\{g_{ij}\}$ with $g_{ij}\in\Gau_1(U_{ij},\mathfrak{v})$
acting on the representations  by gauge transformation. 
The construction of globally defined BV cohomology classes 
described above requires then that $\varkappa$ be replaced by a  collection $\{\varkappa_i\}$
with $\varkappa_i\in \Omega^n(U_i,\mathrm{CE}(\mathfrak{v}))$ such that 
$\varkappa_i={}^{g_{ij}}\varkappa_j$ on $U_{ij}$ and $\varkappa_i={}^{t(W_{ijk})}\varkappa_i$
on $U_{ijk}$ (cf. eqs. \eqref{2plinfdglob0} and \eqref{bash2}).
Let $E^n(\mathfrak{v})$ be the linear space of all $\mathrm{CE}(\mathfrak{v})$--valued $n$--form local data
satisfying these matching conditions.
The locally defined coboundary operator $d_{\mathrm{CE}(\mathfrak{v})}$ extends to 
a globally defined one $d_{\mathrm{CE}(\mathfrak{v})}$ on $E^n(M,\mathfrak{v})$
\footnote{$\vphantom{\Big[}$ This can be seen as follows. Take $M$ diffeomorphic to 
$\mathbb{R}^3$. Let $\varkappa\in \Omega^n(M,\mathrm{CE}(\mathfrak{v}))$ and 
let $g\in\Gau_1(M,\mathfrak{v})$. 
Since $\delta_{\mathrm{BV}}{}^g \mathbfs{p}={}^g \delta_{\mathrm{BV}}\mathbfs{p}$, 
$\delta_{\mathrm{BV}}{}^g \mathbfs{P}={}^g \delta_{\mathrm{BV}}\mathbfs{P}$
(cf. subsect. \ref{subsec:linfvmod}), we have 
$\delta_{\mathrm{BV}}\mathbfs{\mathcal{O}}_\varkappa(\mathbfs{p};\mathbfs{P})=
\delta_{\mathrm{BV}}\mathbfs{\mathcal{O}}_{{}^g \varkappa}({}^g \mathbfs{p};{}^g \mathbfs{P})
=\mathbfs{d}\mathbfs{\mathcal{O}}_{{}^g \varkappa}({}^g \mathbfs{p};{}^g \mathbfs{P})
-\mathbfs{\mathcal{O}}_{d_{\mathrm{CE}(\mathfrak{v})}{}^g\varkappa}({}^g \mathbfs{p};{}^g \mathbfs{P})$ $\vphantom{\bigg[}$ 
from \eqref{2tligtop4}, \eqref{2tligtop10}. But we also have 
$\delta_{\mathrm{BV}}\mathbfs{\mathcal{O}}_\varkappa(\mathbfs{p};\mathbfs{P})
=\mathbfs{d}\mathbfs{\mathcal{O}}_\varkappa(\mathbfs{p};\mathbfs{P})
-\mathbfs{\mathcal{O}}_{d_{\mathrm{CE}(\mathfrak{v})}\varkappa}(\mathbfs{p};\mathbfs{P})
=\mathbfs{d}\mathbfs{\mathcal{O}}_{{}^g \varkappa}({}^g \mathbfs{p};{}^g \mathbfs{P})
-\mathbfs{\mathcal{O}}_{{}^gd_{\mathrm{CE}(\mathfrak{v})}\varkappa}({}^g \mathbfs{p};{}^g \mathbfs{P})$ 
again by \eqref{2tligtop4}, \eqref{2tligtop10}. 
Hence, $d_{\mathrm{CE}(\mathfrak{v})}{}^g\varkappa={}^gd_{\mathrm{CE}(\mathfrak{v})}\varkappa$. So, 
on a non trivial $3$--fold $M$, where $\varkappa\in E^n(M,\mathfrak{v})$ 
has only local representations matching by gauge transformation as indicated above,
$d_{\mathrm{CE}(\mathfrak{v})}\varkappa$ is globally defined and $d_{\mathrm{CE}(\mathfrak{v})}
\varkappa\in E^{n+1}(M,\mathfrak{v})$.}. 
Thus, $(E^*(M,\mathfrak{v}),d_{\mathrm{CE}(\mathfrak{v})})$ is a cochain complex. Proceeding as
earlier, one establishes a homomorphism  $H_{d_{\mathrm{CE}(\mathfrak{v})}}{}^*(E^*(M,\mathfrak{v}))$
into $H_{BV}^*$. 

The BV cohomology classes we have obtained in this way in the 
non rectified version of AKSZ $\mathfrak{v}$ BF gauge theory, can be transposed 
with little effort to the rectified one, expressing the 
non rectified superfields in terms of the rectified ones by inverting
\eqref{rectcibfs}, \eqref{rectCibfs}, 
\begin{subequations}
\label{nrectsfields}
\begin{align}
&\mathbfs{q}_{\mathrm{nr}}=\mathbfs{q}_{\mathrm{r}}-\bar{\mathbfs{\omega}},
\vphantom{\Big]}
\label{nrectcibfs}
\\
&\mathbfs{Q}_{\mathrm{nr}}=\mathbfs{Q}_{\mathrm{r}}
+\frac{1}{2}\mathbfs{\lambda}(\mathbfs{q}_{\mathrm{r}},\mathbfs{q}_{\mathrm{r}})
+\mathbfs{\mu}(\mathbfs{q}_{\mathrm{r}})
-\mathbfs{\lambda}(\bar{\mathbfs{\omega}},\bar{\mathbfs{q}}_{\mathrm{r}})
+\bar{\mathbfs{\varOmega}}.
\vphantom{\Big]}
\label{nrectCibfs}
\end{align}
\end{subequations} 
The expressions of the representatives of the 
rectified theory's BV cohomology classes so obtained depend on the gauge 
rectifier $(\lambda,\mu)$ and the background connection doublet 
$(\bar\omega,\bar\varOmega)$. This dependence is however an artifact of the 
reparametri\-zation \eqref{nrectsfields}. For reasons explained at the end of subsect. 
\ref{subsec:linfgaufix}, the non rectified and rectified versions
are fully equivalent and, so, their BV cohomologies
are isomorphic.
Since the cohomology of the former is independent from $(\lambda,\mu)$, 
$(\bar\omega,\bar\varOmega)$, 
that of the latter also is. 

As explained in subsect. \ref{subsec:2tligfix},
the BRST field variation operator $s$ is simply the restriction of 
$\delta_{\mathrm{BV}}$, $\delta_{\mathrm{tBV}}$ to the field sector of the extended BV 
field space of the rectified theory. Since the component fields of the superfields
$\mathbfs{p}$, $\mathbfs{P}$ belong to that sector, the representatives of
BV cohomology classes we have constructed are automatically also representatives of BRST cohomology 
classes. They thus represent topological operators of 
topological $\mathfrak{v}$ BF gauge theory. By standard arguments of topological field 
theory, their correlators are independent from the background data
$(\lambda,\mu)$, $(\bar\omega,\bar\varOmega)$ and $h$ and, so, compute topological invariants.

\subsection{\normalsize \textcolor{blue}{$\mathfrak{v}$ Chern--Simons gauge theory}}
\label{subsec:2tchern}
 
\hspace{.5cm} 
In this final subsection, we shall describe briefly the $2$--term $L_\infty$ algebra 
analog of standard Chern--Simons (CS) theory \cite{Witten:1988hf}. 
It is another example of $2$--term $L_\infty$ algebra gauge theory
and is related closely to the 
BF gauge theory we have 
studied above in great detail. A more thorough analysis of the CS model will be presented elsewhere.

$\mathfrak{v}$ CS gauge theory can be defined 
when $\mathfrak{v}$ is a reduced  $2$--term $L_\infty$ algebra equipped with an 
invariant metric. 
A $2$--term $L_\infty$ algebra $\mathfrak{v}=(\mathfrak{v}_0,\mathfrak{v}_1,\partial,
[\cdot,\cdot],[\cdot,\cdot,\cdot])$ is said {\it reduced} if $\ker\partial=0$. In that case
$\mathfrak{v}_1$ can be considered as a subspace of $\mathfrak{v}_0$ and the indication of 
$\partial$ can be omitted. 
An {\it invariant metric} on a $2$--term $L_\infty$ algebra $\mathfrak{v}=(\mathfrak{v}_0,
\mathfrak{v}_1,\partial,[\cdot,\cdot],[\cdot,\cdot,\cdot])$ is a non singular 
symmetric bilinear map $(\cdot,\cdot):\mathfrak{v}_0\vee \mathfrak{v}_0\mapsto \mathfrak{v}_0$ 
with the invariance properties $(x,[z,y])+([z,x],y)=0$ and $(x,[w,z,y])+([w,z,x],y)=0$. 

The base manifold of $\mathfrak{v}$ CS gauge theory is a $4$--fold
$N$, which we take to be diffeomorphic to $\mathbb{R}^4$ to avoid for the time being 
the subtleties of the global definition of the model.
The fields of classical $\mathfrak{v}$ CS gauge theory constitute a bidegree $(1,0)$ 
connection doublet $(\omega,\varOmega)$.
The classical action is
\begin{equation}
S_{cl}=\int_N\Big[-(\varOmega,f+\frac{1}{2}\varOmega)+\frac{1}{24}
(\omega, [\omega,\omega,\omega])\Big].
\label{2tchern1}
\end{equation}
The field equations of the theory read
\begin{subequations}
\label{2tchern2}
\begin{align}
&f=0,
\vphantom{\Big]}
\label{2tchern2a}
\\
&F=0.
\vphantom{\Big]}
\label{2tchern2b}
\end{align}
\end{subequations}
They imply that the connection doublet $(\omega,\varOmega)$ is flat, as
in standard CS theory and reproduce the field equations \eqref{clbfeqsa}, \eqref{clbfeqsb}
of classical $\mathfrak{v}$ BF gauge theory. 

Like BF theory, classical $\mathfrak{v}$ CS gauge theory enjoys a high amount of gauge symmetry.
The gauge symmetry variations of the fields are expressed in terms of ghost fields
organized in a bidegree $(0,1)$ field doublet $(c,C)$,  
\begin{subequations}
\label{2tchern3}
\begin{align}
&\delta_{\mathrm{cl}}\omega=-Dc,
\vphantom{\Big]}
\label{2tchern3a}
\\
&\delta_{\mathrm{cl}}\varOmega=-DC.
\vphantom{\Big]}
\label{2tchern3b}
\end{align}
\end{subequations}
They coincide in form with the gauge symmetry variations of the corresponding fields
of classical $\mathfrak{v}$ BF gauge theory (cf. eqs. \eqref{clbffvara}, \eqref{clbffvarb}). 
As is straightforward to verify, the action 
is invariant under the symmetry, 
\begin{equation}
\delta_{\mathrm{cl}}S_{\mathrm{cl}}=0.
\label{2tchern4}
\end{equation}

As in BF theory, it should be possible to define 
gauge symmetry variations of the ghost fields 
rendering the gauge field variation operator $\delta_{\mathrm{cl}}$ nilpotent at least on--shell. 
These variations depend on 
the ghost field doublet $(c,C)$ and a further ghost for ghost field seen as a bidegree $(-1,2)$ 
field doublet $(0,\varGamma)$, 
\begin{subequations}
\label{2tchern5}
\begin{align}
&\delta_{\mathrm{cl}}c=-\frac{1}{2}[c,c]+\varGamma,
\vphantom{\Big]}
\label{2tchern5a}
\\
&\delta_{\mathrm{cl}}C=-[c,C]+\frac{1}{2}[\omega,c,c]-D\varGamma,
\vphantom{\Big]}
\label{2tchern5b}
\end{align}
\begin{align}
&\delta_{\mathrm{cl}}\varGamma=-[c,\varGamma]+\frac{1}{6}[c,c,c].
\vphantom{\dot{\dot{\dot{\dot{\dot{\dot{\dot{x}}}}}}}}
\vphantom{\Big]}
\label{2tchern5c}
\end{align}
\end{subequations}
Again, they coincide in form with the gauge symmetry variations of the corre-
sponding fields
of classical $\mathfrak{v}$ BF gauge theory 
(cf. eqs, \eqref{ghclbffvara}, \eqref{ghclbffvarb}, \eqref{ghclbffvard}).
Using \eqref{2tchern3}, \eqref{2tchern5}, we find that $\delta_{\mathrm{cl}}{}^2\mathcal{F}=0$
for all fields and ghost fields $\mathcal{F}$ except for $\varOmega$, in which case one has
\begin{equation}
\delta_{\mathrm{cl}}{}^2\varOmega=\frac{1}{2}[f,c,c]-[f,\varGamma].
\label{2tchern6}
\end{equation}
Again, as in BF theory, $\delta_{\mathrm{cl}}$ is nilpotent but only on--shell
by the field equation \eqref{2tchern2a} (cf. eq. \eqref{sqclbffvarb}).

As $\mathfrak{v}$ BF gauge theory, $\mathfrak{v}$ CS gauge theory 
admits an AKSZ formulation, {\it AKSZ $\mathfrak{v}$ CS gauge theory}. 
The field content of this consists of a $\hat{\mathfrak{v}}_0[1]$--valued 
superfield $\mathbfs{q}$ and a $\hat{\mathfrak{v}}_1[2]$--valued 
superfield $\mathbfs{Q}$. 
The BV symplectic form is given by 
\begin{equation}
\varOmega_{\mathrm{BV}}=\int_{T[1]N}\varrho (\delta\mathbfs{q},\delta\mathbfs{Q}).
\label{2tchern7}
\end{equation}
Then, by BV theory, associated with $\varOmega_{\mathrm{BV}}$ is the BV bracket 
$(\cdot,\cdot)_{\mathrm{BV}}$.

The BV action AKSZ $\mathfrak{v}$ CS gauge theory is 
\begin{equation}
S_{\mathrm{BV}}=\int_{T[1]N}\varrho\Big[(\mathbfs{Q},\mathbfs{d}\mathbfs{q}
-\frac{1}{2}[\mathbfs{q},\mathbfs{q}]+\frac{1}{2}\mathbfs{Q})
+\frac{1}{24}(\mathbfs{q},[\mathbfs{q},\mathbfs{q},\mathbfs{q}])\Big].
\label{2tchern8}
\end{equation}
It is straightforward to verify that 
$S_{\mathrm{BV}}$ satisfies the classical BV master equation \eqref{SWSW=0}
and, so, according to BV theory, the model is consistent and quantizable.

The BV variations 
of the AKSZ $\mathfrak{v}$ CS gauge theory superfields are 
\begin{subequations}
\label{2tchern9}
\begin{align}
&\delta_{\mathrm{BV}}\mathbfs{q}
=\mathbfs{d}\mathbfs{q}-\frac{1}{2}[\mathbfs{q},\mathbfs{q}]+\mathbfs{Q},
\vphantom{\Big]}
\label{2tchern9a}
\\
&\delta_{\mathrm{BV}}\mathbfs{Q}
=\mathbfs{d}\mathbfs{Q}-[\mathbfs{q},\mathbfs{Q}]+\frac{1}{6}[\mathbfs{q},\mathbfs{q},\mathbfs{q}].
\vphantom{\Big]}
\label{2tchern9b}
\end{align}
\end{subequations}
They agree in form with the BV variations of the corresponding superfields 
of BF theory (cf. eqs. \eqref{dWcibfs}, \eqref{dWCibfs}).
Since $S_{\mathrm{BV}}$ solves the BV master equation \eqref{SWSW=0}, 
the BV superfield variation operator $\delta_{\mathrm{BV}}$ satisfies \eqref{dBV2=0}, \pagebreak 
and thus is nilpotent. For the same reason, the BV action 
$S_{\mathrm{BV}}$ satisfies \eqref{dBVsBV=0} and so is 
BV invariant.

Let us now analyze the issue of gauge covariance in AKSZ $\mathfrak{v}$ CS gauge theory. 
As $\mathfrak{v}$ CS gauge theory involves in an essential way an invariant metric
$(\cdot,\cdot)$, the relevant symmetry group of $\mathfrak{v}$ is not the general
$1$--automorphism group $\Aut_1(\mathfrak{v})$ (cf. subsect. \ref{sec:linftyauto})
but its unitary subgroup $\UAut_1(\mathfrak{v})$. By definition, 
$\UAut_1(\mathfrak{v})$ consists of the $1$--automorphisms 
$\phi\in\Aut_1(\mathfrak{v})$ such that
$(\phi_0(x),\phi_0(y))=(x,y)$ \footnote{$\vphantom{\bigg[}$ It can be shown that, 
for $\phi=(\phi_0,\phi_1,\phi_2)\in\UAut_1(\mathfrak{v})$, 
$\phi_1=\phi_0|_{\hat{\mathfrak{v}}_0}$
and $(\phi_1{}^{-1}\phi_2(x,y),z)$ $+(y,\phi_1{}^{-1}\phi_2(x,z))=0$.}.
Correspondingly, only unitary gauge transformations
are to be considered. These form the unitary subgroup 
$\UGau_1(M,\mathfrak{v})$ of $\Gau_1(M,\mathfrak{v})$. 

For any unitary gauge transformation $g\in\UGau_1(N,\mathfrak{v})$,
the gauge transformed basic superfields are 
\begin{subequations}
\label{2tchern10}
\begin{align}
&{}^g\mathbfs{q}=g_0(\mathbfs{q}+\mathbfs{\sigma}_g),
\vphantom{\Big]}
\label{2tchern10a}
\\
&{}^g\mathbfs{Q}
=g_1(\mathbfs{Q}-\mathbfs{\varSigma}_g-\mathbfs{\tau}_g(\mathbfs{q}+\mathbfs{\sigma}_g))
-\frac{1}{2}g_2(\mathbfs{q}+\mathbfs{\sigma}_g,\mathbfs{q}+\mathbfs{\sigma}_g).
\vphantom{\Big]}
\label{2tchern10b}
\end{align}
\end{subequations}  
These expressions are identical in form to those of the corresponding gauge transformed superfields 
of BF theory (cf. eqs. \eqref{gcibfs}, \eqref{gCibfs}).

Having defined the gauge transformation prescription, we can tackle the problem 
of the global definedness of AKSZ $\mathfrak{v}$ CS gauge theory. 
The BV form $\varOmega_{\mathrm{BV}}$ given in \eqref{2tchern7}
turns out to be gauge invariant. Thus, it can be defined globally on a general $4$--fold $N$. 
Conversely, as in ordinary CS theory, the BV action $S_{\mathrm{BV}}$ given
in \eqref{2tchern8} is not gauge invariant. So, $S_{\mathrm{BV}}$ cannot
be defined globally on a general $N$ in  the usual way. Nevertheless, there is 
an alternative way of giving a global meaning to $S_{\mathrm{BV}}$ proceeding as follows. 
Pick a background connection doublet $(\bar\omega,\bar\varOmega)$. 
Consider the superfield 
\begin{equation}
\mathbfs{\Delta \mathcal{L}}=\mathbfs{\mathcal{L}}-\bar{\mathbfs{\mathcal{L}}}
-\mathbfs{d}\mathbfs{\varLambda}, 
\vphantom{\ul{\ul{\ul{\ul{\ul{\ul{x}}}}}}}
\label{2tchern11}
\end{equation}
where $\mathbfs{\mathcal{L}}$ is the integrand superfield in \eqref{2tchern8},
$\bar{\mathbfs{\mathcal{L}}}$ is $\mathbfs{\mathcal{L}}$ with 
$(\mathbfs{q},\mathbfs{Q})=(-\bar{\mathbfs{\omega}},\bar{\mathbfs{\varOmega}})$ 
and 
\begin{equation}
\mathbfs{\varLambda}=\frac{1}{2}(\bar{\mathbfs{\omega}}+\mathbfs{q},
\bar{\mathbfs{\varOmega}}+\mathbfs{Q}
-\frac{1}{6}[\bar{\mathbfs{\omega}}+\mathbfs{q},\bar{\mathbfs{\omega}}+\mathbfs{q}]).
\label{2tchern12}
\end{equation}
Then, thanks to Stokes' theorem,  one has \hphantom{xxxxxxxxxxx}
\begin{equation}
S_{\mathrm{BV}}=\bar S_{\mathrm{BV}}+\int_{T[1]N}\varrho\mathbfs{\Delta \mathcal{L}},
\label{2tchern13}
\end{equation}
where $\bar S_{\mathrm{BV}}$ is \hphantom{xxxxxxxxxxxxxxxxxxxxxxxxxxxx}
\begin{equation}
\bar S_{\mathrm{BV}}=\int_{T[1]N}\varrho\bar{\mathbfs{\mathcal{L}}}.
\label{2tchern14}
\end{equation}
It is verified that ${}^g\mathbfs{\Delta \mathcal{L}}=\Delta \mathbfs{\mathcal{L}}$
for $g\in\UGau_1(N,\mathfrak{v})$. Therefore, the second term in the right hand side 
can be defined globally also on a general $4$--fold $N$. 
$\bar S_{\mathrm{BV}}=\bar S_{\mathrm{cl}}$ is just the  classical action evaluated 
in the background $(\bar\omega,\bar\varOmega)$. Again, 
it cannot be interpreted as an ordinary integral and must be defined by other means.
However, as it is a mere background term, it does not affect the dynamics at the classical level
and, perturbatively, also at the quantum one. In this way, using \eqref{2tchern13}, in the weaker sense
we have explained, $S_{\mathrm{BV}}$ can be globally defined on a general $4$--fold $N$.

It is important to realize that the superfield $\mathbfs{\varLambda}$  given in 
eq. \eqref{2tchern12} is not by itself 
gauge invariant. The exact term $\mathbfs{d\varLambda}$ appearing in the right hand 
side of \eqref{2tchern11}, so,  cannot be dropped without spoiling the gauge invariance of 
$\mathbfs{\Delta \mathcal{L}}$. For the same reason, it cannot eliminated upon integration on $T[1]N$
using Stokes' theorem on a general $4$--fold $N$. Further, a generic variation $\delta\mathbfs{\varLambda}$ 
of $\mathbfs{\varLambda}$ with respect to the superfields $\mathbfs{q}$, $\mathbfs{Q}$
is also not gauge invariant. Hence, upon integration, $\mathbfs{d\varLambda}$ does not 
yield a topological invariant when $N$ is topologically non trivial.

The BV superfield variations \eqref{2tchern9}
are gauge covariant and, so, are also defined globally on a general 
$4$--fold $N$. This is obvious, since the BV superfield variations and the gauge transformation
rules are formally identical to those of BF theory, which are gauge covariant. 

The basic superfields $\mathbfs{q}$, $\mathbfs{Q}$ have a component expansion
of the form
\begin{subequations}
\label{2tchern15}
\begin{align}
&\mathbfs{q}=c-\omega+\varOmega^+-C^++\varGamma^+,
\vphantom{\Big]}
\label{2tchern15a}
\\
&\mathbfs{Q}=\varGamma-C+\varOmega+\omega^+-c^+,
\vphantom{\Big]}
\label{2tchern15b}
\end{align}
\end{subequations}
where the terms in the right hand side are written down in increasing order of 
form degree and decreasing order of ghost number degree with a conventional
choice signs. It is possible to express the BV action and the BV superfield variations 
in component fields. Working with components, it is then possible to gauge fix the model.

\subsection{\normalsize \textcolor{blue}{Relation to other formulations}}\label{sec:akszcs}


\hspace{.5cm} 
We review some of the results of ref. \cite{Fiorenza:2011jr} 
and compare them with those obtained in the present paper. 
Closely related results were obtained also in ref. \cite{Kotov:2007nr}.

A graded smooth manifold $X$ is a smooth space of the form $V[1]$, the $1$ step grade shift of 
 an ordinary graded smooth vector bundle $V=\bigoplus_{i}V_i\rightarrow M$ on a manifold $M$. 
The local coordinates $x^i$ of $X$ are just the local trivialization coordinates of $V[1]$
seen as Grassmann valued. The 
algebra $C^\infty(X)$ of smooth functions on $X$ is defined as 
the graded commutative algebra $\Fun(V[1])$ of smooth fiberwise polynomial functions on the bundle 
$V[1]$ 
.
An ordinary manifold $M$ can be seen as a graded manifold $X=V[1]$,
where $V$ is $M$ viewed as a vector bundle over $M$ of vanishing rank. 

A graded vector field $D$ on a graded manifold $X$
is a graded derivation operator $D:C^\infty(X)\rightarrow C^\infty(X)$.
In local coordinates $x^i$ of $X$, $D$ has thus the familiar expansion $D=D^i\partial/\partial x^i$.
Graded vector fields form a graded Lie algebra $\Vect(X)$ under graded commutation. 

A differential graded manifold is a pair $(X,\varDelta)$, where $X$ is a graded manifold
and $\varDelta\in \Vect(X)$ is a grade $1$ vector field satisfying the nilpotency
condition $[\varDelta,\varDelta]=2\varDelta\circ \varDelta=0$. The graded vector bundle $V$ underlying 
 a differential graded manifold $(X,\varDelta)$ acquires a $L_\infty$--algebroid structure, 
whose Chevalley--Eilenberg algebra bundle and differential 
operator are $\mathrm{CE}(V)\simeq C^\infty(X)$ and $\mathcal{Q}_{\mathrm{CE}(V)}\simeq \varDelta$, respectively, 
by a reason analog to that by which an $L_\infty$ algebra structure on a graded vector space 
$\mathfrak{v}$ is codified in the Chevalley--Eilenberg algebra and differential
$\mathrm{CE}(\mathfrak{v})$ and $\mathcal{Q}_{\mathrm{CE}(\mathfrak{v})}$ (cf. subsect. \ref{subsec:linfweil}).

A graded manifold $X$ is characterized by the graded commutative algebra of differential 
forms $\Omega^*(X)$. By definition, if $X=V[1]$ with $V$ a graded vector bundle,
then $\Omega^*(X)$ is the graded commutative algebra $\Fun(T[1]V[1])$ 
of smooth 
fiberwise polynomial functions on the $1$ step grade shifted tangent bundle $T[1]V[1]$ of $V[1]$
\footnote{$\vphantom{\dot{\dot{\dot{\dot{x}}}}}$ Here and in the following, 
we refer to the fibers of the 
bundle $T[1]V[1]$. However, the notion of smoothness adopted in this context requires also 
polynomiality with respect to the fibers of the bundle $V[1]$.}. 
The form degree of $\Omega^*(X)$ is the fiberwise polynomial degree 
of $\Fun(T[1]V[1])$. 
The de Rham differential of $X$ is the differential operator on $\Fun(T[1]V[1])$
locally given by $d_X=\xi^i\partial/\partial x^i$, where $\xi^i$ are the fiber coordinates
of $T[1]V[1]$. 
The contraction and Lie derivative operators of a graded vector field $D\in \Vect(X)$ on $X$
are the differential operators on $\Fun(T[1]V[1])$ locally given by 
$i_D=D^i\partial/\partial \xi^i$ and 
$l_D=D^i\partial/\partial x^i+(-1)^D\xi^j\partial D^i/\partial x^j\partial/\partial \xi^i$. 
The graded version of the familiar Cartan relations holds.
This supergeometric framework reduces to the standard one 
in the case where $X$ is an ordinary manifold $M$.


If $(X,\varDelta)$ is a differential graded manifold, then $\Omega^*(X)$ comes equipped with the 
grade $1$ nilpotent differential operator $l_{\varDelta}$. If $V$ is the 
$L_\infty$-algebroid underlying $X$, it is natural to define the Weil algebroid bundle
and differential operator of $V$ to be $\mathrm{W}(V)\simeq\Omega^*(X)$ and $\mathcal{Q}_{\mathrm{W}(V)}\simeq d_X+l_{\varDelta}$,
since, by its construction, $\Omega^*(X)$ is locally an extension of $C^\infty(X)$ 
by grade shifted generators, $d_X$ is the associated shift operator and $l_{\varDelta}$
is the extension of the operator $\varDelta$ from $C^\infty(X)$ to $\Omega^*(X)$  anticommuting with $d_X$, 
much as the Weil algebra $\mathrm{W}(\mathfrak{v})$ and differential $\mathcal{Q}_{\mathrm{W}(\mathfrak{v})}$
of an $L_\infty$--algebra $\mathfrak{v}$ extend the Chevalley--Eilenberg algebra 
$\mathrm{CE}(\mathfrak{v})$ and differential $\mathcal{Q}_{\mathrm{CE}(\mathfrak{v})}$
(cf. subsect. \ref{subsec:linfweil}).

A grade $n$ symplectic differential graded manifold is a differential graded manifold 
$(X,\varDelta)$ equipped with a grade $n$ non singular $2$--form $\omega\in \Omega^2(X)$
such that $d_X\omega=0$ and $l_{\varDelta}\omega=0$. In virtue of its properties, $\omega$ is also 
an element of $\mathrm{W}(V)$ satisfying $\mathcal{Q}_{\mathrm{W}(V)}\omega=0$, 
where $V$ is the graded vector bundle underlying $X$, that is a Weil cocycle. It can  be shown that there is an 
element $\varpi\in \mathrm{CE}(V)$ 
and an element $\varLambda\in \mathrm{W}(V)$ with the property that 
$\mathcal{Q}_{\mathrm{CE}(V)}\varpi=0$ and that 
$\mathcal{Q}_{\mathrm{W}(V)}\varLambda=\omega$ and $i^*\varLambda=\varpi$, where 
$i^*:\mathrm{W}(V)\rightarrow\mathrm{CE}(V)$ is the differential graded
commutative algebra morphism corresponding to the natural 
projection $i^*:\Omega^*(X)\rightarrow \Omega^0(X)\simeq C^\infty(X)$.
The Chevalley--Eilenberg cocycle  
$\varpi$ is said to be in transgression with $\omega$ and 
$\varLambda$ is called the transgression or Chern--Simons element of $\omega$.
Explicit expressions of $\varpi$ and $\varLambda$ are available. 
$\varpi$ is given by 
\begin{equation}
\varpi=\frac{1}{n+1}\omega_{ij}\gr(x^i)x^i\varDelta{}^j,
\label{akszcs1}
\end{equation}
while $\varLambda$ is given by \hphantom{xxxxxxxxxxxxxxxxxxxxxxxx}
\begin{equation}
\varLambda=\frac{1}{n}\omega_{ij}\gr(x^i)x^i\mathcal{Q}_{\mathrm{W}(V)}x^j-\varpi.
\label{akszcs2}
\end{equation}
According to the authors of refs. \cite{Fiorenza:2011jr}, 
the triple of data $(\omega,\varLambda,\varpi)$ defines
an AKSZ sigma like model with base space any closed $n+1$ dimensional world volume $\Sigma$
and target space $X$. Field space can be identified with the set of differential graded commutative algebra
morphisms $\varphi:\mathrm{W}(V)\rightarrow \Omega^*(\Sigma)\simeq \Fun(T[1]\Sigma)$.
The AKSZ sigma model classical BV master action is given explicitly by 
\begin{equation}
S_{\mathrm{BV}}(\varphi)=\int_{T[1]\Sigma}\varphi(\varLambda).
\label{akszcs3}
\end{equation}
Now, we are going to show that the models studied in this section 
can be obtained by the above general procedure. \eqref{akszcs3} provides the 
complete superfield version of $S_{\mathrm{BV}}(\varphi)$.
The authors of \cite{Fiorenza:2011jr} concentrate however on the truncation of $S_{\mathrm{BV}}(\varphi)$ 
to the ghost number $0$ sector of field space, which is just the classical action. 
This is fine but it is no longer sufficient when aiming 
to gauge fix the associated field theory, a necessary step in the path toward its quantization,
as we did above. 

Let us now show that the $\mathfrak{v}$ BF gauge theory model of subsect. 
\ref{subsec:linfvmod} is a special case of the general construction of refs. \cite{Fiorenza:2011jr}.
Consider the graded manifold $X=V[1]$, 
where $V$ is the delooping $b\mathfrak{V}$ 
of the graded vector space $\mathfrak{V}=\hat{\mathfrak{v}}_0[0]\oplus\hat{\mathfrak{v}}_0{}^\vee[0]
\oplus\hat{\mathfrak{v}}_1[1]\oplus\hat{\mathfrak{v}}_1{}^\vee[-1]$, that is $\mathfrak{V}$ 
viewed as a vector bundle over the singleton manifold. Denote by $q$, $p$, $Q$, $P$ the 
coordinates of $X$ corresponding to the direct summands of $\mathfrak{V}$ in the given order.
It is straightforwardly verified that the grade $1$ vector field $\varDelta\in\Vect(X)$ of components 
\begin{subequations}
\label{akszcs4}
\begin{align}
\varDelta^p&=[q,p]^\vee-[Q,P]^\vee-\frac{1}{2}[q,q,P]^\vee,
\vphantom{\Big[}
\label{akszcs4a}
\\
\varDelta^q&=\frac{1}{2}[q,q]-\partial Q,
\vphantom{\Big[}
\label{akszcs4b}
\\
\varDelta^P&=[q,P]^\vee-\partial^\vee p,
\vphantom{\Big[}
\label{akszcs4c}
\\
\varDelta^Q&=[q,Q]-\frac{1}{6}[q,q,q]
\vphantom{\Big[}
\label{akszcs4d}
\end{align}
\end{subequations}
is nilpotent. Thus, $(X,\varDelta)$ is a differential graded manifold. 
$X$ is endowed with a natural grade $2$ symplectic form $\varsigma$
invariant under $\varDelta$, viz
\begin{equation}
\varsigma=-\langle dp,dq\rangle+\langle dP,dQ\rangle,
\label{akszcs5}
\end{equation}
so that $(X,\varDelta)$ is a grade $2$ symplectic differential graded manifold.
Using \eqref{akszcs1}, we can easily get the Chevalley--Eilenberg cocycle
$\varpi$ in transgression with $\varsigma$,
\begin{equation}
\varpi=\langle p,-\frac{1}{2}[q,q]+\partial Q \rangle
-\langle P,-[q,Q]+\frac{1}{6}[q,q,q]\rangle
\label{akszcs6}
\end{equation}
and from this, using \eqref{akszcs2}, the associated Chern--Simons element 
$\varLambda$,
\begin{equation}
\varLambda=-\langle p,\mathcal{Q}_{W(V)}q-\frac{1}{2}[q,q]+\partial Q \rangle
+\langle P,\mathcal{Q}_{W(V)}Q-[q,Q]+\frac{1}{6}[q,q,q]\rangle,
\label{akszcs7}
\end{equation}
up to an irrelevant $\mathcal{Q}_{W(V)}$--exact term.
Application of \eqref{akszcs3} yields immediately the BV action of the BF
theory given in \eqref{SW}. 
We conclude by noticing that, as a byproduct, we have shown that there exists an 
$L_\infty$ algebra $\mathfrak{V}$ canonically associated with any given 
$2$--term $L_\infty$ algebra $\mathfrak{v}$. As a graded vector space, $\mathfrak{V}
=\mathfrak{v}\oplus\tilde{\mathfrak{v}}^\vee:=\tilde T^*\mathfrak{v}$, 
where $\tilde{\mathfrak{v}}^\vee$ is the dual space of $\mathfrak{v}$ with sign reversed grading. 

Our $\mathfrak{V}$ can be viewed in yet another way: as a certain Courant algebroid $E$.
$E$ is the trivial bundle $\hat{\mathfrak{v}}_1{}^\vee\times (\hat{\mathfrak{v}}_0\oplus
\hat{\mathfrak{v}}_0{}^\vee)\rightarrow \hat{\mathfrak{v}}_1{}^\vee$. 
The Courant bracket structure on $E$ is that naturally yielded by the 
brackets $[\cdot,\cdot]$, $[\cdot,\cdot]^\vee$ 
on $\hat{\mathfrak{v}}_1{}^\vee\times (\hat{\mathfrak{v}}_0\oplus\hat{\mathfrak{v}}_0{}^\vee)$.
The anchor of $E$ is induced by the bilinear form
$\langle\cdot, \partial\cdot\rangle: \hat{\mathfrak{v}}_0{}^\vee\times \hat{\mathfrak{v}}_1
\rightarrow \mathbb{R}$ extended trivially 
to one $(\hat{\mathfrak{v}}_0\oplus\hat{\mathfrak{v}}_0{}^\vee)\times \hat{\mathfrak{v}}_1
\rightarrow \mathbb{R}$ and then viewed as a mapping 
$\hat{\mathfrak{v}}_1{}^\vee\times (\hat{\mathfrak{v}}_0\oplus\hat{\mathfrak{v}}_0{}^\vee)\rightarrow 
\hat{\mathfrak{v}}_1{}^\vee\times \hat{\mathfrak{v}}_1{}^\vee\simeq T\hat{\mathfrak{v}}_1{}^\vee$. 
The metric of $E$ is just the fiberwise off diagonal symmetric bilinear form
on $\hat{\mathfrak{v}}_0\oplus\hat{\mathfrak{v}}_0{}^\vee$ determined by the duality 
pairing $\langle\cdot, \cdot\rangle: \hat{\mathfrak{v}}_0{}^\vee\times \hat{\mathfrak{v}}_0
\rightarrow \mathbb{R}$. Our BF theory is therefore a particular case of the Courant algebroid
AKSZ sigma model first studied in ref. \cite{Roytenberg:2006qz} and reexamined  in ref.
\cite{Fiorenza:2011jr} in the above framework. In this paper, we went a step beyond those
endeavours by obtaining the full superfield form of the BV action and carrying out the gauge fixing
of the field theory.

Let us now show that the $\mathfrak{v}$ CS gauge theory model formulated in  subsect. 
\ref{subsec:2tchern} also is a special case of the construction of refs. \cite{Fiorenza:2011jr}.
Consider the graded manifold $X=V[1]$, 
where $V$ is the delooping $b\mathfrak{V}$ 
of the graded vector space $\mathfrak{V}=\hat{\mathfrak{v}}_0[0]\oplus\hat{\mathfrak{v}}_1[1]$.
Denote by $q$, $Q$ the 
coordinates of $X$ corresponding to the direct summands of $\mathfrak{V}$ in the given order.
It is readily checked that the grade $1$ vector field $\varDelta\in\Vect(X)$ of components
\begin{subequations}
\label{akszcs8}
\begin{align}
\varDelta^q&=\frac{1}{2}[q,q]-Q,
\vphantom{\Big[}
\label{akszcs8a}
\\
\varDelta^Q&=[q,Q]-\frac{1}{6}[q,q,q]
\vphantom{\Big[}
\label{akszcs8b}
\end{align}
\end{subequations}
is nilpotent. Thus, $(X,\varDelta)$ is a differential graded manifold. 
$X$ is endowed with a natural grade $3$ symplectic form $\varsigma$
invariant under $\varDelta$, viz
\begin{equation}
\varsigma=(dq,dQ),
\label{akszcs9}
\end{equation}
so that $(X,\varDelta)$ is a grade $3$ symplectic differential graded manifold.
Using \eqref{akszcs1}, we can easily get the Chevalley--Eilenberg cocycle
$\varpi$ in transgression with $\varsigma$,
\begin{equation}
\varpi=-(Q,-\frac{1}{2}[q,q]+Q)-\frac{1}{24}(q,[q,q,q])
\label{akszcs10}
\end{equation}
and from this, using \eqref{akszcs2}, obtain the associated Chern--Simons element 
$\varLambda$,
\begin{equation}
\varLambda=(Q,\mathcal{Q}_{W(V)}q-\frac{1}{2}[q,q]+\frac{1}{2}\partial Q),
+\frac{1}{24}(q,[q,q,q]).
\label{akszcs11}
\end{equation}
again up to an irrelevant $\mathcal{Q}_{W(V)}$--exact term.
Application of \eqref{akszcs3} yields immediately the BV action of the CS
theory given in \eqref{2tchern8}. Note that $\mathfrak{V}$ as an $L_\infty$ algebra is just the  
$2$--term $L_\infty$ algebra $\mathfrak{v}$ we started with.

\vfill\eject

\section{\normalsize \textcolor{blue}{Conclusions and outlook}}\label{sec:concl}


\subsection{\normalsize \textcolor{blue}{Summary of results}}
\label{subsec:summary}

\hspace{.5cm} In this paper, we have worked out a version of semistrict higher gauge theory, 
whose symmetry is encoded in a semistrict Lie $2$--algebra. 
This extends previous constructions
which relied instead on differential Lie crossed modules
\cite{Baez:2002jn,Hofman:2002ey,Pfeiffer:2003je,Baez:2010ya}.

In our formulation of semistrict higher gauge theory, the
symmetry is encoded in a finite dimensional $2$--term
$L_\infty$ algebra $\mathfrak{v}$ at infinitesimal level 
and in the automorphism $2$--group
$\Aut(\mathfrak{v})$ of $\mathfrak{v}$ at finite level. The basic
datum is thus the algebra $\mathfrak{v}$. In this way, we avoid 
any reference to any Lie $2$--group $V$ integrating $\mathfrak{v}$, which
may be infinite dimensional or may be something more general than 
a mere coherent $2$--group, and rely instead on the $2$--group 
$\Aut(\mathfrak{v})$, which is always finite dimensional and strict.
Gauge transformations are mappings valued in the $1$--cell group 
$\Aut_1(\mathfrak{v})$ of $\Aut(\mathfrak{v})$ together with a flat connection 
doublet and other form data satisfying a set of relations. 
Gauge transformations on a neighborhood $O$ form an infinite dimensional group
$\Gau_1(O,\mathfrak{v})$, which is the $1$--cell group of a
strict $2$--group $\Gau(O,\mathfrak{v})$. 
A left action of $\Gau_1(O,\mathfrak{v})$ on fields on $O$ is defined
and gauge invariant Lagrangian field theoretic models can be built. 

This approach has its advantages and disadvantages. At the differential 
level, it is very efficient and provides a powerful algorithm for the construction of local 
semistrict higher gauge models in perturbative Lagrangian field
theory. At the integral level, it is not suitable for the study and 
efficient computation of higher parallel transport even in the strict theory 
and thus also for the investigation of non perturbative issues. 

Using the BV quantization approach in the AKSZ geometrical version, we have been able 
to construct semistrict higher gauge theoretic generalizations of BF and Chern-Simons theory.
These field theories are interesting as exemplification of our
methodology and for the relation  they bear with the general AKSZ construction
of refs. \cite{Fiorenza:2011jr,Kotov:2007nr}.
Other field theoretic models with $2$--term $L_\infty$ algebra 
symmetry can conceivably be worked out. Relatedly, a number of issues call for further
investigation. They are reviewed briefly below. 

\subsection{\normalsize \textcolor{blue}{$2$--term $L_\infty$ algebra Yang--Mills gauge theory}}
\label{subsec:mills}


Though we have not investigated this in detail in this paper, it is not very difficult to 
construct a $2$--term $L_\infty$ algebra Yang--Mills gauge theory. Let $\mathfrak{v}$ be a 
$2$--term $L_\infty$ algebra and let $X$ be any closed oriented manifold.
We assume that $\mathfrak{v}$ is reduced and equipped with an invariant metric
$(\cdot,\cdot)$ (cf. subsect. \ref{subsec:2tchern}). We assume further that 
a gauge rectifier $(\lambda,\mu)$ has been chosen and that
$X$ is equipped with a Riemannian metric $h$.
The basic fields of $\mathfrak{v}$  Yang--Mills gauge theory are the components of a 
connection doublet $(\omega,\varOmega)$  with curvature doublet $(f,F)$. 
The action of the model is 
\begin{equation}
S_{\mathrm{YM}}=\int_X\Big[\frac{1}{2}(\tilde f,*\tilde f)
+\frac{1}{2}(\tilde F,*\tilde F)\Big],
\label{concl1}
\end{equation}
where $(\tilde f,\tilde F)$ is the rectified curvature doublet
\begin{subequations}
\label{concl2}
\begin{align}
&\tilde f=f,
\vphantom{\Big]}
\label{concl2a}
\\
&\tilde F=F+\lambda(\omega,f)-\mu(f).
\vphantom{\Big]}
\label{concl2b}
\end{align}
\end{subequations}
By construction, if the connection doublet $(\omega,\varOmega)$  is canonical, 
the action $S_{\mathrm{YM}}$ is gauge invariant and, so, globally definable.
Though $\mathfrak{v}$  Yang--Mills gauge theory is a straightforward generalization 
of customary Yang--Mills gauge theory, 
the field equations derived from $S_{\mathrm{YM}}$ are rather more involved. They read
\begin{subequations}
\label{concl3}
\begin{align}
&d(*\tilde f-\lambda^t(\omega,*\tilde F)+\mu^t(*\tilde F))
+[\omega,*\tilde f-\lambda^t(\omega,*\tilde F)+\mu^t(*\tilde F)]
\vphantom{\Big]}
\label{concl3a}
\\
&\qquad\qquad\qquad\qquad\qquad+[\varOmega,*\tilde F]+\frac{1}{2}[\omega,\omega,*\tilde F]
+\lambda^t(f,*\tilde F)=0,
\vphantom{\Big]}
\nonumber
\\
&d*\tilde F+[\omega,*\tilde F]+*\tilde f-\lambda^t(\omega,*\tilde F)+\mu^t(*\tilde F)=0,
\vphantom{\Big]}
\label{concl3b}
\end{align}
\end{subequations}
where $(\lambda^t,\mu^t)$ is the transposed rectifier of $(\lambda,\mu)$
with respect to the invariant metric and is defined in a way formally analogous to that
of the dual rectifier $(\lambda^\vee,\mu^\vee)$. The complete symmetries of
 $\mathfrak{v}$  Yang--Mills gauge theory are not known to us. It remains to 
be seen whether the model has any physical applications. 

\subsection{\normalsize \textcolor{blue}{Adding matter}}
\label{subsec:matter}


Adding matter in $2$--term $L_\infty$ algebra gauge theory is a delicate issue.
Suppose that the matter fields are valued in some linear space $W$ and that 
the symmetry $2$--term $L_\infty$ algebra $\mathfrak{v}$ acts on matter fields 
linearly. Then, with every $x\in\mathfrak{v}_0$, 
there is associated an element $T_x\in\mathfrak{gl}(W)$ representing the action of $x$ on 
those fields. It is reasonable to suppose that $T$ is a representation,
so that  
\begin{equation}
[T_x,T_y]=T_{[x,y]}.
\label{}
\end{equation}
Since $\mathfrak{gl}(W)$ is an ordinary Lie algebra, the Jacobi identity holds 
in $\mathfrak{gl}(W)$.
By \eqref{2tlinalgc}, this implies that $T_{\partial[x,y,z]}=0$ identically. 
In general, so, we must have 
\begin{equation}
T_{\partial X}=0,
\label{}
\end{equation}
for every $X\in \mathfrak{v}_1$. By \eqref{2tlinalga}, $\im\partial$ is an ideal
of $\mathfrak{v}_0$ and, by \eqref{2tlinalgc},
$\mathfrak{g}=\mathfrak{v}_0/\im \partial$ is an ordinary Lie algebra.
Therefore, a linear action of $\mathfrak{v}$ on matter fields 
reduces to one of the Lie algebra $\mathfrak{g}$.
The non standard features of $\mathfrak{v}$ get in this way lost by the 
representation $T$. The natural question arises about whether
other forms of linear action on matter fields can be defined, which faithfully
reproduce the richer algebraic structure of $\mathfrak{v}$.


\vfill\eject

\appendix

\section{\normalsize \textcolor{blue}{AKSZ $\mathfrak{v}$ BF gauge theory in components}}
\label{app:vakszcomp}
\hspace{.5cm} Substituting the component expansions \eqref{bcBCi} of the superfields into 
the expressions \eqref{SW}, \eqref{dWsuperfields} of the BV action $S_{\mathrm{BV}}$ and 
field variation $\delta_{\mathrm{BV}}$ of AKSZ $\mathfrak{v}$ BF gauge theory, 
we obtain the corresponding expressions 
in components. 

The components organize in a number of doublets:
the connection doublet $(\omega,\varOmega)$, its curvature doublet $(f,F)$,
four field doublets $(b,B)$, $(c,C)$, $(0,\varGamma)$, $(\beta,0)$ of bidegree
$(2,-1)$, $(0,1)$, $(-1,2)$, $(3,-2)$ and five dual field doublets
$(\varOmega^+,\omega^+)$, $(B^+,b^+)$, $(C^+,c^+)$, $(\varGamma^+,0)$, $(0,\beta^+)$, 
of bidegree $(1,-1)$, $(0,0)$, $(2,-2)$, $(3,-3)$, $(-1,1)$, 
respectively, and their covariant derivative doublets (cf. subsect. \ref{subsec:linfdoub}). 

The component expansion of the BV action $S_{\mathrm{BV}}$ is
\begin{align}
S_{\mathrm{BV}}&=\int_M\Big[\langle b^+,f+[c,b]\rangle
-\langle B^+,F+[c,B]+[b,C]+[\beta,\varGamma]-[\omega,c,b]
\vphantom{\Big]}
\label{SBVact}
\\
&\hspace{1.4cm}-\frac{1}{2}[c,c,\beta]\rangle
+\langle\omega^+,Dc\rangle-\langle\varOmega^+,DC+[b,\varGamma]-\frac{1}{2}[c,c,b]\rangle
\vphantom{\Big]}
\nonumber
\\
&\hspace{1.4cm}+\langle c^+,\frac{1}{2}[c,c]-\partial\varGamma\rangle
-\langle C^+,D\varGamma +[c,C]-\frac{1}{2}[\omega,c,c]\rangle
\vphantom{\Big]}
\nonumber
\\
&\hspace{1.4cm}+\langle\beta^+,Db+[c,\beta]\rangle
-\langle \varGamma^+,[c,\varGamma]-\frac{1}{6}[c,c,c]\rangle\Big]
\vphantom{\Big]}
\nonumber
\end{align}

The component expansion of the BV field variation $\delta_{\mathrm{BV}}$ reads 
\begin{subequations}
\label{dBVcmps}
\begin{align}
&\delta_{\mathrm{BV}}\beta^+
=-[c,\beta^+]^\vee+[\varGamma,B^+]^\vee+\frac{1}{2}[c,c,B^+]^\vee,
\vphantom{\Big]}
\label{dBVcmps1}
\\
&\delta_{\mathrm{BV}}b^+
=-D\beta^+-[c,b^+]^\vee+[\varGamma,\varOmega^+]^\vee+[C,B^+]^\vee
\vphantom{\Big]}
\label{dBVcmps2}
\\
&\hspace{6cm}+[\omega,c,B^+]^\vee+\frac{1}{2}[c,c,\varOmega^+]^\vee,
\vphantom{\Big]}
\nonumber
\\
&\delta_{\mathrm{BV}}\omega^+
=-Db^+-[c,\omega^+]^\vee-[b,\beta^+]^\vee+[C,\varOmega^+]^\vee+[\varGamma,C^+]^\vee
\vphantom{\Big]}
\label{dBVcmps3}
\\
&\hspace{3.3cm}+[c,b,B^+]^\vee++[\omega,c,\varOmega^+]^\vee,
+\frac{1}{2}[c,c,C^+]^\vee,
\vphantom{\Big]}
\nonumber
\\
&\delta_{\mathrm{BV}}c^+
=-D\omega^+-[c,c^+]^\vee-[b,b^+]^\vee-[\beta,\beta^+]^\vee+[\varGamma,\varGamma^+]^\vee
\vphantom{\Big]}
\label{dBVcmps4}
\\
&\hspace{2.0cm}+[C,C^+]^\vee+[B,B^+]^\vee+[c,\beta,B^+]^\vee+[\omega,b,B^+]^\vee
\vphantom{\Big]}
\nonumber
\\
&\hspace{3.7cm}+[c,b,\varOmega^+]^\vee+[\omega,c,C^+]^\vee+\frac{1}{2}[c,c,\varGamma^+]^\vee,
\vphantom{\Big]}
\nonumber
\end{align}
\begin{align}
&\delta_{\mathrm{BV}}c
=-\frac{1}{2}[c,c]+\partial\varGamma,
\vphantom{\Big]}
\label{dBVcmps5}
\\
&\delta_{\mathrm{BV}}\omega
=-Dc,
\vphantom{\Big]}
\label{dBVcmps6}
\\
&\delta_{\mathrm{BV}}b
=-f-[c,b],
\vphantom{\Big]}
\label{dBVcmps7}
\\
&\delta_{\mathrm{BV}}\beta
=-Db-[c,\beta],
\vphantom{\Big]}
\label{dBVcmps8}
\\
&\delta_{\mathrm{BV}}B^+
=-[c,B^+]^\vee+\partial^\vee\beta^+,
\vphantom{\Big]}
\label{dBVcmps9}
\\
&\delta_{\mathrm{BV}}\varOmega^+
=-DB^+-[c,\varOmega^+]^\vee,
\vphantom{\Big]}
\label{dBVcmps10}
\\
&\delta_{\mathrm{BV}}C^+
=-D\varOmega^+-[c,C^+]^\vee-[b,B^+]^\vee,
\vphantom{\Big]}
\label{dBVcmps11}
\\
&\delta_{\mathrm{BV}}\varGamma^+
=-DC^+-[c,\varGamma^+]^\vee-[b,\varOmega^+]^\vee-[\beta,B^+]^\vee,
\vphantom{\Big]}
\label{dBVcmps12}
\\
&\delta_{\mathrm{BV}}\varGamma
=-[c,\varGamma]+\frac{1}{6}[c,c,c],
\vphantom{\Big]}
\label{dBVcmps13}
\\
&\delta_{\mathrm{BV}}C
=-D\varGamma-[c,C]+\frac{1}{2}[\omega,c,c],
\vphantom{\Big]}
\label{dBVcmps14}
\\
&\delta_{\mathrm{BV}}\varOmega
=-DC-[b,\varGamma]+\frac{1}{2}[c,c,b],
\vphantom{\Big]}
\label{dBVcmps15}
\\
&\delta_{\mathrm{BV}}B
=-F-[c,B]-[b,C]-[\beta,\varGamma]+[\omega,c,b]+\frac{1}{2}[c,c,\beta].
\vphantom{\Big]}
\label{dBVcmps16}
\end{align}
\end{subequations} 

Under a gauge transformation $g\in\Gau(M,\mathfrak{v})$, the component fields transform as follows
\begin{subequations}
\label{gactcmps}
\begin{align}
&{}^g\beta^+
=g^\vee{}_0(\beta^+)-g^\vee{}_2(g_0(c),B^+), 
\vphantom{\Big]}
\label{gactcmps1}
\\
&{}^gb^+
=g^\vee{}_0(b^+-\tau_g{}^\vee(B^+))-g^\vee{}_2(g_0(\omega-\sigma_g),B^+)
\vphantom{\Big]}
\label{gactcmps2}
\\
&\hspace{8.9cm}-g^\vee{}_2(g_0(c),\varOmega^+),
\vphantom{\Big]}
\nonumber
\\
&{}^g\omega^+
=g^\vee{}_0(\omega^+-\tau_g{}^\vee(\varOmega^+))-g^\vee{}_2(g_0(b),B^+)
\vphantom{\Big]}
\label{gactcmps3}
\\
&\hspace{4.5cm}-g^\vee{}_2(g_0(\omega-\sigma_g),\varOmega^+)-g^\vee{}_2(g_0(c),C^+),
\vphantom{\Big]}
\nonumber
\\
&{}^gc^+
=g^\vee{}_0(c^+-\tau_g{}^\vee(C^+))-g^\vee{}_2(g_0(\beta),B^+)
\vphantom{\Big]}
\label{gactcmps4}
\\
&\hspace{1.2cm}-g^\vee{}_2(g_0(b),\varOmega^+)
-g^\vee{}_2(g_0(\omega-\sigma_g),C^+)-g^\vee{}_2(g_0(c),\varGamma^+),
\vphantom{\Big]}
\nonumber
\\
&{}^gc=g_0(c),
\vphantom{\Big]}
\label{gactcmps5}
\end{align}
\begin{align}
&{}^g\omega=g_0(\omega-\sigma_g),
\vphantom{\Big]}
\label{gactcmps6}
\\
&{}^gb=g_0(b),
\vphantom{\Big]}
\label{gactcmps7}
\\
&{}^g\beta
=g_0(\beta),
\vphantom{\Big]}
\label{gactcmps8}
\\
&{}^gB^+
=g^\vee{}_1(B^+)
\vphantom{\Big]}
\label{gactcmps9}
\\
&{}^g\varOmega^+
=g^\vee{}_1(\varOmega^+)
\vphantom{\Big]}
\label{gactcmps10}
\\
&{}^gC^+
=g^\vee{}_1(C^+)
\vphantom{\Big]}
\label{gactcmps11}
\\
&{}^g\varGamma^+
=g^\vee{}_1(\varGamma^+)
\vphantom{\Big]}
\label{gactcmps12}
\\
&{}^g\varGamma
=g_1(\varGamma)-\frac{1}{2}g_2(c,c), 
\vphantom{\Big]}
\label{gactcmps13}
\\
&{}^gC
=g_1(C+\tau_g(c))-g_2(\omega-\sigma_g,c), 
\vphantom{\Big]}
\label{gactcmps14}
\\
&{}^g\varOmega
=g_1(\varOmega-\varSigma_g+\tau_g(\omega-\sigma_g))-\frac{1}{2}g_2(\omega-\sigma_g,\omega-\sigma_g)
-g_2(b,c),
\vphantom{\Big]}
\label{gactcmps15}
\\
&{}^gB
=g_1(B+\tau_g(b))-g_2(\omega-\sigma_g,b)-g_2(c,\beta).
\vphantom{\Big]}
\label{gactcmps16}
\end{align}
\end{subequations}

\vfil\eject

\section{\normalsize \textcolor{blue}{Rectified
AKSZ $\mathfrak{v}$ BF gauge theory in components}}
\label{app:rvakszcomp}

\hspace{.5cm} Substituting the component expansions \eqref{bcBCi} of the superfields into 
the expressions \eqref{rSW}, \eqref{rdWsuperfields} of the BV action $S_{\mathrm{BV}}$ and 
field variation $\delta_{\mathrm{BV}}$ of rectified AKSZ $\mathfrak{v}$ BF gauge theory, 
we obtain the corresponding expressions in components. 
The components organize in a number of doublets in the same way as in the 
non rectified theory described in app. \ref{app:vakszcomp}. 

The component expansion of the BV action $S_{\mathrm{BV}}$ reads
\begin{align}
S_{\mathrm{BV}}&=\int_M\Big[\langle b^+,\bar f_{\lambda,\mu}+\bar D_{\lambda,\mu}\omega
+\frac{1}{2}[\omega,\omega]_\lambda
+[c,b]_\lambda-\partial\varOmega\rangle
\vphantom{\Big]}
\label{rSBVact}
\\
&\hspace{1.4cm}
-\langle B^+,\bar F_{\lambda,\mu}+\bar D_{\lambda,\mu}\varOmega
+[\omega,\varOmega+\bar\varOmega-\frac{1}{2}\lambda(\bar\omega,\bar\omega)+\mu(\bar\omega)]_\lambda
\vphantom{\Big]}
\nonumber
\\
&\hspace{1.4cm}+[c,B]_\lambda+[b,C]_\lambda+[\beta,\varGamma]_\lambda
-[\omega+\bar\omega,c,b]_\lambda-\frac{1}{2}[c,c,\beta]_\lambda
\vphantom{\Big]}
\nonumber
\\
&\hspace{1.4cm}-\frac{1}{6}[\omega+\bar\omega,\omega+\bar\omega,\omega+\bar\omega]_\lambda
+\frac{1}{6}[\bar\omega,\bar\omega,\bar\omega]_\lambda+v_{\lambda,\mu}(c,b)
\vphantom{\Big]}
\nonumber
\\
&\hspace{1.4cm}+\frac{1}{2}v_{\lambda,\mu}(\omega+\bar\omega,\omega+\bar\omega)
-\frac{1}{2}v_{\lambda,\mu}(\bar\omega,\bar\omega)-w_{\lambda,\mu}(\omega)
\rangle
\vphantom{\Big]}
\nonumber
\\
&\hspace{1.4cm}
+\langle\omega^+,\bar D_{\lambda,\mu}c+[\omega,c]_\lambda-\partial C\rangle
\vphantom{\Big]}
\nonumber
\\
&\hspace{1.4cm}
-\langle\varOmega^+,\bar D_{\lambda,\mu}C+[\omega,C]_\lambda
+[c,\varOmega+\bar\varOmega-\frac{1}{2}\lambda(\bar\omega,\bar\omega)+\mu(\bar\omega)]_\lambda
\vphantom{\Big]}
\nonumber
\\
&\hspace{1.4cm}
+[b,\varGamma]_\lambda-\frac{1}{2}[c,c,b]_\lambda
-\frac{1}{2}[\omega+\bar\omega,\omega+\bar\omega,c]_\lambda
\vphantom{\Big]}
\nonumber
\\
&\hspace{1.4cm}+v_{\lambda,\mu}(\omega+\bar\omega,c)
-w_{\lambda,\mu}(c)\rangle+\langle c^+,\frac{1}{2}[c,c]_\lambda-\partial\varGamma\rangle
\vphantom{\Big]}
\nonumber
\\
&\hspace{1.4cm}
-\langle C^+,\bar D_{\lambda,\mu}\varGamma+[\omega,\varGamma]_\lambda
+[c,C]_\lambda-\frac{1}{2}[\omega+\bar\omega,c,c]_\lambda
\vphantom{\Big]}
\nonumber
\\
&\hspace{1.4cm}+\frac{1}{2}v_{\lambda,\mu}(c,c)\rangle
+\langle\beta^+,\bar D_{\lambda,\mu}b+[\omega,b]_\lambda+[c,\beta]_\lambda
-\partial B\rangle
\vphantom{\Big]}
\nonumber
\\
&\hspace{1.4cm}
-\langle \varGamma^+,[c,\varGamma]_\lambda-\frac{1}{6}[c,c,c]_\lambda\rangle\Big]_\lambda
\vphantom{\Big]}
\nonumber
\end{align}

The component expansion of the BV field variation $\delta_{\mathrm{BV}}$ is 
\begin{subequations}
\label{rdBVcmps}
\begin{align}
&\delta_{\mathrm{BV}}\beta^+
=-[c,\beta^+]_\lambda{}^\vee+[\varGamma,B^+]_\lambda{}^\vee+\frac{1}{2}[c,c,B^+]_\lambda{}^\vee,
\hspace{2.4cm}
\vphantom{\Big]}
\label{rdBVcmps1}
\\
&\delta_{\mathrm{BV}}b^+
=-\bar D_{\lambda,\mu}\beta^+-[\omega,\beta^+]_\lambda{}^\vee
-[c,b^+]_\lambda{}^\vee+[\varGamma,\varOmega^+]_\lambda{}^\vee
\vphantom{\Big]}
\label{rdBVcmps2}
\end{align}
\begin{align}
&\hspace{2cm}+[C,B^+]_\lambda{}^\vee
+[\omega+\bar\omega,c,B^+]_\lambda{}^\vee+\frac{1}{2}[c,c,\varOmega^+]_\lambda{}^\vee
-v_{\lambda,\mu}{}^\vee(c,B^+),
\vphantom{\Big]}
\nonumber
\\
&\delta_{\mathrm{BV}}\omega^+
=-\bar D_{\lambda,\mu}b^+-[\omega,b^+]_\lambda{}^\vee
-[c,\omega^+]_\lambda{}^\vee-[b,\beta^+]_\lambda{}^\vee
\vphantom{\Big]}
\label{rdBVcmps3}
\\
&\hspace{2.cm}
+[\varOmega+\bar\varOmega-\frac{1}{2}\lambda(\bar\omega,\bar\omega)+\mu(\bar\omega),B^+]_\lambda{}^\vee
+[\varGamma,C^+]_\lambda{}^\vee+[C,\varOmega^+]_\lambda{}^\vee
\vphantom{\Big]}
\nonumber
\\
&\hspace{2.cm}+\frac{1}{2}[\omega+\bar\omega,\omega+\bar\omega,B^+]_\lambda{}^\vee
+[c,b,B^+]_\lambda{}^\vee+[\omega+\bar\omega,c,\varOmega^+]_\lambda{}^\vee
\vphantom{\Big]}
\nonumber
\\
&\hspace{2.cm}+\frac{1}{2}[c,c,C^+]_\lambda{}^\vee
-v_{\lambda,\mu}{}^\vee(\omega+\bar\omega,B^+)-v_{\lambda,\mu}{}^\vee(c,\varOmega^+)
+w_{\lambda,\mu}{}^\vee(B^+),
\vphantom{\Big]}
\nonumber
\\
&\delta_{\mathrm{BV}}c^+
=-\bar D_{\lambda,\mu}\omega^+-[\omega,\omega^+]_\lambda{}^\vee-[c,c^+]_\lambda{}^\vee-[b,b^+]_\lambda{}^\vee
-[\beta,\beta^+]_\lambda{}^\vee
\vphantom{\Big]}
\label{rdBVcmps4}
\\
&\hspace{2.0cm}+[\varOmega+\bar\varOmega-\frac{1}{2}\lambda(\bar\omega,\bar\omega)
+\mu(\bar\omega),\varOmega^+]_\lambda{}^\vee+[\varGamma,\varGamma^+]_\lambda{}^\vee+[C,C^+]_\lambda{}^\vee
\vphantom{\Big]}
\nonumber
\\
&\hspace{2.0cm}+[B,B^+]_\lambda{}^\vee+[c,\beta,B^+]_\lambda{}^\vee+[\omega+\bar\omega,b,B^+]_\lambda{}^\vee
+[c,b,\varOmega^+]_\lambda{}^\vee
\vphantom{\Big]}
\nonumber
\\
&\hspace{2.0cm}+\frac{1}{2}[\omega+\bar\omega,\omega+\bar\omega,\varOmega^+]_\lambda{}^\vee
+[\omega+\bar\omega,c,C^+]_\lambda{}^\vee+\frac{1}{2}[c,c,\varGamma^+]_\lambda{}^\vee,
\vphantom{\Big]}
\nonumber
\\
&\hspace{2.0cm}-v_{\lambda,\mu}{}^\vee(b,B^+)-v_{\lambda,\mu}{}^\vee(\omega+\bar\omega,\varOmega^+)
-v_{\lambda,\mu}{}^\vee(c,C^+)+w_{\lambda,\mu}{}^\vee(\varOmega^+),
\vphantom{\Big]}
\nonumber
\\
&\delta_{\mathrm{BV}}c
=-\frac{1}{2}[c,c]_\lambda+\partial\varGamma,
\vphantom{\Big]}
\label{rdBVcmps5}
\\
&\delta_{\mathrm{BV}}\omega
=-\bar D_{\lambda,\mu}c-[\omega,c]_\lambda+\partial C,
\vphantom{\Big]}
\label{rdBVcmps6}
\\
&\delta_{\mathrm{BV}}b
=-\bar f_{\lambda,\mu}-\bar D_{\lambda,\mu}\omega-\frac{1}{2}[\omega,\omega]_\lambda-[c,b]_\lambda
+\partial\varOmega,
\vphantom{\Big]}
\label{rdBVcmps7}
\\
&\delta_{\mathrm{BV}}\beta
=-\bar D_{\lambda,\mu}b-[\omega,b]_\lambda-[c,\beta]_\lambda+\partial B,
\vphantom{\Big]}
\label{rdBVcmps8}
\\
&\delta_{\mathrm{BV}}B^+
=-[c,B^+]_\lambda{}^\vee+\partial^\vee\beta^+,
\vphantom{\Big]}
\label{rdBVcmps9}
\\
&\delta_{\mathrm{BV}}\varOmega^+
=-\bar D_{\lambda,\mu}B^+-[\omega,B^+]_\lambda{}^\vee
-[c,\varOmega^+]_\lambda{}^\vee+\partial^\vee b^+,
\vphantom{\Big]}
\label{rdBVcmps10}
\\
&\delta_{\mathrm{BV}}C^+
=-\bar D_{\lambda,\mu}\varOmega^+-[\omega,\varOmega^+]_\lambda{}^\vee
-[c,C^+]_\lambda{}^\vee-[b,B^+]_\lambda{}^\vee+\partial^\vee\omega^+,
\vphantom{\Big]}
\label{rdBVcmps11}
\\
&\delta_{\mathrm{BV}}\varGamma^+
=-\bar D_{\lambda,\mu}C^+-[\omega,C^+]_\lambda{}^\vee
-[c,\varGamma^+]_\lambda{}^\vee
\vphantom{\Big]}
\label{rdBVcmps12}
\\
&\hspace{5.50cm}-[b,\varOmega^+]_\lambda{}^\vee-[\beta,B^+]_\lambda{}^\vee+\partial^\vee c^+,
\vphantom{\Big]}
\nonumber
\\
&\delta_{\mathrm{BV}}\varGamma
=-[c,\varGamma]_\lambda+\frac{1}{6}[c,c,c]_\lambda,
\vphantom{\Big]}
\label{rdBVcmps13}
\\
&\delta_{\mathrm{BV}}C
=-\bar D_{\lambda,\mu}\varGamma-[\omega,\varGamma]_\lambda
-[c,C]_\lambda+\frac{1}{2}[\omega+\bar\omega,c,c]_\lambda-\frac{1}{2}v_{\lambda,\mu}(c,c),
\vphantom{\Big]}
\label{rdBVcmps14}
\\
&\delta_{\mathrm{BV}}\varOmega
=-\bar D_{\lambda,\mu}C-[\omega,C]_\lambda
-[c,\varOmega+\bar\varOmega-\frac{1}{2}\lambda(\bar\omega,\bar\omega)+\mu(\bar\omega)]_\lambda
\vphantom{\Big]}
\label{rdBVcmps15}
\end{align}
\begin{align}
&\hspace{2cm}
-[b,\varGamma]_\lambda+\frac{1}{2}[c,c,b]_\lambda
+\frac{1}{2}[\omega+\bar\omega,\omega+\bar\omega,c]_\lambda
\vphantom{\Big]}
\nonumber
\\
&\hspace{2cm}-v_{\lambda,\mu}(\omega+\bar\omega,c)+w_{\lambda,\mu}(c),
\vphantom{\Big]}
\nonumber
\\
&\delta_{\mathrm{BV}}B
=-\bar F_{\lambda,\mu}-\bar D_{\lambda,\mu}\varOmega
-[\omega,\varOmega+\bar\varOmega-\frac{1}{2}\lambda(\bar\omega,\bar\omega)+\mu(\bar\omega)]_\lambda
\vphantom{\Big]}
\label{rdBVcmps16}
\\
&\hspace{2cm}-[c,B]_\lambda-[b,C]_\lambda-[\beta,\varGamma]_\lambda
+[\omega+\bar\omega,c,b]_\lambda+\frac{1}{2}[c,c,\beta]_\lambda
\vphantom{\Big]}
\nonumber
\\
&\hspace{2cm}+\frac{1}{6}[\omega+\bar\omega,\omega+\bar\omega,\omega+\bar\omega]_\lambda
-\frac{1}{6}[\bar\omega,\bar\omega,\bar\omega]_\lambda-v_{\lambda,\mu}(c,b)
\hspace{2cm}
\vphantom{\Big]}
\nonumber
\\
&\hspace{2cm}-\frac{1}{2}v_{\lambda,\mu}(\omega+\bar\omega,\omega+\bar\omega)
+\frac{1}{2}v_{\lambda,\mu}(\bar\omega,\bar\omega)+w_{\lambda,\mu}(\omega).
\nonumber
\end{align}
\end{subequations} 

Under a gauge transformation $g\in\Gau(M,\mathfrak{v})$, the component fields transform 
in an obvious way, since they are all rectified fields
(cf. subsect. \ref{subsec:rectf}).

\vfill\eject

\section{\normalsize \textcolor{blue}{The Lagrangian submanifold $\matheul{L}$}}
\label{app:psilagr}

\hspace{.5cm} The Lagrangian submanifold $\matheul{L}$ of the BV field space $\matheul{F}$ 
generated by the gauge fermion $\Psi$ given by \eqref{2tligfix10}
is specified by the constraints 
\begin{subequations}
\label{psilagr}
\begin{align}
&\beta^+=-*\tilde q,
\vphantom{\Big]}
\label{psilagr1}
\\
&b^+=*\bar D_{\lambda,\mu}\tilde\omega,
\vphantom{\Big]}
\label{psilagr2}
\\
&\omega^+=-*\bar D_{\lambda,\mu}\tilde c,
\vphantom{\Big]}
\label{psilagr3}
\\
&c^+=0,
\vphantom{\Big]}
\label{psilagr4}
\\
&B^+=*\tilde F,
\vphantom{\Big]}
\label{psilagr5}
\\
&\varOmega^+=*\bar D_{\lambda,\mu}\tilde C,
\vphantom{\Big]}
\label{psilagr6}
\\
&C^+=-*\bar D_{\lambda,\mu}\tilde\varGamma,
\vphantom{\Big]}
\label{psilagr7}
\\
&\varGamma^+=0, 
\vphantom{\Big]}
\label{psilagr8}
\\
&\tilde c^+=-\bar D_{\lambda,\mu}*\omega
\vphantom{\Big]}
\label{psilagr9}
\\
&\tilde \theta^+=0, 
\vphantom{\Big]}
\label{psilagr10}
\\
&\tilde\omega^+=-\bar D_{\lambda,\mu}*b-*\bar D_{\lambda,\mu}a,
\vphantom{\Big]}
\label{psilagr11}
\\
&\tilde\xi^+=0, 
\vphantom{\Big]}
\label{psilagr12}
\\
&\tilde q^+=*\beta
\vphantom{\Big]}
\label{psilagr13}
\\
&\tilde \chi^+=0, 
\vphantom{\Big]}
\label{psilagr14}
\\
&\tilde\varGamma^+=\bar D_{\lambda,\mu}*C,
\vphantom{\Big]}
\label{psilagr15}
\\
&\tilde H^+=0, 
\vphantom{\Big]}
\label{psilagr16}
\\
&\tilde C^+=\bar D_{\lambda,\mu}*\varOmega-*\bar D_{\lambda,\mu}\varPhi,
\vphantom{\Big]}
\label{psilagr17}
\\
&\tilde \varTheta^+=0,
\vphantom{\Big]}
\label{psilagr18}
\\
&\tilde F^+=-*B,
\vphantom{\Big]}
\label{psilagr19}
\end{align}
\eject\noindent
\begin{align}
&\tilde \varSigma^+=0
\vphantom{\Big]}
\label{psilagr20}
\\
&a^+=-\bar D_{\lambda,\mu}*\tilde\omega
\vphantom{\Big]}
\label{psilagr21}
\\
&\theta^+=0,
\vphantom{\Big]}
\label{psilagr22}
\\
&\varPhi^+=\bar D_{\lambda,\mu}*\tilde C, 
\vphantom{\Big]}
\label{psilagr23}
\\
&H^+=0.
\vphantom{\Big]}
\label{psilagr24}
\end{align}
\end{subequations}

\vfill\eject

\textcolor{blue}{Acknowledgement} We thank the referee of the paper
and C. Wockel for pointing out the mistakes and weaknesses of the first version 
of the paper and for providing useful suggestions for its improvement.
The paper is dedicated to the memory of my mother, Lucia Baviera, who died 
while it was being completed. 

\vfill\eject


\begin{thebibliography}{99}

\bibitem{Polchinski:1998rr}
  J.~Polchinski,
``String theory. Vol. 2: superstring theory and beyond'', 
{\it Cambridge, UK: Univ. Pr. (1998) 531 p.}

\bibitem{Becker:2007zj}
  K.~Becker, M.~Becker and J.~H.~Schwarz,
``String theory and M-theory: a modern introduction'',  
{\it  Cambridge, UK: Univ. Pr. (2007) 739 p.}

\bibitem{Johnson:2003gi}
C.~V.~Johnson,
``D-branes'', available at 
{\it  Cambridge, USA: Univ. Pr. (2003) 548 p.}

\bibitem{Baez:1999sr}
  J.~C.~Baez,
  ``An introduction to spin foam models of BF theory and quantum gravity'',
  Lect.\ Notes Phys.\  {\bf 543} (2000) 25
  [arXiv:gr-qc/9905087].

\bibitem{Rovelli:2004tv}
C.~Rovelli,
``Quantum gravity'',
{\it  Cambridge, UK: Univ. Pr. (2004) 455 p.}

\bibitem{Savit:1977fw}
R.~Savit,
``Topological excitations in U(1) invariant theories'',
  Phys.\ Rev.\ Lett.\  {\bf 39} (1977) 55.

\bibitem{Freedman:1980us}
D.~Z.~Freedman and P.~K.~Townsend,
``Antisymmetric tensor gauge theories and non linear sigma models'',
  Nucl.\ Phys.\  B {\bf 177} (1981) 282.

\bibitem{Nepomechie:1982rb}
 R.~I.~Nepomechie,
``Approaches to a non Abelian antisymmetric tensor gauge field theory'',
  Nucl.\ Phys.\  B {\bf 212} (1983) 301.

\bibitem{Teitelboim:1985ya}
C.~Teitelboim,
``Gauge Invariance for Extended Objects'',
  Phys.\ Lett.\  B {\bf 167} (1986) 63.

\bibitem{Henneaux:1986ht}
M.~Henneaux and C.~Teitelboim,
``$p$ form electrodynamics'',
  Found.\ Phys.\  {\bf 16} (1986) 593.

\bibitem{Blau:1989bq}
  M.~Blau and G.~Thompson,
``Topological gauge theories of antisymmetric tensor fields'',
  Annals Phys.\  {\bf 205} (1991) 130.

\bibitem{Solodukhin:1992jg}
S.~N.~Solodukhin,
``Non Abelian gauge antisymmetric tensor fields'',
  Nuovo Cim.\  B {\bf 108} (1993) 1275
  [arXiv:hep-th/9211046].

\bibitem{Lahiri:2001di}
  A.~Lahiri,
 ``Gauge transformations of the non Abelian two form'',
  Mod.\ Phys.\ Lett.\  A {\bf 17} (2002) 1643
  [arXiv:hep-th/0107104].

\bibitem{Lahiri:2001ks}
  A.~Lahiri,
  ``Local symmetries of the non Abelian two form'',
  J.\ Phys.\ A  {\bf 35} (2002) 8779
  [arXiv:hep-th/0109220].

\bibitem{Chepelev:2001mg}
  I.~Chepelev,
  ``Non Abelian Wilson surfaces'',
  JHEP {\bf 0202} (2002) 013
  [arXiv:hep-th/0111018].

\bibitem{Baez:2002jn}
J.~C.~Baez,
``Higher Yang-Mills theory'',
arXiv:hep-th/0206130.

\bibitem{Hofman:2002ey}
  C.~Hofman,
``Non Abelian 2--forms'',
  arXiv:hep-th/0207017.

\bibitem{Pfeiffer:2003je}
 H.~Pfeiffer,
``Higher gauge theory and a non Abelian generalization of $2$--form electrodynamics'',
 Annals Phys.\  {\bf 308} (2003) 447
 [arXiv:hep-th/0304074].

\bibitem{Baez:2010ya}
J.~C.~Baez and J.~Huerta,
``An invitation to higher gauge theory'',
arXiv: 1003.4485 [hep-th].

\bibitem{Baez5}
J.~Baez and A.~Lauda, 
``Higher dimensional algebra V: 2-groups'', 
Theor.\ Appl.\ Categor.\ {\bf 12} (2004) 423
[arXiv:math.0307200].

\bibitem{Baez:2003fs}
  J.~C.~Baez and A.~S.~Crans,
``Higher dimensional algebra VI: Lie $2$--algebras'',
  Theor.\ Appl.\ Categor.\  {\bf 12} (2004) 492
  [arXiv:math/0307263].

\bibitem{Lada:1992wc}
  T.~Lada and J.~Stasheff,
``Introduction to SH Lie algebras for physicists'',
  Int.\ J.\ Theor.\ Phys.\  {\bf 32} (1993) 1087
  [arXiv:hep-th/9209099].

\bibitem{Lada:1994mn}
T.~Lada and M.~Markl,
``Strongly homotopy Lie algebras'',
Comm.\ Algebra {\bf 23} (1995) 2147
[arXiv:hep-th/9406095].

\bibitem{Brylinski:1993ab}
  J.~L.~Brylinski,
``Loop spaces, characteristic classes and geometric quantization'',
{\it  Boston, USA: Birkhaeuser (1993) (Progress in mathematics, 107) 300 p.}

\bibitem{Breen:2001ie}
  L.~Breen and W.~Messing,
``Differential geometry of gerbes'',
Adv.\ Math.\ {\bf 198} (2005) 732
[arXiv:math/0106083].

\bibitem{Schwarz:1978}
A.~S.~Schwartz, 
``The partition function of degenerate quadratic functionals and Ray-Singer invariants'', 
Lett.\ Math.\ Phys.\ {\bf 2} (1978) 247.

\bibitem{Horowitz:1989ng}
G.~T.~Horowitz,
``Exactly soluble diffeomorphism invariant theories'',
Commun.\ Math.\ Phys.\  {\bf 125} (1989) 417.

\bibitem{Guo:2002yc}
H.~Y.~Guo, Y.~Ling, R.~S.~Tung and Y.~Z.~Zhang,
  ``Chern-Simons term for BF theory and gravity as a generalized topological
  field theory in four-dimensions'',
  Phys.\ Rev.\  D {\bf 66} (2002) 064017
  [arXiv:hep-th/0204059].

\bibitem{Freidel:2005ak}
L.~Freidel and A.~Starodubtsev,
``Quantum gravity in terms of topological observables'',
  arXiv:hep-th/0501191.

\bibitem{Girelli:2003ev}
F.~Girelli and H.~Pfeiffer,
``Higher gauge theory: differential versus integral formulation'',
J.\ Math.\ Phys.\  {\bf 45} (2004) 3949
[arXiv:hep-th/0309173].

\bibitem{Girelli:2007tt}
F.~Girelli, H.~Pfeiffer and E.~M.~Popescu,
``Topological higher gauge theory - from BF to BFCG theory'',
  J.\ Math.\ Phys.\  {\bf 49} (2008) 032503
  [arXiv:0708.3051 [hep-th]].

\bibitem{Baez:2004in}
J.~Baez and U.~Schreiber,
``Higher gauge theory: 2-connections on 2-bundles'',
  arXiv:hep-th/0412325.

\bibitem{Baez:2005qu}
J.~C.~Baez and U.~Schreiber,
``Higher gauge theory'',
in {\it Categories in Algebra, Geometry and Mathematical Physics}, 
eds. A. Davydov et al., Contemp.\ Math.\ {\bf 431} AMS, Providence (USA) (2007) 7
[arXiv:math/0511710].

\bibitem{Aschieri:2003mw}
  P.~Aschieri, L.~Cantini and B.~Jurco,
``NonAbelian bundle gerbes, their differential geometry and gauge theory'',
  Commun.\ Math.\ Phys.\  {\bf 254} (2005) 367
  [hep-th/0312154].

\bibitem{Fiorenza2011} 
D.~Fiorenza, U.~Schreiber and J.~Stasheff
``\v Cech cocycles for differential characteristic classes – 
An $\infty$--Lie theoretic construction'',
arXiv:1011.4735 [math.AT].

\bibitem{Schreiber2011}
U.~Schreiber,
``Differential cohomology in a cohesive $\infty$--topos'',
available at \href{http://ncatlab.org/schreiber/files/cohesivedocumentv032.pdf}
{http://ncatlab.org/schreiber/files/cohesivedocumentv032.pdf} and references therein.

\bibitem{BV1}
I.~A.~Batalin and G.~A.~Vilkovisky,
``Gauge algebra and quantization'',
Phys.\ Lett.\ B {\bf 102} (1981) 27.

\bibitem{BV2}
I.~A.~Batalin and G.~A.~Vilkovisky,
``Quantization of gauge theories with linearly dependent generators'',
Phys.\ Rev.\ D {\bf 28} (1983) 2567
(Erratum-ibid.\ D {\bf 30} (1984) 508).

\bibitem{AKSZ}
M.~Alexandrov, M.~Kontsevich, A.~Schwartz and O.~Zaboronsky,
``The geometry of the master equation and topological quantum field theory'',
Int.\ J.\ Mod.\ Phys.\ A {\bf 12} (1997) 1405
[arXiv:hep-th/9502010].

\bibitem{Ikeda:2012pv} 
  N.~Ikeda,
  ``Lectures on AKSZ Topological Field Theories for Physicists'',
  arXiv:1204.3714 [hep-th].

\bibitem{Zucchini6}
R.~Zucchini,
``The Hitchin Model, Poisson-quasi-Nijenhuis Geometry and Symmetry
Reduction'', JHEP {\bf 0710} (2007) 075
[arXiv:0706.1289 [hep-th]].

\bibitem{Zucchini7}
R.~Zucchini,
``Gauging the Poisson sigma model,''
JHEP {\bf 0805} (2008) 018
[arXiv:0801.0655 [hep-th]].


\bibitem{Witten:1988ze}
  E.~Witten,
``Topological Quantum Field Theory'',
Commun.\ Math.\ Phys.\  {\bf 117} (1988) 353.

\bibitem{Witten:1988hf}
  E.~Witten,
  ``Quantum field theory and the Jones polynomial'',
  Commun.\ Math.\ Phys.\  {\bf 121} (1989) 351.

\bibitem{Fiorenza:2011jr}
  D.~Fiorenza, C.~L.~Rogers and U.~Schreiber,
  ``A Higher Chern-Weil derivation of AKSZ $\sigma$-models'',
  arXiv:1108.4378 [math-ph].

\bibitem{Kotov:2007nr}
  A.~Kotov and T.~Strobl,
  ``Characteristic classes associated to Q-bundles'',
  arXiv:0711.4106 [math.DG].



\bibitem{Stasheff1}
M.~Schlessinger and J.~D.~Stasheff, 
``The Lie algebra structure of tangent cohomology
and deformation theory'', 
J. of Pure and Appl. Algebra {\bf 38} (1985) 313. 

\bibitem{Gerstenhaber:1964}
M.~Gerstenhaber, 
``A uniform cohomology theory for algebras'', 
Proc.\ Nat.\ Acad.\ Sci.\ {\bf 51} (1964) 626.

\bibitem{Dorfman:1987}
I.~Ya.~Dorfman, 
``Dirac structures of integrable evolution equations'', 
Phys.\ Lett.\  A {\bf 125} (1987) 240.

\bibitem{Courant:1990}
T.~Courant, ``Dirac manifolds'', 
Trans.\ Amer.\ Math.\ Soc.\ {\bf 319} (1990) 631.

\bibitem{Roytenberg:1998vn}
D.~Roytenberg and A.~Weinstein,
``Courant algebroids and strongly homotopy Lie algebras'',
Lett.\ Math.\ Phys.\  {\bf 46} (1998) 81
[arXiv:math/9802118].

\bibitem{BCSS}
J.~Baez, A.~Crans, U.~Schreiber and D.~Stevenson, 
``From loop groups to 2-groups'', 
HHA {\bf 9} (2007), 101. 
[arXiv:math.QA/0504123].

\bibitem{CS}
S.~S.~Chern and J.~Simons, 
``Characteristic forms and geometric invariants'', 
Ann. \ Math. \ {\bf 99} (1974) 48.

\bibitem{CheegSim}
J.~Cheeger and J.~Simons, 
``Differential characters and geometric invariants'', 
in {\it Geometry and Topology}, 
eds. J.~Alexander and J.~Harer, 
Lecture Notes in Mathematics {\bf 1167} (1985) 50.

\bibitem{Whitehead:1946}
J. H. C. Whitehead, 
``Note on a previous paper entitled ‘On adding relations to homotopy groups' ’', 
Ann.\ Math.\ {\bf 47} (1946) 80

\bibitem{Brown:1976}
R.~Brown and C.~B.~Spencer, 
``$G$-groupoids, crossed modules, and the classifying space of a topological group'', 
Proc.\ Kon.\ Akad.\ v.\ Wet.\  {\bf 79} (1976) 296. 

\bibitem{Moufang:1935}
R.~Moufang, 
``Zur Struktur von Alternativk\"orpern'', 
Math. \ Ann. \ {\bf 110} (1935) 416.

\bibitem{PressSeg}
A.~Pressley and G.~Segal, ``Loop groups''
{\it Oxford U. K., Oxford University Press (1986)}.

\bibitem{Hen}
A.~Henriques, ``Integrating $L_\infty$ algebras'',
arXiv:math/0603563 .

\bibitem{SchPr}
C.~Schommer--Pries, ``A finite dimensional string 2--group'',
arXiv:0911.2483.


\bibitem{Chevalley1}
C.~Chevalley and S.~Eilenberg, 
``Cohomology theory of Lie groups and Lie algebras'', 
Trans. \ Amer. \ Math. \ Soc. \ {\bf 63} (1948) 85.

\bibitem{Weil1}
A.~Weil, 
``G\'eom\'etrie diff\'erentielle des espaces fibr\'es'', 
{\it Letters to Chevalley and Koszul} (1949).

\bibitem{HCartan2}
H.~Cartan, 
``Cohomologie re\'elle d’un espace fibr\'e principal diff\'erentielle I'',
{\it Sminaire Henri Cartan 1949/50},
CBRM (1950) 19–01.

\bibitem{HCartan1}
H.~Cartan,
``Cohomologie re\'elle d’un espace fibr\'e principal diff\'erentielle II'', 
{\it Sminaire Henri Cartan 1949/50},
CBRM (1950) 20-01.

\bibitem{Ehresmann1}
C.~Ehresmann,
`` Les connexions infinit\'esimales dans un espace fibr\'e diff\'erentiable'',
{\it Colloque de topologie (espaces fibr\'es) Bruxelles 1950}, 
Georges Thone, Li\`ege, (1951) 29.


\bibitem{Waldorf1}
T.~Nikolaus and K.~Waldorf, 
``Four equivalent versions of non-abelian gerbes'',
arXiv:1103.4815 [math.DG].

\bibitem{Schreiber:2005ff} 
U.~Schreiber,
``From loop space mechanics to nonAbelian strings'',
hep-th/0509163.

\bibitem{Nikolaus1}
T.~Nikolaus, U.~Schreiber and D.~Stevenson 
``Principal infinity-bundles - General theory'',
arXiv:1207.0248 [math.AT].

\bibitem{Nikolaus2}
T.~Nikolaus, U.~Schreiber and D.~Stevenson 
``Principal infinity-bundles - Presentations'',
arXiv:1207.0249 [math.AT].

\bibitem{Roytenberg:2006qz}
  D.~Roytenberg,
  ``AKSZ-BV Formalism and Courant Algebroid-induced Topological Field Theories'',
  Lett.\ Math.\ Phys.\  {\bf 79} (2007) 143
  [hep-th/ 0608150].




\end{thebibliography}
\end{document}